\documentclass[reprint,aps,prd,floatfix]{revtex4-2}

\usepackage{afterpage}
\usepackage{microtype}
\microtypesetup{nopatch=item}
\usepackage{hyperref}
\usepackage{graphicx}
\usepackage{booktabs}
\usepackage{enumitem}
\usepackage{adjustbox}
\usepackage[english]{babel}
\usepackage{wasysym}
\usepackage{bm}
\usepackage{textcomp}
\usepackage{latexsym}
\usepackage{amsfonts} 
\usepackage{lineno}
\usepackage{dcolumn}

\usepackage{amsmath,mathtools}
\usepackage{slashed}
\usepackage{physics}

\newcommand{\tcell}[2][0.18\textwidth]{\parbox[t]{#1}{\raggedright #2}}

\begin{document}

\title{$G(2)$ Glueball Boson Stars as Restricted Kerr Mimickers in the Thomas--Fermi Limit}

\author{Nicol\`o Masi}
\email{masin@bo.infn.it}
\affiliation{
	INFN, Physics Department of Bologna University, Via Irnerio 46 Bologna, 40126, Italy}

\begin{abstract}
We investigate whether horizonless compact stars made of dark glueball matter from an exceptional $G(2)$ broken gauge sector can reproduce selected Kerr observables within a strong-coupling Thomas--Fermi effective-fluid description. The heavy composite condensate is conditionally represented by a complex order parameter with an approximately conserved charge. Comparing standard self-interaction families in the ultraheavy-constituent regime shows that the simultaneous requirements of large stellar mass and high compactness force a separation between microscopic and collective scales; the size of a weak-coupling quartic coefficient by itself is therefore not a unique discriminator once such scale separation is allowed. We introduce GC9, a scale-separated $G(2)$-inspired positive nonic-density closure whose high-density stiffness, smooth bag-free surface and Q-ball safety make it a useful compact-star benchmark. The $G(2)$ Casimir/cumulant language motivates the operator channel but does not uniquely derive the power $p=9$, which is selected within the considered positive hierarchy by the required causal stiffness and compactness. After the Thomas--Fermi rescaling, the reduced stellar observables are independent of the microscopic glueball mass $m_G$; $m_G$ enters the field normalization and gradient-control diagnostics, whereas the collective stiffness $\Lambda_T$ fixes the dimensional mass and radius.
Solving the TOV, Hinderer and Hartle--Thorne systems, we obtain a stable maximum-mass compactness $C=0.3247$, tidal deformability $\Lambda=4.37$, the directly computed rotational combination $\bar I C^{3/2}=0.947$ and slow-rotation quadrupole $\bar Q=1.522$. Converting $\bar I C^{3/2}$ into a mass-shedding spin requires an EOS-specific calibration of the factor $\kappa_{\rm ms}$. At the controlled slow-row value $j=0.328$, the same-order ISCO shift is $1.24\%$. The stable endpoint has no light ring, but its exterior Regge--Wheeler barrier supports a genuine static axial mode,
$M\omega=0.463846-0.098881i$, with a damping time about ten percent shorter than Schwarzschild and a real frequency $24.1\%$ higher. The isotropic GC6--9--12 extension changes this static spectrum only at the few-percent level and does not remove the offset from rotating-remnant ringdown benchmarks; compatibility remains a target for the uncomputed rotating coupled spectrum. Cowling calculations additionally identify matter-supported quadrupolar modes.
The broader scalar-burst, seeded-collapse and cosmological discussions are explicitly phenomenological. A single fixed $\Lambda_T$ produces a compact Kerr-comparison branch only within a factor of a few of its own maximum mass; the multi-sector phenomenology therefore assumes distinct constant-$\Lambda_T$ branches, from the keV supermassive sector to the PeV asteroid-mass sector, whose microscopic origin and possible transitions remain open. Within these qualifications, GC9 defines a falsifiable restricted Kerr-mimicry framework and a concrete target for gravitational-wave, horizon-scale and compact-dark-matter tests.
\end{abstract}

	\maketitle

\section{Introduction: Motivations and Theoretical Context}
\label{sec:intro}

General Relativity (GR) has passed an extraordinary set of strong-field tests and the Kerr family of vacuum solutions has emerged as the standard phenomenological template for interpreting compact-object observations across vastly different scales \citep{Abbott2016GW150914,Abbott2022TestsGR,BertiEtAl2015TestingGR,Cardoso2019BHmimickers,DeLaurentisPani2025TestingNature}. X-ray spectroscopy and timing of accreting sources, stellar dynamics around galactic nuclei, black hole (BH) shadow imaging and gravitational-wave (GW) measurements of binary coalescences have collectively built a remarkably consistent picture in which many astrophysical compact objects behave, to current accuracy, as if their exterior spacetime were well approximated by a Kerr metric characterized by mass $M$ and angular momentum $J$ \citep{EHT2019M87I,EHTSgrA2022I,Abbott2016GW150914,Abbott2022TestsGR,BertiEtAl2015TestingGR,Cardoso2019BHmimickers}.

At the same time, the classical black hole interpretation carries a conceptual tension: if GR is taken literally down to arbitrarily small scales, sufficiently compact configurations generically lead to the formation of an event horizon and a spacetime singularity \citep{Cardoso2019BHmimickers,DeLaurentisPani2025TestingNature}. This raises the well-known incompatibility between unitary quantum evolution and horizon-based information loss, and more broadly the question of whether the ``Kerr black hole'' is a fundamental endpoint or an effective coarse-grained description of a deeper microphysical state. Independently of quantum gravity, the classical picture also implies that horizon formation is an \emph{absorbing inner boundary condition} for dynamics: matter and radiation crossing the horizon do not re-emerge and therefore do not feed back on the environment in ways that could be accessible to observations \citep{Cardoso2019BHmimickers,DeLaurentisPani2025TestingNature}.
Although the astrophysical evidence for compact dark objects is consistent with the Kerr BH paradigm, the question of whether a fraction (or even the totality) of the observed population could consist of exotic compact objects (ECOs) remains open and is actively constrained by GWs and very-long-baseline interferometry (VLBI) data \citep{Cardoso2019BHmimickers,EHT2019M87I,EHTSgrA2022I}.
More generally, Buchdahl's theorem \citep{Buchdahl1959} provides a useful guide to the ECO landscape: it limits the compactness of self-gravitating matter configurations, but the underlying assumptions can be evaded in the presence of anisotropy, multi-component matter, non-perfect-fluid sectors or non-standard stress tensors \citep{BowersLiang1974,CattoenFaberVisser2005,
	RaposoPaniBezaresPalenzuelaCardoso2019,AlhoNatarioPaniRaposo2022,
	AlhoNatarioPaniRaposo2024}, precisely the regime in which several ECO models (such as boson stars, gravastars, fermion-boson stars, ultracompact anisotropic stars, elastic stars, regular-black hole constructions, frozen-star/string-fluid configurations and wormholes \citep{MazurMottola2001,MazurMottola2004,MottolaVaulin2006,DelGrossoFrancioliniPaniUrbano2023,LanYangGuoMiao2023RegularBHReview,BrusteinMedved2024FrozenStars,MorrisThorne1988,DeLaurentisPani2025TestingNature}) become relevant.
Among ECO candidates, boson stars are particularly attractive because they provide explicit, horizonless solutions within well-controlled field theories \citep{Kaup1968,RuffiniBonazzola1969,Liebling2017,SchunckMielke2003,DeLaurentisPani2025TestingNature}.
Self-interactions can dramatically increase the maximum mass and compactness \citep{ColpiShapiroWasserman1986} and specialized potentials can generate ultra-compact configurations exhibiting light rings and strong lensing, thereby challenging the uniqueness of imaging and ringdown-based tests \citep{CunhaBertiHerdeiro2017LRstability,CunhaHerdeiroRadu2017Shadows,DiFilippo2024InnerLR,CardosoFranzinPani2016ECO}.
Concrete boson star (BS) and BH-mimicker case studies already illustrate this degeneracy at the level of horizon-scale imaging and orbital observables, from galactic-center modeling and EHT-compatible shadows to deformed rotating mimickers \citep{VincentEtAl2021M87,LiEtAl2022DeltaKerrShadow,Shaikh2023SgrAEHT}.

A useful modern way to frame the compact-object problem is to separate three progressively stronger questions: whether compact objects other than black holes and neutron stars (NS) exist in nature, whether all BH candidates conform to the Kerr solution of GR, and whether classical black holes exist at all as fundamental objects \citep{DeLaurentisPani2025TestingNature}. This hierarchy is particularly relevant in the present discussion. Our proposal does not begin by modifying the exterior success of the Kerr paradigm, but by questioning the nature of the interior boundary condition and the microscopic composition of ultracompact objects. In this sense, the aim is not to deny the observational effectiveness of Kerr, but to investigate whether some fraction of the compact-object population could instead belong to the broader class of ECOs, while remaining compatible with present gravitational-wave and horizon-scale imaging data \citep{Cardoso2019BHmimickers,DeLaurentisPani2025TestingNature}.
In practice, the relevant observables can be organized according to the physical layer that they probe. The exterior geometry is constrained by spin measurements, multipolar tests and searches for non-Kerr
charges or dipoles, all of which affect orbital motion, lensing and the approximate integrability of geodesics \citep{YunesPretorius2009,BertiEtAl2015TestingGR,
	Cardoso2019BHmimickers}.  The material nature of the object is instead tested by tidal deformability and tidal heating: while
Kerr black holes in GR have vanishing static Love numbers, horizonless objects generically retain a matter-dependent response \citep{Cardoso2019BHmimickers,JohnsonMcDanielEtAl2018,
	PacilioEtAl2021}.  Finally, the inner boundary condition is probed dynamically through merger remnants,
post-merger ringdown, possible echoes or late-time internal modes and through accretion-powered surface, electromagnetic or neutrino signatures \citep{BertiEtAl2015TestingGR,Cardoso2019BHmimickers,
	Abbott2021AreaTest,DeLaurentisPani2025TestingNature}. A viable Kerr mimicker must therefore reproduce the leading
geodesic observables of a black hole and, where a ringdown calculation exists, remain compatible with measured ringdown data, while keeping matter-sensitive deviations below current bounds.
	
The paper explores the viability of the following scenario: the possibility that the compact objects currently interpreted as black holes may instead belong to a special family of \emph{horizonless Kerr mimickers}---compact objects whose exterior spacetime is Kerr-like to high precision while the interior is a regular, macroscopic fluid of dark matter (DM). In this special case it is DM that produces BH mimickers rather than BHs making up DM. In this view, the observational success of Kerr is preserved as an \emph{effective exterior description}, while the inner boundary condition is radically modified: horizons and singularities are replaced by a material core with dynamics, normal modes and non-trivial responses to accretion and mergers. In particular, these Kerr--mimicking boson stars are made of dark composite bosons coming from a non-Abelian theory for dark matter.
This perspective is particularly well suited to the present $G(2)$-glueball construction based on \citep{Masi2024G2Resurgence}: the model is sufficiently structured to make concrete predictions, yet sufficiently flexible to be confronted with the broader ECO-testing program now emerging from gravitational-wave astronomy, BH imaging and precision X-ray spectroscopy \citep{EHT2019M87I,EHTSgrA2022I}.
Furthermore, if black holes are reinterpreted as boson stars made of dark glueball matter, the standard cosmological narrative also changes. In that case dark matter is no longer a diffuse collisionless component: it forms compact objects over a broad mass range, potentially assembling hierarchically into galactic nuclei and contributing to halo substructure. 
 Any compact-DM reinterpretation must be confronted with cosmological and astrophysical bounds, including cosmic microwave background (CMB) constraints, microlensing, dynamical heating and compact-object population studies \citep{Planck2018,Carr2020PBH,Tisserand2007EROS2,Alcock2000MACHO,MonroyRodriguez2014,Brandt2016EridanusII,Abbott2022TestsGR}. This will offer new predictions for future GW detectors and precision measurements of compact-object tidal response and ringdown spectra.

\subsection{Kerr mimickers and what observations can actually constrain}
The observational status of astrophysical BH candidates is basically shaped by gravitational-wave detections, horizon-scale imaging, X-ray and stellar-dynamical measurements and strong-field consistency tests: representative references include \citep{Abbott2016GW150914,Abbott2022TestsGR,Abbott2021AreaTest,EHT2019M87I,EHTSgrA2022I,Cardoso2019BHmimickers,CunhaBertiHerdeiro2017LRstability,CunhaHerdeiroRadu2017Shadows,CardosoFranzinPani2016ECO,Isi2021AreaLaw,Shenar2022VFTS243,Panuzzo2024GaiaBH3,DeLaurentisPani2025TestingNature,BertiEtAl2015TestingGR,Olivares2020,VincentEtAl2021M87}.

An astrophysical compact object that aims to replace a Kerr BH must reproduce, at minimum, the following classes of phenomenology:
\begin{itemize}
	\item \textbf{Exterior geodesic structure:} near-horizon redshift, strong lensing and the location of characteristic orbits (e.g. innermost
	stable circular orbit (ISCO)-like scales) relevant for accretion and emission.
	\item \textbf{Compactness and ``surface'' effects:} the object must be sufficiently compact that electromagnetic signatures do not reveal a hard surface at radii far above $2GM$; conversely, any non-horizon reflectivity or emission must remain consistent with observational bounds.
	\item \textbf{Rotation:} realistic sources are spinning. A viable mimicker must admit rapidly rotating configurations with dimensionless spin $j$ comparable to those inferred for black hole candidates and must remain stable against rotational and non-axisymmetric instabilities.
	\item \textbf{Tidal response:} in GR, for Kerr black holes, static tidal Love numbers vanish. Horizonless objects generally develop nonzero Love numbers and mode excitations; observational constraints therefore bound the deviation from the Kerr response.
	\item \textbf{Merger and ringdown:} GW observations are sensitive to the post-merger spectrum. Kerr mimickers must reproduce an effective ringdown consistent with measured quasi-normal mode (QNM) content, while allowing for potential additional late-time features (e.g.\ echoes or mode mixing) that must remain below current constraints.
\end{itemize}
In short, ``mimicking Kerr'' is a quantitative target: the exterior must be sufficiently Kerr-like and the non-Kerr signatures must lie within present bounds while yielding testable predictions for future instruments. Recent examples include shadow-based constraints on deformed rotating mimickers, direct EHT comparisons for Sgr~A$^*$-like alternatives and waveform-level analyses showing that some BS mergers can still be partially degenerate with present GW templates \citep{LiEtAl2022DeltaKerrShadow,Shaikh2023SgrAEHT,EvstafyevaSperhakeRomeroShawAgathos2024GWDA}.

\subsection{Boson stars as natural arena for horizonless compact objects}
Boson stars and related self-gravitating solitons have a long history, from the original mini-BS constructions to self-interacting, rotating, Proca, axion-like and Q-ball-inspired generalizations; useful reviews and representative examples include \citep{Kaup1968,RuffiniBonazzola1969,Jetzer1992Review,SchunckMielke2003,Liebling2017,ColpiShapiroWasserman1986,Gleiser1988Stability,GleiserWatkins1989,FriedbergLeePang1987,Ryan1997Spinning,KleihausKunzListSchaffer2005,BritoCardosoHerdeiroRadu2016,HerdeiroRadu2017,Visinelli2021Review,Guerra2019AxionBosonStars,Coleman1985Qballs,Kusenko1997Qball}. More recent work has sharpened the phenomenology of this class by analyzing binary orbital dynamics and merger remnants, the multipolar structure of rotating configurations, mixed fermion-boson and fermion-soliton stars and even stellar orbits around BS candidates \citep{PalenzuelaLehnerLiebling2007,BezaresPalenzuela2017,SiemonsenEast2023BinaryBosonStars,VaglioPacilioMaselliPani2022Multipoles,DiGiovanniEtAl2021FermionBoson,MourelleAdamBustillo2024RotatingFermionBoson,BertiDeLucaDelGrossoPani2024FermionSolitonLove,PomboSaltas2023SunLike}.

Among the many candidate horizonless compact objects, boson stars occupy a particularly compelling niche. They are self-gravitating configurations of bosonic fields whose stability can be protected by conserved charges and/or self-interactions. In particular, they have to manifest the following features:
\begin{itemize}
	\item \textbf{Horizonless ultra-compactness:} with sufficiently stiff effective pressure at high density (typically provided by repulsive self-interactions), boson stars can achieve large compactness without forming an event horizon.
	\item \textbf{Rotation with quantized angular momentum:} for a complex bosonic field with a global $U(1)$ symmetry, stationary rotating solutions admit a conserved Noether charge $N$ and a discrete winding number $m\in\mathbb{Z}$ such that one can define an angular momentum and a spin
	\begin{equation}
		J = m\,N,
		\qquad
		j \equiv \frac{J}{M^2}.
		\label{eq:JmN_intro}
	\end{equation}
	This provides a clean, non-perturbative handle on the spin content and on families of stationary rotating solutions. In the present $G(2)$-motivated framework this relation should be understood as the effective rotating-sector description associated with the approximately conserved complex scalar degree of freedom.
	\item \textbf{Controlled effective theory:} the macroscopic properties (mass--radius relation, compactness, stability bands, mode spectra) can be traced back to the form of an effective potential $V(|\Phi|^2)$ or, equivalently, to an effective equation of state (EOS) in the fluid limit.
\end{itemize}
The central challenge is then \emph{model-building}: one must identify a realistic particle-physics origin for a bosonic field (preferably complex or effectively complex) with the correct mass scale and self-interactions to yield compact Kerr-like objects across the astrophysical mass range, while remaining consistent with cosmology and laboratory constraints.

\subsection{Non-Abelian dark sectors and glueball matter}
Glueball dark matter and hidden confining sectors have been studied in a variety of contexts, including pure Yang--Mills hidden valleys, self-interacting glueball relics and dark-confinement cosmology \citep{Juknevich2009,SoniZhang2016,ForestellMorrissey2017GlueballDM,Acharya2017GlueballDM,Carlson1992Cannibal,Hochberg2014SIMP,TulinYu2018SIDM,Bruno2024GlueballGW,Masi2024G2Resurgence}.

Strongly coupled hidden sectors provide a natural origin for macroscopic bosonic matter. A confining non-Abelian gauge theory generically produces a spectrum of bound states, including scalar glueballs. When such states are stable or long-lived on cosmological timescales, they can serve as dark matter candidates. In dense environments, their self-interactions can generate a stiff EOS, enabling the formation of compact self-gravitating objects.

However, the simplest confining theories often produce a \emph{real} lightest scalar glueball (e.g.\ a $0^{++}$), which does not automatically carry a conserved number. In that case, the most generic compact objects are time-dependent oscillatory configurations (oscillatons), and the construction of stationary, rapidly rotating families becomes less viable. Whereas, for the present program---aimed at testing restricted Kerr-like observables including rotation---a key desideratum is an effective complex scalar description with an approximate $U(1)$ symmetry, so that the charge $N$ is meaningful and Eq.~\eqref{eq:JmN_intro} governs rotation.

This assumption is load-bearing and should not be inferred from coset counting alone.  The existence of several real broken-sector degrees of freedom does not by itself prove an approximately conserved complex charge in the confined many-body state.  In the present paper the complex field is therefore a conditional infrared order parameter.  All stationary-spin, quadrupole and ISCO results require number-changing reactions to be slow on the stellar timescale,
Therefore, the compact-star calculation performed below does not include explicit cannibal number-changing reactions, such as \(3\to2\), \(4\to2\) or more general \(n\to2\) processes, as local collision terms in the
stellar dynamics.  Such reactions are instead part of the early-universe dark-sector chemistry: in hidden confining sectors they can determine the glueball relic abundance, heat the dark bath during the nonrelativistic
``cannibal'' phase and delay the cooling of the secluded sector
\citep{HochbergEtAl2014SIMP,HochbergEtAl2014SIMPlest,
	PappadopuloRudermanTrevisan2016,FarinaEtAl2016Cannibal,
	ForestellMorrisseySigurdson2017Glueball,CarenzaEtAl2022GlueballDM}.
In the late-time compact object we assume that these reactions have frozen out, are phase-space suppressed or have been absorbed into the
zero-temperature effective GC9 equation of state.  Equivalently, the macroscopic configuration is treated as an approximately number-conserving complex matter on the dynamical timescale,
\begin{equation}
	\Gamma_{3\to2}
	\sim n_\chi^2\langle\sigma v^2\rangle_{3\to2}
	\ll
	t_{\rm dyn}^{-1}
	\sim
	\sqrt{G\rho},
\end{equation}
with analogous inequalities for higher number-changing channels. Here \(n_\chi\) is the number density of the dark glueball constituents,
\(\langle\sigma v^2\rangle_{3\to2}\) is the effective three-body
number-changing reaction coefficient, \(\Gamma_{3\to2}\) is the
corresponding per-particle cannibalization rate \citep{PappadopuloRudermanTrevisan2016,FarinaEtAl2016Cannibal,ForestellMorrisseySigurdson2017Glueball} and
\(t_{\rm dyn}\sim(G\rho)^{-1/2}\) is the gravitational dynamical time of
a configuration with characteristic density \(\rho\) \citep{BinneyTremaine2008}.  The condition
\(\Gamma_{3\to2}\ll t_{\rm dyn}^{-1}\) states that number-changing
processes are frozen out on the timescale relevant for the stellar
equilibrium. If
this condition were violated, the appropriate description would no
longer be the equilibrium TOV--Hartle--Hinderer problem used here, but a
finite-temperature dissipative kinetic treatment of dense glueball
matter.

No first-principles compact-star calculation of $\langle\sigma v^2\rangle_{3\to2}$ or of the higher number-changing rates is supplied here. Consequently, the stationary complex-field, rotation, quadrupole and ISCO results are conditional on the inequality above; cosmological freeze-out alone does not prove that it remains valid at stellar densities.

In the unification picture proposed in \citep{Masi_2021, Masi2024G2Resurgence,Masi2026E6G2}, the central point is a special non-regular embedding
\begin{equation}
E_6 \to G(2)\times SU(3)_A,
\end{equation}
in which the exceptional factor \(G(2)\) plays the role of a hidden strong sector, while \(SU(3)_A\) acts as the ancestor of the electroweak gauge group. Subsequent breaking steps
\begin{equation}
G(2)\to SU(3)_C,
\qquad
SU(3)_A\to SU(2)_L\times U(1)_A,
\end{equation}
together with the additional $U(1)_X$ abelian factor from an $E_7$ uplift to reproduce hypercharges, recover the Standard Model (SM) gauge structure at low energy while preserving a secluded non-abelian dark sector.
The exceptional Lie group $G(2)$ is characterized by several intertwined features relevant for model-building:
\begin{itemize}
	\item It is a simple, anomaly-free gauge group that contains $SU(3)$ as a subgroup, enabling a tight conceptual link to Quantum Chromodynamics (QCD).
	\item A symmetry-breaking pattern $G(2)\to SU(3)_C$ can be realized via a suitable Higgs sector, producing additional massive gauge bosons associated with the broken generators in the coset $S^6 = G(2)/SU(3)$, alongside an unbroken $SU(3)_C$-like confining sector.
	\item In the broken phase, the spectrum and residual symmetries can naturally support \emph{dark relic} bound states of the heavy vector bosons produced by $G(2)$ breaking, whose effective low-energy description need not coincide with the simplest ``real glueball'' of a pure confining theory. In particular, the broken-sector composites can admit a \emph{complex scalar} description. 
	\item The dark glueball mass, according to the $\alpha$ couplings running analysis performed in \citep{Masi2024G2Resurgence,Masi2026E6G2}, should lie approximately in the range $m\sim10^{9}$--$10^{14}\,\mathrm{GeV}$, so above the EeV scale. This reflects the high energy scale of the $G(2)$ symmetry breaking. If one imposes this transition to occur above the electroweak unification, this bound becomes $m\gtrsim10^{13}\,\mathrm{GeV}$.
\end{itemize}
Hereafter we adopt this $G(2)\to SU(3)_C$ motivated viewpoint as our ultraviolet (UV) anchor. The goal is not merely to posit an ad hoc bosonic field, but to embed the required macroscopic phenomenology in a coherent non-Abelian gauge framework with a controlled symmetry-breaking origin, which is embedded in the grand unification scenario described in \citep{Masi2024G2Resurgence,Masi2026E6G2}. In this sense, the GC9 boson star scenario should be viewed as the strong-gravity, finite-density realization of a broader \(E_6\)-motivated dark sector. The role of the compact-object analysis is then to ask whether the dense glueball medium descending from this exceptional construction can support horizonless, Kerr-mimicking configurations across the astrophysical mass range.

The $E_6\to G(2)\times SU(3)_A$ embedding and the numerical mass window $m\sim10^9$--$10^{14}\,{\rm GeV}$ are imported ultraviolet priors from \citep{Masi2024G2Resurgence,Masi2026E6G2}, not independent results of the present compact-star calculation.  Existing $G(2)$ lattice studies establish valuable facts about confinement, screening and sign-problem-free dense matter, but they do not presently determine this mass window, the approximate $U(1)$, or the GC9 Wilson coefficients.  The reduced stellar calculation is rescalable and should therefore be read conditionally on these UV inputs rather than as an independent validation of them.

For notation to be unambiguous, $SU(3)_C$ in this manuscript denotes
Standard Model color. A long-lived broken-sector relic must therefore be
an $SU(3)_C$ singlet after confinement, and its stability and approximate
conserved number require an additional residual symmetry or dynamical
suppression that is not derived from the coset $G(2)/SU(3)_C$ alone.

\paragraph{Microscopic stability and the meaning of the complex field.}
A further distinction is required between the microscopic stability of the
broken-sector composite and the approximate number conservation used in the
macroscopic boson star description.  In the present \(G(2)\to SU(3)_C\)
interpretation, the six massive coset vectors decompose as
\(X\sim{\bf3}_C\oplus\overline{\bf3}_C\) and therefore carry ordinary QCD
color before confinement \citep{Masi_2021,Masi2024G2Resurgence,Masi2026E6G2}.
We retain as a concrete microscopic reference the lightest color-singlet
two-vector configuration
\begin{equation}
	{\cal G}_X\sim[X\overline X]_{{\bf1}_C},
	\qquad
	{\bf3}_C\otimes\overline{\bf3}_C
	={\bf1}_C\oplus{\bf8}_C ,
	\label{eq:GX_reference_constituent}
\end{equation}
which provides the natural \(J=0\) benchmark considered in the
original broken-\(G(2)\) construction \citep{Masi_2021}.  This identification
must, however, be accompanied by a genuine stability selection rule.
Color singletness alone does not prevent \(X\overline X\) annihilation into
ordinary QCD gluons.  For an S-wave Coulombic heavy pair, dimensional
NRQCD/pNRQCD power counting gives parametrically
\begin{equation}
	\Gamma_{\rm ann}^{(0)}
	\sim
	\frac{\alpha_s^2}{m_X^2}|\psi(0)|^2
	\sim
	\alpha_s^5m_X ,
	\label{eq:GX_unprotected_width}
\end{equation}
up to spin- and channel-dependent coefficients
\citep{BodwinBraatenLepage1995,BrambillaPinedaSotoVairo2005}.
For the values \(m_X\simeq5.6\times10^{13}\,{\rm GeV}\) and \(\alpha_s\simeq0.03\) in \citep{Masi2026E6G2}, this scale is
\begin{equation}
	\Gamma_{\rm ann}^{(0)}
	\sim1.4\times10^{6}\ {\rm GeV},
	\qquad
	\tau_{\rm ann}^{(0)}
	\sim5\times10^{-31}\ {\rm s}.
	\label{eq:GX_unprotected_lifetime}
\end{equation}
The numerical coefficient in Eq.~\eqref{eq:GX_unprotected_width} is not
computed for the massive vector state and these values should therefore be
read only as a parametric instability estimate; nevertheless the hierarchy
is so large that ordinary color singletness cannot provide cosmological
stability.

Accordingly, we assume an additional exact or extremely
accurate selection rule \({\cal S}_X\), of the type contemplated in the
original \(G(2)\) construction through an accidental constituent number,
a discrete symmetry, or a generalized \(G\)-parity
\citep{Masi_2021}.  For the specific \(X\overline X\) realization this
symmetry must act on the physical composite in such a way that
\begin{equation}
	\langle gg,\ldots|\,H\,|{\cal G}_X\rangle\simeq0 .
	\label{eq:GX_stability_rule}
\end{equation}

The effective complex field \(\Phi\) used below should not be identified with a fundamental complex gauge field, nor does \(\Phi^\dagger\) have to
represent a distinct microscopic anti-glueball.  Rather, \(\Phi\) is the condensate order parameter of a long-lived neutral
composite whose effective particle number is approximately conserved in
the cold phase.  We write
\begin{equation}
	\Phi(x)=\sqrt{s(x)}\,e^{i\theta(x)},
	\qquad
	s(x)\equiv\Phi^\dagger\Phi=|\Phi|^2 ,
	\label{eq:Phi_emergent_phase}
\end{equation}
where \(s\) is the local condensate density variable used throughout the
GC9 effective potential and \(\theta\) is its macroscopic phase.  The
would-be Noether current associated with the emergent approximate
global \(U(1)\) is taken to be
\begin{equation}
	j_D^\mu
	\equiv
	-i\Phi^\dagger
	\overleftrightarrow{\nabla^\mu}\Phi.
	\label{eq:GC9_Noether_current}
	\end{equation}
Thus two logically independent conditions enter the present construction:
\begin{equation}
	\Gamma_{{\cal G}_X}\,t_{\rm age}\ll1,
	\qquad
	\Gamma_{\Delta N}\,t_{\rm dyn}\ll1 ,
	\label{eq:GC9_two_stability_conditions}
\end{equation}
where \(\Gamma_{{\cal G}_X}\) is the total decay or annihilation rate of
an individual microscopic color-singlet composite
\({\cal G}_X\sim[X\overline X]_1\), while \(t_{\rm age}\) is the required
survival time of the dark constituent, conservatively of order the age
of the compact object or, for cosmological DM stability, the age of the Universe. The first protects the microscopic composite against decay or annihilation and the second suppresses the aforementioned number-changing reactions among
otherwise stable composites.  
For quasi-equilibrium stellar structure the second inequality need only
hold on \(t_{\rm dyn}\); interpreting \(N\) as a conserved charge over the
full lifetime of an old GC9 object requires the stronger condition
\(\Gamma_{\Delta N}t_{\rm age}\ll1\).
Only the second condition gives the emergent approximate \(U(1)\) underlying
the Noether charge \(N\) and the rotating-field relation \(J=mN\).  

\paragraph{Alternative candidates.}
In the $G(2)\to SU(3)$ gauge--Higgs theory the protecting discrete transformation acts as charge conjugation on the gauge sector, so the ordinary scalar meson $X_\mu\bar X^\mu$ has $J^{PC}=0^{++}$ and is not protected.  
The protected $C=-$ sector nevertheless contains gauge-invariant composite channels with different spin assignments. In particular, using the $G(2)\to SU(3)$ decomposition and charge-conjugation action of Ref.~\citep{ButtazzoEtAl2019}, a leading two-vector projection can occur in the $J^{PC}=1^{+-}$ channel; an S-wave pair of massive spin-one constituents generically admits $J^{P}=0^{+},1^{+},2^{+}$
\citep{KeLiuLi2022}. The natural local three-vector combination $XXX+\bar X\bar X\bar X$ also admits a scalar $J^{PC}=0^{--}$ realization. Such $0^{--}$
three-vector/gluon channels are independently known from constituent-spin and gauge-invariant interpolating-current constructions \citep{BoulangerEtAl2008,PimikovEtAl2017}. If the exact odd-sector parity assumed above is realized by the ultraviolet completion, it stabilizes the lightest state in this sector; the detailed ordering of the $1^{+-}$ and $0^{--}$ levels is then a nonperturbative spectroscopy question. The compact-star calculation does not itself prove that ultraviolet parity.  This microscopic uncertainty does not require abandoning the scalar GC9 compact-star construction.  The GC9 field can be interpreted conservatively as a macroscopic scalar order parameter, or equivalently as an effective isotropic equation-of-state variable, for the stable composite medium rather than as a literal elementary constituent field.  If the lightest protected state is $0^{--}$ the scalar interpretation is direct; if instead a $1^{+-}$ state is lighter, the same static GC9 fluid description can remain a useful leading approximation when the vector polarizations are unpolarized and the medium is effectively isotropic, while spin-sensitive rotational and perturbative observables may eventually require a vector-field refinement. The case of a GC9-Vector star will be discussed in \ref{subsec:GC9_SV_comparison}. We therefore regard the precise microscopic $J^{PC}$ assignment as an open input to be fixed by future nonperturbative spectroscopy.

\subsection{From microphysics to an EOS}
The need for stiff but controllable repulsion has been emphasized in both self-interacting BS studies and scalar-field DM effective field theory (EFT) analyses: representative discussions include \citep{LiddleMadsen1992,MielkeSchunck2000,LeeKoh1996,Schive2014Cosmic,Ferreira2021UltraLight,SuarezRoblesMatos2014}.
Kerr-mimicking compactness at large masses requires \emph{stiff high-density pressure} without spoiling low-density cosmology. A purely quartic repulsion,
\begin{equation}
	V(|\Phi|)=m^2|\Phi|^2 + \frac{\lambda}{2}|\Phi|^4,
	\label{eq:quartic_intro}
\end{equation}
is an instructive benchmark, but it often forces a tension: making the star sufficiently massive and compact can require large effective self-couplings, potentially conflicting with a naive low-energy truncation and with microscopic expectations (e.g.\ large-$N$ suppression of glueball couplings in simple confining theories).
In \emph{saturating repulsive} potentials \citep{Chavanis2011Log,ColemanWeinberg1973,GildenerWeinberg1976,Masi2024G2Resurgence,Arcadi2020} the interaction can stiffen the EOS at high density while remaining perturbatively mild at low density. They can support ultra-compact configurations across a wide mass range, although the heavy-constituent supermassive regime still exposes severe scale-hierarchy, tuning or Planck-ceiling issues, as discussed in Section II.
A primary motivation for seeking stiff, saturating EOS families is the astrophysical breadth of BH candidates. Observed compact objects include: i) stellar-mass sources in X-ray binaries and GW events; ii) intermediate-mass candidates in dense environments; iii) supermassive compact objects in galactic nuclei. This wide dynamical range is precisely what makes the BS alternative both attractive and difficult, as emphasized in early galactic-center studies, compact-object surveys and image-based discrimination analyses \citep{Torres2000,LeeKoh1996,BroderickNarayan2006,BroderickNarayan2007,Vincent2016,Olivares2020}.

A viable horizonless replacement must therefore admit stable configurations across many orders of magnitude in mass. In the BS picture, the achievable macroscopic masses depend sensitively on the constituent mass $m$ and the interaction scales. This broad phenomenological reach is also what allows bosonic alternatives to be discussed in settings as diverse as galactic-center supermassive objects, Gaia-like dark companions and even heavy GW events such as GW190521 interpreted through vector boson star mergers \citep{Torres2000,PomboSaltas2023SunLike,CalderonBustilloEtAl2021GW190521Proca}. 

Our analysis emphasizes that a new class of scale-separated positive-cumulant potentials can dramatically enlarge the parameter space, allowing even very heavy constituents (including ultra-heavy glueball-like candidates) to form massive compact objects.

\subsection{Roadmap of the paper}
The paper follows the standard logic of boson star phenomenology: microphysical motivation, equilibrium structure, relativistic observables, dynamical signatures and astrophysical/cosmological features and constraints \citep{SchunckMielke2003,Liebling2017,Visinelli2021Review,Barack2019BHSurvey}.
First, the heavy-glueball plus supermassive-object requirement is used to filter the space of BS potentials and shows that scale separation is required; among the scale-separated closures considered here, the $G(2)$-inspired GC9 family is retained primarily because its high-density stiffness yields the desired compactness while preserving causality, smoothness, Q-ball safety and a bag-free surface. Second, the resulting compact branch is shown to remain sufficiently Kerr-like in the set of leading observables, with an order-unity quadrupole estimate and a percent-level ISCO shift in the controlled slow-rotation regime. Photon-ring and shadow predictions remain open because the computed surface lies outside the external photon-sphere radius. A static axial $\ell=2$ spacetime mode is computed directly, while its rotating continuation, polar coupling and merger excitation remain open. Third, the same framework identifies structured deviations --- in tidal response, merger relaxation, matter-supported internal oscillations, seeded collapse and hybrid compact DM cosmology --- that make the scenario falsifiable. 

In Sec.~\ref{sec:potentials} we compare the main classes of boson-star self-interaction potentials and use the TON--618 mass bound to expose the scale hierarchy required in the heavy-glueball regime. The mass bound alone does not select GC9 uniquely: once an independent collective scale is admitted, other scale-separated closures can reproduce the dimensional mass scaling. GC9 is retained within the restricted design criteria adopted here because the resulting $c_s^2\to4/5$ stiffness gives substantially better compactness and tidal performance than the standard quartic benchmark while remaining causal, smooth, Q-ball safe and bag free. 

In Sec.~\ref{sec:GC9} we introduce the GC9 potential itself, discuss its microscopic motivation from the
dark \(G(2)\) glueball sector and analyze its vacuum structure, scaling properties and physical interpretation, showing how a boson star can mimic a Kerr BH and how it can be distinguished.

In Sec.~\ref{sec:observables} we develop the compact-object structure of the GC9 construction, introducing an effective EOS and solving the TOV--Hartle--Hinderer system. We also derive the second-order Hartle--Thorne quadrupole of the compact GC9 branch, using
it to compute the slow-rotation multipole and ISCO diagnostics. We also record the formal exterior continuation of the photon-region formulas, while showing explicitly why it is not a physical light ring for the present stellar radius. A GC9 Proca vector glueball alternative is also discussed and compared with GC9 scalar solution.

In Sec.~\ref{sec:SB} we turn to scalar bursts, their energetics and their prospective detectability, emphasizing their role as a distinctive consequence of horizonless post-merger relaxation. 

In Sec.~\ref{sec:seeded} we introduce a minimal seeded-collapse
framework for forming GC9 boson star remnants in stellar collapse, including post-collapse bookkeeping,
re-landing conditions and the connection to failed supernovae and disappearing-star searches. Secondly, we propose the shock-revival mechanism in core-collapse supernovae with a horizonless GC9 boson core, formulate the modified criticality condition and discuss the resulting stellar and compact-object constraints.

In Sec.~\ref{sec:cosmo} we widen the perspective to population and cosmological implications. There we discuss the hybrid
GC9 population, the compact-object versus fluid crossover, the role of asteroid-window primordial GC9s, the remnant-dominated stellar-mass sector, viable GC9 cores inside neutron stars and the hierarchical growth
channel leading to supermassive GC9 objects. 

We conclude in Sec.~\ref{sec:conclusions} with a summary of the main results, their current limitations and the most relevant future tests.

In sum, the guiding principle of this paper is conservative in its gravitational exterior (Kerr-like where observations demand it) but radical in its interior microphysics: we explore whether compact objects currently modeled as black holes could instead be described by boson stars made of non-Abelian dark matter anchored in an exceptional-group strong sector and we derive the corresponding testable astrophysical consequences.

\section{Comparison of Self-Interaction Potentials}
\label{sec:potentials}

The literature on boson-star self-interactions is broad and
heterogeneous and different potentials encode rather different physical
assumptions. The canonical starting points are the free massive scalar
and the repulsive quartic model, which respectively define the Kaup
limit and the Colpi--Shapiro--Wasserman large-self-interaction scaling
\citep{Kaup1968,RuffiniBonazzola1969,ColpiShapiroWasserman1986,
	LiddleMadsen1992,SchunckLiddle1997}. Extensions based on higher-order
polynomial interactions, Q-ball potentials and solitonic or flat-top
self-interactions can generate self-bound or thin-wall configurations,
but usually at the price of additional scales, tuned signs or explicit
charge-stabilizing assumptions
\citep{FriedbergLeeSirlin1986,Coleman1985Qballs,Kusenko1997Qball,
	KusenkoShaposhnikov1998,LeeKoh1996,MultamakiVilja2000,ArodzLis2010,
	HartmannKleihausKunzSchaffer2012,KleihausKunzShnir2010}.
A separate class is provided by axion and axion-like potentials, where
the periodic or cosine structure produces dilute and dense branches with
phenomenology that is strongly tied to the decay constant and to the
attractive nature of the leading self-interaction
\citep{BarrancoBernal2011,BraatenMohapatraZhang2016,Eby2015Axion,
	Eby2016Axion,Helfer2017Axion,Visinelli2018Axion,LevkovPaninTkachev2018,
	BraatenEtAl2018,Guerra2019AxionBosonStars}. Other generalizations
change the matter content rather than only the potential: boson--fermion
stars, multistate configurations and mixed dark-sector objects introduce
additional components, while vector or Proca stars replace the scalar by
a massive spin-one field and can lead to more compact or qualitatively
different branches
\citep{HenriquesLiddleMoorhouse1990,BernalBecerrilGuzman2010,
	UrenaLopezBernal2010,ValdezAlcubierreUrena2013,Brito2016,
	HerdeiroRadu2016ProcaHair,SanchisGual2019ProcaBinary,
	HerdeiroPanotopoulosRadu2020}. Finally, EFT and
modified-kinetic constructions introduce derivative operators,
noncanonical kinetic terms or Horndeski/Proca-like extensions, making the stiffness and stability properties depend not only on the potential
but also on the kinetic geometry of field space
\citep{BrihayeMinkovHartmann2020,BrihayeHartmannKunz2022HorndeskiProca,
	Minamitsuji2018,Babichev2018,KlimasLoginovShnirZhilin2022,
	Mendonca2021}.
These families have distinct consequences for the maximum mass,
compactness, tidal response, multipolar structure and rotational support
of the resulting object
\citep{GleiserWatkins1989,FriedbergLeePang1987,Chavanis2011Log,
	BritoCardosoHerdeiroRadu2016,JohnsonMcDanielEtAl2018,PacilioEtAl2021,
	ChenSunHuangWang2024}.

The macroscopic properties of boson stars---their maximum mass, compactness, stability and ability to mimic Kerr black holes---are controlled almost entirely by the form of the self-interaction potential $V(|\Phi|^2)$ of the underlying bosonic field. In the present framework the bosons are not fundamental scalars, but heavy composite states emerging from a confining $G(2)$ gauge sector; throughout, however, the mapping to an effective complex scalar and to a barotropic EOS should be understood as an assumed low-energy description rather than as a first-principles derivation from the microscopic gauge theory. Their interactions therefore encode non-perturbative strong dynamics and cannot be reliably approximated by a low-order polynomial expansion.
Not every BS self-interaction potential developed in literature is suitable for describing compact objects made of dark \(G(2)\) glueball matter. Standard polynomial, quartic, logarithmic, Q-ball and plateau-like potentials are useful phenomenological templates, but they generally encode microscopic assumptions --- such as elementary weakly coupled scalars or ad hoc saturation scales --- that are not naturally aligned with a strongly coupled composite glueball sector. Moreover, in the ultra-heavy regime relevant here, many of these families either remain tied to the usual \(M_{\max}\propto m^{-2}\) suppression or require tuned dimensionless couplings or Planck-sized scales in order to reach supermassive compact-object masses.

Two conflicting requirements must be reconciled, \textit{i.e.} low-density consistency and high-density stiffness. At small field amplitude the effective theory must reduce to a massive scalar EFT with perturbative self-interactions around the vacuum, so that dilute dark matter remains approximately free and standard large-scale DM phenomenology is not spoiled. At large field amplitude the equation of state must become sufficiently stiff to support highly compact configurations with masses comparable to those of observed supermassive black holes (SMBHs), while remaining causal and avoiding a self-bound bag surface.
Generic low-order polynomial interactions often fail to satisfy these requirements simultaneously: they either remain too soft at high density or require very large coefficients that are difficult to justify in a controlled glueball EFT. 

The construction pursued below is instead a constrained high-order positive density polynomial with an independent emergent stiffness scale. Its repulsive contribution grows monotonically, while its Thomas--Fermi equation of state approaches the finite causal ratio $P/\rho\to4/5$ and $c_s^2\to4/5$. Thus GC9 is not a pressure plateau and not a non-polynomial saturation ansatz; it is a scale-separated density-cumulant closure. This distinction is essential when it is compared with exponential or $\tanh$ plateau potentials, for which the interaction energy itself tends to a constant, as described below.

The microscopic motivation for focusing on ultra-heavy dark glueballs is not ad hoc. In the \(E_6\to G(2)\times SU(3)_A\) unification framework proposed in \citep{Masi2026E6G2}, the exceptional factor \(G(2)\) plays the role of a hidden strong sector, while subsequent breaking \(G(2)\to SU(3)_C\) leaves a secluded confining dark gauge sector whose non-SM bosonic remnants are naturally lifted to very high scales. In that setting, the dark \(G(2)\) vectors associated with the broken generators become massive at the intermediate breaking scale and can later bind into heavy glueball states once the residual non-Abelian sector confines. The resulting dark matter candidates are therefore naturally expected to be ultra-heavy composites rather than light axion-like or weak-scale bosons. This UV picture is precisely what makes the compact-object problem nontrivial: the BS potential must support highly compact equilibria approaching the photon-sphere threshold while the constituent mass can lie many orders of magnitude above the scales usually considered in the BS literature.

Furthermore, the need to mimic SMBHs is conceptually decisive for any BS alternative. Stellar-mass compact objects already provide important tests, but they do not by themselves guarantee that the model can reproduce the full observed BH population. By contrast, SMBH candidates impose the most severe combined requirements on the theory: extremely large masses, sustained compactness over cosmological timescales, compatibility with accretion and imaging observables and a formation pathway consistent with galactic assembly. For this reason, the supermassive sector is the natural place where many otherwise viable BS potentials fail. In the present work, the requirement of mimicking SMBHs --- and in particular the TON--618 benchmark --- is therefore used as one of the main discriminators among candidate self-interaction potentials.

For these reasons, before constructing the GC9 model we reserve a dedicated section to a systematic comparison of potential families. The point is not merely taxonomic, but to identify which effective descriptions could plausibly represent a dense, strongly interacting glueball fluid arising from a $G(2)$ gauge sector while satisfying: (i) smooth repulsive high-density stiffening with a finite subluminal asymptotic sound speed; (ii) an ultraheavy constituent mass $m\sim10^{9}$--$10^{14}\,\mathrm{GeV}$, as considered in \citep{Masi2024G2Resurgence,Masi2026E6G2}; and (iii) supermassive equilibria compatible with the TON--618 benchmark $M\sim10^{10}M_\odot$. These conditions severely constrain, but do not mathematically exhaust, the class of viable effective potentials.

\subsection{TON--618 bound as a decisive selection criterion}
\label{subsec:ton618_selection}

Before comparing specific self-interaction potentials, it is useful to stress why the SMBH TON--618 mass can be exploited as a discriminator. Indeed, if we require that the boson star made of dark $G(2)$ glueball matter covers the whole known range of black holes, up to ultra-massive quasars, the requirement for its maximum mass
\begin{equation}
	M_{\max} \ge M_{\rm TON} \sim 10^{10} M_\odot
	\label{eq:mton_bound_clarify}
\end{equation}
acts as an exceptionally strong discriminator for the present analysis. TON--618 is widely cited as one of the most massive BH candidates known, with a virial mass estimate of order $6.6\times10^{10}\,M_\odot$ \citep{Shemmer2004,Aggarwal2025}. In a framework based on ultra-heavy \(G(2)\)-sector glueball
constituents described in \citep{Masi2024G2Resurgence,Masi2026E6G2}, this bound becomes highly restrictive. A viable effective potential must simultaneously support very heavy microscopic bosons and equilibrium configurations extending all the way to the observed
supermassive range, considering a dark matter scale $M \sim 10^{9}$--$10^{14}\,\mathrm{GeV}$.

This requirement sharply constrains any one-scale realization. Free or weakly self-interacting models force the constituent mass into the ultralight regime, while standard quartic parameterizations require extremely strong dimensionless coupling if the microscopic mass itself is kept ultraheavy; plateau-like families can instead push auxiliary scales toward Planckian values or effectively reintroduce a bag-like vacuum plateau \citep{Schechter1980TraceAnomaly,MottolaVaulin2006}. The important lesson is therefore not that a large numerical coupling uniquely excludes every alternative, but that the heavy-particle and supermassive-star requirements demand an additional collective scale. The potential comparison below then asks which scale-separated closures also provide the stiffness, surface structure and stability properties required for restricted Kerr mimicry.

Moreover, the importance of the TON--618 bound is not limited to the maximum mass alone. The same
high-density stiffness required to satisfy Eq.~\eqref{eq:mton_bound_clarify} is also what later drives the GC9 branch toward a highly compact, restricted Kerr-comparison regime. The TON--618 bound is therefore one useful high-mass filter when comparing potential families in the following subsections. It reveals the scale hierarchy that every ultraheavy-constituent model must address, but it does not by itself prove that GC9 is unique. More precisely, this filter selects for \emph{scale separation} between the dark sector's condensation scale and the compact-object mass scale, not for a unique keV value of the stiffness $\Lambda_T$ introduced below; any sufficiently scale-separated closure passes the same filter, and Sec.~\ref{sec:sector} makes this scale-dependence explicit rather than leaving it as an implicit uniqueness claim (see also Sec.~\ref{sec:lambdaT-frozen}).
Established analytic maximum-mass relations exist for the free, quartic and solitonic benchmark families, whereas other potentials are mostly characterized in the literature through numerical GR solutions and heuristic high-density extrapolations rather than by a universal closed \(M_{\max}\) formula \citep{Kaup1968,RuffiniBonazzola1969, LeeKoh1996,Cardoso2019BHmimickers,ChoiHeSchiappacasse2019,ColpiShapiroWasserman1986}.

\subsection{Free massive scalar field}
The free mini-BS limit remains the canonical reference point for essentially all later constructions \citep{Kaup1968,RuffiniBonazzola1969,LiddleMadsen1992,KusmartsevMielkeSchunck1991,Visinelli2021Review}.
The simplest BS model is based on a free, complex scalar field with potential
\begin{equation}
	V_{\rm free}(s) = m^2 s, \qquad s = |\Phi|^2.
\end{equation}
which contains only the $m$ mass term. In this case the equilibrium configurations are supported purely by gradient (quantum) pressure. The maximum mass is the well known Kaup limit:
\begin{equation}
	M_{\max}^{\rm free} \simeq 0.633 \,\frac{M_{\rm Pl}^2}{m},
	\label{eq:free_mass}
\end{equation}
where $M_{\rm Pl}$ is the Planck mass, while the characteristic radius along the sequence is of order
\begin{equation}
	R \sim \frac{1}{m},
\end{equation}
up to order-one factors that depend on where the configuration lies along the equilibrium branch. Imposing $M_{\max} \gtrsim M_{\rm TON} \sim 10^{10} M_\odot$ immediately yields
\begin{equation}
	m \lesssim 8 \times 10^{-21}\,\mathrm{eV},
\end{equation}
which places the constituent in the fuzzy DM regime \citep{Ferreira2021UltraLight,Schive2014Cosmic}. This is incompatible with our $G(2)$ glueball interpretation, where the natural mass scale is many orders of magnitude larger. Consequently, free boson stars are excluded for our purposes.

\subsection{Quartic self-interaction potentials (Colpi–Shapiro–Wasserman)}
Repulsive quartic stars provide the standard self-interacting benchmark and are often used as the first step beyond the mini-BS regime \citep{ColpiShapiroWasserman1986,Colpi1986,LeeKoh1996,ChavanisDelfini2011,JohnsonMcDanielEtAl2018}.
A widely studied extension introduces a repulsive quartic interaction,
\begin{equation}
	V_{\lambda}(s)=m^2s+\frac{\lambda}{2}s^2.
\end{equation}
leading, in the Thomas--Fermi (TF) regime, to the Colpi--Shapiro--Wasserman scaling
\begin{equation}
	M_{\max}^{\lambda} \sim c_\lambda \sqrt{\lambda}\, \frac{M_{\rm Pl}^3}{m^2},
	\qquad
	R \sim c_\lambda \sqrt{\lambda}\, \frac{M_{\rm Pl}}{m^2}.
	\label{eq:quartic_mass}
\end{equation}
Requiring $M_{\max} \gtrsim M_{\rm TON}$ implies
\begin{equation}
	\lambda\;\sim\;\left(\frac{M_{\rm TON}\,m^2}{c_\lambda\,M_{\rm Pl}^3}\right)^2.
	\label{eq:lambda_req}
\end{equation}
Numerically, taking the constant $c_\lambda=0.1$ as a representative order-one benchmark,
\begin{align}
	m=10~{\rm TeV}:\qquad &\lambda\sim 4\times 10^{37},\\
	m=1~{\rm PeV}:\qquad &\lambda\sim 4\times 10^{45},\\
	m=1~{\rm EeV}:\qquad &\lambda\sim 4\times 10^{57}.
\end{align}
In the ordinary renormalizable quartic interpretation these values are far outside weak-coupling control.  Coupling size alone, however, is not a parametrization-independent discriminator once one permits the same scale separation later used in GC9.  Indeed, define a collective quartic scale $\Lambda_q$ by
\begin{equation}
 V_\lambda=m^2s+\Lambda_q^4\left(\frac{s}{s_q}\right)^2,
 \qquad
 s_q=\frac{\Lambda_q^4}{m^2},
 \qquad
 \lambda=2\left(\frac{m}{\Lambda_q}\right)^4 .
 \label{eq:quartic_scale_separation}
\end{equation}
The $m=10$~TeV TON--618 benchmark corresponds to $\Lambda_q\simeq4.8$~keV, i.e.\ to the same kind of microscopic--collective hierarchy that appears in GC9.  We therefore do \emph{not} reject a scale-separated quartic closure on naturalness grounds alone.  Its robust disadvantage for the present purpose is dynamical: in the strong-coupling limit the quartic EOS approaches $c_s^2\to1/3$ and the resulting branch remains substantially less compact and more tidally deformable than GC9.  The selection made below is consequently based on the combined stiffness, compactness, tidal, surface and stability requirements, not on the numerical magnitude of $\lambda$ by itself.

In addition, the most general polynomial scalar glueball effective potential, in the large $N$ limit of a $SU(N)$ gauge theory, may contain not only a quartic interaction, but also the cubic and higher order terms \citep{SONI2017379,Yamanaka:2019yek,SoniZhang2016}, in the form
\begin{equation}\label{effpot}
	V(\Phi) = \sum_{i=2}^{\infty}{\frac{a_{i}}{i!}{\Big(\frac{4\pi}{N}\Big)}^{i-2} m^{4-i}\Phi^i}
\end{equation}
where the coefficients are $a_i\approx 1$, which could be obtained from lattice computations. The trilinear interaction might generate an attractive Yukawa-like potential \citep{Yamanaka:2019yek}, whereas the quartic one may be repulsive, according to the sign of the coupling; the fifth and higher terms might be suppressed by the mass scale and the decreasing couplings. 

\subsection{Higher-order polynomial potentials}
\label{subsec:GC9_higher_order_polynomial}
Higher-order polynomial truncations appear in nontopological-soliton, Q-ball and phenomenological EFT settings, but their regime of validity is often delicate once the large-field sector is probed \citep{FriedbergLeeSirlin1976,FriedbergLeeSirlin1986,LeePang1992,Coleman1985Qballs,TsumagariCopelandSaffin2008}.
One may attempt to alleviate the need for large quartic couplings by including higher-order terms,
\begin{equation}
	V(s) = m^2 s + \sum_{n\ge 2} \frac{\lambda_n}{n} s^n.
\end{equation}
While this formally allows additional stiffness at high density, it introduces an uncontrolled proliferation of dimensionful couplings $\lambda_n$ that must be tuned against each other to stabilize the star and achieve large masses. Moreover, polynomial truncations are intrinsically unreliable in the strong-field regime $s \gg m^2$, where glueball compositeness and multi-particle effects dominate. From both an EFT and a phenomenological perspective, this approach lacks predictivity.

\paragraph{Generalized power-law closure and relation to GC9.}

A particularly relevant recent preprint by
\citet{PitzSchaffnerBielich2023} considers a generalized interaction
of the form
\begin{equation}
	U_n(s)
	=
	\frac{\lambda_n}{2^{n/2}}s^{n/2},
	\qquad
	s\equiv|\Phi|^2,
	\qquad
	n>2,
	\label{eq:generalized_power_potential}
\end{equation}
in a local-density, strong-self-interaction treatment. Under the same
algebraic map used in the Thomas--Fermi limit, this family gives
\begin{align}
	P_n
	&=
	\left(\frac{n}{2}-1\right)U_n,
	\label{eq:generalized_power_pressure}
	\\
	\rho_n
	&=
	2m^2s+
	\left(\frac{n}{2}+1\right)U_n.
	\label{eq:generalized_power_density}
\end{align}
After a coupling-dependent rescaling of the pressure and energy
density, the corresponding scale-invariant EOS can be written as
\begin{equation}
	\hat\rho
	=
	\hat P^{\,2/n}
	+
	\frac{n+2}{n-2}\hat P.
	\label{eq:generalized_power_EOS}
\end{equation}
The high-density sound-speed limit is
\begin{equation}
	c_s^2
	\longrightarrow
	\frac{n-2}{n+2}.
	\label{eq:generalized_power_sound_speed}
\end{equation}

The mathematical relation with GC9 is direct. For \(n=18\),
Eq.~\eqref{eq:generalized_power_potential} becomes
\begin{equation}
 U_{18}(s)\propto s^9=|\Phi|^{18},
\end{equation}
and Eqs.~\eqref{eq:generalized_power_pressure} and
\eqref{eq:generalized_power_density} reduce to
\begin{equation}
 P_{18}=8U_{18},
 \qquad
 \rho_{18}=2m^2s+10U_{18},
 \label{eq:n18_GC9_algebra}
\end{equation}
with
\begin{equation}
 c_s^2\longrightarrow\frac{4}{5}.
\end{equation}
This is precisely the algebraic structure of the GC9
Thomas--Fermi closure introduced below.

The \(s^9\) fluid relation is therefore not, by itself, unique to the
present construction. The distinction of GC9 lies instead in its
proposed connection with the primitive degree-six invariant of
\(G(2)\), its scale-separated normalization
\(s_T=\Lambda_T^4/m_G^2\), its interpretation in terms of ultraheavy
composite glueball matter and the compact-object observables developed
in this work. The generalized power-law study should consequently be
regarded as direct mathematical prior art for the GC9 EOS, but not as
a microscopic derivation of its \(G(2)\) interpretation.

\subsection{Solitonic and flat-top (plateau) potentials}
Solitonic and flat-top families are widely used because they can generate nearly constant-density cores and enhance compactness relative to quartic stars \citep{FriedbergLeePang1987,GleiserWatkins1989,Brihaye2009Interacting,BrihayeHartmann2009,HartmannKleihausKunzSchaffer2012}, with potentials of the Lee--Koh type
\begin{equation}
	V_{\rm sol}(s) = \mu^2 s \left(1 - \frac{2 s}{\sigma_0^2}\right)^2,
	\label{eq:solitonic}
\end{equation}
where $\mu$ is an effective mass parameter, with a scale $\sigma_0$ controlling the onset of self-interaction and a ``flat'' region enabling a nearly constant-density core. In natural units $(\hbar=c=1)$, one may identify $\mu\sim m$ up to model-dependent normalization.
These models admit large, nearly constant-density cores and can support high masses with comparatively moderate parameters. 
In Table 2 of \citep{Cardoso2019BHmimickers} 
the authors summarize the approximate scaling
\[
\frac{M_{\max}}{M_\odot}
\sim
5\left(\frac{10^{-12}}{\sigma_0}\right)^2
\left(\frac{500\,{\rm GeV}}{m}\right),
\]
in the convention where the scalar field is dimensionless and the formula applies for \(\sigma_0\ll1\).

These potentials can yield large masses and radii without the enormous quartic couplings of simple polynomial repulsion. However, their interpretation is delicate in a glueball EFT. The Lee--Koh potential in is not a strict asymptotic plateau: at very large \(s\) it grows as \(s^3\). Its ``flat-top'' or solitonic character instead refers to the existence of a self-bound, thin-wall branch in which the field sits near a nonzero interior value and returns to the vacuum across a narrow surface layer.
In this thin-wall interpretation the interior energy scale plays a role reminiscent of a bag constant or vacuum-energy offset. 

 In theories where the scalar truly represents the order parameter of an unbroken confining vacuum (as in pure $SU(3)$ glueballs), this constant must be matched to the Yang--Mills trace anomaly \citep{Schechter1980TraceAnomaly,MottolaVaulin2006},
\begin{equation}
	\langle T^\mu_{\ \mu} \rangle \sim \Lambda_{\rm QCD}^4,
\end{equation}
So, if one tried to identify the offset with the vacuum energy of an unbroken pure Yang--Mills glueball EFT, it would be constrained by the trace anomaly. 
This severely constrains $\sigma_0$ and undermines the freedom needed to reach SMBH scales with heavy constituents. This topic will be discussed again for strict plateau families and contrasted with the bag-free GC9 density polynomial.

These solitonic families remain useful as comparison benchmarks in both structure and waveform studies, even when they are not viewed as realistic UV completions \citep{JohnsonMcDanielEtAl2018,PacilioEtAl2021,Sennett2017,Vincent2016,Olivares2020}.

\subsection{Cosh--Gordon potential}
\label{subsec:cosh_gordon}

A literature-supported example of a smooth, non-polynomial and
everywhere repulsive interaction is the Cosh--Gordon potential studied
for complex boson stars in Ref.~\citep{SchunckTorres2000}. In a
normalization in which \(m\) is the physical small-field mass, it may be
written as
\begin{equation}
	V_{\rm CG}(s)
	=
	m^2f^2
	\left[
	\cosh\!\left(\frac{\sqrt{2s}}{f}\right)-1
	\right],
	\qquad
	s\equiv|\Phi|^2,
	\label{eq:cosh_gordon_potential}
\end{equation}
where \(f\) is a field-amplitude scale. This is equivalent, up to
parameter conventions, to the form
\begin{equation}
	U_{\rm cosh}
	=
	\alpha m^2
	\left[
	\cosh\!\left(\beta\sqrt{|\Phi|^2}\right)-1
	\right]
\end{equation}
used in Ref.~\citep{SchunckTorres2000}.

The quadratic mass term is already contained in
Eq.~\eqref{eq:cosh_gordon_potential}; adding a separate term \(m^2s\)
would therefore double count the scalar mass. Indeed, its expansion
about the vacuum is
\begin{equation}
	\begin{split}
		V_{\rm CG}(s)
		={}&
		m^2s
		+\frac{m^2}{6f^2}s^2
		+\frac{m^2}{90f^4}s^3
		+\frac{m^2}{2520f^6}s^4
		+\mathcal O(s^5).
	\end{split}
	\label{eq:cosh_gordon_expansion}
\end{equation}
All nonlinear coefficients are positive. The model consequently
describes an infinite tower of repulsive even-field interactions, with
the leading nonlinear contribution corresponding to an effective
quartic coupling
\begin{equation}
	\lambda_{\rm eff}
	=
	\frac{m^2}{6f^2}
\end{equation}
in the convention
\(V=m^2s+\lambda_{\rm eff}s^2+\cdots\). It does not contain the
attractive term normally required for flat-space Q-ball or
non-topological-soliton solutions. This is consistent with the
classification given in Ref.~\citep{SchunckTorres2000}.

\paragraph{Thomas--Fermi fluid properties.}

The Cosh--Gordon interaction is convex and produces positive pressure
under the same Thomas--Fermi map used throughout this work. Defining
\begin{equation}
	x\equiv\frac{\sqrt{2s}}{f},
	\qquad
	s=\frac{f^2x^2}{2},
\end{equation}
and using
\begin{equation}
	\rho=sV_{,s}+V,
	\qquad
	P=sV_{,s}-V,
\end{equation}
one obtains
\begin{align}
	\rho_{\rm CG}
	&=
	m^2f^2
	\left[
	\frac{x}{2}\sinh x+\cosh x-1
	\right],
	\label{eq:cosh_rho}
	\\
	P_{\rm CG}
	&=
	m^2f^2
	\left[
	\frac{x}{2}\sinh x-\cosh x+1
	\right].
	\label{eq:cosh_pressure}
\end{align}
The pressure is positive for every \(x>0\). To see this, define
\begin{equation}
	F(x)=\frac{x}{2}\sinh x-\cosh x+1.
\end{equation}
Since
\begin{equation}
\begin{aligned}
	F(0)&=0,\\
	F'(x)&=\frac{1}{2}\left(x\cosh x-\sinh x\right)>0
	\qquad (x>0).
\end{aligned}
\end{equation}
one has
\begin{equation}
	P_{\rm CG}>0
	\qquad
	(s>0).
\end{equation}
The low-density expansion is
\begin{align}
	P_{\rm CG}
	&=
	\frac{m^2}{6f^2}s^2
	+\frac{m^2}{45f^4}s^3
	+\mathcal O(s^4),
	\\
	\rho_{\rm CG}
	&=
	2m^2s
	+\frac{m^2}{2f^2}s^2
	+\frac{2m^2}{45f^4}s^3
	+\mathcal O(s^4).
\end{align}

The corresponding adiabatic sound speed is
\begin{equation}
	c_{s,{\rm CG}}^2
	=
	\frac{dP_{\rm CG}}{d\rho_{\rm CG}}
	=
	\frac{x\cosh x-\sinh x}
	{x\cosh x+3\sinh x}.
	\label{eq:cosh_sound_speed}
\end{equation}
It satisfies
\begin{equation}
	0<c_{s,{\rm CG}}^2<1
	\qquad
	\text{for finite }x>0,
\end{equation}
with limiting behavior
\begin{align}
	c_{s,{\rm CG}}^2
	&=
	\frac{x^2}{12}
	+\mathcal O(x^4),
	&& x\ll1,
	\\
	c_{s,{\rm CG}}^2
	&\longrightarrow1,
	&& x\gg1.
\end{align}
The Cosh--Gordon fluid is therefore mechanically stable and causal, but
its high-density limit approaches the maximally stiff relation
\begin{equation}
	P_{\rm CG}\simeq\rho_{\rm CG}.
\end{equation}
More explicitly,
\begin{align}
	\rho_{\rm CG}
	&\simeq
	\frac{m^2f^2}{4}(x+2)e^x,
	\\
	P_{\rm CG}
	&\simeq
	\frac{m^2f^2}{4}(x-2)e^x,
\end{align}
and hence
\begin{equation}
	\frac{P_{\rm CG}}{\rho_{\rm CG}}
	\simeq
	\frac{x-2}{x+2}
	\longrightarrow1.
\end{equation}
Unlike a bounded interaction plateau, the Cosh--Gordon potential grows
exponentially and supplies increasingly strong positive pressure.

\paragraph{Characteristic mass scale.}

The Thomas--Fermi energy-density scale of the model is
\begin{equation}
	\rho_{\rm CG,0}=m^2f^2.
\end{equation}
It is useful to introduce the collective scale
\begin{equation}
	\Lambda_{\rm CG}
	\equiv
	\sqrt{mf},
	\qquad
	\rho_{\rm CG,0}=\Lambda_{\rm CG}^4.
\end{equation}
Dimensional rescaling of the TOV equations then gives
\begin{align}
	M_{\max}^{\rm CG}
	&=
	\mathcal C_{\rm CG}
	\frac{M_{\rm Pl}^3}{mf}
	=
	\mathcal C_{\rm CG}
	\frac{M_{\rm Pl}^3}{\Lambda_{\rm CG}^2},
	\label{eq:cosh_mass_scaling}
	\\
	R_{\max}^{\rm CG}
	&=
	\mathcal D_{\rm CG}
	\frac{M_{\rm Pl}}{mf}
	=
	\mathcal D_{\rm CG}
	\frac{M_{\rm Pl}}{\Lambda_{\rm CG}^2},
	\label{eq:cosh_radius_scaling}
\end{align}
where the dimensionless coefficients
\(\mathcal C_{\rm CG}\) and \(\mathcal D_{\rm CG}\) must be determined
from the equilibrium sequence and depend on whether the full
Einstein--Klein--Gordon system or its Thomas--Fermi limit is used.

Reaching the TON--618 benchmark would therefore require
\begin{equation}
	\Lambda_{\rm CG}
	\lesssim
	4.97\,{\rm keV}\,
	\sqrt{\mathcal C_{\rm CG}}
	\left(
	\frac{6.6\times10^{10}M_\odot}{M_{\rm obs}}
	\right)^{1/2},
	\label{eq:cosh_collective_scale}
\end{equation}
or, equivalently,
\begin{equation}
	\begin{split}
		f
		\lesssim{}&
		2.47\times10^{-20}\,{\rm GeV}\,
		\mathcal C_{\rm CG}
		\left(\frac{10^9\,{\rm GeV}}{m}\right)
		\\
		&\times
		\left(
		\frac{6.6\times10^{10}M_\odot}{M_{\rm obs}}
		\right).
	\end{split}
	\label{eq:cosh_f_bound}
\end{equation}
For \(m\simeq10^9\)--\(10^{14}\,{\rm GeV}\), this corresponds
parametrically to
\begin{equation}
	f\lesssim
	10^{-20}\text{--}10^{-25}\,{\rm GeV}
\end{equation}
up to the sequence-dependent coefficient
\(\mathcal C_{\rm CG}\).

The appearance of a keV collective scale is not, by itself, a reason to
discard the model. It is the same dimensional trade that will appear in
GC9 potential, where the supermassive scale is controlled by \(\Lambda_T\) rather
than directly by the ultraheavy constituent mass. Correspondingly, the
small value of \(f\) reflects the relation
\begin{equation}
	f=\frac{\Lambda_{\rm CG}^2}{m},
\end{equation}
which is analogous to the small condensate-amplitude scale appearing in
the scale-separated GC9 construction.

\paragraph{Why it is not selected for the \(G(2)\) construction.}

The Cosh--Gordon potential is a mathematically consistent boson-star
interaction. It is convex, produces positive pressure, has
\(V_{\rm CG}(0)=0\), does not require an explicit bag constant and
remains causal. It is therefore not excluded by a basic stability or
surface argument. Nevertheless, it is not adopted as the final
\(G(2)\)-glueball closure for the following model-specific reasons.

First, its operator content is not connected to the exceptional-group
invariant structure motivating GC9. Equation
\eqref{eq:cosh_gordon_expansion} begins with a quartic interaction
\(s^2\) and contains every higher power \(s^n\), with coefficients fixed
by the factorial expansion of the hyperbolic cosine. By contrast, the
GC9 ansatz will be designed to isolate a positive degree-six density
cumulant,
\begin{equation}
	U_{\rm GC9}(s)\propto s^9,
\end{equation}
motivated phenomenologically by the primitive degree-six invariant of
\(G(2)\). Adopting the Cosh--Gordon form would reintroduce unsuppressed
quartic, nonic and all higher interactions without a known
group-theoretic or lattice derivation for their relative coefficients.

Second, the Cosh--Gordon form is an entire analytic function. Its
factorial tower is a convenient phenomenological resummation, but it
does not encode thresholds, connected-cumulant selection or other
nontrivial analytic structures expected from a strongly coupled
composite glueball sector. 

Third, the high-density equation of state approaches the causal
boundary,
\begin{equation}
	c_{s,{\rm CG}}^2\longrightarrow1.
\end{equation}
Although this limit is not formally acausal, it leaves no asymptotic
margin below the causal bound and makes the compact branch depend on an
extremely stiff exponential regime in which infinitely many operators
are simultaneously important. 

Fourth, the existing full Einstein--Klein--Gordon study found only a
modest enhancement of the maximum mass for its representative
Cosh--Gordon parameters. For the benchmark considered in
Ref.~\citep{SchunckTorres2000},
\begin{equation}
	M_{\max}^{\rm CG}
	\simeq
	0.638\,\frac{M_{\rm Pl}^2}{m},
\end{equation}
approximately \(0.8\%\) above the corresponding free-field value. The
first equilibrium branch was stable, while the subsequent branches
were unstable. Thus, the Cosh--Gordon potential is known to support
ordinary boson stars, but the existing calculation does not establish
a stable ultracompact or Kerr-mimicking branch capable of spanning the
supermassive sector for ultraheavy constituents.

Finally, making the interaction sufficiently strong through
\(f\ll m\) produces an enormous low-density quartic coefficient,
\begin{equation}
	\lambda_{\rm eff}
	=
	\frac{m^2}{6f^2}.
\end{equation}
Such a regime is not inconsistent for a composite effective field, but
it requires a nonperturbative microscopic explanation for the complete
infinite tower of correlated Wilson coefficients. 

We therefore retain the Cosh--Gordon family as a useful
literature-backed benchmark demonstrating that non-polynomial convex
potentials can generate stable positive-pressure boson stars. It is not
selected as the final model because its leading quartic interaction,
fixed all-orders exponential tower, asymptotically maximally stiff
equation of state and absence of a specific \(G(2)\) matching argument
do not satisfy the restricted microscopic and phenomenological design
criteria imposed on GC9.

\subsection{Non-polynomial logarithmic and bounded interaction families}

Non-polynomial scalar potentials have been studied as phenomenological
resummations of higher-order self-interactions and as possible effective
descriptions of composite or strongly coupled bosonic matter. Their
physical implications depend crucially on their convexity: a potential
whose interaction energy saturates to a constant does not necessarily
generate a positive saturating pressure.

The $G(2)$ construction motivates considering non-polynomial effective
interactions for composite dark glueball matter
\citep{Masi2024G2Resurgence}, but it does not fix their functional form.
Two non-polynomial potentials that have been studied directly in the
boson-star literature are the modified Liouville and logarithmic
potentials \citep{SchunckTorres2000,ChoiHeSchiappacasse2019},
\begin{align}
	V_{\rm L}(s)
	&=
	m^2 f^2
	\left[
	\exp\!\left(\frac{s}{f^2}\right)-1
	\right],
	\label{eq:Liouville_potential_correct}
	\\
	V_{\log}(s)
	&=
	m^2 f^2
	\ln\!\left(1+\frac{s}{f^2}\right),
	\label{eq:log_potential_correct}
\end{align}
where $s=|\Phi|^2$. In both cases the coefficient of the leading
small-field term is $m^2$, so that $m$ is the scalar pole mass.

\paragraph{Liouville exponential potential.}

The Liouville form has the expansion
\begin{equation}
	V_{\rm L}(s)
	=
	m^2s
	+\frac{m^2}{2f^2}s^2
	+\frac{m^2}{6f^4}s^3
	+\mathcal O(s^4),
\end{equation}
and therefore generates an infinite series of positive self-interaction
coefficients. It is a convex, exponentially growing interaction rather
than a saturating plateau.

Defining
\begin{equation}
	x\equiv\frac{s}{f^2},
\end{equation}
the Thomas--Fermi map
\begin{equation}
	\rho=sV_{,s}+V,
	\qquad
	P=sV_{,s}-V
\end{equation}
gives
\begin{align}
	\rho_{\rm L}
	&=
	m^2f^2
	\left[
	(x+1)e^x-1
	\right],
	\\
	P_{\rm L}
	&=
	m^2f^2
	\left[
	(x-1)e^x+1
	\right].
\end{align}
For $x>0$ one has $P_{\rm L}>0$. The corresponding sound speed is
\begin{equation}
	c_{s,{\rm L}}^2
	=
	\frac{dP_{\rm L}}{d\rho_{\rm L}}
	=
	\frac{x}{x+2},
\end{equation}
so that
\begin{equation}
	c_{s,{\rm L}}^2\longrightarrow1
	\qquad (x\longrightarrow\infty).
\end{equation}
The Liouville interaction is therefore genuinely repulsive in the
Thomas--Fermi sense, but its asymptotic fluid approaches the causal
limit. Its compact-star properties must be obtained from either the full
Einstein--Klein--Gordon system or a controlled Thomas--Fermi reduction;
there is no universal plateau-controlled maximum-mass formula.

The Liouville potential of Eq.~\eqref{eq:Liouville_potential_correct} is
genuinely convex and generates a causal, monotonically increasing
Thomas--Fermi pressure, so it cannot be excluded on the same grounds as
the concave families above. We nonetheless prefer the GC9 closure, for
three reasons.

\begin{itemize}
	\item \emph{Causal margin.} The Liouville sound speed saturates the
	causal bound itself, $c_{s,{\rm L}}^2\to1$ as $x\to\infty$, with no
	margin left before superluminal propagation. The GC9 branch instead
	approaches $c_s^2\to4/5$, a deliberately conservative compromise
	between compactness and causality.
	
	\item \emph{Closed form versus truncated series.} $V_{\rm L}(s)$ is an
	infinite, non-terminating series in $s$ with no finite resummation
	beyond the exponential itself; practical implementations in the
	literature accordingly rely on low-order truncation (up to $s^3$ in
	\citep{ChoiHeSchiappacasse2019}). The GC9 nonlinear term
	$\Lambda_T^4(s/s_T)^9$ is instead a single closed term, which is what
	allows the TOV--Hinderer--Hartle--Thorne pipeline used throughout this
	work to remain exact rather than series-truncated.
	
	\item \emph{Microscopic motivation.} The Liouville form in
	\citep{SchunckTorres2000,ChoiHeSchiappacasse2019} is introduced as a
	generic effective resummation -- in the latter case motivated by
	Planck-scale quantum-gravity corrections -- without a specific tie to
	the $G(2)$ confining sector. The GC9 exponent is instead tied to the
	Casimir-six third cumulant of the dark glueball condensate, i.e.\
	to a specific gauge-invariant operator of the microscopic theory
	rather than a generic phenomenological choice.
\end{itemize}

At a purely phenomenological level, the compactness reached by the
Liouville branch in the regime explored in
\citep{ChoiHeSchiappacasse2019} is $C_{\max}^{\rm L}\simeq0.176$ at
their strongest quoted coupling ($f\simeq0.1\,M_{\rm Pl}$), reaching up
to $C_{\max}\sim0.18$ in the strong-coupling limit reported there --
well below the $C_{\max}=0.337118$ obtained here for the post-turning GC9 continuation (Table~\ref{tab:GC9_first_order}).
Since the Liouville causal limit ($c_s^2\to1$) is nominally stiffer
than that of GC9 ($c_s^2\to4/5$), this comparison should not be read as
a proof that the Liouville branch cannot reach comparable compactness
at more extreme coupling than explored so far; it indicates only that,
in the regime studied in the literature to date, it does not.

\paragraph{Logarithmic potential.}

The logarithmic potential has the alternating expansion
\begin{equation}
	V_{\log}(s)
	=
	m^2s
	-\frac{m^2}{2f^2}s^2
	+\frac{m^2}{3f^4}s^3
	-\frac{m^2}{4f^6}s^4
	+\cdots.
\end{equation}
Its leading nonlinear term is therefore attractive, in contrast with
the Liouville model.

Under the same Thomas--Fermi map,
\begin{align}
	\rho_{\log}
	&=
	m^2f^2
	\left[
	\frac{x}{1+x}+\ln(1+x)
	\right],
	\\
	P_{\log}
	&=
	m^2f^2
	\left[
	\frac{x}{1+x}-\ln(1+x)
	\right].
\end{align}
Since
\begin{equation}
	\ln(1+x)>\frac{x}{1+x}
	\qquad (x>0),
\end{equation}
one obtains
\begin{equation}
	P_{\log}<0.
\end{equation}
Indeed,
\begin{equation}
	c_{s,\log}^2
	=
	\frac{dP_{\log}}{d\rho_{\log}}
	=
	-\frac{x}{x+2}<0.
\end{equation}
Thus the logarithmic model studied in the full boson-star literature
does not define a stable repulsive Thomas--Fermi fluid closure by itself.
Localized solutions found from the full Einstein--Klein--Gordon system
retain gradient and harmonic-field contributions that are discarded in
the algebraic Thomas--Fermi map. They should not be interpreted as
positive-pressure plateau stars.

\paragraph{Logarithmic nonlinearities in Gross--Pitaevskii dark matter.}

A distinct use of logarithmic interactions has recently been studied
by \citet{HaghaniHarko2026} in the context of Newtonian
Bose--Einstein-condensate dark matter. In that construction the
logarithm enters the Gross--Pitaevskii equation as a nonlinear
wave-mechanical term. After a Madelung transformation and a
Thomas--Fermi reduction, the resulting hydrodynamic equation of state
has an isothermal or ideal-gas-like form,
\begin{equation}
	P_{\log{\rm BEC}}\propto\rho.
	\label{eq:log_BEC_linear_EOS}
\end{equation}
The model was developed for galactic density profiles and rotation
curves rather than for relativistic compact boson stars.

This result does not contradict the relativistic treatment adopted
here, Eq.~\eqref{eq:log_relativistic_potential}. The
Gross--Pitaevskii logarithmic nonlinearity, its nonrelativistic
Hamiltonian density and the fluid variables obtained after the
Madelung transformation are not identical to the relativistic
potential
\begin{equation}
	U_{\log}(s)
	=
	\Lambda^4
	\ln\!\left(1+\frac{s}{\sigma_0^2}\right)
	\label{eq:log_relativistic_potential}
\end{equation}
inserted into the algebraic relation
\(P=sU_{,s}-U\). Logarithmic models must therefore be compared at the
level of their complete action and reduction procedure, rather than
being grouped together solely because they contain a logarithm.

The recent Gross--Pitaevskii construction provides a useful example of
a viable logarithmic nonlinearity in a different physical regime, but
it does not supply positive-pressure support for the relativistic
concave interaction considered here.

\paragraph{Hyperbolic and exponential plateaus in real scalar stars.}

Hyperbolic and bounded exponential shapes have appeared in the
literature on real, coherently oscillating scalar stars. In particular,
\citet{MuiaEtAl2019DenseScalarStars} studied full general-relativistic
evolutions for the \(\alpha\)-attractor T- and E-model potentials
\begin{align}
	V_T(\varphi)
	&=
	V_0
	\tanh^{2n}\!\left(\frac{\varphi}{\Lambda}\right),
	\label{eq:alpha_T_model}
	\\
	V_E(\varphi)
	&=
	V_0
	\left[
	1-\exp\!\left(-\frac{\varphi}{\Lambda}\right)
	\right]^2.
	\label{eq:alpha_E_model}
\end{align}
Here \(\varphi\) is a real scalar field. For the simplest T-model,
\(n=1\), the small-field expansion is
\begin{equation}
	V_T(\varphi)
	=
	V_0
	\left[
	x^2-\frac{2}{3}x^4
	+\frac{17}{45}x^6
	+\mathcal O(x^8)
	\right],
	\qquad
	x\equiv\frac{\varphi}{\Lambda},
	\label{eq:T_model_expansion}
\end{equation}
whereas the E-model gives
\begin{equation}
	V_E(\varphi)
	=
	V_0
	\left[
	x^2-x^3+\frac{7}{12}x^4
	+\mathcal O(x^5)
	\right].
	\label{eq:E_model_expansion}
\end{equation}

These examples establish published precedents for \(\tanh\)-type and
bounded exponential shapes in self-gravitating scalar configurations.
They are not, however, direct precedents for
\begin{align}
	U_{\tanh}(s)
	&=
	\Sigma^4\tanh\!\left(\frac{s}{\sigma_0^2}\right),\\
	U_{\rm bexp}(s)
	&=
	\Sigma^4\left(1-e^{-s/\sigma_0^2}\right).
\end{align}
The T- and E-model objects are real scalar oscillons or oscillatons,
not stationary \(U(1)\)-charged complex boson stars. Moreover,
\(V_T\) contains an even power of \(\tanh\), while \(V_E\) is
asymmetric about its minimum and contains a cubic term. Their
equilibrium and stability therefore follow from the time-dependent
Einstein--Klein--Gordon problem and cannot be inferred from the
complex-field Thomas--Fermi map \(P=sV_{,s}-V\).

Consequently, Eqs.~\eqref{eq:alpha_T_model} and
\eqref{eq:alpha_E_model} can be cited as contextual precedents for
plateau-shaped scalar potentials, but not as sources for the particular
bounded functions of \(s=|\Phi|^2\) introduced above.

\paragraph{Effective saturation after nonrelativistic averaging.}

A more direct published example of effective saturating repulsion was
studied by \citet{GalazoGarciaBraxValageas2025}. Their relativistic
real-scalar potential is
\begin{equation}
	V(\varphi)
	=
	\frac{1}{2}m^2\varphi^2+V_I(\varphi),
	\label{eq:Galazo_total_potential}
\end{equation}
with
\begin{equation}
	V_I(\varphi)
	=
	M_I^4
	\left[
	\cos\!\left(\frac{\varphi}{\Lambda}\right)
	-1
	+\frac{\varphi^2}{2\Lambda^2}
	\right].
	\label{eq:Galazo_interaction}
\end{equation}
The subtraction of the quadratic contribution makes the interaction
quartic at small amplitude:
\begin{equation}
	V_I(\varphi)
	=
	M_I^4
	\left[
	\frac{\varphi^4}{24\Lambda^4}
	-\frac{\varphi^6}{720\Lambda^6}
	+\mathcal O(\varphi^8)
	\right].
	\label{eq:Galazo_small_field}
\end{equation}
Thus, its leading nonlinear term is repulsive, while the alternating
higher-order series weakens the interaction relative to a pure
quartic law.

The saturation discussed in
Ref.~\citep{GalazoGarciaBraxValageas2025} does not correspond to a
positive constant plateau in the original relativistic potential.
It appears after averaging the rapidly oscillating real scalar in the
nonrelativistic regime. In the notation of that work, the resulting
density-dependent interaction contribution can be written as
\begin{equation}
	\Phi_I^{\rm NR}(\rho)
	=
	\frac{8\rho_b}{\rho_a}
	\left[
	1-
	\frac{
		2J_1\!\left(\sqrt{\rho/\rho_b}\right)
	}{
		\sqrt{\rho/\rho_b}
	}
	\right],
	\label{eq:Galazo_averaged_interaction}
\end{equation}
where \(J_1\) is a Bessel function and
\begin{equation}
	\rho_a
	=
	\frac{8m^4\Lambda^4}{M_I^4},
	\qquad
	\rho_b
	=
	\frac{m^2\Lambda^2}{2}.
	\label{eq:Galazo_density_scales}
\end{equation}
Its limiting behavior is
\begin{align}
	\Phi_I^{\rm NR}(\rho)
	&\simeq
	\frac{\rho}{\rho_a},
	&&
	\rho\ll\rho_b,
	\label{eq:Galazo_low_density}
	\\
	\Phi_I^{\rm NR}(\rho)
	&\longrightarrow
	\frac{8\rho_b}{\rho_a},
	&&
	\rho\gg\rho_b.
	\label{eq:Galazo_high_density}
\end{align}
The corresponding hydrodynamic pressure is obtained from
\begin{equation}
	P_I(\rho)
	=
	\int_0^\rho
	d\tilde\rho\,
	\tilde\rho\,
	\frac{d\Phi_I^{\rm NR}}{d\tilde\rho}.
	\label{eq:Galazo_pressure_relation}
\end{equation}

This construction is a genuine example of an interaction that is
repulsive and quartic at low density but whose nonrelativistic
effective force saturates at high density. It is therefore a more
appropriate literature precedent for the phrase ``saturating
repulsion'' than a bounded positive function inserted directly into a
relativistic complex-field potential.

Nevertheless, the construction differs from GC9 in several essential
respects. It concerns a real oscillating scalar and
Schr\"odinger--Poisson solitons, its saturation emerges only after
time averaging, and wave pressure becomes important outside the
semiclassical regime. The relativistic relation
\begin{equation}
	P=sV_{,s}-V
\end{equation}
must not be applied directly to
\(\Phi_I^{\rm NR}(\rho)\), because the latter is already an averaged
density-dependent quantity rather than the original relativistic
potential.

The model is consequently valuable as evidence that controlled
effective saturation mechanisms exist, but it is not a stationary
complex-boson-star alternative to the GC9 closure and does not provide
the required ultraheavy-\(G(2)\) matching.

\paragraph{Bounded phenomenological ans\"atze.}

For comparison, one may introduce the bounded interactions
\begin{align}
	V_{\tanh}(s)
	&=
	m_0^2s+
	\Sigma^4\tanh\!\left(\frac{s}{\sigma_0^2}\right),
	\label{eq:tanh_phenomenological}
	\\
	V_{\rm bexp}(s)
	&=
	m_0^2s+
	\Sigma^4
	\left[
	1-\exp\!\left(-\frac{s}{\sigma_0^2}\right)
	\right].
	\label{eq:bexp_phenomenological}
\end{align}
Equations~\eqref{eq:tanh_phenomenological} and
\eqref{eq:bexp_phenomenological} are used here only as illustrative
phenomenological ans\"atze; they are not the Liouville exponential
potential of Refs.~\citep{SchunckTorres2000,ChoiHeSchiappacasse2019}.

The physical quadratic coefficient of both models is
\begin{equation}
	m_{\rm pole}^2
	=
	m_0^2+\frac{\Sigma^4}{\sigma_0^2},
\end{equation}
so $m_0$ should not be identified directly with the constituent mass
unless the linear contribution from the bounded interaction has already
been absorbed into its definition.

Writing $x=s/\sigma_0^2$, their interaction pressures are
\begin{align}
	P_{\tanh}^{\rm int}
	&=
	\Sigma^4
	\left[
	x\,\sech^2x-\tanh x
	\right],
	\\
	P_{\rm bexp}^{\rm int}
	&=
	\Sigma^4
	\left[
	(1+x)e^{-x}-1
	\right].
\end{align}
Both are strictly negative for $x>0$. At high density,
\begin{equation}
	\rho_{\rm int}\longrightarrow\Sigma^4,
	\qquad
	P_{\rm int}\longrightarrow-\Sigma^4.
\end{equation}
The bounded interaction plateau therefore behaves as a positive
vacuum-energy contribution with $P=-\rho$, not as a positive repulsive
pressure.

Consequently, the bounded potentials
\eqref{eq:tanh_phenomenological} and
\eqref{eq:bexp_phenomenological} do not possess a positive-pressure
plateau-dominated Thomas--Fermi branch. In particular, one cannot infer
from these potentials the scalings
\begin{equation}
	M_{\max}\sim\frac{M_{\rm Pl}^3}{\Sigma^2},
	\qquad
	R\sim\frac{M_{\rm Pl}}{\Sigma^2},
\end{equation}
or use them to derive a TON--618 bound on $\Sigma$. Such scalings apply
to genuinely self-bound equations of state containing both a density
scale and an independent positive-pressure sector, not to a constant
positive interaction plateau alone.

\subsection{Axion / cosine potential}
Axion-star literature provides the main reference class for bounded periodic potentials and their dilute-versus-dense branches \citep{BarrancoBernal2011,Visinelli2018Axion,Visinelli2021Review,Eby2015Axion,Eby2016Axion,Helfer2017Axion,LevkovPaninTkachev2018,ChenSunHuangWang2024}.
Another deeply studied potential in literature is the canonical axion potential:

\begin{equation}
	V(\Phi)=m_a^2 f_a^2
	\left[1-\cos\!\left(\frac{\sqrt{2}|\Phi|}{f_a}\right)\right].
\end{equation}
where $f_a$ is the axion decay constant. This is mainly exploited for ultralight candidates. At large field amplitude the potential remains bounded and oscillatory and therefore does not generate a monotonic saturation plateau. Expanding for small field amplitudes gives
\begin{equation}
	V \simeq
	m_a^2|\Phi|^2
	-\frac{m_a^2}{6f_a^2}|\Phi|^4
	+\frac{m_a^2}{90f_a^4}|\Phi|^6+\cdots
\end{equation}
and the dilute axion-star branch has the approximate maximum-mass scaling
\begin{equation}
	M_{\max}^{\rm axion}
	\sim
	\frac{f_a M_{\rm Pl}}{m_a}.
\end{equation}
up to convention-dependent numerical coefficients. This should not be confused with a universal maximum mass of the full cosine potential: relativistic axion boson stars with the complete periodic potential exhibit branch-dependent behavior and are generally treated numerically.
For heavy particle masses this leads to extremely small maximum masses for the boson star, incompatible with the supermassive regime we want to reach.
Furthermore, axion stars typically describe weakly interacting, dilute configurations rather than a dense repulsive fluid. These features render axion potentials unsuitable for modeling strongly interacting glueballs.

\paragraph{Full chiral QCD axion potential.}

The simple cosine form is the standard analytical benchmark for a
generic axion-like particle, but the leading-order QCD axion potential
has a more structured non-polynomial form. Writing
\begin{equation}
	z\equiv\frac{m_u}{m_d},
\end{equation}
the zero-temperature chiral potential can be expressed as
\begin{equation}
	V_\chi(a)
	=
	m_\pi^2f_\pi^2
	\left[
	1-
	\sqrt{
		1-
		\frac{4z}{(1+z)^2}
		\sin^2\!\left(\frac{a}{2f_a}\right)
	}
	\right]
	\label{eq:full_chiral_axion_potential}
\end{equation}
\citep{GrillidiCortonaEtAl2016}. Using
\begin{equation}
	m_a^2f_a^2
	=
	m_\pi^2f_\pi^2
	\frac{z}{(1+z)^2},
\end{equation}
an equivalent representation is
\begin{equation}
	V_\chi(a)
	=
	m_a^2f_a^2
	\frac{1+z}{z}
	\left[
	1+z-
	\sqrt{
		1+z^2+2z\cos(a/f_a)
	}
	\right].
	\label{eq:full_chiral_axion_equivalent}
\end{equation}

Near the vacuum,
\begin{equation}
	V_\chi(a)
	=
	\frac{1}{2}m_a^2a^2
	-
	\frac{m_a^2}{4!f_a^2}
	\frac{1-z+z^2}{(1+z)^2}
	a^4
	+
	\mathcal O(a^6).
	\label{eq:chiral_axion_expansion}
\end{equation}
The leading quartic interaction is attractive, as in the simple cosine
model, but its coefficient is modified by the light-quark mass ratio.

The square-root structure makes the full chiral potential an
instructive example of a non-polynomial interaction that is genuinely
derived from strongly coupled gauge dynamics. Dense configurations can
probe higher harmonics and departures from the dilute cosine
approximation \citep{BraatenMohapatraZhang2016,
	Guerra2019AxionBosonStars}. Nevertheless, the interaction is periodic
and predominantly attractive near the vacuum. It does not generate a
monotonic positive-pressure repulsive closure of the GC9 type, and its
dense branches must be obtained from the full field equations rather
than from a single-power Thomas--Fermi EOS.

The chiral axion potential is therefore relevant as a microscopic
counterexample to simple polynomial truncations, but it is unsuitable
for the present ultraheavy-glueball construction because it relies on
a pseudo-Nambu--Goldstone periodicity and an attractive dilute
interaction that have no known analogue in the \(G(2)\) glueball
sector.

\subsection{Quartic--nonic $Q$-ball potential}
Quartic--nonic and related $Q$-ball models remain the canonical examples of non-topological solitons stabilized by a global charge \citep{Coleman1985Qballs,LeePang1992,Kusenko1997Qball,KusenkoShaposhnikov1998,MultamakiVilja2000,TsumagariCopelandSaffin2008}:
\begin{equation}
	V(|\Phi|)=
	\mu^2 |\Phi|^2
	-\lambda |\Phi|^4
	+\beta |\Phi|^6 ,
	\qquad
	\lambda>0,\quad \beta>0.
\end{equation}
The flat-space $Q$-ball existence condition is obtained from
\begin{equation}
	\frac{V(\phi)}{\phi^2}=\mu^2-\lambda \phi^2+\beta \phi^4.
\end{equation}
To build up a Q–ball, an attractive quartic self–interaction potential is needed, opposite
to the standard quartic boson star case. In the present sign convention, the quartic interaction is attractive because it enters as \(-\lambda |\Phi|^4\), with \(\lambda>0\), while the nonic term with \(\beta>0\) stabilizes the potential at large field amplitude.
The minimum occurs at $\phi^2=\frac{\lambda}{2\beta}$ and the minimum frequency is $\omega_{\min}^2=\mu^2-\frac{\lambda^2}{4\beta}$. Thus localized solutions exist for $\omega_{\min}<\omega<\mu$,
which requires $\lambda^2 < 4\beta\mu^2$. 

For the present \(G(2)\)-glueball framework, the quartic--nonic Q-ball potential is disfavoured for two related reasons. First, it is structurally tied to a fundamental global \(U(1)\) charge, whereas in our setup the complex scalar and its associated number should be understood only as an \emph{effective infrared description} of the composite glueball ensemble, which is sufficient for a stationary rotating boson star treatment, not as a microscopic symmetry of the underlying confining theory. Second, even in the strong self-interaction regime the maximum BS mass remains parametrically of the form \cite{LeePang1992}
\begin{equation}
M_{\max}^{Q{\rm -ball\ BS}}
\sim
\mathcal{C}_{Q}\,
\frac{M_{\rm Pl}^4}{\mu^3}\
\end{equation}
so the heavy-mass suppression \(M_{\max}\propto \mu^{-2}\) is not truly removed. In this sense, the Q-ball construction does not naturally solve the ultra-heavy glueball problem addressed here. Moreover, the large-mass Q-ball regime is typically the \emph{thin-wall regime}, in which the scalar field is nearly constant in the interior and returns to the vacuum only across a narrow boundary layer, so that the configuration behaves more like a charge-supported droplet than a dense fluid star. For the present glueball scenario, the bag-free density-polynomial GC9 closure is structurally closer to a fluid star than a thin-wall Q-ball ansatz. This is a model-selection criterion, not a derivation of its hierarchy.

\subsection{Compact signum-Gordon potential}
V-shaped and signum-Gordon interactions are interesting precisely because they produce compact-support objects rather than exponentially localized stars \citep{ArodzLis2008,ArodzLis2009,ArodzLis2010,HartmannKleihausKunzSchaffer2012}:

\begin{equation}
	V(\Phi)=\lambda_{SG} |\Phi|
\end{equation}
which exhibit sharply bounded configurations and large compactness.
A massive variant is
\begin{equation}
	V(|\Phi|)=m^2 |\Phi|^2+\lambda_{SG} |\Phi|.
\end{equation}
A distinctive feature of these models is that the scalar field has compact support $\phi(r)=0$ for $r>R_*$, hence these objects are compactons rather than exponentially decaying stars. 
In fact, the potential is non-analytic at $\Phi=0$ and does not arise from a standard effective field theory description of composite degrees of freedom.  The corresponding EOS is effectively incompressible and lacks a physically motivated
saturation mechanism.  In addition, the absence of a smooth high-density plateau prevents a controlled description of repulsive interactions in a strongly coupled glueball fluid.
Dimensional arguments suggest a characteristic maximum-mass scaling
\begin{equation}
	M_{\max}^{\rm signum}
	\sim
	\mathcal{C}_{SG}
	\frac{M_{\rm Pl}^2}{\lambda_{SG}^{1/3}}.\\
\end{equation}

\subsection{Coleman--Weinberg potentials and radiative symmetry breaking}
\label{subsec:Coleman_Weinberg}

Coleman--Weinberg interactions provide a qualitatively distinct
non-polynomial family in which logarithms arise from radiative
corrections rather than from an assumed high-density saturation
mechanism \citep{ColemanWeinberg1973,GildenerWeinberg1976}. Their
characteristic large-field behavior is
\begin{equation}
	V_{\rm CW}(\phi)
	\sim
	\phi^4
	\ln\!\left(\frac{\phi^2}{\mu^2}\right).
\end{equation}

A direct application to complex boson stars was recently considered by
\citep{EidizadeGhaffarnejad2025}. In the convention used in that work,
the total scalar potential consists of a separate quadratic mass term
and a Coleman--Weinberg interaction,
\begin{equation}
	V_{\rm tot}(s)
	=
	m^2s+U_{\rm CW}(s),
	\qquad
	s\equiv|\psi|^2,
	\label{eq:CW_total_potential}
\end{equation}
where
\begin{equation}
	U_{\rm CW}(s)
	=
	\lambda_{\rm CW}
	\left[
	\frac{1}{2}s^2
	\ln\!\left(\frac{s}{v^2}\right)
	-\frac{1}{4}s^2
	+\frac{1}{4}v^4
	\right].
	\label{eq:CW_direct_boson_star}
\end{equation}
Equivalently, in terms of the field amplitude,
\begin{equation}
	U_{\rm CW}(|\psi|)
	=
	\lambda_{\rm CW}
	\left[
	|\psi|^4
	\ln\!\left(\frac{|\psi|}{v}\right)
	-\frac{1}{4}|\psi|^4
	+\frac{1}{4}v^4
	\right].
\end{equation}
Different factors of two sometimes used in the literature amount to a
redefinition of \(\lambda_{\rm CW}\).

The interaction satisfies
\begin{equation}
	U_{{\rm CW},s}
	=
	\lambda_{\rm CW}
	s\ln\!\left(\frac{s}{v^2}\right),
\end{equation}
and has a stationary symmetry-breaking minimum at
\begin{equation}
	s=v^2,
	\qquad
	U_{\rm CW}(v^2)=0.
\end{equation}
However,
\begin{equation}
	U_{\rm CW}(0)
	=
	\frac{\lambda_{\rm CW}v^4}{4}.
	\label{eq:CW_origin_vacuum_energy}
\end{equation}
This constant is physically important in a gravitating system. If a
boson-star solution is required to satisfy \(s\to0\) at spatial
infinity, the potential in
Eq.~\eqref{eq:CW_direct_boson_star} approaches a nonzero vacuum-energy
density. An asymptotically flat construction must therefore specify
either a subtraction that sets \(V_{\rm tot}(0)=0\), a different
asymptotic vacuum, or an exterior cosmological term. The normalization
cannot be ignored when interpreting the solution as an isolated
compact object.

At large field amplitude,
\begin{equation}
	U_{\rm CW}(s)
	\simeq
	\frac{\lambda_{\rm CW}}{2}
	s^2\ln\!\left(\frac{s}{v^2}\right),
	\qquad
	s\gg v^2.
\end{equation}
The interaction is therefore unbounded and asymptotically
quartic--logarithmic; it does not saturate. Its characteristic stellar
mass remains controlled by the microscopic mass and the radiative
coupling and vacuum scales, rather than by a universal finite
high-density plateau.

Coleman--Weinberg boson stars are valuable as a published example of
radiatively generated non-polynomial structure and nontrivial vacuum
physics. They are not selected here because no \(G(2)\) matching
calculation motivates the logarithmic coefficient relations, their
vacuum subtraction requires separate gravitational treatment and their
large-field behavior does not yield the finite-power causal limit
\begin{equation}
	c_s^2\longrightarrow\frac{4}{5}
\end{equation}
that characterizes the GC9 closure.

\subsection{Nonlinear sigma-model and kinetic generalizations}
Kinetic and derivative-sector deformations are well motivated in effective field theory, Horndeski-like vector-tensor constructions and nonlinear-sigma-model descriptions, although their stellar interpretation is more involved than that of ordinary potential deformations \citep{Minamitsuji2018,Babichev2018,BrihayeHartmannKunz2022HorndeskiProca,KlimasLoginovShnirZhilin2022}. An alternative approach modifies the kinetic term rather than the potential,often motivated by effective field theory or nonlinear sigma-model constructions. The general structure is:
\begin{equation}
	\mathcal{L} \sim K(|\Phi|^2) |\partial \Phi|^2 - V(|\Phi|^2).
\end{equation}
or includes higher-order derivative operators such as
\begin{equation}
	\mathcal{L} \supset \alpha (|\partial\Phi|^2)^2 + \cdots.
\end{equation}
While such models can alter the propagation speed and stability properties, they introduce additional complexity.
As a result the effective stiffness is configuration-dependent and stability may be affected by higher-derivative instabilities.

\subsection{Multi-field and symmetry-breaking constructions}
Multi-field models and symmetry-breaking scenarios have repeatedly been explored as a route to richer equilibrium branches and effective saturation phenomena \citep{BernalBecerrilGuzman2010,UrenaLopezBernal2010,HenriquesLiddleMoorhouse1989,HenriquesLiddleMoorhouse1990,ValdezAlcubierreUrena2013}. A representative example involves interacting scalar sectors,
\begin{equation}
	V = V_1(\phi_1) + V_2(\phi_2) + \lambda\,\phi_1^2\phi_2^2,
\end{equation}
or boson--fermion mixtures. In these systems, effective saturation can emerge dynamically from inter-field interactions. However, they typically involve additional degrees of freedom and fine-tuning and do not provide a minimal or universal description of glueball matter.

\subsection{Vector Boson Stars (Proca stars)}
Vector boson stars and Proca-hair configurations provide the closest relativistic competitor to scalar boson stars in terms of compactness and Kerr-like phenomenology \citep{Brito2016,HerdeiroRadu2016ProcaHair,HerdeiroPanotopoulosRadu2020,SanchisGual2017Proca,SanchisGual2019ProcaBinary,Mendonca2021}.
Boson stars built from massive vector fields have received significant attention in recent years as promising BH mimickers. 
Their dynamics is governed by
\begin{equation}
	\mathcal{L}_{\rm Proca} = -\frac{1}{4}F_{\mu\nu}F^{\mu\nu} + \frac{m^2}{2} A_\mu A^\mu,
\end{equation}
leading to equilibrium configurations that are generically more compact than scalar boson stars.
From the standpoint of relativistic observables, Proca stars exhibit improved Kerr-mimicking properties, including higher compactness and richer multipole structure. However, their maximum mass remains parametrically tied to the particle mass,
\begin{equation}
	M_{\max}^{\rm Proca} \sim \mathcal{O}(1)\,\frac{M_{\rm Pl}^2}{m}.
\end{equation}
Consequently, for ultra-heavy constituents (e.g. $m\gtrsim \mathrm{TeV}$), the achievable macroscopic masses are far below the supermassive regime.\\

\begin{table*}[p]
	\centering
	\fontsize{6.5}{7.1}\selectfont
	\renewcommand{\arraystretch}{1.02}
	\setlength{\tabcolsep}{2.0pt}
	\begin{tabular}{llllll}
		\hline
		\textbf{Family} &
		\textbf{Representative interaction} &
		\textbf{High-density behavior} &
		\textbf{\(M_{\max}\) scaling/status} &
		\textbf{\(G(2)\)-glueball fit} &
		\textbf{Main TON--618 obstruction} \\
		\hline
		
		\tcell[0.115\textwidth]{Free} &
		\tcell[0.205\textwidth]{\(V(s)=m^2s\)} &
		\tcell[0.105\textwidth]{No interaction pressure} &
		\tcell[0.155\textwidth]{
			\(M_{\max}\simeq
			0.633\,M_{\rm Pl}^2/m\)
		} &
		\tcell[0.075\textwidth]{No} &
		\tcell[0.205\textwidth]{
			Reaching TON--618 requires
			\(m\lesssim8\times10^{-21}\,{\rm eV}\),
			incompatible with ultraheavy glueballs.
		} \\
		
		\hline
		
		\tcell[0.115\textwidth]{Quartic} &
		\tcell[0.205\textwidth]{
			\(V(s)=m^2s+\lambda s^2/2\)
		} &
		\tcell[0.105\textwidth]{
			\(V_{\rm int}\propto s^2\);
			\(c_s^2\to1/3\)
		} &
		\tcell[0.155\textwidth]{
			\(M_{\max}\propto
			\sqrt{\lambda}\,
			M_{\rm Pl}^3/m^2\)
		} &
		\tcell[0.075\textwidth]{Weak} &
		\tcell[0.205\textwidth]{
			Requires an enormous, nonperturbative
			\(\lambda\) when
			\(m\sim10^9\)--\(10^{14}\,{\rm GeV}\).
		} \\
		
		\hline
		
		\tcell[0.115\textwidth]{Generic polynomial truncation} &
		\tcell[0.205\textwidth]{
			\(V(s)=m^2s+
			\sum_{k\geq2}\lambda_k s^k/k\)
		} &
		\tcell[0.105\textwidth]{
			Set by the highest retained power
		} &
		\tcell[0.155\textwidth]{
			Model-dependent; multiple dimensional
			couplings
		} &
		\tcell[0.075\textwidth]{Weak} &
		\tcell[0.205\textwidth]{
			Truncation ambiguity, correlated tuning and
			loss of EFT control at large field amplitude.
		} \\
		
		\hline
		
		\tcell[0.115\textwidth]{
			Generalized power law
			\citep{PitzSchaffnerBielich2023}
		} &
		\tcell[0.205\textwidth]{
			\(V(s)=m^2s+\alpha_n s^{n/2}\),
			\(n>2\)
		} &
		\tcell[0.105\textwidth]{
			\(V_{\rm int}\propto s^{n/2}\);
			\(c_s^2\to(n-2)/(n+2)\)
		} &
		\tcell[0.155\textwidth]{
			Scale-invariant TF EOS; normalization
			depends on \(m\) and \(\alpha_n\).
		} &
		\tcell[0.075\textwidth]{Moderate} &
		\tcell[0.205\textwidth]{
			The power law alone does not explain the
			keV collective scale and its radiative stability.
		} \\
		
		\hline
		
		\tcell[0.115\textwidth]{Solitonic / flat-top} &
		\tcell[0.205\textwidth]{
			\(V(s)=m^2s(1-s/s_0)^2\)
		} &
		\tcell[0.105\textwidth]{
			Degenerate-vacuum, thin-wall and self-bound
			branches
		} &
		\tcell[0.155\textwidth]{
			Numerical; thin-wall scaling depends on
			\((m,s_0)\) and field conventions
		} &
		\tcell[0.075\textwidth]{Moderate} &
		\tcell[0.205\textwidth]{
			Requires a specially arranged second vacuum,
			thin-wall dynamics and a sufficiently robust
			conserved \(U(1)\).
		} \\
		
		\hline
		
		\tcell[0.115\textwidth]{
			Liouville exponential
			\citep{SchunckTorres2000,
				ChoiHeSchiappacasse2019}
		} &
		\tcell[0.205\textwidth]{
			\(V_{\rm L}(s)=
			m^2f^2(e^{s/f^2}-1)\)
		} &
		\tcell[0.105\textwidth]{
			Convex exponential growth;
			\(P>0\), \(c_s^2\to1\)
		} &
		\tcell[0.155\textwidth]{
			Full EKG numerical;
			formal strong-coupling TF scale
			\(\sim M_{\rm Pl}^3/(mf)\)
		} &
		\tcell[0.075\textwidth]{Weak--mod.} &
		\tcell[0.205\textwidth]{
			Ultraheavy constituents require an extreme
			\(m/f\) hierarchy; no \(G(2)\) argument fixes
			the complete factorial interaction tower.
		} \\
		
		\hline
		
		\tcell[0.115\textwidth]{
			Cosh--Gordon
			\citep{SchunckTorres2000}
		} &
		\tcell[0.205\textwidth]{
			\(V_{\rm CG}(s)=
			m^2f^2[
			\cosh(\sqrt{2s}/f)-1]\)
		} &
		\tcell[0.105\textwidth]{
			Convex exponential growth;
			\(P>0\), \(c_s^2\to1\)
		} &
		\tcell[0.155\textwidth]{
			Full EKG numerical;
			formal TF scale
			\(\sim M_{\rm Pl}^3/(mf)\)
		} &
		\tcell[0.075\textwidth]{Moderate} &
		\tcell[0.205\textwidth]{
			No demonstrated stable ultraheavy-SMBH
			branch; begins with an unsuppressed quartic
			and contains an unmotivated all-orders tower.
		} \\
		
		\hline
		
		\tcell[0.115\textwidth]{
			Logarithmic
			\citep{ChoiHeSchiappacasse2019}
		} &
		\tcell[0.205\textwidth]{
			\(V_{\log}(s)=
			m^2f^2\ln(1+s/f^2)\)
		} &
		\tcell[0.105\textwidth]{
			Concave; \(P_{\rm TF}<0\) and
			\(c_s^2<0\)
		} &
		\tcell[0.155\textwidth]{
			Full EKG solutions are numerical;
			no stable repulsive TF mass law
		} &
		\tcell[0.075\textwidth]{Poor in TF} &
		\tcell[0.205\textwidth]{
			Does not provide positive-pressure
			high-density support.
		} \\
		
		\hline
		
		\tcell[0.115\textwidth]{
			Bounded exponential ansatz
		} &
		\tcell[0.205\textwidth]{
			\(V(s)=m_0^2s+
			\Sigma^4(1-e^{-s/\sigma_0^2})\)
		} &
		\tcell[0.105\textwidth]{
			\(U\to\Sigma^4\);
			\(P_{\rm int}\to-\Sigma^4\)
		} &
		\tcell[0.155\textwidth]{
			No positive-pressure plateau branch and no
			\(M_{\max}\propto M_{\rm Pl}^3/\Sigma^2\)
			law
		} &
		\tcell[0.075\textwidth]{No; phenomen.} &
		\tcell[0.205\textwidth]{
			The positive plateau is vacuum-like, not
			repulsive. TON--618 therefore gives no
			consistent bound on \(\Sigma\).
		} \\
		
		\hline
		
		\tcell[0.115\textwidth]{
			Bounded \(\tanh\) ansatz
		} &
		\tcell[0.205\textwidth]{
			\(V(s)=m_0^2s+
			\Sigma^4\tanh(s/\sigma_0^2)\)
		} &
		\tcell[0.105\textwidth]{
			\(U\to\Sigma^4\);
			\(P_{\rm int}\to-\Sigma^4\)
		} &
		\tcell[0.155\textwidth]{
			No positive-pressure plateau branch and no
			universal maximum-mass law
		} &
		\tcell[0.075\textwidth]{No; phenomen.} &
		\tcell[0.205\textwidth]{
			Concavity gives negative pressure and
			negative compressibility.
		} \\
		
		\hline
		
		\tcell[0.115\textwidth]{
			Real-scalar T/E plateaus
			\citep{MuiaEtAl2019DenseScalarStars}
		} &
		\tcell[0.205\textwidth]{
			\(V_T(\varphi)=
			V_0\tanh^{2n}(\varphi/\Lambda)\);
			\(V_E(\varphi)=
			V_0(1-e^{-\varphi/\Lambda})^2\)
		} &
		\tcell[0.105\textwidth]{
			Bounded plateau for large real field
		} &
		\tcell[0.155\textwidth]{
			Full-GR oscillaton evolution;
			no stationary complex-\(U(1)\)
			\(M_{\max}\) law
		} &
		\tcell[0.075\textwidth]{Context only} &
		\tcell[0.205\textwidth]{
			These are real, time-dependent scalar stars,
			not stationary complex glueball condensates;
			the TF map in \(s=|\Phi|^2\) does not apply.
		} \\
		
		\hline
		
		\tcell[0.115\textwidth]{
			Effective NR saturation
			\citep{GalazoGarciaBraxValageas2025}
		} &
		\tcell[0.205\textwidth]{
			\(\frac12m^2\varphi^2+
			M_I^4[
			\cos(\varphi/\Lambda)-1+
			\varphi^2/(2\Lambda^2)]\)
		} &
		\tcell[0.105\textwidth]{
			Quartic at low density; averaged
			nonrelativistic force saturates
		} &
		\tcell[0.155\textwidth]{
			Schr\"odinger--Poisson TF/FDM solitons;
			no relativistic compact-star mass law
		} &
		\tcell[0.075\textwidth]{Context only} &
		\tcell[0.205\textwidth]{
			Saturation appears only after oscillation
			averaging; wave pressure and real-field
			dynamics remain essential.
		} \\
		
		\hline
		
		\tcell[0.115\textwidth]{
			Axion / chiral periodic
			\citep{GrillidiCortonaEtAl2016,
				Guerra2019AxionBosonStars}
		} &
		\tcell[0.205\textwidth]{
			\(m_a^2f_a^2[
			1-\cos(\sqrt{2s}/f_a)]\), or the full
			chiral square-root potential
		} &
		\tcell[0.105\textwidth]{
			Periodic and bounded; leading nonlinear
			term attractive
		} &
		\tcell[0.155\textwidth]{
			Dilute scaling
			\(\sim f_aM_{\rm Pl}/m_a\);
			dense branches numerical
		} &
		\tcell[0.075\textwidth]{Poor} &
		\tcell[0.205\textwidth]{
			No monotonic repulsive TF closure;
			periodic pseudo-Goldstone microphysics has no
			known \(G(2)\)-glueball analogue.
		} \\
		
		\hline
		
		\tcell[0.115\textwidth]{
			\(Q\)-ball nonic
		} &
		\tcell[0.205\textwidth]{
			\(V(s)=m^2s-\lambda s^2+\beta s^3\)
		} &
		\tcell[0.105\textwidth]{
			Attractive quartic plus stabilizing nonic;
			thin-wall possible
		} &
		\tcell[0.155\textwidth]{
			Parametric and numerical; depends on
			\(\lambda\) and the dimensionless
			combination \(\beta m^2\)
		} &
		\tcell[0.075\textwidth]{Poor--mod.} &
		\tcell[0.205\textwidth]{
			Requires tuned attraction, higher-order
			stabilization and a robust conserved
			\(U(1)\) charge.
		} \\
		
		\hline
		
		\tcell[0.115\textwidth]{Signum--Gordon} &
		\tcell[0.205\textwidth]{
			\(V(s)=\lambda_{\rm SG}\sqrt{s}\), or
			\(m^2s+\lambda_{\rm SG}\sqrt{s}\)
		} &
		\tcell[0.105\textwidth]{
			Nonanalytic at the vacuum; compact-support
			solutions possible
		} &
		\tcell[0.155\textwidth]{
			Toy-model and coefficient-dependent;
			no universal \(M_{\max}\) formula
		} &
		\tcell[0.075\textwidth]{Poor} &
		\tcell[0.205\textwidth]{
			No natural derivation from a smooth
			ultraheavy composite-glueball EFT.
		} \\
		
		\hline
		
		\tcell[0.115\textwidth]{
			Coleman--Weinberg
			\citep{EidizadeGhaffarnejad2025}
		} &
		\tcell[0.205\textwidth]{
			\(V=m^2s+\lambda[
			\tfrac12s^2\ln(s/v^2)
			-\tfrac14s^2+\tfrac14v^4]\)
		} &
		\tcell[0.105\textwidth]{
			Unbounded
			\(s^2\ln s\) growth
		} &
		\tcell[0.155\textwidth]{
			Full EKG numerical;
			no universal closed maximum-mass law
		} &
		\tcell[0.075\textwidth]{Weak--mod.} &
		\tcell[0.205\textwidth]{
			Requires a specified vacuum subtraction or
			exterior cosmological term, plus a tuned
			radiative scale and \(G(2)\) matching.
		} \\
		
		\hline
		
		\tcell[0.115\textwidth]{
			Noncanonical kinetic / sigma model
		} &
		\tcell[0.205\textwidth]{
			\(\mathcal L=
			K(s)|\partial\Phi|^2-V(s)\)
		} &
		\tcell[0.105\textwidth]{
			Controlled by field-space geometry and
			derivative operators
		} &
		\tcell[0.155\textwidth]{
			Model-dependent; determined by both
			\(K(s)\) and \(V(s)\)
		} &
		\tcell[0.075\textwidth]{Moderate} &
		\tcell[0.205\textwidth]{
			No universal mass scaling; ghosts,
			hyperbolicity and EFT cutoff must be checked
			model by model.
		} \\

		\hline
	\end{tabular}
	\caption{Comparison of benchmark self-interaction BS families. Here $s=|\Phi|^2$, except for real-field entries. Model-specific definitions, pressure signs and limitations are detailed in Sec.~II.}
	\label{tab:master_potential_comparison}
\end{table*}

Table~\ref{tab:master_potential_comparison} synthesizes the overall comparison of boson star potential families, showing $M_{max}$ scaling, glueball suitability and main limitations related to the aforementioned capability to produce SMBH such as TON--618. The table makes explicit a central point: all traditional potential families require either ultra-light constituents, astronomically large dimensionless couplings or interaction scales pushed to (or beyond) the Planck scale, once TON--618 is imposed for heavy $m$. 

The above analysis highlights a gap in the standard benchmark families considered here: none simultaneously provides a transparent confining non-Abelian interpretation, strong causal repulsion at high density and a clean separation between the microscopic particle mass and the macroscopic condensate-stiffness scale.

\section{The scale-separated $G(2)$-inspired Casimir-six third-cumulant potential}
\label{sec:GC9}

The comparison above shows that the obstacle is not the mere presence of self-interactions, but the way in which microscopic and macroscopic scales are tied together. A low-order scalar polynomial normalized by the heavy glueball mass tends either to remain too soft, to introduce self-bound or $Q$-ball branches, or to leave the maximum stellar mass suppressed by the microscopic scale. Conversely, a purely phenomenological density scale can reproduce supermassive objects, but it risks looking like a hidden bag constant unless its role is carefully separated from the microscopic vacuum trace anomaly.

We therefore define the final potential directly in the form used below. We call this the \emph{scale-separated $G(2)$-inspired Casimir-six third-cumulant} closure, abbreviated GC9. The name is an interpretive label for the microscopic channel, not a claim that $G(2)$ uniquely derives the nonic density power. GC9 is intended as a low-energy, lattice-matchable glueball-condensate EFT whose exponent is selected phenomenologically within a positive cumulant hierarchy by the compactness and causality requirements.

\subsection{Definition and scale separation}
\label{subsec:GC9_definition}

Let $s\equiv |\Phi|^2$ be the scalar density of the effective complex glueball condensate. The GC9 potential is
\begin{equation}
V_{\rm GC9}(s)=m_G^2s+\Lambda_T^4\left(\frac{s}{s_T}\right)^9,
\qquad s_T=\frac{\Lambda_T^4}{m_G^2}.
\label{eq:Vgc9_origin}
\end{equation}
Here $m_G$ is the microscopic mass of the relevant dark glueball-like composite, while $\Lambda_T$ is an infrared condensate-stiffness scale. The two scales have different physical meanings. The first fixes the pole mass of the excitation in the dilute regime; the second fixes the macroscopic pressure and density normalization of the coherent condensed medium.
Mathematically, the nonlinear term in
Eq.~\eqref{eq:Vgc9_origin} is the \(n=18\) member of the generalized
power-law family \(U_n\propto|\Phi|^n\) considered in
Ref.~\citep{PitzSchaffnerBielich2023}. Its Thomas--Fermi algebra,
\begin{equation}
 P_{\rm GC9}=8U_{\rm GC9},
 \qquad
 \rho_{\rm GC9}=2m_G^2s+10U_{\rm GC9},
\end{equation}
is therefore a specific realization of a known generalized-power
closure. The new hypothesis made here is not the mathematical
existence of an \(|\Phi|^{18}\) interaction, but its proposed
identification with a scale-separated positive third connected cumulant
motivated by the degree-six invariant structure of the \(G(2)\)
glueball sector.

\begin{figure}[t]\centering\includegraphics[width=\columnwidth]{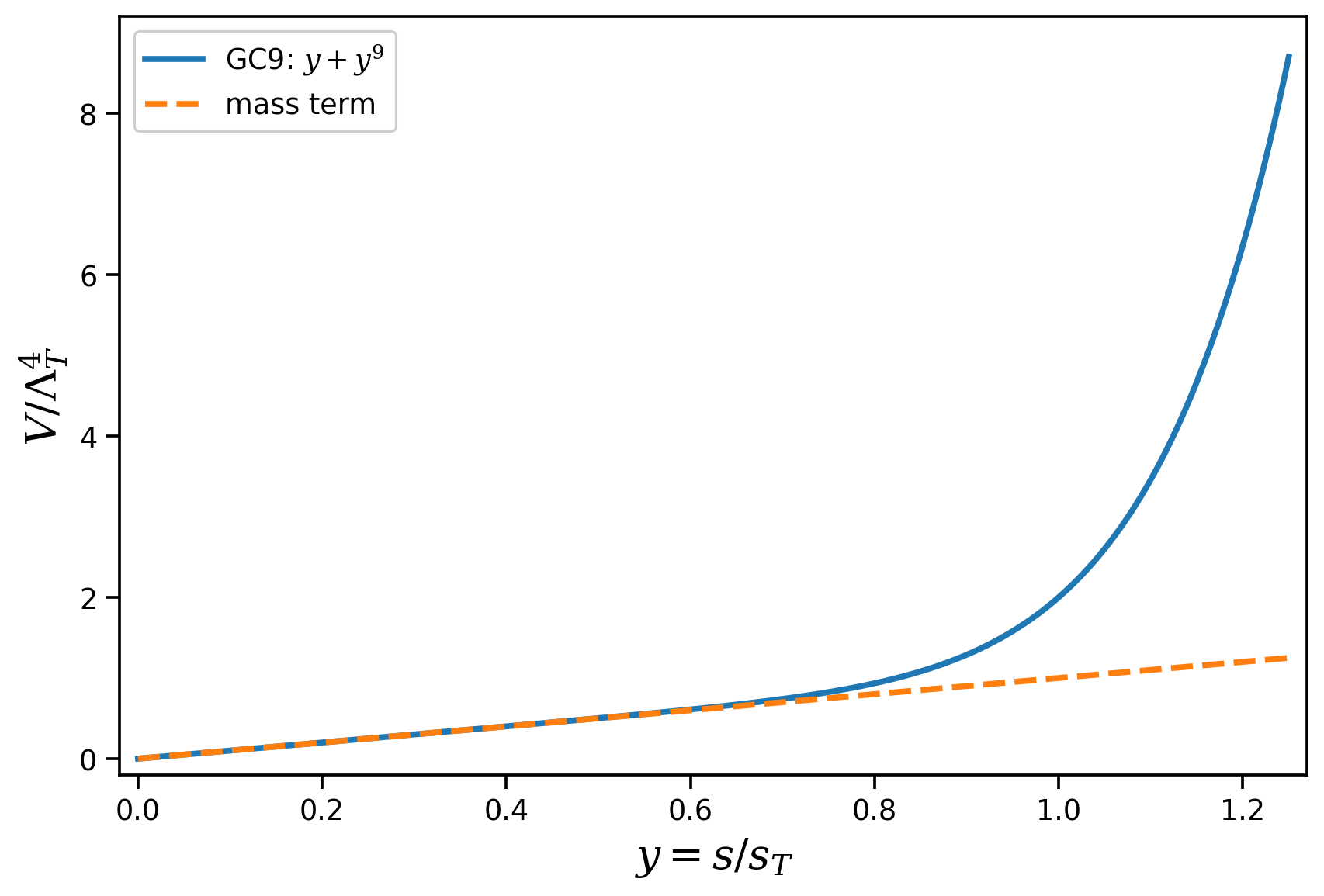}\caption{Reduced GC9 potential $V/\Lambda_T^4=y+y^9$.}\label{fig:GC9_potential}\end{figure}

The small-field expansion is
\begin{equation}
V_{\rm GC9}(s)=m_G^2s+\mathcal O(s^9),
\label{eq:gc9_small_field}
\end{equation}
so the microscopic glueball mass is not erased. At the same time, defining $y=s/s_T$, one obtains $m_G^2s=\Lambda_T^4y$ and therefore
\begin{equation}
V_{\rm GC9}=\Lambda_T^4(y+y^9).
\label{eq:gc9_reduced_potential}
\end{equation}
The reduced potential is thus
\begin{equation}
    f(y)=y+y^9,
\label{eq:gc9_reduced_f}
\end{equation}
shown in Figure \ref{fig:GC9_potential}. An important scope consequence is immediate: once the Thomas--Fermi problem is written in terms of $y$, the reduced TOV sequence and all dimensionless observables computed from it---$C$, $k_2$, the tidal deformability, $\bar I$, $\bar Q$, the reduced ISCO diagnostics, the static axial $M\omega$ and the Cowling $M\omega$ ladder---are independent of $m_G$.  The microscopic mass is therefore an ultraviolet input, not a quantity measured by the stellar calculation. It re-enters through the field normalization $s_T$, the finite-gradient control parameter, the dilute microscopic excitation scale and vector-specific hard modes.  The dimensional stellar mass and radius are controlled by $\Lambda_T$.
This simple expression is the reason the model remains analytically transparent. The dimensional normalization is no longer $m_G^2\sigma_0^2$, so the maximum star mass is not forced to scale as $m_G^{-2}$. Instead, $\Lambda_T$ is introduced as the IR stiffness of the coherent condensate. The term ``scale-separated'' labels this separation of microscopic and macroscopic scales; no renormalization-group or many-body derivation of its value is claimed in the present paper.

The potential is selected by simultaneous prescriptions: correct microscopic pole, no direct $m_G^{-2}$ maximum-mass suppression, smoothness, Q-ball safety, absence of a finite-density bag surface, causality, high compactness, a heuristic high-spin proxy near $0.9$, an order-unity slow-rotation quadrupole and lattice matchability. The overall features are displayed in Table \ref{tab:GC9_prescriptions}.

\begin{table*}
\caption{Structural prescriptions satisfied by the GC9 closure. The term ``density polynomial'' emphasizes that the nonlinear piece is a high-order polynomial in $s=|\Phi|^2$, not an ordinary low-order scalar polynomial used as a weak-coupling truncation. A few Kerr-mimicking and maximum-mass numerical results obtained later are anticipated.}
\label{tab:GC9_prescriptions}
\begin{ruledtabular}
\begin{tabular}{lll}
Prescription & GC9 realization & Consequence \\
\hline
Microscopic mass & $V=m_G^2s+\mathcal O(s^9)$ & glueball pole preserved \\
Scale separation & $s_T=\Lambda_T^4/m_G^2$ & $M_{\max}$ controlled by $\Lambda_T$ \\
Smoothness & $V\in C^\infty$ for $s\ge0$ & no wall, cusp or branch cut \\
Q-ball safety & $V/s=m_G^2(1+y^8)$ & $\min V/s\ge m_G^2$ \\
Bag-free surface & $P=0\Leftrightarrow \rho=0$ & no finite-density edge \\
Causality & $c_s^2\to4/5$ & high-density stiffness remains subluminal \\
Kerr mimicking & $C_{\max}^{\rm stable}=0.325$, $\bar Q=1.52$ & few-percent geodesic shifts \\
SMBH scaling & $M_{\max}=0.0642M_{\rm Pl}^3/\Lambda_T^2$ & no direct $m_G^{-2}$ suppression \\
$G(2)$ interpretation & $I_6\to\langle I_6^3\rangle_c\to y^9$ & positive Casimir-six third cumulant \\
\end{tabular}
\end{ruledtabular}
\end{table*}

\subsection{Why this does not contradict the rejection of generic polynomial potentials}
\label{subsec:GC9_polynomial_reconciliation}

The precise obstruction found in Sec.~\ref{sec:potentials} applies to generic low-order scalar-polynomial potentials when they are treated as ordinary one-scale weak-coupling truncations tied directly to the heavy microscopic mass. A quartic interaction can itself be rewritten with an independent collective scale, Eq.~\eqref{eq:quartic_scale_separation}, so scale separation is not unique to GC9. What distinguishes the GC9 benchmark in the present comparison is the combination of its much stiffer causal high-density EOS, high compactness, small tidal response, Q-ball safety and bag-free surface.

The GC9 potential is polynomial only in a restricted and physically different sense. It is a positive high-order polynomial in the coarse-grained glueball density $s=|\Phi|^2$, namely $s^9=|\Phi|^{18}$, with a scale-separated normalization. Its role is closer to a connected density-cumulant closure than to an ordinary low-energy Taylor expansion. Thus the constraints do not merely reject polynomiality; they select a special positive density polynomial with an independent condensate-stiffness scale.

\subsection{Vacuum structure, Q-ball safety and absence of a bag surface}
\label{subsec:GC9_vacuum}

The potential is nonnegative for $s\ge0$ and has its unique vacuum at $s=0$. The ratio relevant for flat-space $Q$-ball formation is
\begin{equation}
	\frac{V_{\rm GC9}(s)}{s}=m_G^2+\Lambda_T^4\frac{s^8}{s_T^9}=m_G^2(1+y^8)\ge m_G^2.
	\label{eq:GC9_qball_ratio}
\end{equation}
Thus the Coleman condition $\min_{s>0}V(s)/s<m_G^2$ is never satisfied. The model is therefore not a $Q$-ball or thin-wall soliton potential. The stability of the star is gravitational and charge-supported, not a consequence of a flat-space self-bound minimum.

\subsection{$G(2)$ and the Casimir-six third cumulant}
\label{subsec:GC9_G2_interpretation}

The $G(2)$ origin matters in three ways. First, $G(2)$ is a simple exceptional group with an $SU(3)$ subgroup, and the adjoint decomposes schematically as $\mathbf{14}\to \mathbf{8}\oplus\mathbf{3}\oplus\bar{\mathbf 3}$. This motivates a language in which the dark gauge sector contains an $SU(3)$-like confined component together with additional coset channels associated with $G(2)/SU(3)$. Second, the broken-sector composites need not coincide with the simplest real $0^{++}$ glueball of a pure Yang--Mills theory; in the effective rotating sector they can be represented by a complex condensate field with an approximately conserved number. Third, dense glueball matter should be described by connected multi-glueball correlation functions rather than by an elementary weakly coupled scalar polynomial.

The acronym ``GC9'' denotes a $G(2)$-inspired Casimir-six third-cumulant closure. The numeral refers to the resulting power $y^9$ of the coarse-grained density, not to a Casimir of degree nine: $G(2)$ has exactly two primitive Casimir invariants, of degrees $2$ and $6$. Neither the degree-six invariant nor connected-cumulant counting uniquely selects the third cumulant: the neighboring $y^6$ and $y^{12}$ terms considered below are equally admissible at the level of this schematic operator language. The choice $p=9$ is therefore phenomenological within that hierarchy, because it gives the comparatively stiff but still comfortably causal limit $c_{s,\infty}^2=4/5$ and the required compactness. With this status made explicit, the schematic scalar-density projection $y^3\leftrightarrow I_6$ provides a microscopic interpretation of the selected channel, so that $\langle I_6^3\rangle_c\to y^9$. Schematically,
\begin{equation}
    I_6\longrightarrow \langle I_6^3\rangle_c\longrightarrow \left(\frac{|\Phi|^2}{s_T}\right)^9,
\label{eq:GC9_cumulant_mapping}
\end{equation}
where $I_6$ denotes a higher gauge-invariant glueball channel whose scalar-density projection contributes positively to the effective energy. We do not claim that Eq.~\eqref{eq:GC9_cumulant_mapping} is a lattice derivation or an operator-dimension identity. Rather, it is an interpretive projection for the positive channel selected by the macroscopic constraints. A first-principles calculation would have to determine whether such a channel is dominant, or whether a resummed family with nearby exponents and coefficients is preferred.
One may test whether an even higher power would work. As discussed later, a family $f(y)=y+y^p$ gives
\begin{equation}
	\hat\rho=2y+(p+1)y^p,
	\qquad
	\hat P=(p-1)y^p,
\end{equation}
if the same Thomas-Fermi map is used with the appropriate scalar-density power. The high-density sound speed tends to
\begin{equation}
	c_{s,\infty}^2={p-1\over p+1}.
\end{equation}
As $p$ increases, the EOS approaches the causal limit.  The GC9 choice $p=9$ gives $4/5$, which is stiff but not dangerously close to unity.  It is therefore a conservative compromise between compactness and causality.
A resummed version of GC9 such as
\begin{equation}
	f(y)=y+{(1+\alpha y^9)^q-1\over \alpha q},
\end{equation}
could in principle improve compactness and quadrupole values. However, the minimal GC9 branch is preferable because it is analytically transparent and avoids adding an extra shape parameter $q$. The resummation alternative could be the subject of a future extension.

\subsection{Possible GC6--9--12A extension}
\label{subsec:GC6912A_extension}

The GC9 closure can be embedded, without changing any result derived in this
paper, in a broader positive connected-cumulant hierarchy. A minimal
three-term extension is
\begin{equation}
\begin{aligned}
 V_{6\text{--}9\text{--}12}(s)
 &=\Lambda_T^4 f_{6\text{--}9\text{--}12}(y),\\
 f_{6\text{--}9\text{--}12}(y)
 &=y+a_6y^6+a_9y^9+a_{12}y^{12}.
\end{aligned}
 \label{eq:GC6912_potential}
\end{equation}
with $y=s/s_T$, $s_T=\Lambda_T^4/m_G^2$ and non-negative coefficients
$a_6,a_9,a_{12}$. Under the schematic projection $y^3\leftrightarrow I_6$,
the terms $y^6$, $y^9$ and $y^{12}$ may be read respectively as positive
second-, third- and fourth-connected-cumulant contributions of the degree-six
channel. The corresponding isotropic Thomas--Fermi closure is
\begin{align}
 \hat\rho&=2y+7a_6y^6+10a_9y^9+13a_{12}y^{12},\\
 \hat P&=5a_6y^6+8a_9y^9+11a_{12}y^{12},
 \label{eq:GC6912_eos}
\end{align}
and hence
\begin{equation}
 c_r^2=
 \frac{30a_6y^5+72a_9y^8+132a_{12}y^{11}}
 {2+42a_6y^5+90a_9y^8+156a_{12}y^{11}}.
 \label{eq:GC6912_sound_speed}
\end{equation}
For positive coefficients the vacuum remains at $y=0$, the pressure vanishes
only at zero density and
\begin{equation}
 \frac{V_{6\text{--}9\text{--}12}}{s}
 =m_G^2\left(1+a_6y^5+a_9y^8+a_{12}y^{11}\right)\ge m_G^2,
\end{equation}
so the same bag-free and flat-space $Q$-ball-safety criteria are preserved.
When the density-twelve term dominates, $P/\rho$ and $c_r^2$ tend to $11/13$;
when it is absent and the density-nine term dominates, the GC9 asymptote
$4/5$ is recovered.

The suffix ``A'' denotes an optional anisotropic completion of this composite
interaction, rather than an additional power in the scalar potential. One
possible phenomenological closure is
\begin{equation}
 P_t=P_r+\Delta_A,
 \qquad
 \Delta_A=\frac{\lambda_A}{3}
 \frac{(\rho+P_r)(\rho+3P_r)r^2}{1-2m(r)/r},
 \label{eq:GC6912_anisotropy}
\end{equation}
which is of Bowers--Liang type \citep{BowersLiang1974}. The model denoted
GC6--9--12A is therefore Eq.~\eqref{eq:GC6912_potential} supplemented by
Eq.~\eqref{eq:GC6912_anisotropy}. The GC9 model analyzed throughout this
paper is the exact isotropic special case
\begin{equation}
 a_6=a_{12}=0,\qquad a_9=1,\qquad \lambda_A=0.
 \label{eq:GC9_special_case}
\end{equation}
The label GC6--9--12 refers to the density powers retained in the positive-cumulant
hierarchy and does not denote a pure-GC6 benchmark.

Positive anisotropy provides additional tangential support and can raise the
maximum compactness, but the useful range is bounded from above rather than
open-ended. Once a smooth horizonless configuration develops a light-ring
pair, the stable member can support long-lived trapped perturbations and
introduces the known nonlinear-instability concern, potentially leading to
migration or collapse
\citep{CunhaBertiHerdeiro2017LRstability,CunhaHerdeiroRaduSanchisGual2023LRInstability}.
This is the mechanism that Sec.~\ref{subsec:GC9_qnm_complete} shows to be
absent on the static GC9 endpoint. Crossing $C=1/3$ is therefore treated here
as a warning boundary of the extension rather than as a target. The most
interesting regime is the narrow interval $3<R/M<3.281$, in which the peak of
the exterior $\ell=2$ scattering barrier remains outside the star while the
vacuum photon sphere at $3M$ is absent.

To separate the effect of the connected-cumulant interaction from that of
anisotropy, we first solve the isotropic $6$--$9$--$12$ closure with
$a_6=a_9=1$ and $a_{12}=10$. The value $a_{12}=10$ is chosen only to make the
fourth-cumulant contribution numerically visible; it is neither lattice
matched nor theoretically preferred. The isotropic maximum-mass point is
\begin{equation}
\begin{aligned}
 M_{\max,\mathrm{iso}}^{\rm red}&=0.0720561,\\
 C_{\mathrm{iso}}&=0.323053,
 \qquad R/M=3.09547 .
\end{aligned}
 \label{eq:GC6912_isotropic_benchmark}
\end{equation}
Thus the extended interaction by itself raises the reduced maximum-mass
coefficient by about $12\%$ relative to GC9, but does not raise the
maximum-mass compactness.

We next add the Bowers--Liang completion of
Eq.~\eqref{eq:GC6912_anisotropy}. We choose the moderate illustrative value
$\lambda_A=0.3$ phenomenologically so that the configuration remains outside
the photon-sphere threshold. This number is not derived from, and must not be
identified with, the vortex coefficient $\alpha_{\rm vort}$ in
Eq.~\eqref{eq:delta}; the two closures have different functional dependence
and normalization. The resulting maximum-mass benchmark is
\begin{equation}
\begin{aligned}
 M_{\max,A}^{\rm red}&=0.0739113,\\
 C_A&=0.327990,
 \qquad R/M=3.04888 .
\end{aligned}
 \label{eq:GC6912A_benchmark}
\end{equation}
The anisotropic completion therefore supplies the compactness increase,
while the neighboring cumulants primarily increase the reduced mass scale.
The three benchmarks are summarized in Table~\ref{tab:GC6912A_comparison}.

\begin{table}[t]
\caption{Separation of the interaction and anisotropy effects in the
GC6--9--12A outlook benchmark. The quoted points are the first maximum of the
reduced mass sequence.}
\label{tab:GC6912A_comparison}
\begin{ruledtabular}
\begin{tabular}{lcccc}
Model & $M_{\max}^{\rm red}$ & $C$ & $R/M$ & $\chi_{\max}$\\
\hline
GC9, isotropic & 0.064157 & 0.324694 & 3.07983 & 0.975563\\
GC6--9--12, isotropic & 0.072056 & 0.323053 & 3.09547 & 0.973348\\
GC6--9--12A, $\lambda_A=0.3$ & 0.073911 & 0.327990 & 3.04888 & 0.988204
\end{tabular}
\end{ruledtabular}
\end{table}

Because $R>3M$ excludes only an external vacuum photon sphere, we also apply
the exact interior null-orbit diagnostic to the anisotropic solution. For a
static spherical stress tensor the circular-null-orbit condition remains
\begin{equation}
 \chi(r)\equiv\frac{3m(r)}{r}+4\pi r^2P_r(r)=1 .
 \label{eq:GC6912A_chi}
\end{equation}
For the benchmark of Eq.~\eqref{eq:GC6912A_benchmark} we find
\begin{equation}
 \chi_{\max}=0.988204<1,
 \qquad r(\chi_{\max})/R=0.97321,
 \label{eq:GC6912A_chimax}
\end{equation}
so it has neither an external nor an internal light ring. The exterior
$\ell=2$ Regge--Wheeler peak at $3.28078M$ remains outside the surface by
$7.6\%$. For comparison, $\lambda_A=1$ gives $R/M\simeq2.94$ and
$\chi_{\max}\simeq1.025$, while $\lambda_A=5$ gives $R/M\simeq2.30$ and
$\chi_{\max}\simeq1.315$; both lie in the light-ring-bearing regime and are
retained only as illustrations of where the phenomenological extension
ceases to be conservative.

None of these values is used elsewhere in the manuscript or presented as a
validated stellar prediction. The extension still requires independent
checks of radial and tangential causality, energy conditions, radial and
non-radial stability, cracking, light-ring trapping, ergoregion stability and
full rotating equilibria. Equation~\eqref{eq:GC6912_anisotropy} and the vortex
closure of Eq.~\eqref{eq:delta} should be regarded as distinct
phenomenological parametrizations of possible superfluid stresses; matching
them requires a microscopic calculation.

Finally, the multi-coefficient form of Eq.~\eqref{eq:GC6912_potential} should
not be read as a return to the generic low-order polynomial truncations discussed in
Sec.~\ref{subsec:GC9_higher_order_polynomial}. All retained coefficients are
non-negative, and the family therefore does not satisfy the standard canonical
flat-space $Q$-ball criterion used in this work. The terms may be interpreted as
phenomenological projections of successive connected cumulants of the same
degree-six $G(2)$ channel; until a microscopic or lattice matching is performed,
however, $a_6$, $a_9$ and $a_{12}$ remain independent effective coefficients.
The macroscopic mass scale remains controlled by $\Lambda_T$ rather than by the
microscopic glueball mass. The role of the hierarchy is to exhibit a systematic,
lattice-matchable neighborhood of GC9 whose coefficients and anisotropic stresses
could be selected by future dense-$G(2)$ calculations.

\subsection{Trace anomaly, bag constants and the role of $\Lambda_T$}
\label{subsec:GC9_trace_anomaly}

Pure $G(2)$ Yang--Mills has the trace-anomaly operator relation
\begin{equation}
T^\mu_{\ \mu}=\frac{\beta(g_{G(2)})}{2g_{G(2)}}F^A_{\mu\nu}F^{A\mu\nu}.
\label{eq:GC9_trace_anomaly}
\end{equation}
At one loop,
\begin{equation}
\begin{aligned}
\beta(g_{G(2)})&=-\frac{b_0}{16\pi^2}g_{G(2)}^3+\cdots,\\
b_0&=\frac{11}{3}C_2(G(2))=\frac{44}{3}.
\end{aligned}
\label{eq:GC9_G2_beta}
\end{equation}
because $C_2(G(2))=4$. This is a genuine first-principles group-theory constraint, but the matrix element $\langle F^2\rangle$ is nonperturbative.

It would be dangerous to identify $\Lambda_T^4$ directly with a microscopic vacuum trace-anomaly density. Such an identification would make the model resemble a bag-constant construction and would typically restore the unwanted heavy-mass scaling. The GC9 interpretation is instead
\begin{equation}
\begin{aligned}
\Lambda_T^4&=Z_{\rm IR}\,\mathcal A_D,\qquad Z_{\rm IR}>0,\\
\mathcal A_D&\equiv\left|\frac14\left\langle
\frac{\beta(g_D)}{2g_D}F^A_{\mu\nu}F^{A\mu\nu}
\right\rangle_{G(2)}\right|.
\end{aligned}
\label{eq:GC9_IR_anomaly}
\end{equation}
The coefficient $Z_{\rm IR}$ encodes coarse graining, in-medium renormalization and many-body dilution of the coherent glueball condensate. The compact object is not a bag of microscopic dark vacuum energy; it is a self-gravitating coherent state whose stiffness is normalized by, but not equal to, the microscopic anomaly.

\subsection{Lattice-matchable EFT status}
\label{subsec:GC9_lattice_matchable}

A first-principles determination would require nonperturbative $G(2)$ lattice input. The natural observables are the scalar glueball mass, the normalization of the $0^{++}$ operator, the trace-anomaly matrix element, low-energy glueball scattering amplitudes and connected higher glueball-density cumulants. A Wilson action in the seven-dimensional fundamental representation is
\begin{equation}
S_{\rm lat}=\beta\sum_p\left[1-\frac17{\rm Re\,Tr}_{\mathbf 7}U_p\right],
\label{eq:GC9_lattice_action}
\end{equation}
and scalar glueballs are extracted from zero-momentum gauge-invariant operators. Existing $G(2)$ and $G(2)$-QCD lattice studies provide a nonperturbative precedent for such a program, especially because $G(2)$ theories can avoid the usual finite-density sign problem in QCD-like settings \citep{Holland:2003jy,Wellegehausen:2013cya,Wellegehausen:2015iea}. They do not yet compute the pure-glue higher cumulants needed here.

\subsection{Interpretation of the GC9 branch}
\label{subsec:GC9_extended_checks}

The GC9 closure may be viewed as the endpoint of a filtering procedure. Starting from the general scalar-density EFT,
\begin{equation}
	V_{\rm eff}(s)=m_G^2s+\sum_{n\ge2}\lambda_ns^n,
\end{equation}
one imposes positivity of $V(s)/s-m_G^2$, absence of finite-density self-binding, subluminal high-density sound speed, a compactness target near the ultracompact threshold and a maximum-mass formula not dominated by the microscopic glueball mass. These requirements eliminate not only attractive low-order terms, but also many apparently benign repulsive closures. A quartic term gives the familiar Colpi--Shapiro--Wasserman scaling, but it does not naturally produce the desired compactness and Kerr-like multipoles in the ultraheavy glueball regime. A logarithmic or bag-like closure can generate high compactness, but it tends to introduce a finite-density edge or a trace-anomaly-like density scale that behaves as a bag constant. A flat-top potential can mimic compactness but risks a thin-wall or $Q$-ball interpretation.

The density-nine third-cumulant closure avoids these pitfalls because it is repulsive only in the sense needed at high density, while the linear term keeps the vacuum and dilute field theory conventional. The absence of intermediate negative curvature in $V(s)/s$ means that flat-space charge-stabilized solitons are not generated. The pressure grows as $8y^9$, while the energy density contains the larger coefficient $10y^9$, producing the finite asymptotic ratio $P/\rho\to4/5$. In this way the high-density branch is stiff without approaching the causal limit too aggressively.

A useful diagnostic is the logarithmic stiffness,
\begin{equation}
	\Gamma_{\rm eff}\equiv\frac{d\ln P}{d\ln\rho}.
\end{equation}
At low density $P\propto y^9$ and $\rho\propto y$, so $\Gamma_{\rm eff}\to9$. At high density both scale as $y^9$, so $\Gamma_{\rm eff}\to1$ while $P/\rho\to4/5$. The transition from a very steep onset to a linear high-density EOS is precisely what gives the branch a compact stable region without inserting a density discontinuity.

The scale $\Lambda_T$ should be treated as an infrared stiffness, not as the microscopic confinement scale. If one set $\Lambda_T\sim m_G$, the maximum mass would again be microscopic for ultraheavy glueballs. The phenomenologically relevant hierarchy is instead
\begin{equation}
	\Lambda_T\ll m_G,
\end{equation}
which is interpreted as a many-body, coarse-grained, in-medium-renormalized stiffness scale of the coherent condensate. It is structurally distinct from a bag constant because it does not represent a vacuum energy filling the star: it represents the pressure normalization of a macroscopic condensed phase whose microscopic origin remains to be calculated.

\subsection{Comparison with ordinary polynomial, logarithmic and bag-like closures}
\label{subsec:GC9_family_comparison}

It is helpful to classify the alternatives by what controls their maximum mass. For a microscopic polynomial normalization one typically obtains
\begin{equation}
	M_{\max}\sim A\left(\frac{\sigma_0}{M_{\rm Pl}}\right)\frac{M_{\rm Pl}^3}{m_G^2},
\end{equation}
so the ultraheavy glueball mass suppresses the stellar scale unless $\sigma_0$ is made extremely large. For a bag-like model 
\begin{equation}
	M_{\max}\sim A\frac{M_{\rm Pl}^3}{\sqrt{B}},
\end{equation}
where $B$ is a vacuum density. This can fit a supermassive object, but it risks identifying the star with a patch of dark vacuum energy. GC9 instead gives
\begin{equation}
	M_{\max}\sim A\frac{M_{\rm Pl}^3}{\Lambda_T^2},
\end{equation}
where $\Lambda_T^4$ is an in-medium stiffness scale. The formula resembles the bag expression dimensionally, but the interpretation is different: $\Lambda_T^4$ is not the vacuum energy difference between phases, but rather the normalization of a positive connected density cumulant in a coherent condensate.

The distinction can be summarized as follows: in a bag model, the density scale exists even in the absence of particles. In GC9, the nonlinear term vanishes with $s$, and the pressure vanishes with the density. Therefore there is no star without the condensate. This is why $P=0$ implies $\rho=0$ at the surface.

\subsection{Maximum-mass scaling and the TON--618 normalization}
\label{subsec:GC9_mass_scaling}

The key dimensional consequence of Eq.~\eqref{eq:Vgc9_origin} is that the physical mass scale is set by $\Lambda_T$, not by $m_G$. Restoring dimensions from the reduced TOV solution gives
\begin{equation}
M_{\max}^{\rm GC9}=A_{\rm GC9}\frac{M_{\rm Pl}^3}{\Lambda_T^2},
\qquad A_{\rm GC9}=0.064157 .
\label{eq:Mmax_GC9_scaling}
\end{equation}
Here and below $M_{\rm Pl}=G^{-1/2}=1.2209\times10^{19}\,{\rm GeV}$ is the unreduced Planck mass. The associated radius at the maximum-mass turning point scales as
\begin{equation}
R(M_{\max})=B_{\rm GC9}\frac{M_{\rm Pl}}{\Lambda_T^2},
\qquad B_{\rm GC9}=0.197591 .
\label{eq:Rmax_GC9_scaling}
\end{equation}
The TON--618-scale mass $M_{\rm TON}\simeq 6.6 \times10^{10}M_\odot$ corresponds to
\begin{equation}
\Lambda_T=\left(\frac{0.064157\,M_{\rm Pl}^3}{M_{\rm TON}}\right)^{1/2}\simeq 1.26\times10^{-6}{\rm GeV}.
\label{eq:GC9_TON_LambdaT}
\end{equation}
The same normalization gives
\begin{equation}
\begin{aligned}
R(M_{\max})&\simeq 3.00\times10^{11}\,{\rm km},\\
\frac{R}{GM/c^2}&=\frac{B_{\rm GC9}}{A_{\rm GC9}}\simeq3.07983 .
\end{aligned}
\label{eq:GC9_TON_radius}
\end{equation}
Thus the mass and radius are fixed consistently by the same value of $\Lambda_T$, they are not independently tuned. This is a keV-scale stiffness, not a keV-scale particle mass. The crucial point is that in the GC9 scaling the maximum mass and the associated radius are controlled by the emergent stiffness scale $\Lambda_T$, whereas the microscopic dark-glueball mass $m_G$ only enters the density normalization through $s_T=\Lambda_T^4/m_G^2$. Therefore there is no contradiction between obtaining a TON--618-scale compact object and keeping ultraheavy microscopic constituents, for example in the range $m_G\sim 10^9$--$10^{14}\,{\rm GeV}$ considered in the dark glueball scenario. In other words, the requirement $\Lambda_T\sim 10^{-6}\,{\rm GeV}$ should be interpreted as a statement about the effective many-body stiffness of the condensate, not as a statement that the constituent glueball itself must be keV-scale. With this interpretation, the GC9 construction can still reproduce both the TON--618 mass scale and the associated radius scale while remaining compatible with ultraheavy dark matter constituents.

Numerically, the hierarchy between $\Lambda_T$ and an EeV--GUT-scale glueball is as severe as the hierarchy between the plateau scale and the constituent mass in the discarded saturating models. But GC9 improves the surface physics---the nonlinear term vanishes with $s$, $P=0$ implies $\rho=0$, and no vacuum bag fills the star---but it does not derive the small value of $\Lambda_T$ \textit{a priori}. Until $Z_{\rm IR}$ or an equivalent many-body matching coefficient is calculated from $G(2)$ dynamics, $\Lambda_T$ must be treated as a phenomenological collective scale fixed by the target macroscopic mass, not as a solved hierarchy problem.

\subsection{Compactness band at fixed $\Lambda_T$ and phenomenological branch-dependent
	normalization}
\label{sec:sector}

Equation~\eqref{eq:Mmax_GC9_scaling} fixes the maximum mass, but compactness is not uniform
along the equilibrium sequence: only configurations within a factor of
a few of $M_{\max}(\Lambda_T)$ approach the compact endpoint. Well below the
turning point the GC9 fluid enters its dilute regime, in which the
equation of state reduces to the Newtonian polytrope
$P\simeq (1/64)\,\rho^{9}/\rho_s^{8}$ with $\rho_s=\Lambda_T^{4}$ and
polytropic index $n=1/8$ (the derivation is given in Sec.~\ref{subsec:GC9_dilute_crossover}). The
Lane--Emden scalings then give, anchored at the turning point and to
within order-unity factors confirmed by the reduced numerical sequence,
\begin{equation}
 C(M)\simeq C(M_{\max})
 \left(\frac{M}{M_{\max}(\Lambda_T)}\right)^{16/23},
 \label{eq:diluteband0}
\end{equation}
and
\begin{equation}
 R(M)\simeq \frac{B_{\rm GC9}}{A_{\rm GC9}}\,
 \frac{G M_{\max}}{c^{2}}
 \left(\frac{M}{M_{\max}}\right)^{7/23},
 \label{eq:diluteband}
\end{equation}
for $M\ll M_{\max}$, with $B_{\rm GC9}/A_{\rm GC9}=3.07983$.

A single universal stiffness therefore cannot render the entire
astrophysical mass hierarchy simultaneously compact. With the
TON--618 normalization $\Lambda_T\simeq 1.3$~keV,
Eqs.~\eqref{eq:diluteband0} and \eqref{eq:diluteband} give
$C\simeq3.8\times10^{-4}$ and $R\simeq104$~AU at
$M=4\times10^{6}M_\odot$, $C\simeq4.8\times10^{-8}$ and
$R\simeq2.06$~AU at $M=10M_\odot$, and $C\simeq3.8\times10^{-18}$,
$R\simeq1.17\times10^4$~km at the asteroid-window benchmark
$M=3\times10^{-14}M_\odot$: none of these configurations is
a black-hole mimicker.

In the phenomenological sections of this paper we therefore adopt a
\emph{sector-dependent} normalization,
\begin{equation}
	\Lambda_T^{\rm bench}(M_\star)
	=\left(\frac{A_{\rm GC9}\,M_{\rm Pl}^{3}}{M_\star}\right)^{1/2}
	\simeq 0.32~{\rm GeV}
	\left(\frac{M_\odot}{M_\star}\right)^{1/2},
	\label{eq:sectorLambda}
\end{equation}
so that each compact population lies near its own maximum-mass
turning point. Representative values are collected in
Table~\ref{tab:sectors}.

Equation~\eqref{eq:sectorLambda} is a benchmark inversion of the turning-point mass formula, not a local constitutive law and not a prediction that a fluid coupling can depend on the final total mass of a star.  Each row of Table~\ref{tab:sectors} therefore represents a separate constant-$\Lambda_T$ effective-EOS branch.  The stellar mass remains an output of the equilibrium problem within each branch.

Three properties make this branch-by-branch rescaling mathematically self-consistent.
First, the reduced-sequence observables computed in Sec.~\ref{sec:observables}
($C$, $k_2$, $\Lambda$, $\bar I$, $\bar Q$, the ISCO shifts and the
Cowling frequencies $M\omega$) are dimensionless functions of the
reduced central density only: they are $\Lambda_T$-independent and
apply verbatim within every sector, with only the dimensional
normalization changing. Second, the bulk gradient-suppression
parameter of Eq.~\eqref{eq:GC9_epsilon_TF} remains tiny in all sectors,
$\epsilon_{\rm TF}=\Lambda_T^{2}/(m_G M_{\rm Pl})\lesssim
3\times10^{-16}$ even for the most demanding combination
($\Lambda_T\simeq 1.9$~PeV, $m_G=10^{9}$~GeV), so the Thomas--Fermi
reduction retains its parametric justification, and the scale
separation $\Lambda_T\ll m_G$ persists (at worst
$\Lambda_T/m_G\sim 2\times10^{-3}$). Third, the structural properties
of the closure --- causality with $c_s^2\to 4/5$, Q-ball safety,
smoothness and the bag-free surface --- are $\Lambda_T$-independent.

The sector dependence of $\Lambda_T$ is a phenomenological input, not
a derived result. Within the interpretation of Eq.~\eqref{eq:GC9_IR_anomaly} it
corresponds to an environment- or formation-history-dependent
coarse-graining coefficient $Z_{\rm IR}$; distinct condensation
epochs or branches of the dark sector could provide alternative
realizations. We note that for the lighter sectors the required
hierarchy is substantially milder than in the supermassive case:
$\Lambda_T/m_G$ ranges from $\sim10^{-3}$--$10^{-8}$ (PeV stiffness
versus $m_G=10^{9}$--$10^{14}$~GeV) rather than
$\sim10^{-15}$--$10^{-20}$. Conversely, sector dependence introduces
a mass dependence of $\Lambda_T$ that a first-principles calculation
would ultimately have to explain, and it imposes a nontrivial
consistency condition on hierarchical growth, discussed in
Sec.~\ref{sec:hierarchical}. A speculative formation-scale interpretation of this
mass-dependence is explored in Sec.~\ref{sec:lambdaT-frozen}
immediately below. A single-$\Lambda_T$ version of the framework remains
viable, but its compact, Kerr-comparison claims are then restricted
to the mass band within a factor of a few of the corresponding
$M_{\max}$.

\begin{table}[t]
	\caption{Sector-dependent stiffness normalizations from
		Eq.~\eqref{eq:sectorLambda} and the corresponding turning-point
		radius $R=3.07983\,GM_\star/c^{2}$.}
	\label{tab:sectors}
	\begin{ruledtabular}
		\begin{tabular}{lccc}
			Sector & $M_\star$ & $\Lambda_T^{\rm sector}$\\
			\hline
			SMBH (TON--618) & $6.6\times10^{10}M_\odot$ & $1.3$ keV\\
			Galactic center & $4\times10^{6}M_\odot$ & $1.6\times10^{2}$ keV\\
			Stellar remnant & $10\,M_\odot$ & $0.10$ GeV\\
			Lunar mass & $10^{-7}M_\odot$ & $1.0$ TeV\\
			Asteroid window & $3\times10^{-14}M_\odot$ & $1.9$ PeV\\
		\end{tabular}
	\end{ruledtabular}
\end{table}

\subsection{Kerr Mimicking}
The strong-field discrimination problem for horizonless compact objects has been analyzed in literature using shadows, multipolar structure, light rings, ringdown spectra and tidal responses in a wide range of parametrized and explicit models \citep{Cardoso2019BHmimickers,Olivares2020,DeLaurentisPani2025TestingNature,BertiCardosoWill2006,JohannsenPsaltis2010,RezzollaZhidenko2014,Bambi2017,Grandclement2014,Vincent2016,Olivares2020,Pani2015,YagiYunes2013,Cunha2017,DiFilippo2024,EHT2019M87I}.
\label{sec:kerr}
In the present work, the central question is whether compact boson stars supported by the GC9 interaction can reproduce the exterior phenomenology usually attributed to Kerr black holes, while still
remaining horizonless and dynamically distinct in their interior response. The aim of this subsection is therefore
introductory: we identify the classes of observables that matter for Kerr mimicry (such as the multipolar structure, rotational features, tidal response, geodesic properties and merger behavior), explain why they are relevant for GC9 boson stars and set up the logic of the explicit calculations carried out in the following section.

\paragraph{Exterior spacetime and multipole moments.}
A Kerr BH is fully characterized by mass and angular momentum and its entire multipolar structure is
fixed by the relation
\begin{equation}
	M_\ell+iJ_\ell = M (ia)^\ell,
	\qquad a\equiv \frac{J}{M}.
	\label{eq:kerr_multipole_relation_intro}
\end{equation}
By contrast, a boson star is not determined by \(M\) and \(J\) alone; its higher multipoles retain information
about the matter distribution and self-interaction potential \citep{YagiYunes2013,Visinelli2021Review}. 
For the quadrupole moment
\begin{equation}
	M_2 = - a^2 M + \delta M_2, \qquad
	J_3 = - a^3 M + \delta J_3, \quad \ldots
\end{equation}
where the deviations $\delta M_\ell,\delta J_\ell$ encode the internal structure. Its multipoles deviate:
$$\delta M_\ell \equiv M_\ell-M_\ell^{\rm Kerr}(M,J),\qquad
\delta S_\ell \equiv S_\ell-S_\ell^{\rm Kerr}(M,J).$$
The leading observable departure from Kerr is generically the mass quadrupole,
\begin{equation}
	M_2\equiv Q,
	\qquad
	\Delta Q \equiv Q-Q_{\rm Kerr} = Q + \frac{J^2}{M},
	\label{eq:deltaQ_intro}
\end{equation}
which controls, at lowest nontrivial order, the deformation of the exterior spacetime away from the Kerr
prediction at fixed \((M,J)\). In the GC9 case, this quantity will be computed explicitly below from the
second-order Hartle--Thorne expansion and then propagated into geodesic and spectroscopic proxies. At the present stage, the important point is simply that the GC9 branch provides a concrete horizonless model in
which the non-Kerr multipoles are calculable rather than left as free deformation parameters
\citep{Hartle1967SlowRotation,Cardoso2019BHmimickers}.

To first approximation we will describe slow rotation using the Hartle--Thorne expansion \citep{Hartle1967SlowRotation}.
This allows one to compute the moment of inertia, the spin-induced quadrupole and the leading corrections to
the vacuum exterior in a gauge-invariant way. For a Kerr mimicker, the relevant question is not whether these
quantities equal the Kerr values exactly, but whether their deviations remain small enough to evade present
constraints from gravitational waves \citep{Maggiore2020}, horizon-scale imaging and precision X-ray probes \citep{Cardoso2019BHmimickers,EHT2019M87I,EHT2022SgrA}.
The detailed GC9 quadrupole extraction and the corresponding reduced quadrupole \(\bar Q\) are deferred to the
later observable section, where they are obtained consistently from the corrected second-order computation.

\paragraph{Geodesics, ISCO and photon region.}
A second class of observables is determined by the orbital structure of the exterior geometry. The innermost
stable circular orbit controls the characteristic inner scale of accretion flows, while the photon region
and equatorial light rings govern strong lensing, image morphology and shadow-scale observables
\citep{CunhaBertiHerdeiro2017LRstability,CunhaHerdeiroRadu2017Shadows,Olivares2020}.
Ultra-compact horizonless objects can possess light rings and, in sufficiently compact configurations, even
multiple light rings may occur; in that case one must also confront the possible dynamical implications of
stable photon orbits \citep{CunhaBertiHerdeiro2017LRstability,DiFilippo2024InnerLR}.

A Kerr black hole has a well-defined ISCO,
\begin{equation}
	r_{\rm ISCO}^{\rm Kerr}(a)
	=
	M\left[3+Z_2-\sqrt{(3-Z_1)(3+Z_1+2Z_2)}\right],
\end{equation}
with $Z_1,Z_2$ standard functions of $a$.
The associate fractional deviation
\begin{equation}
	\frac{\Delta r_{\rm ISCO}}{r_{\rm ISCO}^{\rm Kerr}}
	\equiv
	\frac{r_{\rm ISCO}^{\rm GC9}-r_{\rm ISCO}^{\rm Kerr}}{r_{\rm ISCO}^{\rm Kerr}}
\end{equation}
is typically $\lesssim 5\%$ in the Kerr-mimicking branch.
Similarly, the photon sphere and shadow radius should differ only at the percent level, consistent with current Event Horizon Telescope (EHT) constraints on M87$^\ast$ and Sgr~A$^\ast$ \citep{VincentEtAl2021M87,EHT2019M87I,EHTSgrA2022I,EHT2022SgrA}.

For GC9 boson stars, these issues are especially relevant because the same saturating EOS that supports large masses can also drive the compact branch close to the photon-ring threshold. In the following section we compute and discuss a controlled set of layer-2 observables using the multipolar data.

\paragraph{Tidal Love numbers.}
A third discriminator is the tidal response. In classical general relativity, the static quadrupolar Love numbers
of Kerr black holes vanish
\begin{equation}
	k_2^{\rm Kerr} = 0,
\end{equation}
whereas horizonless compact objects generally develop nonzero tidal deformability
and matter-induced resonances \citep{Sennett2017,PacilioEtAl2021,Pani2015}.
This makes tidal observables particularly valuable because they are sensitive to the presence or absence of a
material core even when the exterior metric is close to Kerr. In the GC9 framework, the tidal response depends
on both compactness and EOS stiffness and therefore provides an independent consistency check on any claim
of Kerr mimicry. 
Horizonless objects generically have $k_2\neq 0$. 
Current gravitational-wave observations remain consistent with the BH expectation \(k_2=0\), with present uncertainties still too large to exclude small nonzero deviations \citep{Abbott2022TestsGR,JohnsonMcDanielEtAl2018}. Future detectors such as Einstein Telescope, Cosmic Explorer and LISA are expected to improve the sensitivity to quadrupolar finite-size effects substantially \citep{Shterenberg2025FiniteSize}.

\paragraph{High-spin requirement.}
Kerr mimicry is meaningful only if the candidate object can rotate rapidly and remain dynamically
viable. Boson stars can support substantial spin, but the accessible range is model dependent and may be
limited by mass shedding, ergoregion formation, or nonlinear instabilities \citep{Liebling2017,BritoCardosoHerdeiroRadu2016,CunhaBertiHerdeiro2017LRstability}.
For GC9 stars, the relevant issue is whether the saturating interaction can support compact, rapidly rotating
configurations in the same part of parameter space in which the exterior remains approximately Kerr-like.
This question connects directly to the merger, scalar-burst and seeded-collapse phenomenology developed in later sections.

Electromagnetic spin measurements of BH candidates, mainly from continuum fitting and relativistic X-ray reflection spectroscopy, indicate that at least a subset of both stellar-mass systems and SMBH candidates are rapidly rotating, with several sources favoring $a_*\gtrsim 0.9$ and a few canonical cases consistent with near-extremal values $a_*\sim 0.98\text{--}0.99$ \citep{Reynolds2020Spin,McClintock2006GRS1915,Brenneman2011SMBHSpin}. For the present purpose, the important implication is not that every compact object must sit arbitrarily close to the Kerr bound, but that any viable horizonless Kerr mimicker must be able to reproduce at least the high-spin tail of the observed population. In the rotating boson star description this requirement translates into the need for stationary configurations with sufficiently large $j \equiv \frac{J}{M^2}$, with $J=mN$, where $N$ is the effective conserved number associated with the rotating sector.

\paragraph{Ringdown and internal modes.}
The post-merger and perturbed dynamics provide a fourth class of signatures. Even when the early ringdown is
dominated by the exterior photon region and can partially mimic a Kerr response, horizonless objects generically
admit additional matter modes, mode mixing and possible late-time deviations from BH behavior
\citep{CardosoFranzinPani2016ECO,BertiCardosoWill2006,Cardoso2016}.

After a merger, Kerr black holes relax via QNMs with frequencies
\begin{equation}
	\omega_{\ell m n}^{\rm Kerr} = \omega_{\ell m n}(M,a).
\end{equation}
Thus the relevant question is quantitative: whether the GC9 spacetime-dominated response remains close enough to the black-hole spectrum at current sensitivity while preserving distinctive interior dynamics. In the observable section below, we retain the light-ring/eikonal expressions only as formal comparison numbers and solve the static axial $\ell=2$ perturbation problem directly. The rotating polar--axial spectrum and merger excitation amplitudes remain separate tasks.

These features are currently unconstrained but constitute prime targets for future gravitational-wave spectroscopy \citep{BertiEtAl2016BHSpec,Bhagwat2020Spectroscopy}. In particular, LISA observations of extreme-mass-ratio inspirals and massive BH mergers are expected to improve access to multi-mode ringdown measurements and consistency tests of the Kerr QNM spectrum \citep{LISA2017Mission,BertiSesanaBarausseCardosoBelczynski2016,Maselli2023LISAEMRI}.

This establishes the GC9 construction as a concrete and testable Kerr-mimicker framework, while keeping the quantitative results in the sections where they are explicitly derived.

\paragraph{Merger dynamics: gravitational cooling and re-landing.}
\label{sec:mergers}
A static compact object can be tuned to reproduce several Kerr-like observables, but the real dynamical test of
any horizonless BH alternative is what happens during and after a merger. In the GC9 framework, the
central question is whether a compact BS remnant can absorb the violent perturbation of coalescence,
shed the excess energy and angular momentum generated during the event and settle back onto a stable
ultracompact branch without forming a classical horizon \citep{Cardoso2019BHmimickers,Liebling2017,PalenzuelaLehnerLiebling2007,BezaresPalenzuela2017}.

The candidate regulating mechanism is the boson-star analogue of
\emph{gravitational cooling}.  In the successful horizonless re-landing channel
considered here, excess energy and angular momentum are not disposed of by
horizon absorption; they must instead be redistributed among outgoing
radiation, bound oscillations, surrounding matter or other allowed channels.
This does not guarantee re-landing: if the losses and redistribution are
insufficient, collapse or migration outside the controlled GC9 branch remains
an allowed endpoint.  In practice, a post-merger object can lose energy and
angular momentum through
gravitational waves, scalar radiation \citep{SeidelSuen1994,HawleyChoptuik2000,Liebling2017} and, in the most nonlinear regime, partial mass shedding or ejection of
bosonic lumps \citep{YoshidaEriguchi1994,PalenzuelaLehnerLiebling2007,
	BezaresPalenzuela2017}. This converts the merger problem into a dynamical self-regulation problem: an initially
overmassive or overspinning remnant can still remain viable provided it radiatively relaxes quickly enough
toward the stable branch \citep{Liebling2017,PalenzuelaLehnerLiebling2007}.

At the level of global charges, this relaxation can be described schematically by
\begin{equation}
	\frac{dM}{dt}=-\dot E_{\rm GW}-\dot E_{\Phi}^{\infty},\;
	\frac{dJ}{dt}=-\dot J_{\rm GW}-\dot J_{\Phi}^{\infty},\;
	\frac{dN}{dt}=-\dot N_{\Phi}^{\infty},
\end{equation}
where \(\dot E_{\Phi}^{\infty}\), \(\dot J_{\Phi}^{\infty}\), and \(\dot N_{\Phi}^{\infty}\) denote the scalar fluxes carried to
infinity. The physically relevant endpoint is then a \emph{re-landing} condition: after the transient relaxation,
the remnant must lie again on a stable rotating GC9 sequence, which may be expressed either in microscopic
form,
\begin{equation}
	M_0-\Delta E_{\rm rad}=M(\omega_f,m_f),\;
	J_0-\Delta J_{\rm rad}=m_f N(\omega_f,m_f),
\end{equation}
or more phenomenologically as
\begin{equation}
	M_f \le M_{\max}(J_f).
\end{equation}
This is the same dynamical logic that underlies BS formation and post-merger relaxation more generally:
a horizonless remnant survives only if nonlinear cooling drives it back toward the stable branch
\citep{SeidelSuen1994,YoshidaEriguchi1994,Liebling2017,PalenzuelaLehnerLiebling2007,BezaresPalenzuela2017}.

This point is crucial for the rest of the paper. Once gravitational cooling is viewed as the mechanism that
enforces re-landing, two immediate consequences follow. First, the relaxation need not be silent: if the
remnant temporarily overshoots the stable branch, part of the excess energy can emerge as coherent scalar
bursts. Second, the same post-merger relaxation process determines the spectral structure of the ringdown:
a rapidly damped spacetime mode can coexist with a frequency shift from the black-hole spectrum, while the horizonless bosonic core
can support additional matter-sensitive late-time modes. In this sense, scalar bursts and bosonic ringdown are
not separate add-ons, but two complementary manifestations of the same merger-regulation dynamics
\citep{PalenzuelaLehnerLiebling2007,BezaresPalenzuela2017,EvstafyevaSperhakeRomeroShawAgathos2024GWDA}.

\section{Constructing the GC9 boson star observables}
\label{sec:observables}

We now construct the compact-object observables of the GC9 branch exploiting the nonrotating Tolman--Oppenheimer--Volkoff background equations, the first-order slow-rotation formalism of Hartle and the relativistic tidal perturbation treatment of Hinderer \citep{OppenheimerVolkoff1939,Hartle1967SlowRotation,Hinderer2008Love}.

The calculations use the same hierarchy as in the standard self-interacting boson-star literature: first the scalar potential is mapped to a Thomas--Fermi equation of state, then the static equilibrium is computed with TOV, the tidal response with the Hinderer system, the moment of inertia with first-order Hartle rotation and the quadrupole with the second-order Hartle--Thorne equations. The rapidly rotating endpoint is not claimed to be a full two-dimensional rotating solution; it is a heuristic mass-shedding estimate based on the compactness and moment-of-inertia sequence.

The analysis should be interpreted not as a substitute for full rotating GR simulations, but as the quantitatively controlled first stage of predictions.
The observables shown hereafter were not all computed at the same level of approximation.
The mass, radius, compactness, Love number, tidal deformability and dimensionless moment of inertia were obtained directly from a common relativistic pipeline based on the TOV equations, the Hinderer tidal perturbation equation and the first-order Hartle slow-rotation formalism. The spin ceiling was then estimated using a Keplerian mass-shedding estimate for the maximum angular velocity.
The spin-induced quadrupole was computed separately from the corrected second-order Hartle--Thorne expansion, yielding the reduced quadrupole \(\bar Q\) and the non-Kerr quadrupolar deviation \(\Delta q\). These data were propagated into the exterior ISCO and into formal light-ring/eikonal comparison quantities. Because the formal photon orbit lies inside the material surface, those latter quantities are not physical stellar observables. The static axial $\ell=2$ mode is instead obtained from a separate Regge--Wheeler eigenvalue problem on the TOV background. Thus the TOV--Hartle--Hinderer observables, the Hartle--Thorne geodesic diagnostics, the formal eikonal reference and the physical static axial QNM are kept as distinct approximation layers.

\subsection{Thomas--Fermi closure}
\label{subsec:GC9_TF}

For $V(s)=\Lambda_T^4 f(y)$ with $y=s/s_T$, the Thomas--Fermi energy density and pressure are
\begin{equation}
\hat\rho=yf'(y)+f(y),\qquad \hat P=yf'(y)-f(y).
\label{eq:GC9_TF_general}
\end{equation}
With $f(y)=y+y^9$ one obtains
\begin{equation}
\hat\rho=2y+10y^9,\qquad \hat P=8y^9 .
\label{eq:GC9_EOS}
\end{equation}
The sound speed is
\begin{equation}
c_s^2=\frac{d\hat P}{d\hat\rho}=\frac{72y^8}{2+90y^8}.
\label{eq:GC9_sound_speed}
\end{equation}
The two asymptotic regimes are
\begin{align}
	y\ll1:&\qquad \hat\rho\simeq2y,
	\qquad \hat P\simeq8y^9,
	\qquad c_s^2\simeq36y^8,\\
	y\gg1:&\qquad \hat\rho\simeq10y^9,
	\qquad \hat P\simeq8y^9,
	\qquad c_s^2\to4/5.
\end{align}
The high-density EOS approaches $\hat P\simeq(4/5)\hat\rho$, stiff enough to support highly compact configurations while remaining causal.

\begin{figure}[t]\centering\includegraphics[width=\columnwidth]{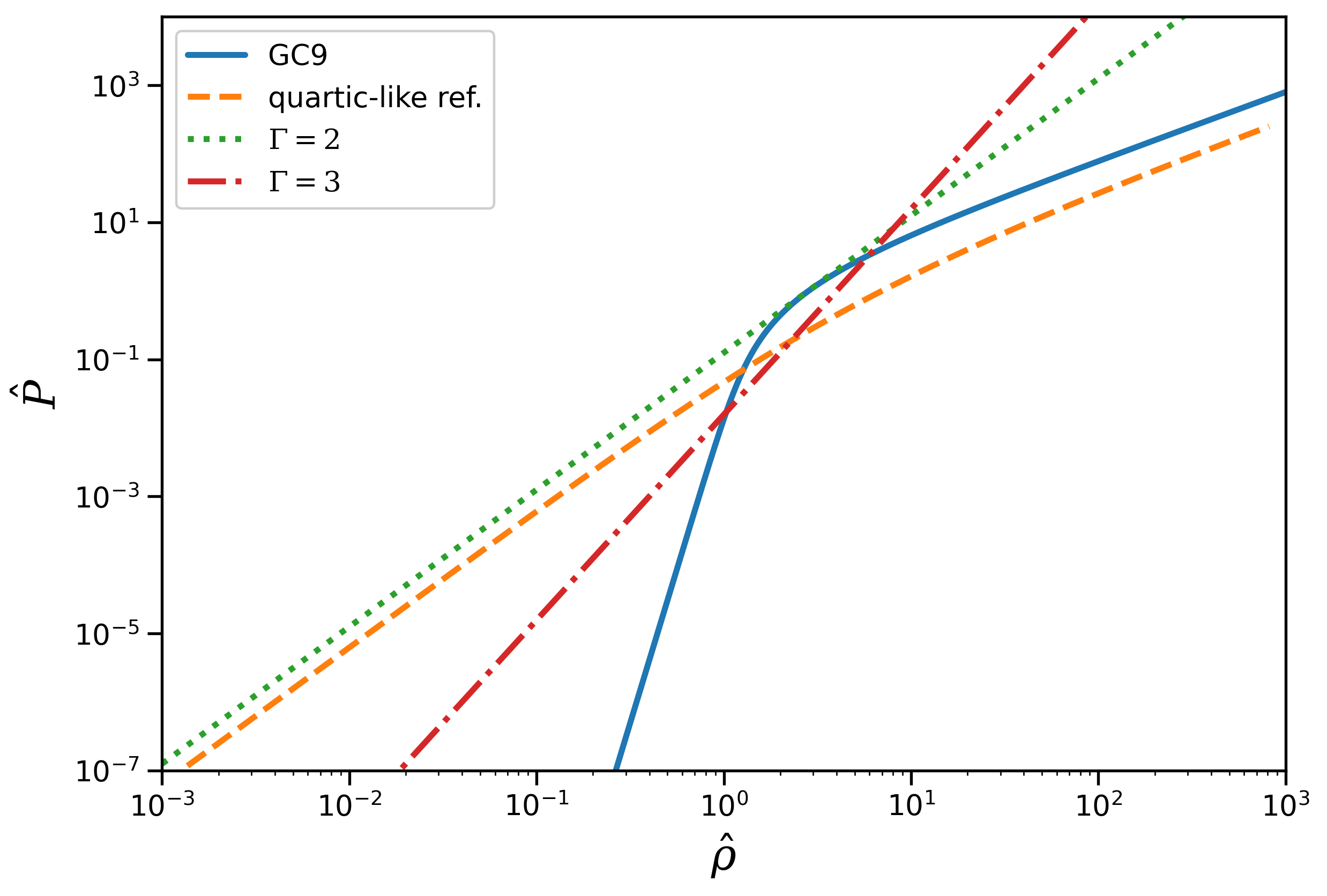}\caption{Reduced equation-of-state comparison, $\hat P(\hat\rho)$, between the GC9 closure, the quartic boson-star benchmark, and representative relativistic $\Gamma=2$ and $\Gamma=3$ polytropic references. GC9 is substantially stiffer at high density while retaining a smooth, bag-free and causal asymptote.}\label{fig:GC9_EOS}\end{figure}

\begin{figure}[t]\centering\includegraphics[width=\columnwidth]{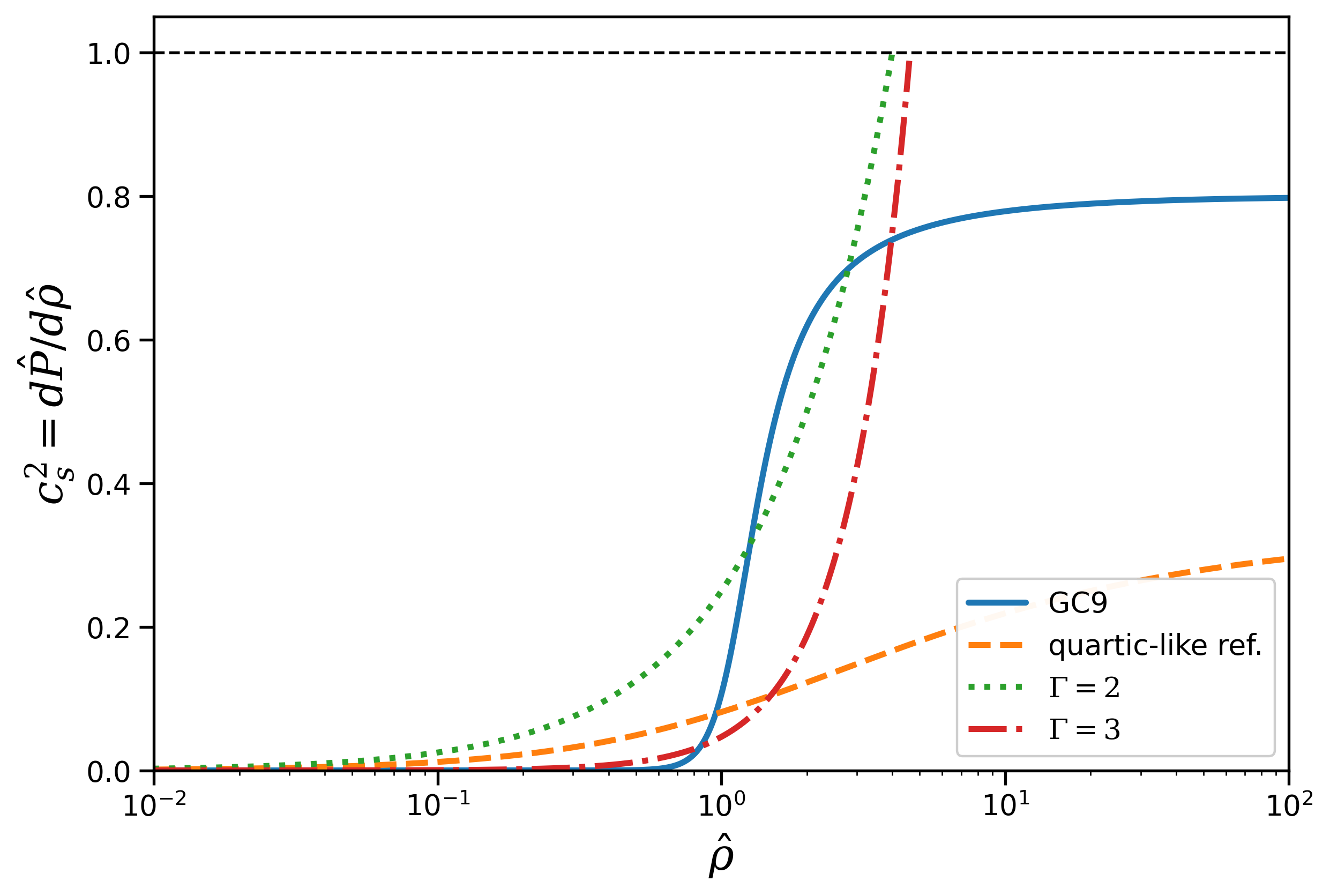}\caption{Reduced sound-speed comparison, $c_s^2(\hat\rho)$. The horizontal dashed line marks the causal limit $c_s^2=1$. The quartic benchmark asymptotes to $c_s^2\to1/3$, whereas GC9 approaches the stiffer but still causal limit $c_s^2\to4/5$. The representative $\Gamma=2$ and $\Gamma=3$ polytropic reference curves are truncated when they reach the causal bound and are not continued into their superluminal regime.}\label{fig:GC9_sound_speed}\end{figure}

Figs.~\ref{fig:GC9_EOS}--\ref{fig:GC9_ILove} compare the GC9 closure with the same quartic-like benchmark and representative relativistic $\Gamma=2$ and $\Gamma=3$ polytropic references used in the GC6 draft. In the sequence plots, solid segments denote the stable branch up to the maximum-mass turning point, while the lighter continuation marks the unstable branch. In the tidal-response diagnostics we impose the compactness cut $C>0.1$, which removes the dilute numerically sensitive tail and isolates the compact comparison sector.


The target parameter hierarchy supplies a useful bulk expansion parameter.  The microscopic Compton length is $\ell_G\sim m_G^{-1}$, whereas the macroscopic GC9 length scale is $L_T\sim M_{\rm Pl}/\Lambda_T^2$.  A parametric measure of bulk gradient suppression is therefore
\begin{equation}
\epsilon_{\rm TF}\equiv {\Lambda_T^2\over m_G M_{\rm Pl}} .
\label{eq:GC9_epsilon_TF}
\end{equation}
For $\Lambda_T=(1.26$--$1.61)\times10^{-6}\,{\rm GeV}$ and $m_G=10^9$--$10^{14}\,{\rm GeV}$,
\begin{equation}
1.3\times10^{-45}\lesssim\epsilon_{\rm TF}\lesssim2.1\times10^{-40} .
\label{eq:GC9_epsilon_TF_range}
\end{equation}
Away from the surface, terms quadratic in smooth gradients are consequently expected to be parametrically suppressed by $\epsilon_{\rm TF}^2$.  
The estimate in Eq.~\eqref{eq:GC9_epsilon_TF_range} is a bulk scale-separation diagnostic, not a numerical convergence theorem.  
A decisive validation requires a TF-seeded stiff boundary-value solver, a virial or scaling-identity residual and a matched bulk--surface boundary-layer analysis.

\subsection{Building the Observables}
\paragraph{Static Structure: TOV Equations.}
The equilibrium structure of the star is obtained by solving the TOV equations:
\begin{align}
	\frac{dP}{dr}
	&=
	-\frac{(\rho+P)\left[m(r)+4\pi r^3 P\right]}{r\left[r-2m(r)\right]},
	\\
	\frac{dm}{dr}
	&=
	4\pi r^2 \rho(r).
\end{align}
The system is integrated outward from the center with initial conditions
\begin{equation}
	m(0)=0,\qquad \rho(0)=\rho_c,\qquad P(0)=P(\rho_c),
\end{equation}
where $\rho_c$ is the central density.
The stellar radius $R$ is defined by the vanishing of the pressure $P(R)=0$ and the total mass is $M = m(R)$.
The compactness is then
\begin{equation}
	C = \frac{M}{R}.
\end{equation}
These quantities are obtained directly from the TOV integration.
The maximum mass is determined by the usual turning-point criterion
\begin{equation}
	\frac{dM}{d\rho_c}=0,
\end{equation}
which, under the standard one-parameter equilibrium-sequence assumptions, marks the onset of the first radial instability

\paragraph{Tidal Deformability and Love Number.}
The quadrupolar tidal Love number $k_2$ is computed using the relativistic perturbation formalism developed by Hinderer.
Introducing the metric perturbation variable $y(r)$, one solves the first-order differential equation
\begin{equation}
	r\frac{dy}{dr}
	+
	y^2
	+
	F(r)\,y
	+
	r^2 Q(r)
	=
	0,
\end{equation}
where $F(r)$ and $Q(r)$ depend on the TOV background and on the EOS through the sound speed $c_s^2=dP/d\rho$.
The regularity condition at the center is $y(0)=2$.
Evaluating the perturbation variable at the stellar surface, $y_R\equiv y(R)$, the Love number is given by
\begin{equation}
	k_2
	=
	\frac{8C^5}{5}(1-2C)^2
	\frac{2+2C(y_R-1)-y_R}{\mathcal D(C,y_R)},
\end{equation}
where $\mathcal D(C,y_R)$ is the standard denominator in the Hinderer formula.
The dimensionless tidal deformability is then
\begin{equation}
	\Lambda = \frac{2}{3}\frac{k_2}{C^5}.
\end{equation}
Thus, both $k_2$ and $\Lambda$ are computed directly from the relativistic perturbation equations.

\paragraph{Moment of Inertia: Slow-Rotation Formalism (Hartle--Thorne).}
The moment of inertia is obtained using the Hartle slow-rotation expansion.
Defining the frame-dragging function $\bar{\omega}(r)=\Omega-\omega(r)$, one solves
\begin{equation}
	\frac{1}{r^4}\frac{d}{dr}
	\left(
	r^4 \mathfrak{j}(r)\frac{d\bar\omega}{dr}
	\right)
	+
	\frac{4}{r}\frac{d\mathfrak{j}}{dr}\bar\omega
	=
	0,
\end{equation}
where
\begin{equation}
	\mathfrak{j}(r)=e^{-(\nu+\lambda)/2}
\end{equation}
is determined from the TOV background.
Matching to the exterior solution yields the moment of inertia
\begin{equation}
	I
	=
	\frac{R^4}{6}
	\left.
	\frac{d\bar\omega/dr}{\Omega}
	\right|_{r=R}.
\end{equation}
The dimensionless moment of inertia is
\begin{equation}
	\bar I = \frac{I}{M^3}.
\end{equation}

\paragraph{Angular Momentum and Spin Parameter.}
At first order in rotation, the angular momentum is given by
\begin{equation}
	J = I\,\Omega.
\end{equation}
However, the maximum allowed angular velocity $\Omega_{\max}$ is not obtained from the slow-rotation formalism. Instead, we estimate it using a mass-shedding (Keplerian) approximation,
\begin{equation}
	\Omega_K \simeq \kappa_{\rm ms}\sqrt{\frac{M}{R^3}},
\end{equation}
where $\kappa_{\rm ms} \sim \mathcal{O}(1)$ accounts for relativistic corrections.
The corresponding spin ceiling is then estimated as
\begin{equation}
	J_{\max} \simeq I\,\Omega_K,
\end{equation}
and the dimensionless spin parameter becomes
\begin{equation}
	j_{\max}
	=
	\frac{J_{\max}}{M^2}
	\simeq
	\kappa_{\rm ms}\,\bar I\,C^{3/2}.
	\label{eq:jmax_formula}
\end{equation}
This expression provides a consistent estimate of the rotational support based on quantities computed within the relativistic framework. To increase $j_{\max}$ further, one can also postulate superfluid features capable of enhancing rotational features.

\paragraph{Anisotropy from vortex superfluidity.}
A rotating superfluid generically forms a vortex lattice and, after coarse graining, may develop an effective
stress anisotropy with tangential pressure larger than radial pressure,
\begin{equation}
	\Delta \equiv P_t-P_r >0 .
\end{equation}
At the phenomenological level, this effect can be incorporated through the anisotropic TOV equation of
Bowers and Liang~\citep{BowersLiang1974},
\begin{equation}
	\dv{P_r}{r}
	=
	-\frac{(\rho+P_r)(M+4\pi r^3 P_r)}{r(r-2M)}
	+\frac{2}{r}\Delta .
\end{equation}
To model the rotational origin of the anisotropy, we adopt a simple GR-aware closure that vanishes in the
nonrotating limit,
\begin{equation}
	\Delta(r)
	=
	\alpha_{\rm vort}\,(\rho+P_r)\,
	\frac{\Omega^2 r^2}{1-2M/r}\,\frac{r}{R},
	\qquad
	\alpha_{\rm vort}\sim 0.1\text{--}0.3,
	\label{eq:delta}
\end{equation}
where \(\alpha_{\rm vort}\) parametrizes the strength of the coarse-grained vortex-induced stress. This closure
should be understood as an effective ansatz rather than as a first-principles derivation of the superfluid
stress tensor. Positive anisotropy of this type can increase the attainable compactness and moment of inertia
and therefore may raise the spin ceiling relative to the isotropic GC9 branch. Any such enhancement must, however, be checked against the absence of
ergoregions, inner stable light rings and other pathologies of the resulting ultra-compact configurations
\citep{Maggio2018,DiFilippo2024}.

\subsection{Second-order Rotation and Quadrupole Moment}
\label{sec:hartle_thorne_second_order}
To fully characterize the Kerr-mimicking properties of the boson star, the spin-induced quadrupole moment $Q$ must be computed by extending the slow-rotation expansion to second order in the angular velocity. The practical implementation adopted in this work is summarized below, while the detailed interior--exterior matching is described in the methodological discussion that follows.

\paragraph{Second-order Hartle--Thorne formalism and quadrupole extraction.}
To compute the spin-induced quadrupole moment $Q$ consistently, the slow-rotation expansion must be extended to second order in the angular velocity $\Omega$, following the Hartle formalism and its Hartle--Thorne exterior matching. At first order, one solves only for frame dragging and obtains $I$ and $J=I\Omega$. The quadrupole requires the even-parity $\ell=2$ sector at $\mathcal O(\Omega^2)$, which is precisely the content of the second-order Hartle equations.
The metric is written as
\begin{align}
	& ds^2 = -e^{\nu(r)}\left[1+2h_0(r)+2h_2(r)P_2(\cos\theta)\right]dt^2 \\
	& + e^{\lambda(r)}\left[1+\frac{2m_0(r)+2m_2(r)P_2(\cos\theta)}{r-2M(r)}\right]dr^2 \\
	& + r^2\left[1+2K_2(r)P_2(\cos\theta)\right]
	\left[d\theta^2+\sin^2\theta\left(d\phi-\omega(r)dt\right)^2\right],
\end{align}
where $\omega(r)=\mathcal O(\Omega)$ and $\{h_0,h_2,m_0,m_2,K_2\}=\mathcal O(\Omega^2)$.
The background functions $M(r)$, $\nu(r)$ and $\lambda(r)$ are supplied by the TOV solution. At first order in $\Omega$, one solves the standard aforementioned frame-dragging equation for $\bar\omega(r)=\Omega-\omega(r)$.
Matching to the exterior gives the angular momentum $J$ and the moment of inertia $I=J/\Omega$.

At second order, one solves the coupled $\ell=0$ and $\ell=2$ even-parity system. The $\ell=0$ sector $\{m_0,h_0\}$ determines the rotational correction to the total mass and the spherical expansion of the star. The $\ell=2$ sector $\{h_2,m_2,k_2\}$ determines the rotational quadrupolar deformation. The explicit field equations are lengthy and are standard in the Hartle treatment; modern high-order re-derivations also recover the same second-order structure as the first nontrivial rotational order.

The exterior vacuum solution is matched to the Hartle--Thorne metric. The asymptotic expansion of $g_{tt}$ reads
\begin{equation}
g_{tt}^{\rm ACMC}
=-1+\frac{2M}{r}
+\frac{2Q}{r^{3}}P_2(\cos\theta)
+\mathcal O(r^{-4}),
\end{equation}
so the quadrupole moment is extracted by isolating the coefficient of the $P_2(\cos\theta)/r^3$ term in the matched exterior solution. Equivalently, one may write
\begin{equation}
	Q = -\frac{J^2}{M} - \frac{8}{5}K M^3,
\end{equation}
where $K=C_{22}^{\rm ext}$ is the integration constant appearing in the exterior $\ell=2$ solution after matching, for which $Q_{\rm Kerr}=-J^2/M$.
It is convenient to define the dimensionless quadrupole parameter
\begin{equation}
	\bar Q \equiv -\frac{Q M}{J^2},
\end{equation}
which satisfies $\bar Q_{\rm Kerr}=1$ for a Kerr black hole. The quantity
\begin{equation}
\delta\overline Q\equiv\overline Q-1 .
\label{eq:delta_Qbar}
\end{equation}
therefore provides a direct measure of the departure from Kerr multipole structure.
The practical computation pipeline is thus $
P(\rho)\;\to\;\{M(r),\nu(r),\lambda(r)\}\;\to\;\bar\omega(r)\;\to\;\{h_0,h_2,m_0,m_2,k_2\}\;\to\;\{J,Q\}$.

\subsection{Results: reduced sequence and first-order observables}
\label{subsec:GC9_first_order_results}

The reduced sequence has maximum mass
\begin{equation}M_{\max}^{\rm red}=0.064157,\qquad \hat\rho_c(M_{\max})=4.691700,
\end{equation}
with
\begin{equation}R_{\rm red}(M_{\max})=0.197591,\qquad C(M_{\max})=0.324694.
\end{equation}
The maximum compactness along the branch is
\begin{equation}C_{\max}=0.337118,\qquad \hat\rho_c(C_{\max})=10.386752.\end{equation}
At the maximum-mass point,
\begin{equation}k_2=0.0236672,\qquad \Lambda=4.37208,\qquad \bar I=5.116358.
\end{equation}
The static and first-order calculation directly fixes the combination
\begin{equation}
\bar I C^{3/2}=0.9466.
\end{equation}
A mass-shedding estimate would instead be
\begin{equation}
j_{\max}^{\rm proxy}=\kappa_{\rm ms}\,\bar I\,C^{3/2},
\end{equation}
where $\kappa_{\rm ms}$ is EOS- and rotation-model dependent.  The illustrative choice $\kappa_{\rm ms}=1$ gives $j_{\max}^{\rm proxy}\simeq0.95$, but this is not a computed spin ceiling; ordinary fluid-star calibrations would give a smaller value, whereas rapidly rotating boson-star families can reach $j\gtrsim1$.  Only a two-dimensional rotating GC9 sequence can determine the relevant $\kappa_{\rm ms}$.
All quantities are summarized in Table \ref{tab:GC9_first_order}.

\begin{figure}[t]\centering\includegraphics[width=\columnwidth]{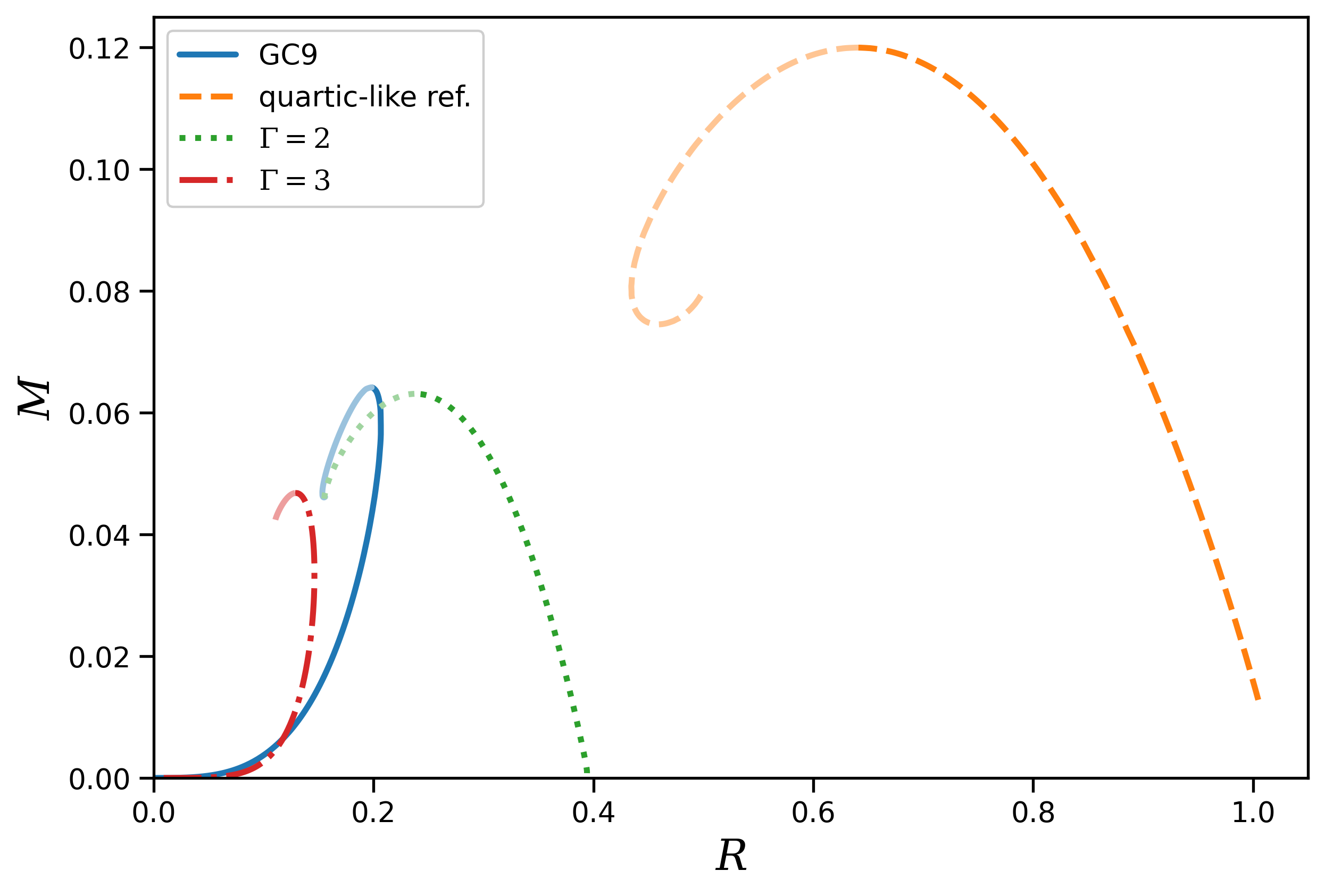}\caption{Reduced mass--radius sequences for the GC9 closure, the quartic boson-star benchmark, and representative relativistic $\Gamma=2$ and $\Gamma=3$ polytropic references. Solid segments indicate the stable branch up to the maximum-mass turning point, while the lighter continuation shows the unstable branch.}\label{fig:GC9_mass_radius}\end{figure}

\begin{figure}[t]\centering\includegraphics[width=\columnwidth]{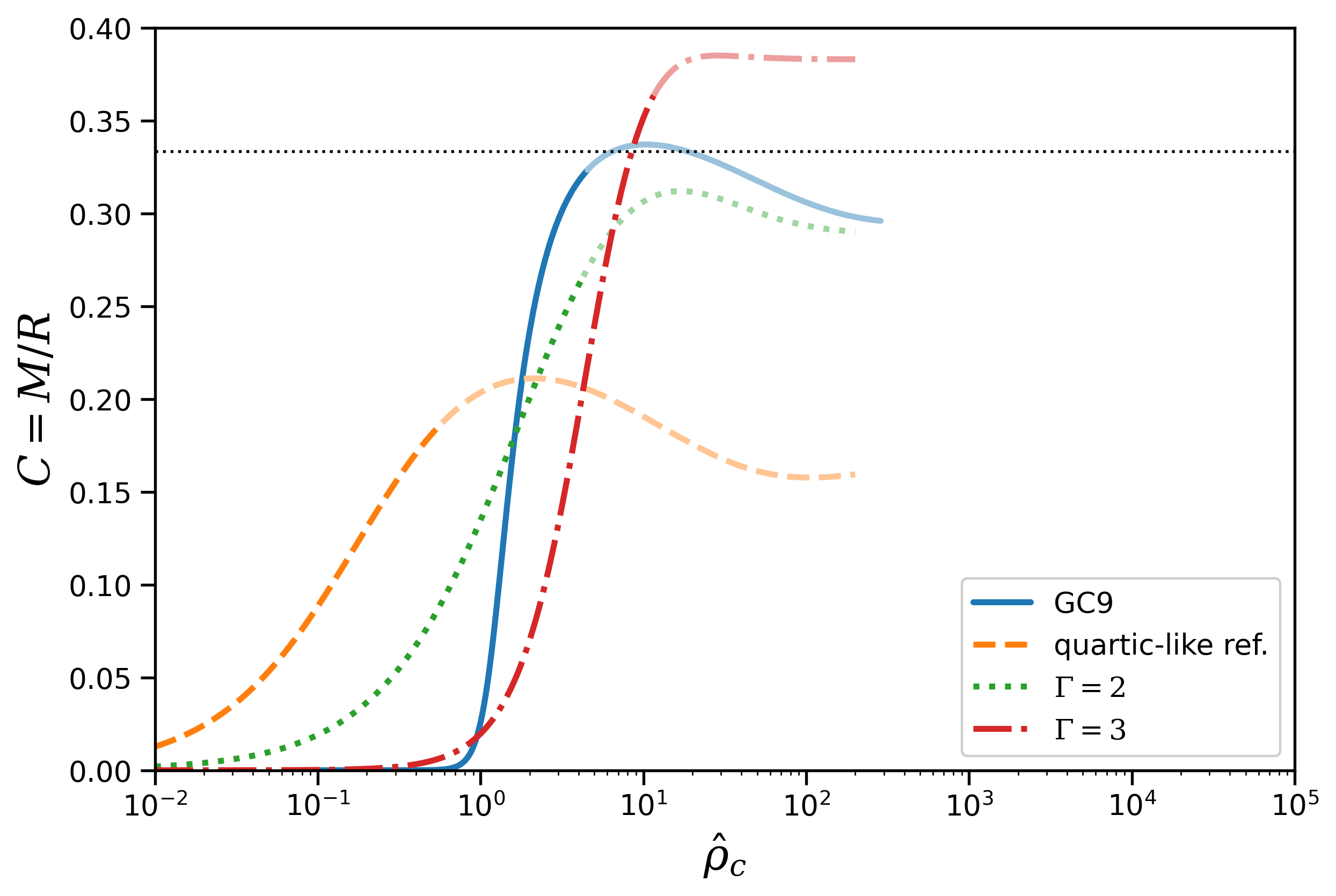}\caption{Reduced compactness $C=M/R$ as a function of central density for the same families shown in Fig.~\ref{fig:GC9_mass_radius}. The horizontal dotted line marks the photon-ring threshold $C=1/3$. The stable GC9 sequence approaches the photon-sphere threshold from below; the post-turning continuation crosses it and reaches $C=0.3371$. The representative quartic and polytropic benchmarks remain well below this threshold on their stable branches. The solid segment is the branch up to the first maximum-mass turning
		point, whose endpoint has $C^{\rm stable}_{\rm max}=0.3247$.  The lighter
		continuation is post-turning-point and contains the larger value
		$C^{\rm sequence}_{\rm max}=0.3371$.}\label{fig:GC9_compactness}\end{figure}

\paragraph{Mass--radius structure and compactness hierarchy.}
The resulting mass--radius relations are shown in Fig.~\ref{fig:GC9_mass_radius}, while the compactness evolution as a function of central density is shown in Fig.~\ref{fig:GC9_compactness}. Quartic and $\Gamma=2$ polytropic solutions remain far below the compactness required for realistic Kerr mimicking.
The GC9 family approaches $C \simeq \frac13$,
namely the spherical photon-sphere compactness threshold, while remaining below it on the stable branch. This shows that the GC9 sequence is not simply ``more compact than quartic,'' but lies near the photon-sphere compactness scale with an asymptotic sound speed $c_s^2=4/5$ below the maximally stiff causal value.

\begin{figure}[t]\centering\includegraphics[width=\columnwidth]{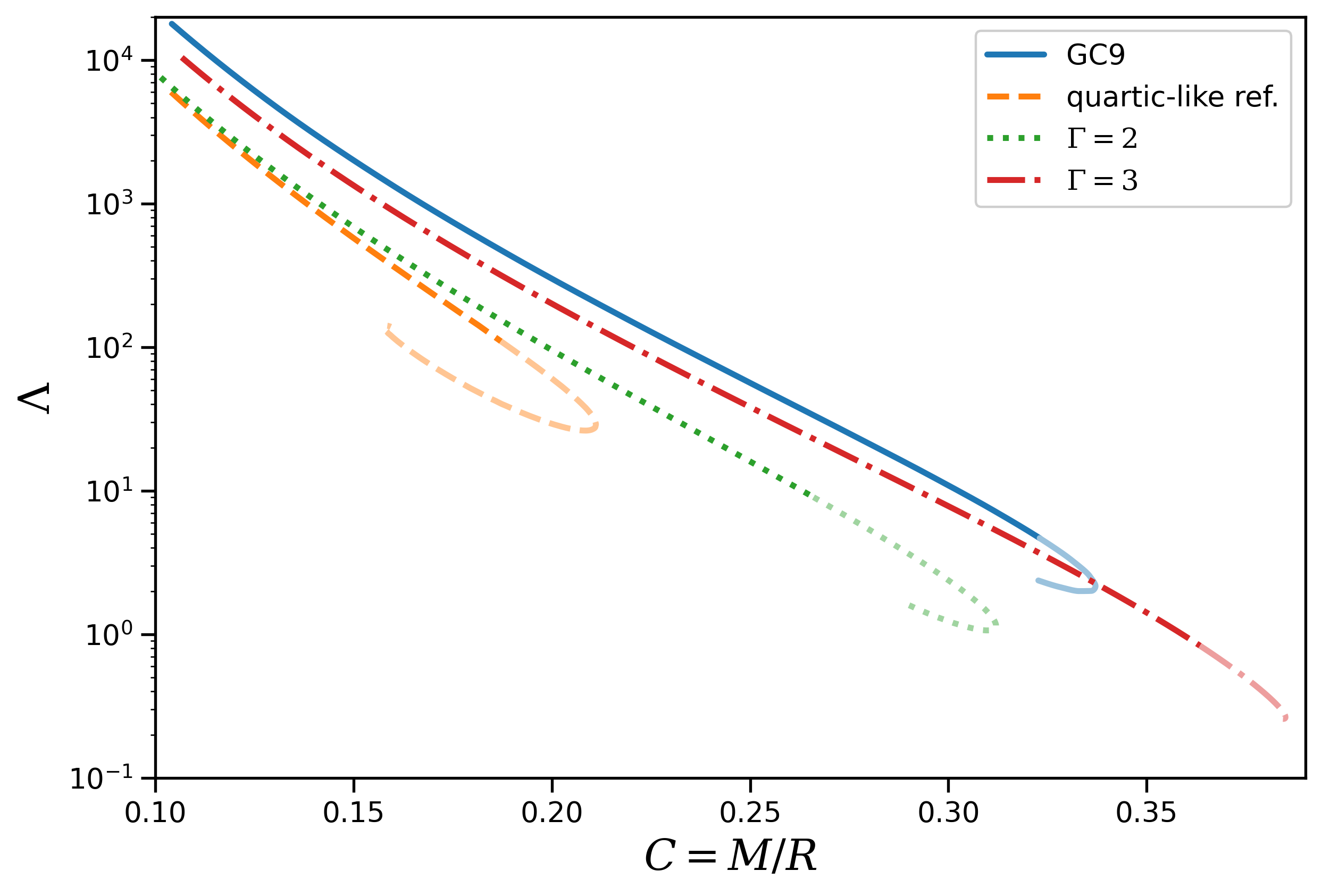}\caption{Dimensionless tidal deformability $\Lambda$ as a function of compactness for GC9, the quartic-like benchmark and representative $\Gamma=2$ and $\Gamma=3$ polytropic references, restricted to $C>0.1$. Solid segments end at the first maximum-mass turning point; lighter continuations are post-turning.}\label{fig:GC9_lambda_compactness}\end{figure}

\begin{figure}[t]\centering\includegraphics[width=\columnwidth]{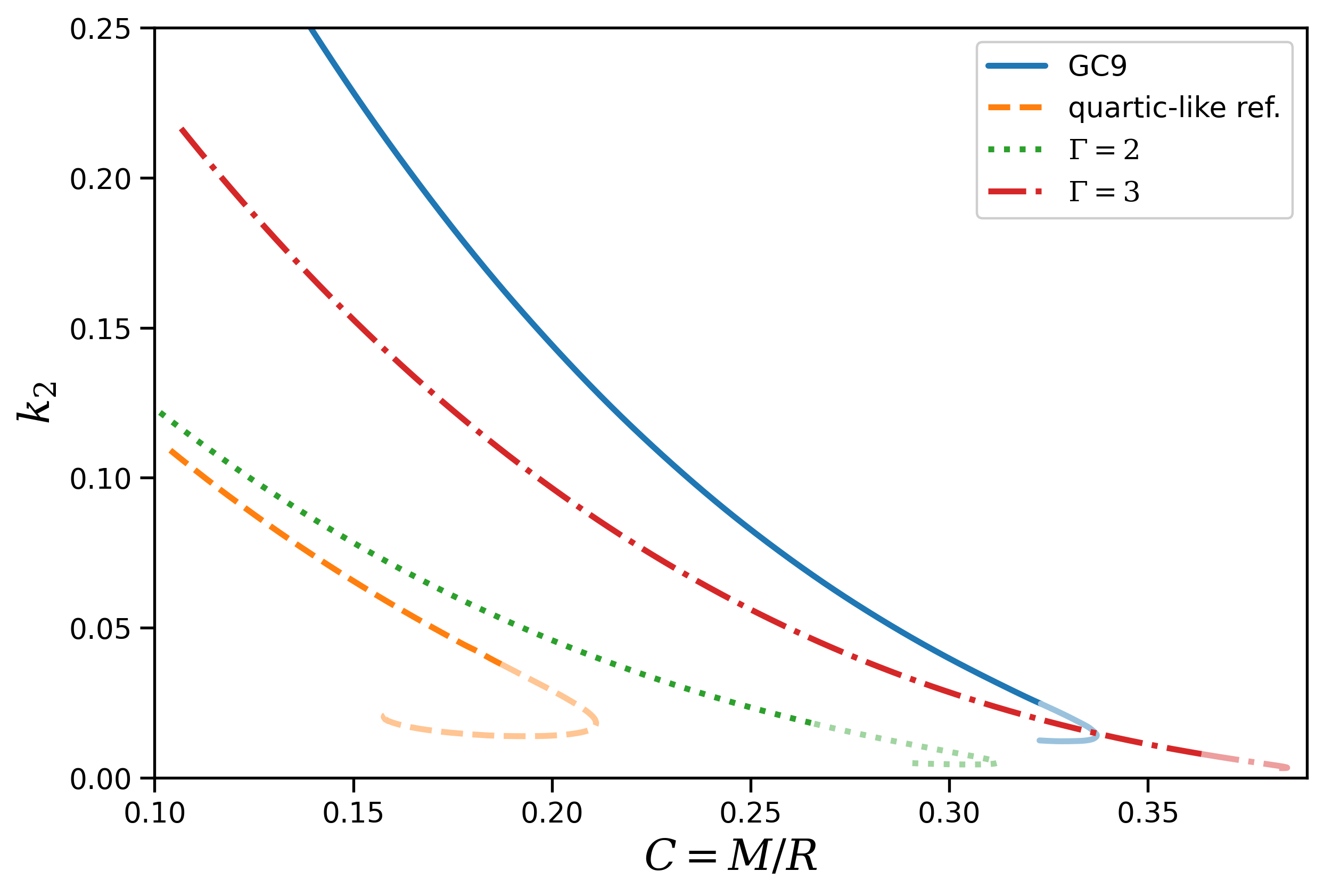}\caption{Quadrupolar Love number $k_2$ as a function of compactness for the same four families, restricted to $C>0.1$. The GC9 branch enters the low-$k_2$, high-compactness corner most relevant for restricted Kerr-mimicking phenomenology.}\label{fig:GC9_k2_compactness}\end{figure}

\begin{figure}[t]\centering\includegraphics[width=\columnwidth]{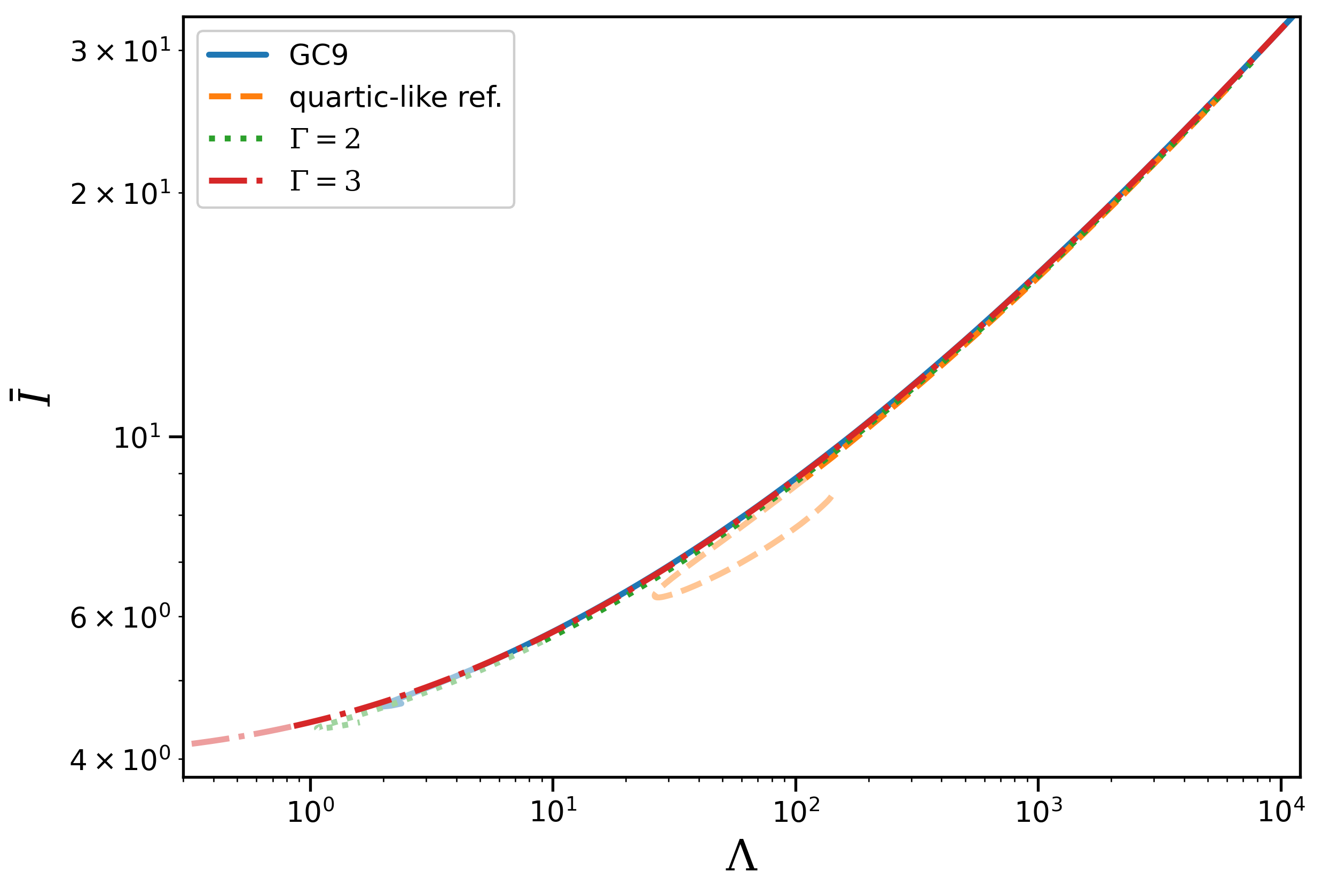}\caption{Reduced I--Love relation, $\bar I(\Lambda)$, for the compact portions $C>0.1$ of GC9 and the three reference families. The common axes and line conventions reproduce the comparison style of the GC6 draft.}\label{fig:GC9_ILove}\end{figure}

\paragraph{Tidal response and I--Love behavior.}
Compactness alone is not sufficient: a realistic Kerr mimicker must also suppress its tidal response. This is tested through the Love number $k_2$ and the dimensionless tidal deformability $\Lambda$, shown in Figs. \ref{fig:GC9_lambda_compactness}, \ref{fig:GC9_k2_compactness} and \ref{fig:GC9_ILove}.
The $\Lambda$ improvement relative to the quartic benchmark is dramatic.
The GC9 family thus reduces the tidal deformability by nearly two orders of magnitude while simultaneously increasing the compactness toward, but not into, the external photon-ring regime.

\begin{table*}\caption{First-order reduced GC9 observables. The tabulated spin endpoint is a heuristic proxy and is rounded accordingly.}\label{tab:GC9_first_order}\begin{ruledtabular}\begin{tabular}{cccccccccc}$M_{\max}^{\rm red}$ & $\hat\rho_c(M_{\max})$ & $R_{\rm red}$ & $C(M_{\max})$ & $C_{\max}$ & $k_2$ & $\Lambda$ & $\bar I$ & $j_{\max}^{\rm proxy}$ & $c_{s,\infty}^2$\\\hline 0.064157 & 4.691698 & 0.197591 & 0.324694 & 0.337118 & 0.023667 & 4.372078 & 5.116358 & 0.95 & 0.800000\end{tabular}\end{ruledtabular}\end{table*}

\paragraph{Heuristic high-spin target.}
Varying the uncalibrated order-unity coefficient over the illustrative interval $\kappa_{\rm ms}=0.9$--$1.1$ gives approximately $j_{\max}^{\rm proxy}\simeq0.85$--$1.04$. The choice $\kappa_{\rm ms}=1$ is not calibrated to a two-dimensional GC9 solution. This number is only an order-of-magnitude target and should not be assigned precision beyond $j_{\max}^{\rm proxy}\sim0.9$. We note that $\kappa_{\rm ms}=1$ is not simply an agnostic midpoint of the illustrative interval above: the empirical Kepler-frequency prefactor for ordinary fluid stars is closer to $\kappa_{\rm ms}\sim0.6$--$0.7$, which would instead favor $j_{\max}^{\rm proxy}\sim0.6$. The higher target adopted here is grounded not in the fluid-star analogy but in the rotating boson-star literature itself, where equal-frequency ($J=mN$) sequences routinely reach $j\gtrsim1$ \citep{Ryan1997Spinning,HerdeiroRadu2017}. Existing two-dimensional rotating boson-star families exhibit useful relations among moment of inertia, spin and quadrupole, but those relations do not calibrate the present EOS without an actual rotating GC9 sequence \citep{AdamEtAl2022RotatingUniversal}.

The spin hierarchy does not explicitly assume vortex-induced superfluid corrections. Such effects may be incorporated phenomenologically as
\begin{equation}
	j_{\max}^{\rm sf}\simeq j_{\max}(1+\delta_{\rm sf}),
	\qquad
	\delta_{\rm sf}\sim 0.05\text{--}0.20,
\end{equation}
depending on the strength of the anisotropic vortex stress.
For GC9 configurations, for which the uncalibrated proxy gives $j_{\max}^{\rm proxy}\sim0.95$, this would formally lift the proxy toward $j\sim1$, but it does not establish a stable extremal rotating solution, whereas softer BS families remain far below the Kerr range even after such corrections.

Electromagnetic spin measurements using reflection spectroscopy and continuum fitting have been reported for roughly a few dozen stellar-mass BHs and a comparable number of SMBHs, while current gravitational-wave measurements remain less constraining on spin magnitudes and orientations. This makes the high-spin end of the GC9 branch especially relevant phenomenologically and motivates testing whether a full rotating sequence can access the near-extremal regime inferred for the fastest astrophysical candidates, without invoking the extreme couplings needed in conventional quartic models. 
Stellar-mass BH candidates have reported near-extremal spins from continuum-fitting and reflection modeling. For instance, continuum-fitting analyses yield \(a_\ast \gtrsim 0.98\) in GRS~1915+105 and \(a_\ast \gtrsim 0.983\) in Cygnus~X-1, as conservative lower bounds in those works \citep{McClintock2006GRS1915,Gou2014CygX1}. Analogous SMBH spin estimates from X-ray reflection spectroscopy are also often high, although substantially more model-dependent in detail \citep{Reynolds2013Spin,Reynolds2020Spin}.


\subsection{Results: Second-order Hartle--Thorne quadrupole}
\label{subsec:GC9_second_order}
To compute the quadrupole moment beyond phenomenological scaling arguments, we implemented the full second-order Hartle--Thorne system using the approach developed in \citep{c3jx-5487}. Starting from the TOV background for a given GC9 closure, we first solved the \(O(\Omega)\) frame-dragging equation and extracted \(J\) by matching to the asymptotically flat exterior solution, then the full \(O(\Omega^2)\), \(\ell=2\) interior sector, including the auxiliary pressure-displacement contribution, using the exact regular-center expansion rather than a naive direct integration. The interior solution was written in a regular basis and matched linearly at the stellar surface to the Hartle--Thorne exterior quadrupolar metric, thereby determining the exterior constant \(C_{22}^{\rm ext}\). The mass quadrupole was then obtained from
\begin{equation}
	Q = -\frac{J^2}{M} - \frac{8}{5}M^3\,C_{22}^{\rm ext},
\end{equation}
and summarized through the reduced quantity \(\bar Q=-QM/J^2\). Technical details are discussed in Appendix A.

The GC9 closure gives
\begin{equation}
 J=1.351094\times10^{-3},\qquad
 C_{22}^{\rm ext}=0.0351243,
\end{equation}
and
\begin{equation}
 Q_{\rm GC9}=-4.32937\times10^{-5},
 \qquad
 \bar Q_{\rm GC9}=1.52158.
 \label{eq:GC9_Qbar}
\end{equation}
With
\begin{equation}
 Q_{\rm Kerr}=-\frac{J^2}{M}=-2.84531\times10^{-5},
\end{equation}
one obtains
\begin{equation}
 \Delta Q=-1.48406\times10^{-5},
 \qquad
 \frac{\Delta Q}{M^3}=-5.6199\times10^{-2}.
\end{equation}
Thus the GC9 benchmark is about $52\%$ more oblate than Kerr at fixed
$(M,J)$. The result is stable under variation of the regular-center
starting radius over $r_0/R=10^{-4}$--$3\times10^{-6}$ at the
$10^{-7}$ level in $\bar Q$. As an independent scale check, inserting
$\bar Q=1.52158$ into the standard slow-rotation I--Q fit gives
$\bar I\simeq5.10$, within about $0.4\%$ of the directly integrated
$\bar I=5.11636$ \citep{YagiYunes2013}. The quoted GC9 value uses the direct exterior normalization in Eq.~\eqref{eq:Q_HT_final}; the independent I--Q consistency check above provides the relevant normalization diagnostic.
All quantities are summarized in Table \ref{tab:GC9_quadrupole}.%
\footnote{The benchmark $j=0.328248$ labels the slow-rotation run performed on the same maximum-mass background, $\hat\rho_c(M_{\max})=4.691700$, used for $C$, $k_2$, $\Lambda$ and $\bar I$. Its numerical value reflects the chosen slow-rotation normalization and is not a physical selection principle or a mass-shedding condition. The dimensionless quadrupole $\bar Q=-QM/J^2$ is independent of the overall infinitesimal rotation normalization at this order, whereas the quoted ISCO displacement is specifically evaluated at this representative slow-row $j$.}

\begin{table}
\caption{Second-order Hartle--Thorne quadrupole estimate of the GC9 branch in the stated exterior normalization convention.}
\label{tab:GC9_quadrupole}\label{tab:gc9_quadrupole_comparison}
\centering\scriptsize
\setlength{\tabcolsep}{3pt}
\begin{ruledtabular}
\begin{tabular}{ccccc}
$j$ & $J$ & $C^{\rm ext}_{22}$ & $Q$ & $\bar Q$\\
\hline
0.328248 & $1.35109\times10^{-3}$ & 0.0351243 & $-4.32937\times10^{-5}$ & 1.52158
\end{tabular}
\end{ruledtabular}
\end{table}

\subsection{ISCO and the status of photon-region diagnostics}
\label{subsec:GC9_kerr_observables}\label{sec:gc9_layer2_kerr}

The quantities computed in the previous sections---compactness, spin, tidal deformability and second-order quadrupole moment---already show that the compact GC9 branch is not exactly Kerr, even when it remains sufficiently compact to act as a plausible horizonless BH alternative. A natural next step is therefore to translate the GC9 multipolar deviations into \emph{geodesic} and \emph{spectroscopic} observables. In the absence of a full numerical geodesic integration in the exact exterior metric, the cleanest intermediate step is to use a Hartle--Thorne / multipole-expansion treatment. This provides a controlled approximation for the innermost stable circular orbit and formal photon-region and eikonal comparison quantities. Because the corresponding orbit lies inside the GC9 surface, the latter are not physical stellar observables.

The basic quantities entering this analysis are
\begin{equation}
	j \equiv \frac{J}{M^2},
	\qquad
	\bar Q \equiv -\frac{Q\,M}{J^2},
	\qquad
	q_2 \equiv -\frac{Q}{M^3}=\bar Q\,j^2,
	\label{eq:gc9_q2_def}
\end{equation}
together with the deviation of the quadrupole from the Kerr value,
\begin{equation}
	\Delta q \equiv \frac{Q-Q_{\rm Kerr}}{M^3}
	=
	\frac{Q+J^2/M}{M^3},
	\qquad
	Q_{\rm Kerr}=-\frac{J^2}{M}.
	\label{eq:gc9_deltaq_def}
\end{equation}
For Kerr one has \(\bar Q_{\rm Kerr}=1\), \(q_2=j^2\) and \(\Delta q=0\). Therefore \(\Delta q\) measures directly how far the GC9 compact branch departs from the Kerr quadrupole at fixed \((M,J)\).

\paragraph{ISCO.}
For reference, the exact Kerr prograde ISCO radius is
\begin{equation}
	\frac{r_{\rm ISCO}^{\rm Kerr}}{M}
	=
	3+Z_2-\sqrt{(3-Z_1)(3+Z_1+2Z_2)},
	\label{eq:kerr_isco_exact}
\end{equation}
with
\begin{equation}
	\begin{split}
		&Z_1=1+(1-j^2)^{1/3}\left[(1+j)^{1/3}+(1-j)^{1/3}\right],\\
		&Z_2=\sqrt{3j^2+Z_1^2}.
	\end{split}
\end{equation}
To estimate the GC9 ISCO in the slow-rotation / quadrupole expansion we use the standard Hartle--Thorne compact-star form \citep{ShibataSasaki1998ISCO,BertiStergioulas2004Match}
\begin{equation}
	\frac{r_{\rm ISCO}^{\rm HT2}}{M}
	\simeq
	6\left(
	1
	-0.54433\,j
	-0.22619\,j^2
	+0.17989\,q_2
	\right),
	\label{eq:ht_isco2}
\end{equation}
which contains the full \(O(j^2,q_2)\) correction. To explore the size of the next-order effect while staying within the available multipole data, we also quote the partial cubic extension
\begin{equation}
	\begin{split}
		\frac{r_{\rm ISCO}^{\rm HT3}}{M} &\simeq
		6(1- 0.54433\,j -0.22619\,j^2 +0.17989\,q_2 \\
		& -0.23002\,j^3 +0.26296\,j\,q_2),
		\label{eq:ht_isco3}
	\end{split}
\end{equation}
where the omitted terms would require higher multipoles not yet computed for GC9.

\paragraph{Light ring and shadow proxy.}
The equatorial prograde Kerr light-ring radius is
\begin{equation}
	\frac{r_{\rm LR}^{\rm Kerr}}{M}
	=
	2\left[
	1+\cos\!\left(\frac{2}{3}\arccos(-j)\right)
	\right].
	\label{eq:kerr_lr_exact}
\end{equation}
As a first non-Kerr proxy for the GC9 light ring we use the slow-rotation Kerr--\(Q\) / Hartle--Thorne estimate \citep{AbramowiczEtAl2003HTGeo}
\begin{equation}
	\frac{r_{\rm LR}^{\rm GC9}}{M}
	\simeq
	3
	-\frac{3}{2}\Delta q
	-\frac{2j}{\sqrt{3}}
	-\frac{2j^2}{9}.
	\label{eq:gc9_lr_proxy}
\end{equation}
Although this is not a full null-geodesic integration, it captures the leading displacement of the equatorial photon orbit induced by the excess quadrupole. Since the shadow edge in compact rotating spacetimes is strongly controlled by the unstable photon region, Eq.~\eqref{eq:gc9_lr_proxy} should be interpreted only as a first \emph{shadow-scale proxy}, not as a full shadow computation.

\paragraph{Formal eikonal reference.}
In the eikonal limit of a vacuum photon-region problem, the real and imaginary parts of high-multipole modes are related to the photon-ring orbital frequency and Lyapunov exponent. Here the corresponding orbit lies inside the material surface, so the following expressions are retained only as formal exterior comparison quantities. Defining
\begin{equation}
	c_1\equiv -16+15\ln 3,
	\qquad
	\tilde q \equiv \frac{5}{8}\Delta q,
	\label{eq:c1_qtilde}
\end{equation}
the corresponding Hartle--Thorne light-ring proxies are \citep{AbramowiczEtAl2003HTGeo}
\begin{equation}
	M\Omega_{\rm LR}^{\rm GC9}
	\simeq
	\frac{1}{3\sqrt{3}}
	\left(
	1
	-\frac12\,\tilde q\,c_1
	+\frac{2j}{3\sqrt{3}}
	+\frac{11}{54}j^2
	\right),
	\label{eq:gc9_omega_lr}
\end{equation}
and
\begin{equation}
	M\lambda_{\rm LR}^{\rm GC9}
	\simeq
	\frac{1}{3\sqrt{3}}
	\left(
	1
	+\tilde q\,c_1
	-\frac{2}{27}j^2
	\right).
	\label{eq:gc9_lambda_lr}
\end{equation}
Equation~\eqref{eq:gc9_omega_lr} estimates the formal frequency shift and Eq.~\eqref{eq:gc9_lambda_lr} the formal damping scale of a light-ring-dominated exterior mode. They are not GC9 stellar QNM predictions because the inferred orbit is internal to the surface. The physical static axial mode and the matter-supported spectrum are computed separately below.

\paragraph{Numerical results.}
For the benchmark slow row,
\begin{equation}
q_2=\bar Qj^2=0.163946,
\qquad
\Delta q=j^2(1-\bar Q)=-0.056199.
\end{equation}
The exact Kerr ISCO at the same spin is
\begin{equation}
\frac{r_{\rm ISCO}^{\rm Kerr,BL}}{M}=4.877207,
\qquad
\frac{R_{\rm circ}^{\rm Kerr}}{M}=4.892758,
\label{eq:kerr_isco_circ}
\end{equation}
where the first value is the Boyer--Lindquist radius and the second is the
equatorial circumferential radius. The second-order Hartle--Thorne
formula gives
\begin{equation}
\begin{aligned}
r_{\rm ISCO}^{\rm HT2}(q_2{=}j^2)/M&=4.898016,\\
r_{\rm ISCO}^{\rm HT2}({\rm GC9})/M&=4.958673.
\end{aligned}
\end{equation}
The coordinate-consistent same-order displacement is
\begin{equation}
\Delta_{\rm ISCO}^{\rm same\text{-}order}=1.238\%,
\end{equation}
while comparison with the exact Kerr circumferential radius gives
$1.347\%$. The residual $0.107\%$ difference between the same-order
Kerr expansion and exact Kerr is the Hartle--Thorne truncation offset.
At fixed $\bar Q=1.52158$ the formal fixed-denominator scaling is
\begin{equation}
\Delta_{\rm ISCO}^{\rm formal,fixed\text{-}den}(j)
\simeq11.51\%\,j^2,
\label{eq:isco_shift_general}
\end{equation}
which would be about $9.3\%$ at $j=0.9$ but is outside the controlled
slow-rotation calculation. Retaining the displayed partial-cubic terms
while setting uncomputed higher multipoles to zero gives
\begin{equation}
r_{\rm ISCO}^{\rm cub}/M=4.994769,
\qquad
\Delta_{\rm ISCO}^{\rm HT3}=1.830\%,
\end{equation}
relative to the corresponding Kerr partial-cubic value $4.905006$.
The maximum-mass benchmark has $R/M=3.079827$, so the ISCO remains
safely outside the material surface.

The photon-region status is different. The maximum-mass configuration,
which is the endpoint of the radially stable branch, has
\begin{equation}
C(M_{\max})=0.324694<\frac13,
\qquad R/M=3.079827,
\end{equation}
and therefore no external Schwarzschild photon sphere. Directly
evaluating the exact interior condition
\begin{equation}
\chi(r)\equiv\frac{3m(r)}{r}+4\pi r^2P(r)=1
\label{eq:GC9_interior_LR_condition}
\end{equation}
gives
\begin{equation}
\chi_{\max}=0.975563
\quad\hbox{at}\quad r/R=0.98633
\end{equation}
for this stable endpoint. Thus the entire radially stable branch remains
light-ring free.

The post-turning continuation is qualitatively different. A degenerate
light ring first appears at
\begin{equation}
y_c=0.909105,\qquad C=0.332115,\qquad r/R=0.9677,
\label{eq:GC9_LR_onset}
\end{equation}
which is already beyond the first maximum of $M(y_c)$ and is radially
unstable. The maximum-compactness point is
\begin{equation}
\begin{aligned}
y_c&=0.981131,\qquad C_{\max}=0.337118,\\
R/M&=2.966322.
\end{aligned}
\end{equation}
with $\chi_{\max}=1.026206$ at $r/R=0.8851$; it therefore has an
external unstable photon orbit at $r=3M$ and the paired internal stable
orbit required for a smooth horizonless ultracompact configuration.
Farther along the unstable sequence, at $y_c=1.2$, the two internal
roots occur at
\begin{equation}
 r_{\rm LR}^{(1)}/R=0.356044,
 \qquad r_{\rm LR}^{(2)}/R=0.797758.
\end{equation}
Consequently GC9 has a light-ring-bearing continuation, but not a stable
long-lived light-ring star. These configurations are already radially
unstable and may occur only as transient migration or collapse stages;
establishing their dynamical role requires nonlinear evolution. Their paired
stable orbit retains the usual trapping-instability concern
\citep{CunhaBertiHerdeiro2017LRstability,CunhaHerdeiroRaduSanchisGual2023LRInstability}.

\begin{figure}[t]
	\centering
	\includegraphics[width=\columnwidth]{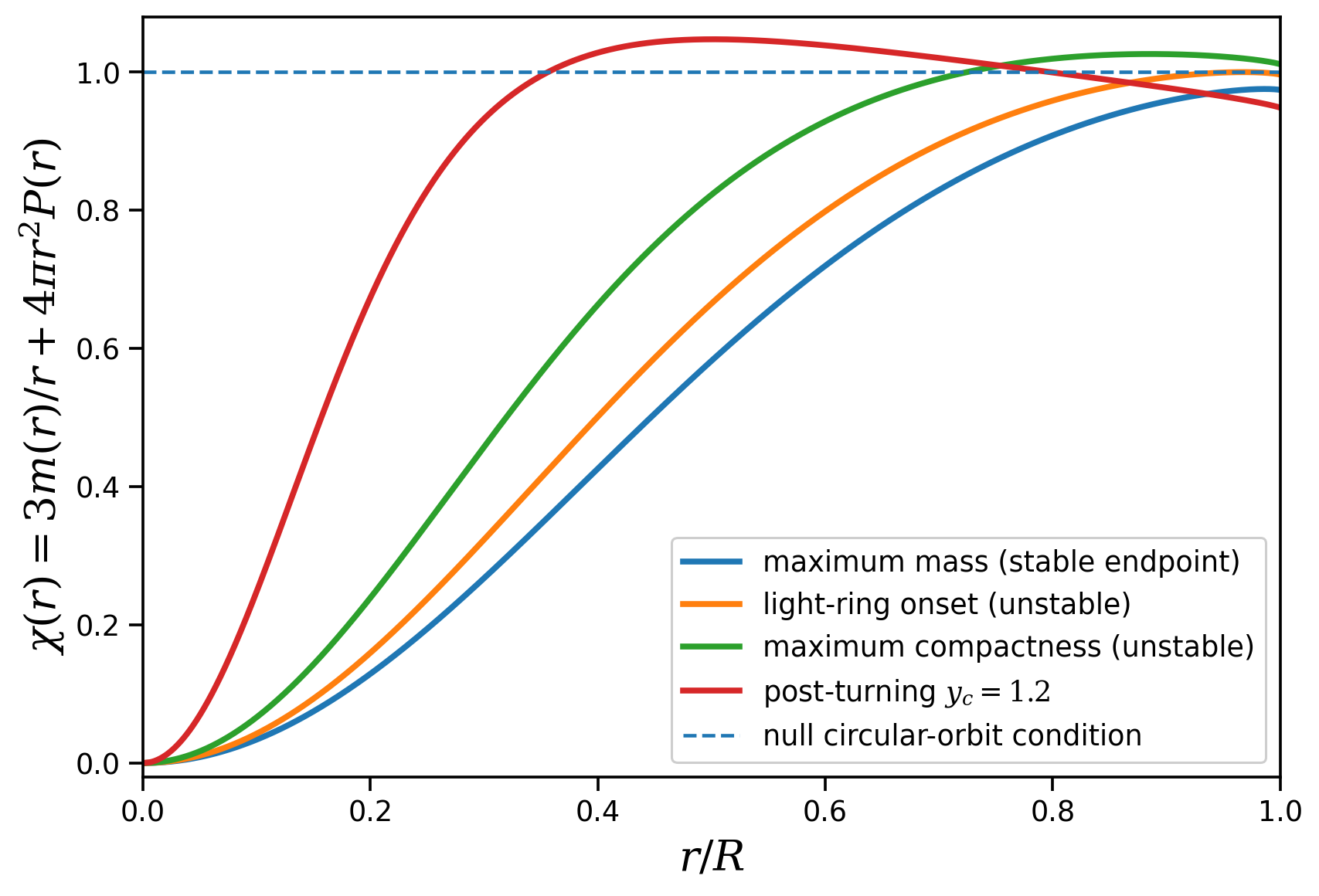}
	\caption{Interior null-orbit diagnostic
		$\chi(r)=3m(r)/r+4\pi r^2P(r)$.  A circular null orbit requires $\chi=1$.
		The maximum-mass stable endpoint remains below threshold. The light-ring onset and maximum-compactness configuration lie on the post-turning radially unstable continuation.}
	\label{fig:GC9_light_ring_diagnostic}
\end{figure}

Formally continuing the Hartle--Thorne exterior formulas inward gives
\begin{equation}
\begin{aligned}
r_{\rm LR}^{\rm formal}/M&=2.681324,\\
M\Omega_{\rm LR}^{\rm formal}&=0.222608,\\
M\lambda_{\rm LR}^{\rm formal}&=0.187675.
\end{aligned}
\label{eq:gc9_formal_lr}
\end{equation}
but $r_{\rm LR}^{\rm formal}<R$ and these quantities are therefore not physical exterior light-ring observables of the star. They are retained only as a multipolar comparison with a formally continued vacuum metric. In particular,
\begin{equation}
\begin{aligned}
M\omega_{220}^{\rm formal}
&\equiv2M\Omega_{\rm LR}^{\rm formal}
-\frac{i}{2}M\lambda_{\rm LR}^{\rm formal}\\
&=0.445217-0.093838i .
\end{aligned}
\label{eq:GC9_formal_eikonal_reference}
\end{equation}
is a reference number, not a predicted QNM of the GC9 star.

\begin{table*}
\caption{Controlled slow-row multipole and ISCO diagnostics. The stellar radius is included to verify that the ISCO lies outside the surface. $\Delta_{\rm ISCO}^{\rm HT2}$ and $\Delta_{\rm ISCO}^{\rm HT3}$ are same-order (HT-vs-HT) shifts, the coordinate-consistent comparison since Eq.~\eqref{eq:ht_isco2}--\eqref{eq:ht_isco3} give circumferential radii \citep{ShibataSasaki1998ISCO}; against the exact Kerr circumferential radius, Eq.~\eqref{eq:kerr_isco_circ}, the HT2 total is $1.347\%$. The cubic column is a partial sensitivity extension in which uncomputed higher multipoles are set to zero.}
\label{tab:GC9_Kerr_observables}\label{tab:gc9_layer2_kerr}
\begin{ruledtabular}
\begin{tabular}{cccccccc}
$j$ & $\bar Q$ & $R/M$ & $r_{\rm ISCO}^{\rm Kerr}/M$ & $r_{\rm ISCO}^{\rm HT2}/M$ & $r_{\rm ISCO}^{\rm HT3}/M$ & $\Delta_{\rm ISCO}^{\rm HT2}$ & $\Delta_{\rm ISCO}^{\rm HT3}$\\
\hline
0.328248 & 1.52158 & 3.079827 & 4.877207 & 4.958673 & 4.994769 & $1.238\%$ & $1.830\%$\\
\end{tabular}
\end{ruledtabular}
\end{table*}

\begin{figure}[t]\centering\includegraphics[width=\columnwidth]{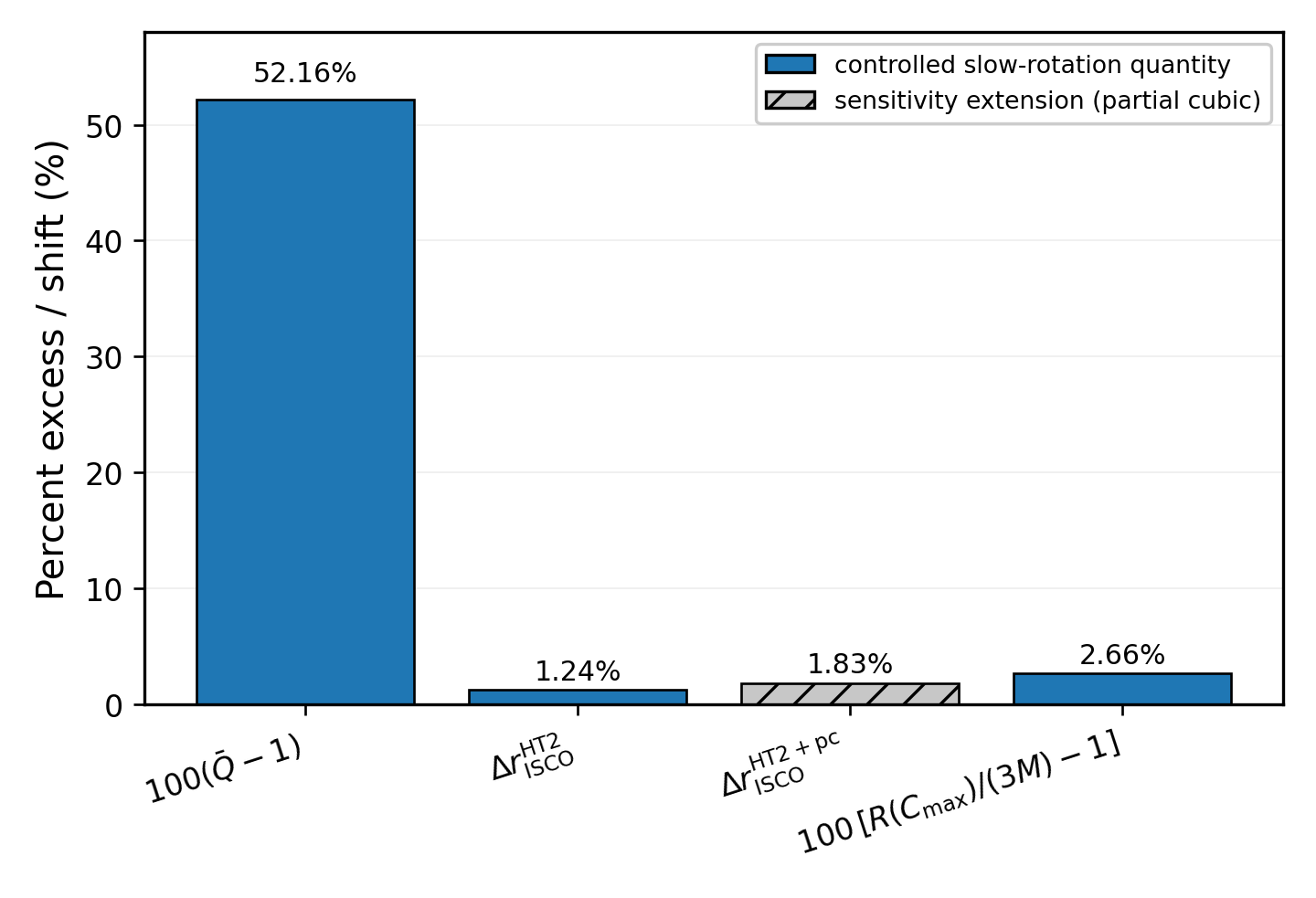}\caption{Percentage Kerr-comparison diagnostics in the controlled slow-rotation row: quadrupole excess, second-order and partial-cubic ISCO shifts, and the positive surface offset from $R=3M$. The partial-cubic bar is a sensitivity extension, not a complete third-order stellar result.}\label{fig:GC9_Kerr_summary}\end{figure}

\subsection{Physical meaning for Kerr mimicking}
\label{subsec:GC9_physical_meaning}

These results sharpen the phenomenological interpretation of the GC9 branch.
The calculated observables \(C\), \(j\), \(k_2\), \(\Lambda\), \(Q\), and
\(\bar Q\) show that the effective-fluid sequence is highly compact and has a
slow-rotation exterior close to, but not exactly equal to, Kerr.  The
controlled quadrupole result,
\begin{equation}
	\bar Q_{\rm GC9}=1.52158,
\end{equation}
corresponds to a quadrupole magnitude approximately \(52\%\) larger than the
Kerr value at fixed \((M,J)\).  Nevertheless, the resulting exterior ISCO
shift is only at the percent level.  GC9 should therefore be described as a
restricted orbital, multipolar, and tidal Kerr mimicker rather than as an exact Kerr substitute.

A definitive characterization requires three further calculations:
\begin{enumerate}
	\item a fully rotating two-dimensional equilibrium sequence, from which the
	mass-shedding limit, stellar surface, ergoregion structure, and multipoles
	\(Q(j)\), \(S_3(j)\), and \(M_4(j)\) can be extracted;
	\item direct timelike and null geodesic integration in the corresponding
	numerical exterior metric, followed by ray tracing through a specified
	accretion flow;
	\item a rotating coupled metric--condensate perturbation calculation yielding
	the polar--axial spectrum, mode mixing and excitation amplitudes.
\end{enumerate}
The present Hartle--Thorne analysis already demonstrates that the GC9 branch
does not differ from Kerr through a large displacement of the exterior ISCO.
The maximum-mass endpoint of the radially stable branch satisfies
\begin{equation}
 C(M_{\max})=0.324694<\frac13,
 \qquad R/M=3.079827,
\end{equation}
and the exact interior diagnostic gives $\chi_{\max}=0.975563<1$;
therefore this stable endpoint has no light ring. The post-turning,
radially unstable continuation instead reaches
\begin{equation}
 C_{\max}=0.337118>\frac13,
 \qquad R/M=2.966322,
\end{equation}
and develops an external unstable light ring together with an internal
stable partner. Thus GC9 supplies a possible transient ultracompact
light-ring stage, but not a demonstrated long-lived stable
light-ring-controlled remnant. This unstable continuation is not needed to
produce the static axial $\ell=2$ mode of the stable endpoint; its possible
role during merger or collapse must instead be tested with nonlinear evolutions.

\paragraph{ISCO-related observables.}
For the controlled slow-rotation configuration,
\begin{equation}
	j=0.328248,
	\qquad
	\bar Q=1.52158,
\end{equation}
the Kerr reference value and the second-order Hartle--Thorne estimate are
\begin{equation}
	\frac{r_{\rm ISCO}^{\rm Kerr,BL}}{M}=4.877207,
	\qquad
	\frac{r_{\rm ISCO}^{\rm HT2}}{M}=4.958673.
\end{equation}
As discussed in Sec.~\ref{sec:gc9_layer2_kerr}, $r_{\rm ISCO}^{\rm Kerr}$
is a Boyer--Lindquist radius while $r_{\rm ISCO}^{\rm HT2}$ is a
circumferential radius \citep{ShibataSasaki1998ISCO}, so the
coordinate-consistent comparison is the same-order shift
\begin{equation}
		\Delta_{\rm ISCO}^{\rm HT2}
		\equiv
		\frac{r_{\rm ISCO}^{\rm HT2}-r_{\rm ISCO}^{\rm HT2}(q_2{=}j^2)}
		{r_{\rm ISCO}^{\rm HT2}(q_2{=}j^2)}
		=
		1.238\%,
\end{equation}
or $1.347\%$ against the exact Kerr value converted to circumferential
radius, Eq.~\eqref{eq:kerr_isco_circ}. A partial cubic sensitivity extension gives
\begin{equation}
	\frac{r_{\rm ISCO}^{\rm HT3}}{M}=4.994769,
	\qquad
	\Delta_{\rm ISCO}^{\rm HT3}=1.830\%,
\end{equation}
but this second number is not a complete third-order stellar result because
the independent higher multipoles have not yet been computed.  The robust
statement is therefore that the GC9 quadrupole changes the exterior ISCO by
$1.238\%$ at the controlled slow-rotation point, while the $\sim9\%$ number obtained from Eq.~\eqref{eq:isco_shift_general} is only a formal fixed-denominator extrapolation and not a controlled high-spin prediction.

The ISCO remains outside the material surface:
\begin{equation}
	\frac{r_{\rm ISCO}^{\rm HT2}}{R}=1.610.
\end{equation}
Thus late circular motion down to the ISCO probes the vacuum exterior rather
than the condensate interior.  Since the ISCO is inferred indirectly through
thermal continuum fitting, reflection spectroscopy, or quasi-periodic
variability, present constraints remain affected by disk thickness,
ionization, emissivity, magnetic stresses, and coronal geometry
\citep{ShashankEtAl2025Reflection}.  A one-percent displacement of the
effective inner-edge scale is therefore not presently expected to be
excluded.  It is more naturally interpreted as a target for future
high-precision accretion measurements and EMRI orbital spectroscopy.

\subsection{Scope of the Kerr-mimicking claim: imaging, ringdown and conditional accretion mapping}
\label{subsec:GC9_mimicry_scope}

It is useful to collect in one place which Kerr observables the present
calculation can and cannot address, organized by the radius each observable
probes. The GC9 stable branch has $R/M=3.079827$ at the maximum-mass benchmark
and $R>3M$ everywhere on the stable sequence. Every diagnostic controlled by
the exterior metric at $r\gtrsim4M$ --- the mass scaling, the slow-rotation
multipoles, the ISCO and its shift, the tidal deformability --- is computed in
vacuum and is a legitimate output of this framework. Every diagnostic that requires an exterior photon region at $r\simeq3M$ --- the photon ring, the associated critical shadow edge and a light-ring/eikonal QNM identification --- is inaccessible. This does not preclude low-multipole stellar spacetime modes, which are treated separately below. The discussion
below makes explicit what that costs observationally. These are not statements
about numerical control: they are structural consequences of $C<1/3$, and they
delimit the claim made in this paper. Table~\ref{tab:mimicry_scope}
summarizes the resulting scope.

\paragraph{Imaging: the missing photon ring is not the binding constraint.}
Current very-long-baseline interferometry does not resolve the higher-order
($n\ge1$) photon subrings, which carry only a small fraction of the total
flux, so the absence of a photon ring is not by itself excluded by an image.
The sharper issue is different, and specific to a dark-sector star: GC9 matter
couples to photons only gravitationally, so the configuration is horizonless
\emph{and} effectively surfaceless in the optical sense, and no mechanism
removes photons from the beam. A central brightness depression can then arise
only from an absence of emitting material, not from absorption.
General-relativistic magnetohydrodynamic simulations followed by radiative
transfer, for accreting nonrotating boson stars, show that baryons accumulate
down into the innermost regions, so that emission is present at all radii and
any dark region is substantially smaller than that of a black hole of the same
mass \citep{Olivares2020}; on this basis stable mini-boson-star configurations
are disfavored as models of Sgr~A$^\ast$, the inner image being bright rather
than depressed \citep{EHT2022SgrAVI}. This is a genuine constraint on the
supermassive end of Table~\ref{tab:sectors} and is not addressed by anything
computed here.

Two considerations keep the question open rather than settled. First, thin-disk
radiative-transfer calculations show that boson stars without photon spheres
can nonetheless display a luminous ring and a central brightness depression,
whose size and contrast depend strongly on the emissivity profile and on how
far the flow penetrates the star
\citep{RosaRubieraGarcia2022NoPhotonSphere}. Second, and more directly relevant
here, an effective shadow can be generated with no light ring at all if the
angular velocity along timelike circular orbits attains a maximum at a finite
radius $R_\Omega$, which then sets the inner edge of the accretion flow and
reproduces the black-hole flow morphology; spherical Proca stars realizing
$R_\Omega=6M$ mimic the Schwarzschild shadow in exactly this way
\citep{HerdeiroEtAl2021ImitationGame}, and the mechanism has been extended to
dynamically robust spinning configurations
\citep{SengoEtAl2024ImitationReloaded}. The GC9 branch does possess an ISCO at
$r_{\rm ISCO}^{\rm HT2}/M=4.958673$, comfortably outside the surface, so an
analogous emission-void mechanism is available in principle. Establishing it
requires dedicated GRMHD and ray tracing in the GC9 spacetime, which is not
attempted here. Until that is done, the imaging status of GC9 at the
supermassive end should be regarded as an open constraint rather than a
satisfied one.

\paragraph{An explicit emission-void scale for GC9.}
The kinematic part of that argument can, however, be settled analytically with
the static solution already in hand, and it turns out to be unusually
favorable. For the static metric $ds^2=-A(r)dt^2+B(r)dr^2+r^2d\Omega^2$ the
coordinate angular velocity of a circular timelike geodesic is
\begin{equation}
\Omega^2(r)=\frac{A'(r)}{2r},
\qquad
A(r)=e^{2\Phi(r)},
\qquad
\Phi'=\frac{m+4\pi r^3P}{r(r-2m)} ,
\label{eq:circ_omega}
\end{equation}
so $\Omega(r)$ follows directly from the TOV solution. Integrating the GC9
closure at the maximum-mass central density $\hat\rho_c=4.691700$ reproduces
$M_{\max}^{\rm red}=0.064157$, $R_{\rm red}=0.197591$ and $R/M=3.079827$ to six
digits, and evaluating Eq.~\eqref{eq:circ_omega} on that background gives
\begin{equation}
\begin{aligned}
M\Omega(0)&=0.132713,\qquad
M\Omega(R_\Omega)=0.185755,\\
M\Omega(R)&=0.185017.
\end{aligned}
\end{equation}
with an interior maximum at
\begin{equation}
R_\Omega/M=3.0068=0.9763\,R .
\label{eq:gc9_Romega}
\end{equation}
A maximum of $\Omega$ away from the origin therefore exists, so the region
$r<R_\Omega$ has $d\Omega/dr>0$ and is magnetorotationally stable: the
Balbus--Hawley mechanism cannot transport a disk inward past $R_\Omega$, which
is precisely the condition identified in
\citet{HerdeiroEtAl2021ImitationGame}.

Two features make the GC9 case stronger than the Proca-star realization. First,
because the slow-row exterior is Hartle--Thorne rather than exactly Schwarzschild, the controlled prograde ISCO is $r_{\rm ISCO}^{\rm HT2}/M=4.958673$, i.e.\ $1.610\,R$; the disk inner edge is therefore at the black-hole value by
construction, rather than being replaced by $R_\Omega$ as in models with stable
circular orbits all the way to the center. Second, the apparent size of the
void is nearly degenerate with a black-hole shadow. The lensed impact
parameter $b(r)=r/\sqrt{A(r)}$ is monotonic throughout, independently
confirming the absence of a light ring (Sec.~\ref{sec:gc9_layer2_kerr}), and
\begin{equation}
b(R_\Omega)=5.1926\,M
\quad\text{versus}\quad
b_{\rm crit}^{\rm Schw}=3\sqrt3\,M=5.1962\,M,
\label{eq:void_apparent_size}
\end{equation}
a difference of $-0.069\%$, with the stellar surface itself at
$b(R)=5.2013\,M$ ($+0.099\%$). This near-coincidence is structural rather than
tuned: because the branch sits just below the light-ring threshold, $b(r)$ is
nearly stationary near $r\simeq3M$, and $b(r)$ remains within $1\%$ of
$b_{\rm crit}^{\rm Schw}$ for all $r/M\in[2.95,3.27]$ --- a window that
comfortably brackets both $R_\Omega$ and $R$. The predicted apparent void size
is therefore insensitive to where exactly inside the star the flow stalls.

One caveat must be stated with the same precision, because it limits what the
above establishes. The MRI criterion bounds where a \emph{disk} can be
maintained, not where matter can go. Integrating the radial turning-point
condition $E^2/A(r)=1+L^2/r^2$ for material plunging from the ISCO
($E^2=8/9$, $L=2\sqrt3\,M$) gives an inner turning point at $r/M=1.63$, well
inside $R_\Omega$, corresponding to an apparent radius of only $4.06\,M$;
lower-angular-momentum material penetrates further, and strictly radial infall
reaches the center. The magnetorotationally stable core is thus a barrier to
viscous inflow, not to ballistic plunge. Whether the resulting configuration is
optically dark depends on how efficiently the plunging component radiates and
on where it settles --- stable circular orbits exist for $r<3.189\,M$ and for
$r>6\,M$, with an unstable band between --- which remains a GRMHD and
radiative-transfer question. What the present calculation does establish is
that the kinematic precondition for an effective shadow is satisfied by GC9,
with a void scale that would appear at $99.93\%$ of the corresponding Schwarzschild shadow radius.

\paragraph{QCD transparency, reprocessing and compositeness of infalling matter.}
The microscopic stability assumption also determines how ordinary
matter can interact with a GC9 star.  Because the breaking is
\(G(2)\to SU(3)_C\), with \(SU(3)_C\) the ordinary QCD color group,
the massive vectors transform as
\begin{equation}
	{\bf14}_{G(2)}
	\longrightarrow
	{\bf8}_C\oplus{\bf3}_C\oplus\overline{\bf3}_C ,
	\label{eq:G2_SU3C_reprocessing}
\end{equation}
and interact with QCD gluons before being confined into the protected
color-singlet composite \({\cal G}_X\)
\citep{Masi_2021,Masi2024G2Resurgence,Masi2026E6G2}.  The unbroken-\(G(2)\)
screening relation
\({\bf7}\otimes{\bf14}^{\otimes3}\supset{\bf1}\) provides a useful
group-theoretic precedent for fermion--gluon hybrids
\citep{Holland:2003jy,Wellegehausen:2013cya}, but it does not imply that
such states can be produced at low energy after \(G(2)\) breaking.

For \(m_X\gg\Lambda_{\rm QCD}\), the protected ground-state reference configuration is extremely compact. Its leading
static singlet potential is
\begin{equation}
	V_{X}(r)\simeq-\frac{C_F\alpha_s}{r},
	\qquad C_F=\frac43 ,
\end{equation}
and therefore
\begin{equation}
	a_X\simeq\frac{2}{C_F\alpha_s m_X},
	\qquad
	E_B\simeq\frac{C_F^2\alpha_s^2m_X}{4}.
	\label{eq:XbarX_compactness}
\end{equation}
For \(m_X\simeq5.6\times10^{13}\,{\rm GeV}\) and
\(\alpha_s\simeq0.03\),
\begin{equation}
	a_X\simeq1.8\times10^{-13}\ {\rm fm},
	\qquad
	E_B\simeq2.2\times10^{10}\ {\rm GeV}.
	\label{eq:XbarX_reference_numbers}
\end{equation}
By comparison, radial infall from rest at infinity onto the stable GC9
endpoint, \(C=0.3247\), gives
\begin{equation}
\begin{aligned}
	\gamma_R&=(1-2C)^{-1/2}\simeq1.689,\\
	K_N&\simeq(\gamma_R-1)m_N\simeq0.65\ {\rm GeV}.
\end{aligned}
	\label{eq:GC9_infall_energy}
\end{equation}
Ordinary accretion therefore cannot dissociate a protected ground-state
composite or create new \(X\) constituents,
\(E_B/K_N\sim3\times10^{10}\).  Broken-phase formation of heavy-light
hybrids can become accessible only through an anomalously small
rearrangement gap or through a pre-existing excited or restructured
component; the unbroken-\(G(2)\) hybrid spectrum does not by itself provide
such a low threshold.

The stellar column density does not compensate this microscopic hierarchy.
Let
\begin{equation}
	n_D(r)\equiv j_D^{\hat 0}(r)
\end{equation}
denote the local Noether-charge density of the effective GC9 condensate
measured in its comoving frame and let
\(n_{D,c}\equiv n_D(0)\) be its central value.  In the
Thomas--Fermi limit,
\begin{equation}
	n_D(y)
	=
	2y\sqrt{1+9y^8}\,
	\frac{\Lambda_T^4}{m_G}.
	\label{eq:GC9_local_number_density}
\end{equation}
Taking \(m_G\simeq2m_X\simeq1.12\times10^{14}\,{\rm GeV}\), the stable
maximum-mass configuration, \(y_c=0.8730\), therefore has
\begin{equation}
	n_{D,c}
	=
	2y_c\sqrt{1+9y_c^8}\,
	\frac{\Lambda_T^4}{m_G}
	\simeq4.1\times10^{23}\ {\rm cm^{-3}}
	\label{eq:GC9_central_number_density}
\end{equation}
for the stellar-sector normalization
\(\Lambda_T=0.10\,{\rm GeV}\).

For an incident baryon, define the GC9 column number density along its
trajectory by
\begin{equation}
	N_D^{\rm col}
	\equiv
	\int_{\rm path} n_D(r)\,d\ell_{\rm prop},
	\label{eq:GC9_column_density_definition}
\end{equation}
where \(d\ell_{\rm prop}\) is the local proper path length through the
star.  For a radial crossing, an order-of-magnitude estimate obtained by
replacing the density profile by its central value and the proper path by
the stellar diameter is
\begin{equation}
	N_D^{\rm col}
	\sim 2R\,n_{D,c}
	\simeq3.7\times10^{30}\ {\rm cm^{-2}},
	\label{eq:GC9_opacity_threshold}
\end{equation}
since the \(10M_\odot\) endpoint has \(R\simeq45.5\) km.
The corresponding baryonic optical depth through the GC9 medium is
\begin{equation}
	\tau_b^{\rm GC9}
	\equiv
	\int_{\rm path}
	n_D(r)\,
	\sigma_{b{\cal G}}(E,r)\,
	d\ell_{\rm prop},
	\label{eq:GC9_baryon_optical_depth}
\end{equation}
where \(\sigma_{b{\cal G}}(E,r)\) denotes the effective total interaction
cross section of an incident baryon with a GC9 composite at the relevant
local energy.  The probability for at least one interaction along the
trajectory is
\begin{equation}
	P_{\rm int}=1-\exp[-\tau_b^{\rm GC9}] .
\end{equation}

As a deliberately conservative dimensional benchmark, assigning the entire
geometrical area of the compact ground-state constituent,
\begin{equation}
	\sigma_{b{\cal G}}\sim\pi a_X^2
	\simeq9.8\times10^{-52}\ {\rm cm^2},
\end{equation}
and using the column estimate of
Eq.~\eqref{eq:GC9_opacity_threshold} gives
\begin{equation}
	\tau_b^{\rm GC9}
	\sim N_D^{\rm col}\pi a_X^2
	\sim4\times10^{-21}.
	\label{eq:GC9_geometric_tau}
\end{equation}
For a compact color-singlet state the actual low-energy QCD interaction is,
by analogy with heavy quarkonium, governed by chromoelectric multipoles and
the corresponding color polarizability rather than by an unscreened
constituent-level QCD cross section
\citep{TarrusCastellaKrein2018,PolyakovSchweitzer2018}.
A separate interaction can arise through projected heavy fermionic states.
Let \(F\) denote such a state, with mass \(M_F\), and define
\begin{equation}
	C_{qF}\equiv
	\langle q_{\rm light}|T_X|F_{\rm heavy}\rangle .
\end{equation}
When \(F\) is far off shell, integrating it out generates a
higher-dimensional interaction.  For a characteristic low-energy scale
or momentum transfer \(E\ll M_F\), the corresponding dimension-six
cross-section estimate is parametrically
\begin{align}
	\sigma_6 &
	\sim
	\frac{g_{G(2)}^4|C_{qF}|^4E^2}
	{16\pi M_F^4}
	\simeq  \\
	& 10^{-82}|C_{qF}|^4\ {\rm cm^2}
	\left(\frac{g_{G(2)}}{0.6}\right)^4
	\left(\frac{E}{1\,{\rm GeV}}\right)^2
	\left(\frac{10^{13}\,{\rm GeV}}{M_F}\right)^4 .
	\label{eq:GC9_offshell_hybridization}
\end{align}
Here \(C_{qF}\) has not been calculated explicitly in the projected
\(E_6\) spectrum; taking \(|C_{qF}|=1\) therefore gives the deliberately
optimistic upper normalization shown above.  Any additional projector or
mixing suppression only decreases the cross section.  Generic off-shell
communication through the high-scale projected fermion sector is therefore
irrelevant for stellar opacity.

Consequently, conditional on the protected compact-singlet
ground-state realization, the baryonic optical depth satisfies
\begin{equation}
\tau_b^{\rm GC9}\ll1,
	\label{eq:GC9_transparency}
\end{equation}
i.e.\ ordinary baryonic matter can penetrate the cold GC9 medium rather
than being automatically absorbed or ``digested'' into it.

A qualitatively different regime is possible only if some microscopic
constituents become extended.  If a fraction \(f_{\rm ext}\) resides in
QCD-sized heavy-light or otherwise restructured states with an illustrative
hadronic cross section \(\sigma_{\rm had}\sim10\,{\rm mb}\), then
\begin{equation}
	f_{\rm ext}^{\rm crit}
	\sim
	\frac{1}{N_D^{\rm col}\sigma_{\rm had}}
	\simeq2.7\times10^{-5},
	\label{eq:GC9_extended_fraction}
\end{equation}
so a tiny extended population could make the medium strongly stopping.
The production and survival of such states is not predicted by the GC9 EOS
and would require a dedicated QCD bound-state and transport calculation.

More fundamentally, the dense GC9 solution must remain compatible with
the existence of the compact glueball constituent.  To quantify this issue, we separate the local GC9 energy density into the quadratic
``rest-mass'' contribution (for a \(X\overline X\) state) and the nonlinear collective contribution,
\begin{equation}
	\rho_{\rm rest}(y)\equiv 2m_G^2s
	=2\Lambda_T^4 y,
	\quad
	\rho_{\rm nl}(y)\equiv
	10\frac{m_G^{18}}{\Lambda_T^{32}}s^9
	=10\Lambda_T^4y^9 ,
	\label{eq:GC9_rest_nonlinear_density}
\end{equation}
where \(y=s/s_T\) and \(s_T=\Lambda_T^4/m_G^2\).
The local condensate chemical potential, equivalently the local phase
frequency in the Thomas--Fermi limit, is
\begin{equation}
	\mu_D(y)\equiv\sqrt{V_{,s}}
	=m_G\sqrt{1+9y^8}.
	\label{eq:GC9_local_chemical_potential}
\end{equation}
At the center of the stable maximum-mass configuration,
\(y_c=0.8730126\), these quantities satisfy
\begin{equation}
	\frac{\mu_{D,c}}{m_G}
	=\sqrt{1+9y_c^8}\simeq2.009,
	\quad
	\frac{\rho_{{\rm nl},c}}{\rho_{{\rm rest},c}}
	=5y_c^8\simeq1.69 .
	\label{eq:GC9_internal_consistency}
\end{equation}
The microscopic binding fraction is only
\begin{equation}
	\frac{E_B}{m_G}\simeq2\times10^{-4}.
\end{equation}
If a fraction \(\epsilon_{\rm int}\) of the collective excess chemical
energy couples to the internal dissociation coordinate, preservation of
compact constituents requires the diagnostic condition
\begin{equation}
	\epsilon_{\rm int}
	\frac{\mu_D-m_G}{E_B}\lesssim1
	\quad\Longrightarrow\quad
\epsilon_{\rm int}\lesssim2\times10^{-4} .
	\label{eq:GC9_Mott_condition}
\end{equation}
If this condition fails, the compact glueball quasiparticle ceases to be a self-consistent microscopic degree of freedom.  By analogy with in-medium heavy-quark bound states, for which color
screening and medium-induced scattering can dissolve a bound state into the continuum, the appropriate description would then be a Mott-like dissociation or a more general restructured QCD-colored
many-body phase rather than the one-component GC9 condensate
\citep{JankowskiBlaschkeGrigorian2010,
	BrambillaGhiglieriPetreczkyVairo2008,
	BrambillaEscobedoGhiglieriVairo2013,LiewLuo2017QCDBoundStates}.
Such a phase could interact efficiently with infalling baryons, but its equation of state, charge transport and radiative properties would all have to be recomputed.
Thus the present model predicts transparency only conditionally on both microscopic composite stability and preservation of the compact-singlet structure.

\paragraph{Ringdown of the static endpoint and the unstable continuation.}
The absence of an external light ring invalidates a direct eikonal or photon-sphere identification, but it does not remove the low-multipole stellar spectrum. Solving the static axial $\ell=2$ eigenproblem directly gives
\begin{equation}
\begin{aligned}
M\omega_{\rm ax}&=0.463846-0.098881i,\\
Q&=2.345,\qquad \tau/M=10.113.
\end{aligned}
\end{equation}
The damping scale is close to Schwarzschild, whereas the real frequency is $24.1\%$ above the Schwarzschild value at the same mass. This is a distinguishable static prediction, not a Kerr degeneracy and not yet a merger waveform: rotation, polar coupling and excitation amplitudes remain uncomputed. The absence of any light ring, the monotonic axial potential and the low-$Q$ mode exclude the specific axial light-ring trapping mechanism for this static endpoint; they do not prove complete nonlinear or rotating stability. The post-turning configurations that do develop a light-ring pair are already radially unstable and may occur only as transient migration or collapse stages. Inspiral observables such as the finite tidal response and quadrupole excess remain independent tests of GC9.

\paragraph{Conditional continuum-fitting mapping.}
For an opaque object whose baryonic disk is truncated at the material surface, the inner edge would satisfy
\begin{equation}
r_{\rm in}=\max(r_{\rm ISCO},R).
\end{equation}
Applying this assumption to the static GC9 radius $R/M=3.079827$ and then fitting the spectrum with a Kerr thin-disk model gives the illustrative mapping
\begin{equation}
a_\ast^{\rm inferred}\lesssim0.766,
\qquad
r_{\rm ISCO}^{\rm Kerr}(0.766)/M=3.0798.
\label{eq:gc9_spin_ceiling}
\end{equation}
This is not a model-independent hard ceiling. 
GC9 matter is optically surfaceless and, in the protected compact-singlet
ground-state realization discussed above, is predicted to be effectively
transparent to baryons despite the QCD color of its microscopic constituents.
The geodesic analysis permits plunging material to penetrate inside both
$R$ and $R_\Omega$; whether an excited or restructured constituent population
instead produces appreciable stopping requires the microphysical transport
calculation described above.
A self-consistent continuum-fitting prediction therefore requires a rotating GC9 equilibrium, GRMHD and radiative transfer. The value $0.766$ should be retained only as a conditional static surface-truncation benchmark, not as a direct exclusion by reported near-extremal Kerr spins.

\begin{table}
\caption{Scope of the GC9 Kerr-mimicking claim. The table separates direct calculations from conditional mappings and observables that require additional rotating, accretion or radiative-transfer modeling.}
\label{tab:mimicry_scope}
\begin{ruledtabular}
\begin{tabular}{lll}
\tcell[0.24\columnwidth]{Observable} & \tcell[0.19\columnwidth]{Probes} & \tcell[0.43\columnwidth]{Status} \\
\hline
\tcell[0.24\columnwidth]{Mass scale, $M_{\max}(\Lambda_T)$} & \tcell[0.19\columnwidth]{global} & \tcell[0.43\columnwidth]{reproduced by construction} \\
\tcell[0.24\columnwidth]{Stellar-dynamical orbits} & \tcell[0.19\columnwidth]{$r\gtrsim10^3M$} & \tcell[0.43\columnwidth]{indistinguishable from Kerr} \\
\tcell[0.24\columnwidth]{Moment of inertia $\bar I$} & \tcell[0.19\columnwidth]{exterior} & \tcell[0.43\columnwidth]{$5.116358$} \\
\tcell[0.24\columnwidth]{Quadrupole $\bar Q$} & \tcell[0.19\columnwidth]{exterior} & \tcell[0.43\columnwidth]{$1.52158$, $52\%$ above Kerr} \\
\tcell[0.24\columnwidth]{ISCO radius and shift} & \tcell[0.19\columnwidth]{$\simeq4.9M$} & \tcell[0.43\columnwidth]{$1.238\%$ (same order)} \\
\tcell[0.24\columnwidth]{Tidal def. $\Lambda$} & \tcell[0.19\columnwidth]{exterior} & \tcell[0.43\columnwidth]{$4.37208$, versus $0$ for a BH} \\
\tcell[0.24\columnwidth]{Effective shadow} & \tcell[0.19\columnwidth]{$R_\Omega=3.0068M$} & \tcell[0.43\columnwidth]{apparent $5.1926M$ versus $5.1962M$ (Schwarzschild)} \\
\hline
\tcell[0.24\columnwidth]{Continuum-fitting spin} & \tcell[0.19\columnwidth]{disk inner edge} & \tcell[0.43\columnwidth]{conditional $a_\ast\lesssim0.77$ surface-truncation mapping} \\
\tcell[0.24\columnwidth]{Photon ring, shadow edge} & \tcell[0.19\columnwidth]{$\simeq3M$} & \tcell[0.43\columnwidth]{no external photon ring; requires GRMHD} \\
\tcell[0.24\columnwidth]{Static axial QNM} & \tcell[0.19\columnwidth]{global $\ell=2$ potential} & \tcell[0.43\columnwidth]{$0.46385-0.098881i$; rotating excitation open} \\
\end{tabular}
\end{ruledtabular}
\end{table}

\paragraph{Accretion, disc winds and non-conservative mass transfer.}
A recent observational example illustrates why luminous accretion and
strong baryonic mass ejection do not by themselves require an absorbing
event horizon.  State-resolved optical spectroscopy of the transient
low-mass X-ray binary Swift~J1727.8$-$1613, conventionally interpreted as
hosting a stellar-mass \(\sim10\,M_\odot\) black hole, reveals during the
late hard-state transition a massive, cool disc outflow with characteristic
velocity \(v_w\simeq750\,{\rm km\,s^{-1}}\) and
\(T\lesssim10^4\,{\rm K}\) \citep{CastroSegura2026J1727}.
Using a Sobolev estimate, Castro Segura et al.\ infer
\begin{equation}
	\dot M_w\gtrsim10^{-9}\,M_\odot\,{\rm yr}^{-1},
	\qquad
	\frac{\dot M_w}{\dot M_{\rm acc}}\sim1 ,
	\label{eq:J1727_wind}
\end{equation}
so that the instantaneous outflow can be comparable to the inward
accretion rate and the binary may undergo strongly non-conservative mass
transfer.  Their order-of-magnitude wind estimate adopts a characteristic
line-forming radius \(R\sim10^{10}\,{\rm cm}\), although the authors stress
that both this radius and the wind geometry are uncertain.  For comparison,
a \(10\,M_\odot\) GC9 configuration near the compact benchmark has
\[
R_{\rm GC9}\simeq3.08\,\frac{GM}{c^2}\simeq4.5\times10^6\,{\rm cm},
\]
so the optical wind is associated with radii of order \(10^3\) GC9
radii, where the gravitational field is essentially determined by the
total mass rather than by the nature of the central boundary.  Such a
disc-wind cycle is therefore compatible, at this level, with an ordinary
donor transferring baryonic matter to a GC9 compact object.

The distinction from a black hole arises only in the innermost flow.
General-relativistic magnetohydrodynamic calculations of accretion onto
surfaceless boson stars show that baryonic plasma need not terminate at a
material surface but can penetrate into and accumulate within the bosonic
configuration \citep{Olivares2020}.  In a GC9 realization, donor material
would likewise carry essentially no dark Noether charge in the minimal
portal-suppressed theory, so baryonic accretion need not correspond to
growth of the GC9 condensate itself.  Matter can instead remain in a
baryonic inner flow or be removed through winds, jets or other feedback
channels.  The powerful outflow observed in Swift~J1727.8$-$1613 therefore
provides a useful phenomenological example of the accretion/ejection
balance that a stellar-mass GC9 binary would also have to reproduce, but
it does not constitute evidence for a horizonless central object.
Discriminating the two cases requires dedicated rotating-GC9 GRMHD and
radiative-transfer calculations of the inner flow, where irreversible
horizon absorption is replaced by interaction with a penetrable,
dynamical dark condensate.

\subsection{Quantitative multipole discriminant for gravitational-wave observations}
\label{subsec:GC9_EMRI_forecast}

At fixed mass and spin, the spin-induced quadrupole may be written in the waveform convention
\begin{equation}
	{Q\over M^3}=-(1+\delta\kappa)j^2 .
\end{equation}
For the controlled slow-row GC9 model,
\begin{equation}
	\delta\kappa=\bar Q-1=0.52158,
\end{equation}
so that
\begin{align}
	{Q_{\rm Kerr}\over M^3}&=-j^2=-0.107747,\\
	{Q_{\rm GC9}\over M^3}&=-\bar Qj^2=-0.163946,
\end{align}
and
\begin{equation}
	\delta q\equiv{Q_{\rm GC9}-Q_{\rm Kerr}\over M^3}=-0.056199 .
	\label{eq:GC9_delta_q_EMRI}
\end{equation}
The quadrupole magnitude is therefore $52.2\%$ larger than Kerr at the same $M$ and $J$.

Existing ground-based inspiral analyses that allow both tidal deformability and a spin-induced quadrupole deviation find the selected GWTC events consistent with the binary-black-hole hypothesis, but their constraints are event- and waveform-dependent and are not a direct exclusion of the present supermassive slow-row model \citep{NarikawaUchikataTanaka2021}.  A more relevant future discriminator is an extreme-mass-ratio inspiral.  As a historical benchmark, the Barack--Cutler Fisher analysis for one year of LISA data at signal-to-noise ratio 100 found indicative uncertainties $\Delta(Q/M^3)\sim10^{-4},10^{-3},10^{-2}$ for central masses $10^{5.5},10^6,10^{6.5}M_\odot$, respectively \citep{BarackCutler2007OffKerr}.  Comparing Eq.~\eqref{eq:GC9_delta_q_EMRI} with those benchmark errors gives the ratios shown in Table~\ref{tab:GC9_EMRI_benchmark}.

\begin{table}
	\caption{Illustrative comparison of the GC9 slow-row quadrupole deviation with published LISA EMRI Fisher errors. These are not bespoke GC9 detection significances; they only show the scale of the effect relative to a standard benchmark.}
	\label{tab:GC9_EMRI_benchmark}
	\begin{ruledtabular}
		\begin{tabular}{ccc}
			$M/M_\odot$ & benchmark $\sigma(Q/M^3)$ & $|\delta q|/\sigma$\\
			\hline
			$10^{5.5}$ & $10^{-4}$ & 562\\
			$10^6$ & $10^{-3}$ & 56.2\\
			$10^{6.5}$ & $10^{-2}$ & 5.62
		\end{tabular}
	\end{ruledtabular}
\end{table}

\paragraph{Photon region and horizon-scale imaging.}
The formal exterior continuation of the slow-rotation metric yields a
light-ring-like radius
\begin{equation}
	\frac{r_{\rm LR}^{\rm formal}}{M}=2.681326.
\end{equation}
However, the GC9 stellar radius at the same benchmark is
\begin{equation}
	\frac{R}{M}=3.079827,
\end{equation}
so that
\begin{equation}
	\frac{r_{\rm LR}^{\rm formal}}{R}=0.871<1.
\end{equation}
The formal orbit therefore lies inside the material configuration and is not
an exterior null circular orbit.  Its associated radius and frequency cannot
be interpreted as physical shadow or eikonal-ringdown observables.

Accordingly, the present GC9 calculation should not be compared directly with
the Kerr critical curve inferred from EHT imaging
\citep{EHT2019M87I,EHTSgrA2022I}.  Absence of an external photon ring does not
by itself preclude GC9 from reproducing selected mass, spin, ISCO, tidal, and
multipolar observables, but it prevents a claim of full optical Kerr
mimicking.  A meaningful horizon-scale comparison requires a fully rotating
surface, numerical null geodesics, and radiative-transfer modeling of the
accretion flow.  Such a calculation would determine whether the image still
develops a sufficiently dark central depression despite the absence of a
vacuum photon ring.

\subsection{Oscillation spectrum and the status of QNM predictions}
\label{subsec:GC9_qnm_complete}

Because the maximum-mass radially stable GC9 endpoint has $R>3M$, it has no external photon ring. The post-turning radially unstable continuation crosses the light-ring threshold, whereas the formally continued prograde Hartle--Thorne light-ring radius of the controlled slow-row endpoint lies inside the material surface. Consequently the light-ring/eikonal correspondence cannot be used to predict a physical stellar QNM, and Eq.~\eqref{eq:GC9_formal_eikonal_reference} is retained only as a comparison number. Within the static perfect-fluid Thomas--Fermi closure, the axial sector decouples from fluid motion at linear order and can be solved as a genuine complex eigenvalue problem. The polar sector, rotating polar--axial coupling and merger excitation amplitudes remain open; the Cowling calculation below supplies approximate matter-mode frequencies only.

\paragraph{The axial $\ell=2$ sector: a static spacetime QNM.}
The eikonal argument above concerns the $\ell\to\infty$ limit. The mode actually
measured in post-merger spectroscopy is $\ell=2$, and its scattering barrier is
not located at the light ring. For Schwarzschild the Regge--Wheeler peak sits at
\begin{equation}
r_{\rm peak}^{(\ell=2)}/M = 3.28078,
\label{eq:rw_peak}
\end{equation}
against the eikonal light ring at $3M$. Since the GC9 maximum-mass surface lies
at $R/M=3.079827$, the $\ell=2$ barrier peak lies in the \emph{vacuum} exterior
with a $6.5\%$ margin, and $82.6\%$ of the barrier's full width at half maximum
($r/M\in[2.261,6.956]$) is exterior to the star. The exterior part of the low-multipole scattering barrier is therefore present even though the eikonal light ring is not.

Within the static effective-fluid model this sector can be solved directly. Axial perturbations of a static perfect
fluid do not couple to fluid motion, so no Cowling-type approximation enters:
with $Z$ the Regge--Wheeler function,
\begin{equation}
\frac{d^2Z}{dr_*^2}+\left[\omega^2-V_{\rm ax}(r)\right]Z=0,
\qquad \frac{dr_*}{dr}=e^{\Lambda_g-\Phi_g},
\label{eq:axial_master}
\end{equation}
\begin{equation}
V_{\rm ax}=e^{2\Phi_g}\!\left[\frac{\ell(\ell+1)}{r^2}+4\pi(\rho-P)-\frac{6m}{r^3}\right],
\label{eq:axial_potential}
\end{equation}
with $Z\sim r^{\ell+1}$ at the centre and purely outgoing behaviour at infinity
\citep{ChandrasekharFerrari1991,KokkotasSchutz1992}. We solved
Eq.~\eqref{eq:axial_master} by Wronskian matching between an interior
integration and an inward integration of the asymptotic outgoing solution.
Applying the upgraded Riccati/logarithmic-derivative machinery to a Schwarzschild background with an ingoing horizon condition returns
$M\omega=0.37367168-0.08896232i$, reproducing the standard Leaver value at the displayed precision \citep{Leaver1985}. For the GC9
maximum-mass configuration,
\begin{equation}
\begin{aligned}
M\omega_{\ell=2}^{\rm axial}&=0.463846-0.098881i,\\
Q&\equiv\left|\frac{\mathrm{Re}\,\omega}{2\,\mathrm{Im}\,\omega}\right|=2.345,
\qquad
\frac{\tau}{M}=10.113 .
\end{aligned}
\label{eq:gc9_axial_qnm}
\end{equation}
The upgraded outgoing-asymptotic/log-derivative solver is internally stable at the $10^{-9}$ level under the audited matching variations; varying the outer matching radius over $35M$--$50M$ changes the quoted root by less than a few parts in $10^9$.

Three consequences follow. First, the maximum-mass static endpoint possesses a rapidly damped, spacetime-dominated quasi-normal mode; the absence of an eikonal light ring does not remove it. Second, its damping scale is black-hole-like: $Q=2.345$ against $Q_{\rm Schw}=2.100$, so the amplitude decays on a timescale corresponding to $0.75$ oscillation periods rather than $0.67$. Third, it is distinguishable from Schwarzschild at the same mass, with a real frequency higher by $24.1\%$. This is a falsifiable static prediction rather than a claim of Kerr degeneracy. In physical units the mode frequency is $f\simeq1.5$~kHz for $10\,M_\odot$ and $3.7$~mHz for $4\times10^6\,M_\odot$; whether a merger excites it with observable amplitude requires a dynamical calculation.

The same computation sharply constrains, but does not settle, the stability question for this static endpoint.
The light-ring instability of ultracompact objects operates by trapping
perturbations in the potential well associated with a \emph{stable} light ring,
where they decay only logarithmically and accumulate until nonlinear
backreaction drives migration or collapse \citep{CunhaBertiHerdeiro2017LRstability,CunhaHerdeiroRaduSanchisGual2023LRInstability}. Its
signature is a family of long-lived modes with $Q\gg1$. The computed maximum-mass static endpoint satisfies none of the preconditions: it has no light ring at all
($\chi_{\max}=0.975563<1$); the axial potential
Eq.~\eqref{eq:axial_potential} is monotonic throughout the interior, with no
turning point and hence no trapping cavity; and the computed mode has
$Q=2.345$, so its energy is radiated to infinity within a single oscillation.
There is no reservoir for the specific axial light-ring trapping mechanism to feed on. This establishes linear axial damping for the computed mode, not complete polar, nonlinear or rotating stability. The light-ring-bearing
configurations identified in Sec.~\ref{sec:gc9_layer2_kerr} lie past the
maximum-mass turning point and are radially unstable for that reason alone,
independently of their photon-region structure.

Two limitations should be stated. The calculation is nonrotating; astrophysical
remnants have $a_\ast\sim0.7$, and the spinning axial problem requires the
two-dimensional equilibria not constructed here. And the polar sector, which
couples to the fluid, may add further modes; the Cowling frequencies below are a
first estimate of that sector and are not the ringdown.

\paragraph{Direct observational benchmark: GW250114.}
The cumulative GWTC--5.0 tests find no evidence for departures from general
relativity in the observed compact-binary population, and the loud event
GW250114 currently supplies the sharpest single-event ringdown constraint
\citep{LVK2026GWTC5TGR,AbacEtAl2025GW250114}. In the full-signal
pSEOBNR analysis, the dominant mode is reconstructed as
\begin{equation}
 f_{220}=251.7^{+5.1}_{-5.0}\ {\rm Hz},
 \qquad
 \tau_{220}=4.09^{+0.42}_{-0.38}\ {\rm ms},
 \label{eq:GW250114_dimensional_qnm}
\end{equation}
at $90\%$ credibility, with the reference redshifted remnant timescale
$t_{M_f}=G(1+z)M_f/c^3=0.337\,{\rm ms}$. The corresponding central
dimensionless complex frequency is
\begin{equation}
 M\omega_{220}^{\rm GW250114}
 =
 2\pi f_{220}t_{M_f}
 -i\,\frac{t_{M_f}}{\tau_{220}}
 \simeq
 0.5330-0.0824\,i .
 \label{eq:GW250114_dimensionless_qnm}
\end{equation}
Used literally at the same dimensionless normalization, the static GC9 mode of
Eq.~\eqref{eq:gc9_axial_qnm} has a real frequency $13.2\%$ below this central
value and a damping time $\tau/M=10.113$, about $16.7\%$ shorter than
$\tau/M=12.14$ inferred for GW250114. Equivalently, a rotating GC9 continuation
would need approximately a $14.9\%$ increase in $\Re(M\omega)$ and a
$20.0\%$ increase in $\tau/M$ relative to the static result to pass through
these central values.

That comparison mixes structural and rotational effects and should not be read
as a direct measure of the departure from general relativity. Inverting
Eq.~\eqref{eq:GW250114_dimensionless_qnm} through the Kerr fundamental mode
gives a remnant spin $a_\ast\simeq0.70$, whereas
Eq.~\eqref{eq:gc9_axial_qnm} is a nonrotating result. Without the rotating GC9
spectrum, the structural and rotational contributions to the offset cannot be
separated. The quantity that LVK actually constrains is the fractional
deviation at the \emph{same} remnant mass and spin. Evaluated at matched
$(M,a_\ast=0)$ against the exact Schwarzschild fundamental
$M\omega=0.373672-0.088962i$,
\begin{equation}
 \delta f_{220}^{\rm static}=+24.1\%,
 \qquad
 \delta\tau_{220}^{\rm static}=-10.0\% ,
 \label{eq:GC9_deltaf_static}
\end{equation}
to be compared with the catalog-combined GWTC--3 bounds
$\delta\hat f_{220}=0.02^{+0.03}_{-0.03}$ (posterior product) and
$0.02^{+0.07}_{-0.07}$ (hierarchical) \citep{Abbott2022TestsGR}. These published
GWTC--3 intervals are retained as a numerical reference, while GWTC--5.0 provides
the latest catalog-level assessment cited above. The static
offset is substantially larger than the fractional deviations allowed by
catalog-level Kerr analyses. If a comparable offset persisted at the spins of
the observed remnants, GC9 would be incompatible with those constraints. The
static result alone, however, cannot be inserted directly into a posterior
constructed from rotating binary remnants.

A complementary static diagnostic is obtained by asking how a Kerr template
would misinterpret the nonrotating GC9 mode. A Kerr ringdown template fitted to
the static GC9 frequency returns
\begin{equation}
 a_\ast^{\rm ringdown}\simeq0.48
 \qquad\text{against a true spin }a_\ast=0 ,
 \label{eq:GC9_spin_misattribution}
\end{equation}
i.e.\ the object would ring as though it were spinning substantially faster
than it is. This illustrates the type of inconsistency targeted by the
inspiral--merger--ringdown test reported by LVK, in which pre-merger and
post-merger mass and spin estimates are found to agree \citep{Abbott2022TestsGR}. Nor can the discrepancy be
absorbed into a spin reassignment, because the damping fails in the opposite
direction: at the frequency-matched Kerr spin $a_\ast\simeq0.48$ one would
expect $\tau/M\simeq11.55$, whereas GC9 gives $10.113$, shorter by $12.5\%$. The
static GC9 mode therefore does not coincide with \emph{any} point of the Kerr
family in the $(\delta f_{220},\delta\tau_{220})$ plane.

\paragraph{Does the GC6--9--12 interaction remove the static discrepancy?}
As a direct check, we repeated the same reduced-fluid TOV plus static axial
calculation for the isotropic connected-cumulant benchmark
$a_6=a_9=1$, $a_{12}=10$ and $\lambda_A=0$. At its first maximum-mass point,
$M_{\max}^{\rm red}=0.072056$, $C=0.323053$ and $R/M=3.09547$, the resulting
fundamental axial estimate is
\begin{equation}
 M\omega_{\ell=2}^{\rm GC6-9-12}
 \simeq 0.463-0.097\,i .
 \label{eq:GC6912_axial_qnm}
\end{equation}
Relative to GC9 the rounded real part changes by well below one percent while the displayed imaginary part differs at the approximately two-percent level.  No significance should be attached to the sign of this small displacement because the GC6--9--12 root is quoted only to three decimal places. Compared with the GW250114 central
reconstruction, this isotropic extension still requires approximately a $15\%$
increase in $\Re(M\omega)$ and an $18\%$ increase in $\tau/M$. Thus the
neighboring positive cumulants raise the reduced maximum-mass coefficient but
do not materially alter the dimensionless static axial spectrum, and they do
not remove the ringdown discrepancy. No exact axial QNM is claimed here for
the anisotropic GC6--9--12A benchmark: its perturbation problem is
model-dependent and must be derived consistently from the anisotropic stress
closure. The remaining possible route to agreement is therefore the full
rotating, coupled polar--axial spectrum rather than a static retuning of the
positive-cumulant coefficients.

The decisive limitation is that Eqs.~\eqref{eq:GC9_deltaf_static} and
\eqref{eq:GC9_spin_misattribution} are static results, while every quantity
entering the LVK ringdown tests refers to a rotating remnant with
$a_\ast\sim0.7$. The two calculations are not the same object. Under rotation
the prograde scattering barrier migrates inward while the equatorial surface
expands, so the fraction of the barrier lying in vacuum --- the quantity that
controls the offset in Eq.~\eqref{eq:GC9_deltaf_static} --- changes in a way
that the present calculation does not determine; in addition the rotating
problem couples the axial and polar sectors, redistributes excitation
amplitudes among overtones and higher multipoles, and modifies the exterior
multipole structure through $\bar Q$ and the higher $S_\ell$, $M_\ell$. Neither
the sign nor the magnitude of the rotating correction to
$\delta f_{220}$ follows from the nonrotating spectrum computed here.

We therefore do not claim compatibility with the LVK ringdown catalog, and we
do not resolve the discrepancy by appealing to a collapsed remnant: the
horizonless endpoint is the object of interest throughout this work, and its
post-merger spectrum is simply not yet computed at the rotation rates the data
probe. Establishing or excluding compatibility requires the rotating
polar--axial GC9 spectrum obtained from a two-dimensional stationary equilibrium
sequence, together with its overtone and higher-multipole content, excitation
amplitudes, and a merger waveform that can be tested directly in the LVK
likelihood. We regard Eq.~\eqref{eq:GC9_deltaf_static} as the sharpest
currently available quantitative target for that calculation, and we identify
it as the principal open problem of the present framework rather than as a settled inconsistency.

\paragraph{A second direction: dissipative interior response.}
Rotation is not the only physical ingredient absent from the static
calculation. The axial eigenvalue problem solved above assumes a real
perfect-fluid stress tensor, a conservative GC9 background and regularity
at the centre. The contrast with a black hole is sharp at the inner end
of the scattering problem: a horizon terminates it absorptively, with the
Regge--Wheeler potential of Eq.~\eqref{eq:axial_potential} decaying to zero
over an infinite tortoise distance, whereas the GC9 star terminates it at
a regular centre, where the centrifugal term
\(\ell(\ell+1)/r^2\) diverges and the wave is returned without loss. The
offset in Eq.~\eqref{eq:GC9_deltaf_static} nevertheless reflects jointly
the stellar interior potential, the finite surface location and this
lossless regular-centre condition, and should not be attributed to either
the inner boundary condition or the equation of state alone.

In analogy with the exotic-compact-object literature, the response of
the entire stellar interior as seen from the exterior barrier may be
summarized phenomenologically by a complex, frequency-dependent effective
reflectivity \(\mathcal R_{\rm eff}(\omega)\)
\citep{CardosoFranzinPani2016ECO,Cardoso2016,Maggio2018,
	Cardoso2019BHmimickers}. The perfect-fluid calculation corresponds to a
lossless interior with \(|\mathcal R_{\rm eff}|=1\), whereas conversion of
axial perturbation energy into microscopic dark-sector excitations would
give \(|\mathcal R_{\rm eff}|<1\). This analogy is not an identification
with a near-horizon reflecting wall: even the formal limit
\(\mathcal R_{\rm eff}=0\) would leave the GC9 surface location and
interior Regge--Wheeler potential different from their Schwarzschild
counterparts. Finite internal dissipation would modify both
\(\Re(\omega)\) and \(\Im(\omega)\), but neither the sign nor the magnitude
of the real-frequency shift follows from \(|\mathcal R_{\rm eff}|\) alone.
The relevant quantity is the full complex function
\(\mathcal R_{\rm eff}(\omega)\), including its phase and coupling to the
GC9 interior potential, so the spectrum must be recomputed rather than
inferred by interpolation between perfectly reflecting and perfectly
absorbing limits. In particular, moving the static GC9 frequency toward
the Schwarzschild value would reduce the quoted \(24.1\%\) static offset,
but would not automatically improve agreement with spinning-remnant
observations.

The present configuration does not form a long-lived echo cavity. The
tortoise separation between the centre and the \(\ell=2\) barrier peak is
\(10.41\,M\), so a round trip takes \(20.83\,M\), compared with the
calculated damping time \(\tau/M=10.113\) in
Eq.~\eqref{eq:gc9_axial_qnm}. The mode therefore survives for only
\(0.49\) round trips, or approximately one one-way interior traversal.
A modification based on weak cumulative losses over many reflections is
therefore unavailable. Any order-unity change in the effective interior
response would have to develop during a single traversal, on a
characteristic timescale of order \(10\,M\). The number-changing channel
invoked in Sec.~\ref{sec:intro} cannot provide such damping. For the
compact benchmark,
\begin{equation}
	t_{\rm dyn}=(G\bar\rho)^{-1/2}=11.06\,M,
	\qquad
	\frac{\tau_{\rm ringdown}}{t_{\rm dyn}}=0.92 ,
	\label{eq:GC9_timescale_coincidence}
\end{equation}
so the freeze-out condition
\(\Gamma_{3\to2}\ll t_{\rm dyn}^{-1}\) also implies
\(\Gamma_{3\to2}\tau_{\rm ringdown}\ll1\). The same slow reaction that
preserves the approximately conserved dark number or charge over a
dynamical time cannot produce order-unity attenuation during one
ringdown.

What remains open is charge-conserving dissipation in the odd-parity
sector, for which the selection rules are restrictive. Ordinary bulk viscosity is not the direct transport coefficient for the nonrotating axial mode, because axial perturbations of a static perfect fluid carry
no leading compressional density or pressure variation---the same
decoupling that permits Eq.~\eqref{eq:axial_master} to be solved without
a Cowling approximation. Possible dissipative channels instead include shear or anisotropic stress, conversion of axial perturbation energy into additional odd-parity dark-sector degrees of freedom, and other microscopic processes that need not rely on compression while preserving
the emergent dark charge. Their relative importance cannot be established
without specifying the underlying dense-\(G(2)\) transport theory. Under
rotation, axial--polar mixing could additionally activate compressional and chemical-relaxation channels, but that belongs to the full rotating coupled-spectrum problem. Quantifying any of these effects requires a causal dissipative extension of
Eqs.~\eqref{eq:axial_master}--\eqref{eq:axial_potential} and a new complex
eigenvalue calculation; a strongly dissipative response would also require checking that the equilibrium and Thomas--Fermi assumptions remain
self-consistent. Finite interior response is therefore a second, independently testable direction alongside rotation.

\paragraph{Quadrupolar internal dark-core modes.}
Within the relativistic Cowling approximation, in which the fluid displacement is perturbed while metric perturbations are frozen, the first four nonrotating $\ell=2$ frequencies of the maximum-mass benchmark are
\begin{equation}
\begin{aligned}
M\omega_n^{(\ell=2)}&\simeq
0.188320,\ 0.558423,\\
&\hspace{1.2cm}0.792306,\ 1.028633,
\qquad n=0,1,2,3 .
\end{aligned}
\label{eq:GC9_core_modes}
\end{equation}
These are approximate real normal-mode frequencies, not complex QNM eigenvalues. They demonstrate that the GC9 fluid background supports a discrete matter-sensitive spectrum, but they do not by themselves determine excitation amplitudes or damping.
The displayed extra digits identify the numerical run and should not be interpreted as a precision estimate; until an independent implementation and convergence table are supplied, only the first few significant figures should be used phenomenologically.
The fundamental matter-supported mode therefore lies well below the formal
exterior comparison frequency:
\begin{equation}
	M\omega_0^{(\ell=2)}
	\simeq0.188
	<
	\Re\!\left(M\omega_{220}^{\rm formal}\right)
	\simeq0.445.
\end{equation}
The first overtone,
\begin{equation}
	M\omega_1^{(\ell=2)}\simeq0.558,
\end{equation}
instead lies close to the same frequency scale.  This proximity suggests that
a future coupled metric--condensate calculation could exhibit mode mixing or
avoided crossings between matter-led and spacetime-led oscillations.  It does
not, however, establish such mixing within the Cowling approximation.

The higher modes at
\begin{equation}
	M\omega_2^{(\ell=2)}\simeq0.792,
	\qquad
	M\omega_3^{(\ell=2)}\simeq1.029
\end{equation}
form a distinct matter-sensitive spectral ladder.  Such an internal spectrum
has no analogue in a vacuum Kerr spacetime and may ultimately provide a more
appropriate gravitational-wave discriminator than a light-ring/eikonal
comparison.

\paragraph{Rotational scale and kinematic splitting.}
For the controlled slow-row solution the angular velocity follows directly from $J=I\Omega$:
\begin{equation}
M\Omega=\frac{j}{\bar I}
=\frac{0.328248}{5.116358}
=0.064157.
\label{eq:GC9_rotation_scale}
\end{equation}
A purely kinematic first-order estimate is then
\begin{equation}
M\omega_{n2m}^{\rm inert}
\simeq
M\omega_{n2}^{(0)}+mM\Omega,
\qquad m=-2,-1,0,1,2.
\label{eq:GC9_mode_splitting}
\end{equation}
Adjacent components are separated by approximately $0.0644$ in $M\omega$, and the full $m=-2$ to $m=2$ span is approximately $0.258$. A true rotational splitting calculation must include a mode-dependent Ledoux coefficient, frame dragging, rotational deformation and polar--axial coupling.

\paragraph{Indicative damping hierarchy.}
The Cowling problem gives real frequencies. If one applies only the dimensional quadrupole-radiation estimate
\begin{equation}
\frac{M}{\tau_{\rm GW}}
\sim
\frac{2}{5}C^2(M\omega)^4,
\label{eq:GC9_tau_scaling}
\end{equation}
with $C=0.324694$, one obtains the indicative values in Table~\ref{tab:GC9_qnm_summary}. These numbers rank the modes by their expected radiative efficiency but are not substitutes for imaginary parts obtained from a complex eigenvalue problem.

\begin{table*}
\caption{Oscillation results for the compact GC9 benchmark. The axial spacetime entry is a genuine static stellar QNM. The formal eikonal entry is only a comparison number because its light-ring radius lies inside the surface. The internal frequencies are nonrotating Cowling eigenvalues; their damping times and quality factors are dimensional quadrupole-radiation estimates.}
\label{tab:GC9_qnm_summary}
\begin{ruledtabular}
\begin{tabular}{cccccc}
sector & $n$ & $\Re(M\omega)$ & $-\Im(M\omega)$ & $\tau/M$ & $Q$ \\
\hline
axial spacetime QNM & -- & 0.4638 & 0.09888 & 10.113 & 2.345 \\
formal eikonal reference & -- & 0.445217 & 0.093838 & 10.66 & 2.372 \\
internal Cowling mode & 0 & 0.188320 & -- & $1.885\times10^{4}$ & $1.775\times10^{3}$ \\
internal Cowling mode & 1 & 0.558423 & -- & 243.86 & 68.09 \\
internal Cowling mode & 2 & 0.792306 & -- & 60.18 & 23.84 \\
internal Cowling mode & 3 & 1.028633 & -- & 21.18 & 10.89
\end{tabular}
\end{ruledtabular}
\end{table*}

The conversion to physical frequency is
\begin{equation}
f=\frac{\Re(M\omega)}{2\pi M}
\simeq
32.3\,{\rm kHz}
\left(\frac{M_\odot}{M}\right)
\Re(M\omega).
\label{eq:GC9_qnm_physical_frequency}
\end{equation}
Thus the internal-mode spectrum moves from the ground-based band for stellar-mass configurations to the millihertz and sub-millihertz domains for massive and supermassive objects.

The static endpoint therefore has both a genuine damped axial spacetime QNM and an approximate matter-supported Cowling ladder. What remains open is the polar complex spectrum, rotational splitting and polar--axial mixing, excitation amplitudes and the nonlinear post-merger waveform. The absence of an exterior photon ring invalidates only a direct light-ring/eikonal identification; it does not invalidate the computed low-multipole axial mode. The post-turning light-ring continuation remains a separate radially unstable transient problem.

\begin{figure}[t]
\centering
\includegraphics[width=\columnwidth]{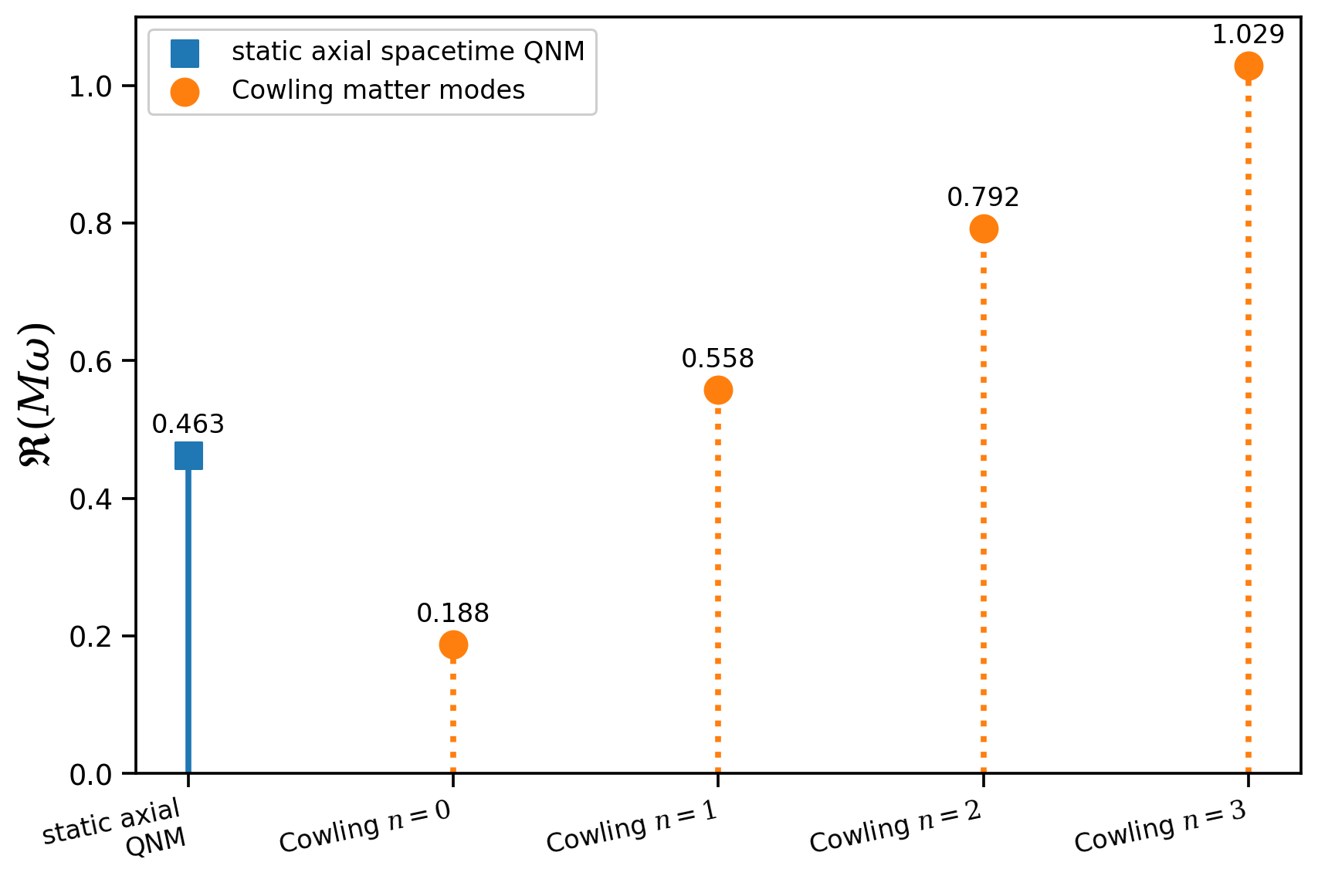}
\caption{Real parts of the genuine static axial spacetime QNM and the first four nonrotating quadrupolar internal-mode frequencies of the GC9 benchmark. The axial entry is obtained from the complex Regge--Wheeler eigenvalue problem, whereas the internal entries are Cowling matter-mode estimates. The separate formal light-ring/eikonal comparison is not plotted because the stable GC9 configuration has no exterior light ring and that number is not a stellar eigenmode.}
\label{fig:GC9_QNM_spectrum}
\end{figure}

\subsubsection{Limitations and future computations}
\label{app:GC9_referee_limitations}

As stressed before, some limitations of this approach are self-evident. First, the GC9 closure is not yet derived from a production $G(2)$ lattice calculation, instead it is a constrained EFT ansatz designed to be lattice-matchable. Second, the compact branch is computed in a Thomas--Fermi fluid approximation. Full finite-coupling EKG solutions should be used to test convergence toward this branch. Third, the high-spin claim is based on a mass-shedding proxy: a full rotating numerical solution is required to determine the exact maximum spin, shape and ergoregion properties. Fourth, although the static axial $\ell=2$ QNM is computed, the polar complex spectrum, rotating mode coupling, excitation amplitudes and nonlinear waveform are not. Fifth, the GC6--9--12A construction of Sec.~\ref{subsec:GC6912A_extension} is an outlook hierarchy only: none of its anisotropic benchmark values enters the GC9 equilibrium, multipole, tidal or QNM results quoted here.

These limitations prevent the fluid branch from being advertised as an established final solution of the $G(2)$ dark matter microscopic scalar theory but they define the next computational program. Nevertheless the GC9 BS passes the analytic and reduced-fluid filters and yields conditional, quantitatively testable observables. 

\subsection{Scalar versus vector realization of the GC9 medium}
\label{subsec:GC9_SV_comparison}

Conditional on the odd-sector selection rule assumed in the microscopic discussion above, two especially relevant candidate protected channels are a scalar $J^{PC}=0^{--}$ channel and a spin--one $J^{PC}=1^{+-}$ channel. The former gives a direct microscopic interpretation of the scalar order parameter used throughout the GC9 construction, whereas the latter suggests that the lightest protected particle may instead be vector-like.  The purpose of this subsection is therefore not to choose between these channels, but to determine how strongly the compact-star predictions depend on that choice.  We denote the scalar realization by GC9-S and a polarization-balanced spin--one realization by GC9-V.
Standard Proca stars provide the relativistic microscopic reference class for gravitating spin--one condensates \citep{BritoCardosoHerdeiroRadu2016,HerdeiroPanotopoulosRadu2020}, while massive-vector perturbations possess additional polarization sectors and rotation-induced axial--polar couplings \citep{PaniEtAl2012ProcaPert}.

\subsubsection{Exact isotropic bulk correspondence}
\label{subsubsec:GC9V_isotropic}

Let $q$ be the positive coarse-grained quadratic occupation invariant of the vector medium and define
\begin{equation}
	U_V(q)=\Lambda_T^4 f(y),\;
	f(y)=y+y^9,\;
	y=\frac{q}{q_T},\;
	q_T=\frac{\Lambda_T^4}{m_V^2}.
	\label{eq:GC9V_potential}
\end{equation}
A convenient local realization of the isotropic limit is an equal-occupation ensemble of the three spin--one polarizations (or, equivalently for the stress tensor, a mutually orthogonal polarization triad).  With the normalization $q=\frac12\sum_a A^{(a)*}_iA^{(a)i}$, the harmonic field equation gives $\omega_c^2=U_{V,q}$.  The summed kinetic and potential stresses are isotropic and reduce to
\begin{equation}
	\rho_V=qU_{V,q}+U_V,\qquad
	P_V=qU_{V,q}-U_V.
	\label{eq:GC9V_legendre}
\end{equation}
Consequently the vector medium has exactly the same Thomas--Fermi bulk closure as GC9-S,
\begin{equation}
		\hat\rho_V=\hat\rho_S=2y+10y^9,\qquad
		\hat P_V=\hat P_S=8y^9,
	\label{eq:GC9V_sameEOS}
\end{equation}
with
\begin{equation}
	c_s^2=\frac{72y^8}{2+90y^8}\longrightarrow\frac45.
	\label{eq:GC9V_cs}
\end{equation}
Equation~\eqref{eq:GC9V_sameEOS} is exact within the polarization-balanced coarse-grained construction.  Hence the isotropic TOV sequence is identical to GC9-S and so are all observables that depend only on the same perfect-fluid background and perturbation closure.  At the maximum-mass point,
\begin{align}
	M_{\max}^{\rm red}&=0.064157, \quad
	R_{\max}^{\rm red}=0.197591, \\ \nonumber
	C(M_{\max})&=0.324694, \quad
	k_2=0.023667, \\ \nonumber
	\Lambda&=4.37208, \quad
	\bar I=5.11636. \nonumber
	\label{eq:GC9V_iso_numbers}
\end{align}
Thus a $1^{+-}$ microscopic ground state does not, by itself, force a change in the leading GC9 mass--radius or tidal construction.  This bulk correspondence is the central reason why the scalar stellar model can remain useful even before the microscopic $0^{--}$--$1^{+-}$ ordering is settled.

\subsubsection{Controlled vector-stress envelope: static structure and first-order rotation}
\label{subsubsec:GC9V_anisotropy}

A coherent or incompletely depolarized spin--one medium can carry residual anisotropic stress.  To measure the sensitivity of GC9 to such effects we adopt the quasi-local Horvat-type envelope \citep{HorvatIlijicMarunovic2011}
\begin{equation}
	\Delta_V(r)\equiv P_t-P_r
	=\eta_V P_r\frac{2m(r)}{r},
	\qquad -0.30\leq\eta_V\leq0.24.
	\label{eq:GC9V_Horvat}
\end{equation}
It vanishes at the center and at the zero-pressure surface.  Equation~\eqref{eq:GC9V_Horvat} is a sensitivity parameterization rather than a derivation of the unique microscopic $1^{+-}$ stress tensor.  It is, however, particularly useful because quasi-local anisotropy can be treated consistently in both static and slow-rotation formalisms \citep{BeltracchiPosada2024,BecerraEtAl2024AnisoHT}. Directly solving the maximum-mass sequence for $C=1/3$ gives
\begin{equation}
\eta_V^{\rm LR}=0.24243,
\label{eq:GC9V_eta_LR}
\end{equation}
so the positive side of the main comparison is conservatively terminated at $\eta_V=0.24$.  This prevents the controlled GC9-V envelope from mixing non-ultracompact stars with backgrounds possessing an external Schwarzschild photon sphere; the latter require a separate multi-root trapped-mode analysis.

For the static background we solve
\begin{align}
	\frac{dm}{dr}&=4\pi r^2\rho,\\
	\frac{dP_r}{dr}&=-\frac{(\rho+P_r)(m+4\pi r^3P_r)}{r(r-2m)}
	+\frac{2\Delta_V}{r}.
	\label{eq:GC9V_anisoTOV}
\end{align}
At first order in rotation the frame-dragging equation becomes
\begin{equation}
	\frac{1}{r^4}\frac{d}{dr}\left(r^4j\frac{d\varpi}{dr}\right)
	+\frac{4j'}{r}\frac{\rho+P_t}{\rho+P_r}\,\varpi=0,
	\qquad \varpi\equiv\Omega-\omega,
	\label{eq:GC9V_anisoHartle1}
\end{equation}
which reduces to the ordinary Hartle equation at $P_t=P_r$.  Matching to $\varpi_{\rm ext}=\Omega-2J/r^3$ gives $I=J/\Omega$.

\begin{table*}[t]
	\centering
	\scriptsize
	\setlength{\tabcolsep}{3.7pt}
	\caption{GC9-S/GC9-V maximum-mass comparison within the non-ultracompact envelope.  The $\eta_V=0$ row is the exact isotropic GC9-S/GC9-V bulk limit. The negative endpoint reaches about $19\%$ local pressure anisotropy, while the positive endpoint $\eta_V=0.24$ reaches about $16\%$ and remains just below the $C=1/3$ threshold of Eq.~\eqref{eq:GC9V_eta_LR}.  $\bar Q_{\rm rapid}$ is the calibrated second-order estimate of Eq.~\eqref{eq:GC9V_Qbar_scaling}.  $M\omega_w$ is the directly solved frozen-vector/effective-fluid axial spacetime mode described below.}
	\label{tab:GC9SV_vector_envelope}
	\begin{ruledtabular}
		\begin{tabular}{c c c c c c c c c}
			$\eta_V$ & $\max|\Delta_V|/P_r$ & $M_{\max}^{\rm red}$ & $R_{\max}^{\rm red}$ & $C$ & $\bar I$ & $\bar I C^{3/2}$ & $\bar Q_{\rm rapid}$ & $M\omega_w$\\
			\hline
			$-0.30$ & 0.188 & 0.061271 & 0.195324 & 0.313687 & 5.3100 & 0.9329 & 1.6785 & $0.46460-0.11481i$\\
			$-0.15$ & 0.096 & 0.062711 & 0.196449 & 0.319225 & 5.2092 & 0.9395 & 1.5979 & $0.46453-0.10677i$\\
			$0$     & 0     & 0.064157 & 0.197591 & 0.324694 & 5.1164 & 0.9466 & 1.5216 & $0.46385-0.09888i$\\
			$+0.15$ & 0.099 & 0.065603 & 0.198752 & 0.330072 & 5.0311 & 0.9541 & 1.4495 & $0.46258-0.09121i$\\
			$+0.24$ & 0.160 & 0.066469 & 0.199458 & 0.333248 & 4.9834 & 0.9587 & 1.4083 & $0.46155-0.08675i$\\
		\end{tabular}
	\end{ruledtabular}
\end{table*}

The bulk response is small even over this deliberately broad envelope.  At $|\eta_V|=0.15$, corresponding to a maximum pressure anisotropy of about ten per cent, $M_{\max}$ changes by $\simeq2.25\%$, $R_{\max}$ by only $\simeq0.6\%$, and $C$ by $\simeq1.7\%$.  Across the adopted non-ultracompact interval, the negative endpoint gives shifts of $-4.50\%$, $-1.15\%$ and $-3.39\%$ in $M_{\max}$, $R_{\max}$ and $C$, whereas the positive $\eta_V=0.24$ endpoint gives $+3.60\%$, $+0.94\%$ and $+2.63\%$.  The dimensionless moment of inertia changes by less than four per cent over the same interval.

The cancellation in the high-spin proxy is still stronger.  The quantity actually fixed by the static plus first-order calculation is
\begin{equation}
	\frac{j_{\max}^{\rm proxy}}{\kappa_{\rm ms}}
	=\bar I C^{3/2},
	\label{eq:GC9V_jproxy}
\end{equation}
which changes only from $0.9466$ in GC9-S to $0.9329$--$0.9587$ over the adopted vector envelope.  Thus the vector correction to this directly computed combination is at most about $1.5\%$, smaller than the present uncertainty associated with calibrating $\kappa_{\rm ms}$ itself.

\subsubsection{Second-order quadrupole envelope}
\label{subsubsec:GC9V_Qbar}

A complete GC9-V quadrupole requires the anisotropic $\ell=2$ Hartle--Thorne displacement system.  Such second-order anisotropic formalisms exist and demonstrate that anisotropy can affect the quadrupole substantially when it is large \citep{BeltracchiPosada2024,BecerraEtAl2024AnisoHT}.  Before undertaking that full calculation, a controlled structural estimate follows from
\begin{equation}
	Q_{\rm rot}\simeq-\kappa_Q\Omega^2R^5,
	\qquad I=k_I MR^2,
\end{equation}
so that
\begin{equation}
	\bar Q\equiv-\frac{QM}{J^2}
	\simeq\frac{\kappa_Q}{k_I^2 C}.
	\label{eq:GC9V_Qbar_scaling}
\end{equation}
We calibrate the single dimensionless shape coefficient $\kappa_Q$ to the direct scalar GC9 Hartle--Thorne result $\bar Q_S=1.52158$ and insert the independently calculated GC9-V values of $C$ and $k_I=I/(MR^2)$.  This yields
\begin{equation}
1.408\lesssim\bar Q_V^{\rm rapid}\lesssim1.678
	\qquad (-0.30\leq\eta_V\leq0.24).
	\label{eq:GC9V_Qbar_range}
\end{equation}
The change is only about $5\%$ for $|\eta_V|=0.15$ and remains within about $10\%$ over the adopted non-ultracompact envelope.  Positive tangential stress moves the estimate toward the Kerr value $\bar Q=1$, whereas radial-dominated stress moves it away from Kerr.  Equation~\eqref{eq:GC9V_Qbar_range} is therefore a second-order sensitivity envelope, not a substitute for the eventual full anisotropic Hartle--Thorne solution.

\subsubsection{Direct effective-fluid axial spacetime mode}
\label{subsubsec:GC9V_axialQNM}

The scalar GC9 maximum-mass configuration has the directly computed static axial mode
\begin{equation}
	M\omega_{w,S}=0.46385-0.098881i,
	\; \frac{\tau}{M}=10.113,
	\; Q=2.345.
	\label{eq:GC9S_axialmode}
\end{equation}
For an anisotropic fluid the odd-parity metric-led sector obeys the generalized Regge--Wheeler equation \citep{Mondal2026AxialW}
\begin{equation}
	\frac{d^2Z}{dr_*^2}+\left[\omega^2-V_{\rm ax}^{(V)}(r)\right]Z=0,
	\label{eq:GC9V_RW}
\end{equation}
with
\begin{equation}
	V_{\rm ax}^{(V)}=e^{2\Phi}
	\left[\frac{\ell(\ell+1)}{r^2}+4\pi(\rho-P_r)-\frac{6m}{r^3}\right].
	\label{eq:GC9V_RWpotential}
\end{equation}
For $\eta_V\neq0$, Eq.~\eqref{eq:GC9V_RW} should not be confused with the
most general perturbation system of an anisotropic star.  Recent consistent
derivations show that pressure anisotropy can induce direct axial
matter--metric coupling and that spacetime-mode frequencies can depend on the
perturbed anisotropy prescription \citep{RodriguezRuizEtAl2026Axial,YuEtAl2026AnisoW}.
We therefore impose a deliberately restricted \emph{frozen-anisotropy}
closure: axial perturbations of the anisotropy tensor and of the microscopic
vector polarization are set to zero, while the anisotropic background
$\{\rho,P_r,m,\Phi\}$ is retained in Eq.~\eqref{eq:GC9V_RW}.  Under this
metric-led diagnostic closure we solve the eigenproblem on each GC9-V
maximum-mass background by two-sided logarithmic-derivative
(Riccati/Wronskian) matching: the regular-centre solution is integrated
outward and matched at the surface to an inward integration initialized by a
high-order purely outgoing asymptotic series.  No frequency calibration is
applied.  The approximation becomes the ordinary perfect-fluid axial problem
at $\eta_V=0$, where the background is exactly the GC9-S maximum-mass
configuration and the calculation gives $M\omega=0.46385-0.098881i$.
As an independent validation of the upgraded matching, the same exterior
machinery applied to the Schwarzschild $\ell=2$ problem gives
$M\omega=0.37367168-0.08896232i$, reproducing the standard Leaver value at
the displayed precision \citep{Leaver1985}.

Varying the outer matching radius from $35M$ to $50M$ changes the roots by less than $2\times10^{-9}$ at the audited reference points; at the new positive endpoint $\eta_V=0.24$ the corresponding variation is below $3\times10^{-11}$. This provides a direct convergence check of the effective-fluid eigenproblem.  The uncalibrated GC9-V frequencies are the last column of Table~\ref{tab:GC9SV_vector_envelope}.  Their most important feature is that the oscillation frequency is considerably more robust than the damping rate:
\begin{equation}
		0.46155\lesssim {\rm Re}(M\omega_{w,V})\lesssim0.46460,
	\label{eq:GC9V_w_Re_range}
\end{equation}
whereas
\begin{equation}
		0.08675\lesssim -{\rm Im}(M\omega_{w,V})\lesssim0.11481.
	\label{eq:GC9V_w_Im_range}
\end{equation}
Thus the real part moves by less than $0.5\%$ across the adopted non-ultracompact envelope, while the damping rate changes from about $-12.3\%$ to $+16.1\%$.  Equivalently,
\begin{equation}
	8.71\lesssim\tau/M\lesssim11.53,
	\qquad
	2.02\lesssim Q\lesssim2.66.
	\label{eq:GC9V_w_Qrange}
\end{equation}
Within the frozen-anisotropy closure this is stronger than a fit-based
extrapolation because Eq.~\eqref{eq:GC9V_RW} is solved directly on the GC9-V
backgrounds.  It is not, however, the fully consistent anisotropic axial
spectrum: the latter can contain matter--metric coupling
\citep{RodriguezRuizEtAl2026Axial,YuEtAl2026AnisoW}, and a microscopic vector
completion adds explicit polarization perturbations.  The quoted roots are
therefore a background-sensitivity diagnostic for the metric-led spacetime
mode, not the full Einstein--Proca--GC9 QNM spectrum.

\subsubsection{Matter-supported modes}
\label{subsubsec:GC9V_mattermodes}

The scalar GC9 nonrotating Cowling ladder is
\begin{equation}
	M\omega_{n,S}=(0.188320,\,0.558423,\,0.792306,\,1.028633).
	\label{eq:GC9S_Cowling}
\end{equation}
A rapid structural estimate for the vector backgrounds follows from the dynamical-frequency scaling $M\omega\propto C^{3/2}$.  It gives
\begin{align}
	\eta_V=-0.30:&\quad
	M\omega_n\simeq(0.1788,\,0.5303,\,0.7524,\,0.9768),\\
	\eta_V=+0.24:&\quad
	M\omega_n\simeq(0.1958,\,0.5806,\,0.8238,\,1.0696).
	\label{eq:GC9V_Cowling_range}
\end{align}
The corresponding structural shift ranges from about $-5.0\%$ at $\eta_V=-0.30$ to $+4.0\%$ at $\eta_V=+0.24$, and is about $2.5\%$ for $|\eta_V|=0.15$.  These numbers remain Cowling estimates and do not include direct polarization perturbations.

\subsubsection{Vector-specific polarization and amplitude scales}
\label{subsubsec:GC9V_vector_modes}

A genuine spin--one medium has degrees of freedom absent from GC9-S.  The GC9 barotropic closure alone, however, does not uniquely fix their gap because it contains only a density interaction.  This is closely analogous to the nonrelativistic Proca EFT, in which the density coupling and spin--spin coupling are independent; when the spin--spin coupling vanishes, the theory enters a symmetry-enhanced sector with additional polarization and multi-frequency structure \citep{ChavezNamboEtAl2025NRProca}.  Recent linear analysis finds the nonrelativistic ground state mode-stable under generic perturbations and also identifies stable polarization-dependent excited configurations \citep{ChavezNamboEtAl2026Stability}.

It is nevertheless useful to establish the scale separation implied by the minimal density-only GC9-V surrogate.  In a local homogeneous patch let
\begin{equation}
	\omega_c^2=U_{V,q}=m_V^2(1+9y^8),
	\qquad
	a\equiv qU_{V,qq}=72m_V^2y^8.
	\label{eq:GC9V_local_parameters}
\end{equation}
Linearization of the density/amplitude sector gives the local determinant
\begin{equation}
	(\Omega^2-k^2-a)^2-4\omega_c^2\Omega^2-a^2=0.
	\label{eq:GC9V_density_dispersion}
\end{equation}
The gapless branch has
\begin{equation}
	c_s^2=\frac{a}{a+2\omega_c^2},
	\label{eq:GC9V_local_cs}
\end{equation}
which reproduces the GC9 sound speed, while the local amplitude branch has
\begin{equation}
	\Omega_H^2(k=0)=4\omega_c^2+2a.
	\label{eq:GC9V_Higgs_gap}
\end{equation}
In the symmetry-enhanced polarization limit the orthogonal reorientation sector has the schematic relativistic dispersion
\begin{equation}
	\Omega_{\rm pol,\pm}(k)=\sqrt{\omega_c^2+k^2}\pm\omega_c,
	\label{eq:GC9V_pol_dispersion}
\end{equation}
so that the lower branch is type-II,
\begin{equation}
	\Omega_{\rm pol,-}(k)\simeq\frac{k^2}{2\omega_c},
	\qquad k\ll\omega_c,
	\label{eq:GC9V_soft_spin}
\end{equation}
whereas the upper branch is microscopic, $\Omega_{\rm pol,+}(0)=2\omega_c$.

At the GC9-S maximum-mass central point $y_c=0.8730126$ one obtains
\begin{equation}
	\frac{\omega_c}{m_V}=2.00916,
	\qquad
	\frac{a}{m_V^2}=24.2939,
	\qquad
	c_s^2=0.75057,
	\label{eq:GC9V_local_numbers1}
\end{equation}
\begin{equation}
	\frac{\Omega_{\rm pol,+}(0)}{m_V}=4.01833,
	\qquad
	\frac{\Omega_H(0)}{m_V}=8.04579.
	\label{eq:GC9V_local_numbers2}
\end{equation}
The lower polarization branch should be interpreted as a symmetry-enhanced near-flat direction rather than as a predicted stellar spectral line.  Its finite-star frequency depends on boundary conditions and on spin--spin or polarization-locking operators that are absent from the density-only closure; accordingly we do not assign an astrophysical frequency to Eq.~\eqref{eq:GC9V_soft_spin}.

By contrast, the hard local scales over $m_V=10^9$--$10^{14}\,$GeV are
\begin{equation}
	f_{\rm pol,+}\simeq9.7\times10^{32}\text{--}9.7\times10^{37}\,{\rm Hz},
\end{equation}
and 
\begin{equation}
	f_H\simeq1.95\times10^{33}\text{--}1.95\times10^{38}\,{\rm Hz}.
	\label{eq:GC9V_hard_numeric}
\end{equation}
These hard frequencies are microscopic scales, not stellar QNMs, and lie outside the intended coarse-grained GC9 regime.  Conversely, the soft branch of Eq.~\eqref{eq:GC9V_soft_spin} is a consequence of the symmetry-enhanced density-only surrogate and is not universal.  A spin--spin interaction, polarization locking, or a canonical single-field Proca completion can lift the mode and introduce a new gap that is not determined by the barotropic GC9 EOS.  
Fundamental self-interacting Proca realizations require independent
consistency checks of their perturbative degrees of freedom: sufficiently
compact quartically self-interacting Proca stars can encounter ghost
pathologies \citep{AokiMinamitsuji2023CompactProca}, while suitable
UV completions, such as the Proca--Higgs construction, can remove the
hyperbolicity pathology of the effective vector theory
\citep{HerdeiroRaduCostaFilho2023ProcaHiggs}.

\subsubsection{Combined scalar--vector comparison and interpretation}
\label{subsubsec:GC9SV_verdict}

\begin{table*}[t]
	\centering
	\tiny
	\setlength{\tabcolsep}{3pt}
	\caption{Status of the GC9-S/GC9-V comparison.  ``Exact'' refers to the polarization-balanced coarse-grained GC9-V construction.}
	\label{tab:GC9SV_status}
	\begin{ruledtabular}
		\begin{tabular}{l l l l}
			Observable/sector & GC9-S & GC9-V result & Status\\
			\hline
			Bulk EOS & $\hat\rho=2y+10y^9$, $\hat P=8y^9$ & identical in isotropic limit & exact in GC9-V bulk\\
			$M_{\max},R,C$ & baseline & $\lesssim4.5\%,1.2\%,3.4\%$ shifts over adopted envelope & direct TOV envelope\\
			$\bar I$ & $5.1164$ & $4.9834$--$5.3100$ & direct first-order Hartle\\
			$\bar I C^{3/2}$ & $0.9466$ & $0.9329$--$0.9587$ & direct first-order Hartle\\
			$\bar Q$ & $1.5216$ & $1.4083$--$1.6785$ & calibrated rapid HT2 envelope\\
			axial $w$ mode & $0.46385-0.09888i$ & Re shift $<0.5\%$, damping shift $-12.3\%$ to $+16.1\%$ & direct effective-fluid eigenproblem\\
			Cowling ladder & Eq.~\eqref{eq:GC9S_Cowling} & structural shift $\lesssim5\%$ & dynamical-frequency proxy\\
			vector polarization & absent & ultra-soft in density-only symmetry limit & model-dependent\\
			vector hard/amplitude modes & absent & $\Omega\sim m_V$ & microscopic, outside stellar EFT\\
			full rotating spectrum & open & extra axial/polar polarization couplings & requires Einstein--vector solve\\
		\end{tabular}
	\end{ruledtabular}
\end{table*}

The comparison therefore has a simple interpretation: if the protected microscopic ground state is $0^{--}$, GC9-S has a direct scalar microscopic realization.  If instead the $1^{+-}$ state is lighter, the polarization-balanced GC9-V medium reproduces the same bulk EOS exactly and even deliberately large residual vector stresses modify the static and first-order observables only at the few-per-cent level.  The rapid quadrupole remains order unity and, for tangentially dominated stress, actually moves toward Kerr.  The directly solved metric-led axial spacetime frequency remains pinned near ${\rm Re}(M\omega)\simeq0.46$, although its damping time is more sensitive to polarization anisotropy.

The genuinely discriminating observables are therefore not the existence of the compact branch or its leading mass--radius relation, but the fully self-consistent second-order multipoles and the additional vector perturbation content.  Canonical Proca stars are known to possess distinct scalar-versus-vector dynamical properties, especially in rotation \citep{SanchisGualEtAl2019Spinning}, and a fundamental self-interacting Proca realization can introduce consistency constraints absent from the scalar theory \citep{AokiMinamitsuji2023CompactProca,HerdeiroRaduCostaFilho2023ProcaHiggs}.  

Accordingly, the present GC9-S solutions remain a consistent and economical baseline for the compact-star phenomenology even if future microscopic spectroscopy selects $1^{+-}$: in that case GC9-S should be read as the scalar coarse-grained order parameter of the isotropized protected medium, while GC9-V quantifies the leading systematic uncertainty and identifies the vector-specific observables that must ultimately be recomputed.

\section{Phenomenological outlook I: scalar bursts and the Galactic Center}

The following three sections are phenomenological outlooks and are not used to establish the existence or equilibrium properties of the compact GC9 branch. Their order-unity efficiencies, threshold functions and population fractions are illustrative placeholders rather than calibrated fits. They are retained to identify future calculations and observables, while the submission-level evidence for GC9 rests on the conditional Thomas--Fermi equilibrium, tidal, slow-rotation and static axial-QNM analyses of Sec.~\ref{sec:observables}.

\label{sec:SB}
Burst-like phenomena, gravitational cooling and violent relaxation have analogues in boson star, oscillaton and axion-star dynamics, consequently providing a useful phenomenological bridge between compact-object mergers and particle-like scalar bursts \citep{SeidelSuen1994,HawleyChoptuik2000,PalenzuelaLehnerLiebling2007,BezaresPalenzuela2017,Eby2016Axion,Helfer2017Axion,LevkovPaninTkachev2018}.

A possible consequence of horizonless GC9 merger relaxation is episodic, non-destructive mass--energy ejection in the form of \emph{scalar bursts}; no event rate or detector-level signal is predicted in the present work. These events have no direct analogue in classical BH physics and arise from the interplay between strong self-interactions, relativistic gravity and the absence of an event horizon. 
Numerical studies of dynamical boson stars in scalar-tensor gravity show that transitions between equilibrium branches can be accompanied by a burst of scalar radiation, followed by a long-lived oscillatory
remnant: in the simulations by Ruiz et al. \citep{Scalarization012}, the radiated scalar-field
energy can reach the \(\sim10^{-2}\)--\(10^{-1}\) level of the mass scale of the configuration, with dominant late-time scalar frequencies of order \(10^{-3}\)--\(10^{-2}\) in dimensionless boson-star units.

In GC9 boson stars, the equation of state asymptotically approaches a stiff but causal linear regime $P/\rho\to4/5$, while remaining analytic and non-singular. As a consequence, the mass--radius relation admits configurations close to the relativistic stability boundary. When such objects undergo accretion, mergers, or tidal perturbations, they can temporarily exceed the maximum mass or angular momentum allowed for stationary equilibrium. The black-hole-substitution channel considered here consists of events for which energy and angular-momentum losses drive the remnant back onto a stable horizonless branch. If that re-landing condition fails, the present effective-fluid model does not determine the endpoint; such events lie outside the demonstrated GC9 substitution channel. Scalar radiation is therefore a candidate loss channel rather than an established generic outcome for GC9.

This putative relaxation mechanism is a nonlinear analogue of the well-known process of \emph{gravitational cooling} observed in boson-star formation \citep{SeidelSuen1994}; if an efficient asymptotically propagating loss channel exists in GC9, the same strong-field regime could produce observable transient signatures.

Scalar bursts may be triggered by several astrophysical processes:
\begin{itemize}
	\item \textbf{Accretion-induced overmass}: gradual accumulation of baryonic or dark matter pushes the star beyond the local stability branch.
	\item \textbf{Post-merger relaxation}: binary GC9 boson star mergers can produce a transient overmassive or rapidly rotating configuration whose endpoint depends on the nonlinear dynamics.
	\item \textbf{Spin-up to extremality}: angular momentum transfer (e.g.\ via disks or mergers) drives the star beyond its maximal stable spin $j_{\max}$.
	\item \textbf{Tidal excitation}: close encounters or EMRIs excite non-linear scalar modes.
\end{itemize}
In all cases, the defining property is the absence of a horizon, which allows the excess energy to be radiated away rather than hidden.

\subsection{Burst energetics}
Numerical studies of BS formation, collapse and mergers show that strongly perturbed configurations can relax through gravitational cooling, scalar-field emission and gravitational-wave losses. In particular, when a remnant is driven beyond the stability threshold by merger or rapid accretion, part of the excess energy must be shed before the system can re-land on a stable branch of solutions \citep{SeidelSuen1994,Liebling2017,BezaresPalenzuela2017}.
We call the excess mass above the stable branch by $\Delta M$; the scalar burst is then assumed to tap a fraction \(f_{\rm sb}\) of this overshoot energy,
\begin{equation}
	E_{\rm sb}\sim f_{\rm sb}\,\Delta M\,c^2,
	\qquad
	0<f_{\rm sb}\le 1.
\end{equation}
Accordingly, the burst efficiency relative to the total remnant mass is
\begin{equation}
	\epsilon_{\rm sb}
	\equiv \frac{E_{\rm sb}}{M c^2}
	\sim
	f_{\rm sb}\,\frac{\Delta M}{M}.
	\label{eq:eps_burst_from_gap}
\end{equation}
For fractional overshoots \(\Delta M/M\sim10^{-3}\text{--}10^{-1}\) and order-unity release fractions \(f_{\rm sb}\sim0.1\text{--}1\), this gives the phenomenologically reasonable range $\epsilon_{\rm sb}\sim 10^{-3}\text{--}10^{-1}$, where the lower end corresponds to mild overcompression and weak scalar leakage, while the upper end represents strongly nonlinear post-merger relaxation. 
This range should be interpreted as an informed estimate rather than as a simulation-level prediction for the GC9 model.
Immediately after a merger or rapid accretion episode, the remnant is expected to relax by exciting internal bosonic-core modes (which are absent in a classical Kerr spacetime) in addition to whatever spacetime-dominated response emerges from the full coupled problem. In the present GC9 framework, this means that scalar-burst emission should be viewed as the matter-sector counterpart of post-merger relaxation: part of the excess energy is first deposited into the compact bosonic core and is then shed through nonlinear scalar emission as the remnant re-lands on the stable branch. 
The Cowling analysis indicates internal quadrupolar bosonic modes that can participate in burst-assisted relaxation, although their coupling to the spacetime waveform has not yet been computed.
These collective frequencies do not by themselves prove the existence of an asymptotically propagating scalar channel: for an ultraheavy carrier field, the emitted disturbance must be shown to contain on-shell quanta, sufficiently high nonlinear harmonics, a gapless collective mode, or an indirect gravitational/portal channel.  This kinematic question is left open.
For supermassive GC9 boson stars with
\[
M\sim 10^6\text{--}10^{10}M_\odot,
\]
the corresponding burst energy scale is
\begin{equation}
	E_{\rm sb}\sim \epsilon_{\rm sb}\,Mc^2
	\sim 10^{57}\text{--}10^{63}\ {\rm erg},
\end{equation}
whereas lower-mass remnants would populate proportionally lower ranges. 

A more GC9-specific estimate may be obtained by characterizing the transient overcompression of the remnant core. Let \(\rho_{\rm eq}\) denote the equilibrium core density and let the peak density reached during the merger be
\begin{equation}
	\rho_{\rm peak}=\eta\,\rho_{\rm eq},
	\qquad \eta>1.
\end{equation}
The excess scalar energy stored in a core of radius \(R_c\) is then
\begin{equation}
	E_{\rm sb}
	\sim
	\int d^3x\,(\rho_{\rm peak}-\rho_{\rm eq})
	\sim
	\frac{4\pi}{3}R_c^3(\eta-1)\rho_{\rm eq}.
\end{equation}
Defining the equilibrium core mass
\begin{equation}
	M_c\sim \frac{4\pi}{3}R_c^3\rho_{\rm eq},
\end{equation}
one finds, in natural units,
\begin{equation}
	E_{\rm sb}\sim (\eta-1)M_c.
	\label{eq:scalar_burst_core}
\end{equation}
If the core contains a fraction \(f_c\) of the remnant mass $M_c=f_c M_f$, then
\begin{equation}
	E_{\rm sb}\sim (\eta-1)f_c M_f.
	\label{eq:scalar_burst_fraction}
\end{equation}
This shows that the scalar-burst efficiency is controlled mainly by the degree of overcompression and by the core mass fraction participating in the nonlinear relaxation.

A characteristic feature of the GC9 interaction is that the local potential remains positive and analytic while the Thomas--Fermi pressure grows as a controlled density-nine third-cumulant closure.  At high density the reduced EOS approaches
\begin{equation}
\hat P\simeq {4\over5}\hat\rho,
\qquad
 c_s^2\to {4\over5}<1,
\end{equation}
rather than a bag-like constant-density state or a superluminal hard core.  The interaction energy grows sufficiently rapidly to support the modeled stable branch, while the fluid response remains causal and smooth.  This property permits a smooth repulsive response during overcompressed post-merger relaxation; the magnitude and efficiency of any scalar-energy release nevertheless require dedicated nonlinear simulations.

An immediate observational consequence is that part of the post-merger energy budget is transferred into the scalar sector rather than into gravitational waves. Relative to a BH merger with the same initial masses, a GC9 merger may therefore exhibit an apparent deficit in the emitted GW energy,
\begin{equation}
	\Delta E_{\rm GW}\propto E_{\rm sb},
\end{equation}
together with secondary signatures in the late-time relaxation. In particular, scalar bursts may leave imprints through mode shifts in the gravitational-wave signal, partial suppression of the standard Kerr-like ringdown amplitude and possibly indirect electromagnetic counterparts through environmental reprocessing \citep{Liebling2017,BezaresPalenzuela2017}.

Scalar bursts would differ from the expectations of classical vacuum-BH dynamics. In a horizonless GC9 remnant, excess field energy can in principle be radiated or redistributed within the macroscopic bosonic configuration rather than being absorbed by a horizon. A robust observation of episodic post-merger scalar-sector energy release from an otherwise compact source would therefore provide evidence for additional horizonless matter dynamics, although connecting such a signal uniquely to GC9 would require waveform modeling.

Depending on the excitation efficiency and waveform, LISA could probe post-merger relaxation in the supermassive regime, while third-generation ground-based detectors (such as Einstein Telescope, Cosmic Explorer) may be sensitive to stellar-mass analogues. 
Scalar bursts could therefore be a potential discriminant of horizonless matter dynamics, although identifying them specifically with GC9 requires waveform and propagation calculations.

A crucial theoretical consistency requirement is that post-burst configurations re-land on the stable branch rather than entering a runaway sequence. In the GC9 model this may be favored by the monotonic positive repulsion and by the absence of a vacuum-energy offset or bag constant. Existing analytical arguments and numerical studies of boson-star dynamics
show that radiative cooling can drive perturbed configurations toward new
equilibria in appropriate regions of parameter space
\citep{SeidelSuen1994,Liebling2017,BezaresPalenzuela2017}.  They do not establish
that this outcome is generic for GC9; dedicated \(3+1\) relativistic
simulations remain necessary.

\subsection{Detectability of Scalar Bursts with LISA and Connections to High-Energy Anomalies}
\label{subsec:lisa_scalar_bursts}

Scalar bursts are a possible consequence of a gravitational-cooling channel capable of regulating overmassive or overspinning GC9 remnants \citep{SeidelSuen1994,Liebling2017,Bezares2024ECO}. In the compact GC9 branch the relevant structural parameters are already known to lie in the range
\begin{equation}
\begin{aligned}
C(M_{\max})&=0.324694,\\
C_{\rm post\mbox{-}turn}^{\max}&=0.337118,\\
j_{\max}^{\rm proxy}&\simeq0.95,\qquad
\bar Q_{\rm slow}\simeq1.522.
\end{aligned}
\end{equation}
so one can sharpen the burst energetics beyond a purely phenomenological prior. A useful bookkeeping scale for a successful re-landing episode is
\begin{equation}
	\Delta E_{\rm rel}\sim
	\max\!\left[\Delta M\,c^2,\,\Omega_f\,\Delta J\right],
	\label{eq:sb_lower_bound}
\end{equation}
where $\Delta M\equiv M_0-M_{\rm stab}(J_f)$ is the overshoot above the
stable branch and $\Omega_f\Delta J$ is the corresponding rotational-work
scale.  Equation~\eqref{eq:sb_lower_bound} is not a rigorous lower bound on
the energy of one radiative channel: the actual energy-to-angular-momentum
ratio depends on the emitted modes and part of the excess can remain bound or
be transferred to surrounding matter.  We therefore use
\begin{equation}
	\epsilon_{\rm rel}\equiv \frac{\Delta E_{\rm rel}}{M_0c^2}
	\sim
	\max\!\left[\frac{\Delta M}{M_0},\,\frac{\Omega_f\Delta J}{M_0c^2}\right]
	\label{eq:epssb_def}
\end{equation}
only as a characteristic re-landing budget.  A scalar-burst fraction
$\epsilon_{\rm sb}\leq\epsilon_{\rm rel}$ is an additional dynamical input.
For near-threshold overshoot, a characteristic re-landing budget of a few
$10^{-3}$--$10^{-2}$ of the total mass is plausible, while more violent
episodes can involve a few percent.  Only an unknown fraction of that budget
need emerge specifically as propagating scalar radiation.  Accordingly,
\begin{equation}
	\epsilon_{\rm sb}\sim 10^{-3}\text{--}{\rm few}\times10^{-2}
	\label{eq:epssb_mild}
\end{equation}
is retained as an illustrative scalar-emission range rather than a consequence
of Eqs.~\eqref{eq:sb_lower_bound}--\eqref{eq:epssb_def}; values near
$10^{-1}$ are stressed upper edges, not generic expectations.
Using the controlled slow-row observables, the angular velocity follows from $J=I\Omega$ rather than from the order-unity estimate $I\sim MR^2$:
\begin{equation}
M\Omega=\frac{j}{\bar I}=0.064157.
\end{equation}
Thus the rotational work scale is naturally only a few percent of the total mass-energy. This supports the previous more restrictive burst-efficiency estimate.
The physical static axial QNM, formal eikonal comparison and nonrotating Cowling lines are
\begin{align}
M\omega_{\rm ax}&\simeq0.4638-0.09888i,\\
M\omega_{220}^{\rm formal}&\simeq0.445217-0.093838 i,\\
M\omega_n^{(\ell=2)}&\simeq0.188320,\ 0.558423,\ 0.792306,\ 1.028633.
\end{align}
The first is a genuine static axial stellar QNM, the second is not because its formal light ring is inside the surface, and the Cowling lines provide approximate internal dynamical scales. 
The scalar-burst scenarios considered here concern the deeply relativistic
compact branch, for which any macroscopic relaxation signal is naturally
organized around the dynamical crossing time of the emitting core.
At branch level, we estimate the emitting core radius by the compact mass-containing region of the remnant, taking \(R_c\sim R_{99}\) as a conservative proxy. Since the GC9 compact benchmarks have \(C\simeq0.3247\text{--}0.3371\), this implies
\begin{equation}
	\frac{R_c}{M}\sim \frac{R_{99}}{M}\sim \mathcal{O}(3\text{--}4),
\end{equation}
so the burst originates from a region of only a few gravitational radii.
Writing \(R_c=\alpha GM/c^2\) with \(\alpha\sim \mathcal O(3\text{--}4)\), the burst timescale is
\begin{equation}
	\tau_{\rm sb}\sim \frac{R_c}{c_s}\sim \frac{\alpha GM}{c^3},
\end{equation}
This implies a characteristic burst frequency
\begin{equation}
	f_{\rm sb}
	\sim \frac{1}{2\pi\tau_{\rm sb}}
	\sim \frac{c^3}{2\pi\alpha GM}
	\simeq
	8\times10^{-5}\text{--}1.1\,{\rm Hz},
\end{equation}
for $M \sim 10^{4}\text{--}10^{8} M_\odot$, which spans frequencies below, within and above LISA's most sensitive band; only the intermediate part of this mass range is a plausible LISA target. This selects populations normalized near their sector maximum mass in the sense of Sec.~\ref{sec:sector}. Under a single universal $\Lambda_T$, only remnants within a factor of a few of the corresponding $M_{\max}$ are compact enough to
radiate in this band.
Although the scalar field itself does not couple directly to the detector, rapid rearrangement of the mass quadrupole and excitation of spacetime modes induce a tensor counterpart. 
The point is not that the scalar channel is itself observed, but that the scalar-driven relaxation modifies the quadrupolar dynamics on precisely the timescales to which LISA is most sensitive.

A crucial question for testability is the event rate. Let $n_{\rm GC9}(M,z)$ denote the comoving number density of GC9 boson stars of mass $M$ at redshift $z$. The differential burst rate per comoving volume can be written as
\begin{equation}
	\frac{d\Gamma_{\rm sb}}{dV} =
	\int dM \, n_{\rm GC9}(M,z)\, \lambda_{\rm sb}(M,z),
\end{equation}
where $\lambda_{\rm sb}$ is the average burst rate per object. In the GC9 framework, bursts are triggered by accretion, mergers, or angular-momentum transfer, suggesting the scaling
\begin{equation}
	\lambda_{\rm sb}(M) \sim \eta_{\rm sb}\, \frac{1}{\tau_{\rm sb}}
	\sim \eta_{\rm sb}\, \frac{c^3}{10\,GM}.
\end{equation}
The quantity $\eta_{\rm sb}$ should be interpreted as a duty cycle, not as
an energy efficiency; the illustrative scalar-emission range in
Eq.~\eqref{eq:epssb_mild} is a separate phenomenological input.  The observable
rate can be written as
\small
\begin{equation}
	\begin{split}
	&\Gamma_{\rm obs} =
	\int dz \, \frac{dV}{dz}\frac{1}{1+z}\times \\
	&\int dM \, n_{\rm GC9}(M,z)\,
	\lambda_{\rm sb}(M,z)\,
	\Theta({\rm SNR}(M,z)-{\rm SNR}_{\rm thr}),
	\end{split}
\end{equation}
\normalsize
where $\Theta$ enforces detectability by LISA.  Equation above is the useful
result at the present level: neither the GC9 population function
$n_{\rm GC9}(M,z)$ nor the burst duty cycle $\eta_{\rm sb}$ has been derived,
so no reliable event-rate normalization is claimed.  If a supermassive GC9
population and the proposed relaxation channel are realized, the part of the
population whose dynamical frequencies overlap the LISA band is expected to
lie roughly in the range
\[
10^{5} M_\odot \lesssim M \lesssim 10^{7} M_\odot .
\]
A detector-level rate requires population synthesis, a merger/trigger model
and a computed scalar-to-observable transfer function.

In addition, while the scalar field is dark, scalar bursts can indirectly power electromagnetic and high-energy signals through several channels: sudden changes in the gravitational potential can perturb surrounding disks; time-dependent curvature can accelerate plasma; and if the $G(2)$ glueball scalar couples to heavy mediators, scalar self-interactions can seed SM cascades. In particular,
\begin{equation}
	E_\nu \sim \alpha_\nu E_{\rm sb},
	\qquad
	\alpha_\nu \ll 1,
\end{equation}
can still provide a large available energy scale for high-energy secondaries
if a concrete portal and conversion mechanism exist.  The absence of a horizon
permits, but does not guarantee, outward energy release or repeated activity;
both require dynamical coupling to the environment.

The cleanest candidate discriminants are therefore:
\begin{enumerate}
	\item \textbf{Low-frequency relaxation bursts} in the LISA band, if nonlinear merger simulations confirm an appreciable scalar-radiation efficiency.
	\item \textbf{Repeated transient activity} from the same compact source, indicating a nonabsorbing matter sector rather than a vacuum exterior alone.
	\item \textbf{A non-Kerr post-merger spectrum} containing the shifted static axial mode and matter-supported internal modes; the rotating spectrum and excitation amplitudes remain to be computed.
	\item \textbf{Possible multi-messenger correlations} with high-energy neutrinos or gamma-ray flares, conditional on a specified portal to Standard Model particles.
	\item \textbf{Rapid rotation without a horizon}, subject to confirmation by a full two-dimensional rotating-equilibrium calculation rather than the present spin proxy.
\end{enumerate}
A correlated observation of several such features would motivate a horizonless matter interpretation, but discriminating GC9 specifically would require dedicated waveforms, ray tracing and portal modeling.

\subsection{GC9 boson stars versus fermionic DM core models for the Galactic center}
\label{sec:GC_relation}

We compare the GC9 \(G(2)\) glueball boson star framework with recent Galactic-center studies \citep{Crespi2026GCFermions,Pelle2024FermionCoreImaging} in which Sgr~A$^\ast$ is modeled as a horizonless compact object made of fermionic dark matter. The point of the comparison is twofold: first, to clarify whether the GC9 scenario is compatible with current observations of S-stars and G-objects; second, to identify the observational regimes in which bosonic and fermionic compact-core models begin to differ. So the comparison concerns the \emph{nature of Sgr~A$^\ast$} while the S-stars and G-objects are treated as dynamical tracers of its gravitational field.

The relevant present-day constraints come from stellar dynamics and horizon-scale imaging. The S2 orbit has now provided both a detection of the relativistic gravitational redshift near pericenter and a measurement of Schwarzschild precession, fixing the central mass at
\begin{equation}
M_{\rm GC}\simeq 4\times 10^6\,M_\odot
\end{equation}
and probing radii of order
\begin{equation}
r \sim 10^3\text{--}10^4\,r_g,
\qquad
r_g\equiv \frac{GM_{\rm GC}}{c^2},
\end{equation}
while the EHT image of Sgr~A$^\ast$ constrains the horizon-scale brightness depression and characteristic angular size \citep{Gravity2018S2Redshift,Gravity2020S2Precession,EHTSgrA2022I}. At these distances, the dominant observables are the Keplerian mass, relativistic pericenter precession, gravitational redshift and possible small extended-mass effects. Current data constrain the enclosed mass profile tightly, but they do not uniquely select an event horizon over a sufficiently compact horizonless alternative.

This point has been demonstrated explicitly in recent fermionic DM core analyses \citep{Crespi2026GCFermions,Pelle2024FermionCoreImaging}. In those models, Sgr~A$^\ast$ is described as a compact degenerate fermion core embedded in a larger core--halo dark matter distribution. Recent fits to S2 and several G-objects show that such horizonless configurations can reproduce the observed orbital data with differences from the BH case below the percent level, provided the compact core radius satisfies
\begin{equation}
R_c \ll r_{\rm peri}(S2)
\end{equation}
and the enclosed extended mass remains sufficiently small in the orbital region \citep{Crespi2026GCFermions,Pelle2024FermionCoreImaging}. The key lesson is therefore robust: \emph{present Galactic-center data do not by themselves require the existence of an event horizon.} 
\begin{table}[t]
	\centering
	\small
	\renewcommand{\arraystretch}{1.14}
	\setlength{\tabcolsep}{4pt}
	\begin{tabular}{ll}
		\hline
		\tcell[0.32\linewidth]{\textbf{Quantity}} &
		\tcell[0.55\linewidth]{\textbf{Benchmark value / GC9 implication}} \\
		\hline
		
		\tcell[0.32\linewidth]{Central mass of Sgr~A$^\ast$} &
		\tcell[0.55\linewidth]{\(M_{\rm GC}\simeq4\times10^6\,M_\odot\); fixes the target mass for a GC9 Galactic-center core.} \\
		
		\tcell[0.32\linewidth]{Gravitational radius} &
		\tcell[0.55\linewidth]{\(r_g=GM_{\rm GC}/c^2\simeq5.9\times10^6\,{\rm km}\simeq0.039\,{\rm AU}\).} \\
		
		\tcell[0.32\linewidth]{S2 pericenter scale} &
		\tcell[0.55\linewidth]{\(r_{\rm peri}(S2)\sim10^3\text{--}10^4\,r_g\); current stellar probes remain far outside the horizon-scale core.} \\
		
		\tcell[0.32\linewidth]{Compact GC9 core radius} &
		\tcell[0.55\linewidth]{For \(C\simeq0.3247\text{--}0.3371\), \(R_{\rm GC9}/M\simeq2.97\text{--}3.08\).} \\
		
		\tcell[0.32\linewidth]{Ratio to S2 pericenter} &
		\tcell[0.55\linewidth]{\(R_{\rm GC9}/r_{\rm peri}(S2)\sim10^{-3}\); present S-star astrometry cannot resolve the GC9 interior.} \\
		
		\tcell[0.32\linewidth]{EHT horizon-scale image} &
		\tcell[0.55\linewidth]{The observed Sgr~A$^\ast$ ring/shadow scale is \(\sim50\,\mu{\rm as}\); this is the relevant scale for Kerr--GC9 differences.} \\
		
		\tcell[0.32\linewidth]{Kerr-mimicking shifts} &
		\tcell[0.55\linewidth]{No external photon ring is present on the radially stable branch; the post-turning unstable continuation develops one, but dedicated ray tracing and nonlinear stability analysis are required before an image-plane claim can be made.} \\
		
		\tcell[0.32\linewidth]{Prompt ringdown sector} &
		\tcell[0.55\linewidth]{The static endpoint has an axial spacetime mode $M\omega=0.4638-0.09888i$ with $Q=2.345$. It is $24.1\%$ above Schwarzschild in real frequency at fixed mass; the rotating spectrum and merger excitation remain open.} \\
		
		\hline
	\end{tabular}
	\caption{Benchmark comparison for interpreting Sgr~A$^\ast$ as a compact GC9 boson-star core. The GC9 branch remains horizon-scale, \(R_{\rm GC9}\sim{\cal O}(r_g)\), while observed S-stars probe radii of order \(10^3\text{--}10^4\,r_g\).}
	\label{tab:sgrA_gc9_benchmark}
\end{table}
The overall properties for Sgr~A$^\ast$ are displayed in the Table \ref{tab:sgrA_gc9_benchmark}.
The GC9 framework lies in the same broad phenomenological class allowed by current S-star/G-object data. In our case, the central object is a self-gravitating boson star supported not by fermionic degeneracy pressure but by repulsive self-interactions of an effective complex glueball field governed by the GC9 potential. In the Thomas--Fermi regime, the resulting EOS approaches $P_{\rm GC9}\simeq(4/5)\rho$ with $c_s^2\to4/5$, allowing a highly compact branch whose radius is much smaller than the S2 pericenter scale once the macroscopic stiffness is fixed. Whenever
\begin{equation}
R_{\rm GC9} \ll r_{\rm peri}(S2),
\end{equation}
the exterior spacetime sampled by the currently observed S-stars is effectively point-mass dominated and the model is therefore expected to be observationally degenerate with the Kerr BH and fermionic-core descriptions at present precision. In this sense, the GC9 scenario is plausibly compatible with current S-star and G-object data, although a dedicated Sgr~A$^\ast$ fit remains to be carried out. The relation is satisfied in the Galactic-center-sector
normalization of Sec.~\ref{sec:sector}. Setting
$M_{\max}=M_{\rm GC}\simeq4\times10^{6}M_\odot$ requires
$\Lambda_T^{\rm GC}\simeq1.6\times10^{-4}$~GeV, for which
$R_{\rm GC9}\simeq3.07983\,GM_{\rm GC}/c^{2}\simeq1.82\times10^{7}$~km
$\simeq0.13$~AU, i.e.\
$R_{\rm GC9}/r_{\rm peri}({\rm S2})\sim10^{-3}$.

Fermionic DM cores and GC9 boson stars differ in their microphysics, rotational structure and horizon-scale lensing phenomenology. Fermionic models are typically embedded in a core--halo framework and can produce compact horizonless centers together with extended DM halos. In the GC9 framework, by contrast, the dark \(G(2)\) sector can be realized as a mixture of diffuse glueball matter and a population of compact GC9 objects, so the extended mass fraction near Sgr~A$^\ast$ is not fixed by a single equilibrium core--halo solution and can be tuned to remain below current observational limits,
\begin{equation}
\frac{M_{\rm ext}(r<r_{\rm peri})}{M_{\rm GC}}\lesssim 10^{-2}.
\end{equation}

Most importantly, GC9 boson stars are designed as restricted Kerr mimickers. Their compact branch supports rotation, an order-unity quadrupole estimate and a percent-level ISCO shift, but the computed surface remains outside the external photon-sphere radius. Thus, at the level of current Galactic-center astrometry, they remain degenerate with both black holes and fermionic cores, but they predict a different pattern of deviations once observables sensitive to spin, multipoles, photon rings, or tidal response become accessible. In particular:
\begin{itemize}
\item at S-star radii, all sufficiently compact models are effectively point-mass-like and remain hard to distinguish;
\item at horizon-scale radii, EHT-like imaging and future infrared interferometry can test the absence of a standard external photon ring through dedicated ray-traced brightness and lensing predictions;
\item outside the Galactic-center setting, GC9 boson stars motivate distinctive merger phenomenology --- gravitational cooling, scalar bursts and matter-supported internal oscillations --- which has no direct analogue in the same form for fermionic core models.
\end{itemize}

The recent fermionic DM core literature provides a useful comparison class: it shows that sufficiently compact horizonless mass distributions can remain compatible with present Galactic-center observations at the resolved scales. Our GC9 \(G(2)\) glueball boson star scenario belongs to the same observationally viable class at S-star scales, while differing fundamentally in its microphysics, rotational sector, merger behavior and cosmological embedding. Present stellar-orbit data alone do not establish the nature of the central boundary condition; distinguishing among these possibilities will require moving beyond stellar orbits toward horizon-scale imaging, strong-field timing, gravitational-wave spectroscopy and high-energy transient probes \citep{Crespi2026GCFermions,Pelle2024FermionCoreImaging,EHTSgrA2022I}.

\section{Phenomenological outlook II: seeded stellar collapse in a minimal model}
\label{sec:seeded}
A central question for the present framework is whether an ordinary core-collapse supernova can create a \(G(2)\) glueball boson star directly from baryonic matter, thereby allowing stellar-mass BH candidates to be reinterpreted as GC9 dark boson stars. For the ultraheavy glueball masses relevant here, the answer is generically negative. We therefore focus on a more plausible alternative: \emph{seeded collapse}, in which a compact dark seed is already present in the progenitor and steers the collapse outcome toward a horizonless GC9 remnant.

The basic obstruction to \emph{in situ} formation is kinematic and dynamical. A core-collapse supernova reaches temperatures
\begin{equation}
	T \sim 10\text{--}50\,{\rm MeV}
\end{equation}
and baryon chemical potentials
\begin{equation}
	\mu_B\sim \mathcal{O}(100\,{\rm MeV}),
\end{equation}
well below the threshold required to produce on-shell glueballs of mass $m_\phi \gtrsim 10^9\,{\rm GeV}$
from SM degrees of freedom \citep{Janka2012,Burrows2013}. Even in the presence of Higgs-portal or higher-dimensional gluonic portal operators (which are forbidden or suppressed in the present theory \citep{Masi2026E6G2}), production rates are strongly suppressed by phase space and by the small effective couplings allowed by collider cosmological and supernova-cooling constraints \citep{Arcadi2020,Bringmann2018}. There is also no rapid conserved-charge flow capable of converting \(\gg M_\odot\) of baryons into ultraheavy bosons on millisecond--second timescales.

Therefore, a pure \(G(2)\) glueball boson star remnant cannot be created simply by re-labeling baryons during collapse. The BS equation of state requires a macroscopic dark configuration that must either pre-exist in the progenitor or be assembled over much longer timescales than core collapse itself.

\subsection{Seeded collapse hypothesis}

We therefore consider a progenitor star of mass \(M_\star\) and radius \(R_\star\) containing a compact dark seed of mass \(M_{\rm seed}\), with fractional abundance
\begin{equation}
	\epsilon \equiv \frac{M_{\rm seed}}{M_\star}\ll 1.
	\label{eq:eps}
\end{equation}
For \(\epsilon\ll 1\), ordinary stellar evolution is essentially unchanged: the seed perturbs the gravitational potential only weakly,
\begin{equation}
	\Phi_N(r)\simeq -\frac{G M_b(r)}{r}-\frac{G M_{\rm seed}}{r},
	\label{eq:potential}
\end{equation}
where \(M_b(r)\) is the enclosed baryonic mass \citep{Kippenhahn2012}. In this regime, the seed remains dynamically negligible through most of stellar life, but can become important once the baryonic core contracts during collapse.

A minimal bookkeeping for the incidence of seeds at collapse time introduces a seed probability
\begin{equation}
	f_{\rm seed}=1-\exp\!\left[-\int_0^{t_\star}\Gamma_{\rm seed}(t)\,dt\right],
	\label{eq:fseed}
\end{equation}
with the schematic decomposition
\begin{equation}
	\Gamma_{\rm seed}=\Gamma_{\rm form}+\Gamma_{\rm cap\,seed}+\Gamma_{\rm cap\,DM}.
	\label{eq:GammaDecomp}
\end{equation}
Here \(\Gamma_{\rm form}\) describes star formation around a pre-existing compact seed in the natal cloud, \(\Gamma_{\rm cap\,seed}\) the dynamical capture of compact seeds in dense environments and \(\Gamma_{\rm cap\,DM}\) the gradual accumulation of individual dark particles. 
So star forms around a seed  (dominant if seeds are common in the interstellar medium (ISM)):
\begin{equation}
	\Gamma_{\rm form}\sim \frac{n_{\rm seed}}{n_{\rm gas}}\,\frac{1}{t_{\rm SF}},
\end{equation}
where $n_{\rm seed}$ is the number density of pre-existing compact seeds in the star-forming cloud, $t_{\rm SF}$ a star-formation time.
A simple estimate for compact-seed capture is
\begin{equation}
\begin{aligned}
\Gamma_{\rm cap\,seed}&\sim n_{\rm seed}\,\pi b_{\max}^2\,v\,P_{\rm diss},\\
b_{\max}&\simeq R_\star\sqrt{1+\frac{2GM_\star}{R_\star v^2}}.
\end{aligned}
\end{equation}
where \(n_{\rm seed}\) is the seed number density and \(v\) the relative velocity.
Here $P_{\rm diss}$ is the probability that an encounter loses enough orbital energy for permanent capture; without drag, scattering or another dissipative interaction, the geometrical expression alone is only an encounter rate and $P_{\rm diss}$ can be very small.  Likewise, $\Gamma_{\rm form}$ is an incidence proxy rather than a rate derived from star-formation theory.
For ultraheavy glueballs, the last channel is typically negligible because capture rates scale schematically as
\begin{equation}
	C_{\rm cap}\propto \frac{\rho_{\rm DM}}{m_\phi}\,\sigma_{\chi b}\,v^{-1}\,N_b\,F_{\rm foc},
	\label{eq:Ccap}
\end{equation}
where $\sigma_{\chi b}$ is the portal-induced scattering cross section on baryons, $v$ is the relative velocity and
$F_{\rm foc}$ accounts for gravitational focusing \citep{Gould1987}. This is strongly suppressed at large \(m_\Phi\).

\paragraph{Post-collapse bookkeeping and re-landing.}
Let \(M_\phi\) and \(J_\phi\) denote the gravitating mass and angular momentum of the compact dark core. The final BS endpoint must satisfy the stability condition
\begin{equation}
	M_\phi^{\,f}
	\le
	M_{\max}\!\left(J_\phi^{\,f};N_\phi^{\,f}\right),
	\qquad
	\left(N_\phi^{\,f},J_\phi^{\,f}\right)
	\in\mathcal D_{\rm stable},
	\label{eq:stability}
\end{equation}
where \(M_{\max}(J;N)\) denotes the phenomenological rotating stability
surface at fixed dark charge $N_\phi$, and \(\mathcal D_{\rm stable}\) is the
domain in the \((N,J)\) plane admitting a stable GC9 equilibrium.
This condition is not automatic: if the collapsing system overshoots the stability curve, the excess energy and angular momentum must be shed through gravitational cooling and scalar bursts before a stable horizonless remnant can be reached \citep{SeidelSuen1994,Liebling2017,Bezares2024ECO}.

A minimal bookkeeping writes
\begin{align}
	M_\phi^{\,f} &= M_{\rm seed}+M_{\rm coh}-\frac{E_{\rm sb}}{c^2},
	\label{eq:MassBook}\\
	J_\phi^{\,f} &= J_{\rm seed}+J_{\rm acc}-J_{\rm sb}-J_{\rm GW},
	\label{eq:AngBook}
\end{align}
where \(M_{\rm coh}\) denotes the amount of energy coherently stored in the pre-existing dark configuration during collapse, \(E_{\rm sb}\) and \(J_{\rm sb}\) are the scalar-burst losses and \(J_{\rm acc}\) is the angular momentum deposited by fallback/accretion. Importantly, \(M_{\rm coh}\) should not be interpreted as newly produced dark rest mass from baryons, but as the fraction of collapse binding energy that is transferred into the pre-existing dark core through coherent gravitational excitation. This distinction has a sharp consequence: if the dark sector carries the approximately conserved charge $N$ discussed in the Introduction, gravitational excitation can grow $M_{\rm coh}$ but cannot grow $N$. A subsolar seed with correspondingly small $N_{\rm seed}$ therefore cannot become a several-solar-mass \emph{pure} GC9 remnant through binding-energy deposition alone: the mechanism above only regulates re-landing for a seed whose charge already supports the target remnant mass, $M_{\max}(N_{\rm seed},J)\gtrsim M_\phi^{\,f}$. Absent an external charge reservoir or an explicit portal supplying additional $N$, an under-charged seed's natural endpoint is a hybrid baryon--GC9 object with a subdominant dark component rather than a pure-GC9 remnant, and would require a two-fluid treatment not attempted here.

A simple phenomenological evolution model is
\begin{align}
	\dot M_{\rm tot}(t) &=
	\dot M_{\rm acc}(t)-
	\frac{\dot E_{\rm GW}(t)+\dot E_\nu(t)+\dot E_{\rm sb}(t)}{c^2},
	\label{eq:Mdot_tot}\\
	\dot M_\phi(t) &=
	\eta_{\rm coh}\,\frac{\dot E_{\rm bind}(t)}{c^2}
	-\frac{\dot E_{\rm sb}(t)}{c^2},
	\label{eq:Mdot_phi}
\end{align}
where \(\eta_{\rm coh}\in[0,1]\) parametrizes the efficiency with which a fraction of the gravitational binding-energy release is stored in the pre-existing dark configuration. In Eq.~\eqref{eq:Mdot_phi}, $M_\phi$ denotes the total energy of the fixed-charge dark configuration, not newly created dark rest mass; the equation cannot turn an under-charged seed into a several-solar-mass pure GC9 star. 
We can also consider a possible dark scalar charge
\begin{equation}
	\dot N_\phi(t)
	=
	\dot N_{\rm acc}^{\rm dark}(t)
	-
	\dot N_{\rm sb}(t)
	-
	\dot N_{\Delta N}(t)
	\label{eq:Ndot_phi}
\end{equation}
with baryonic fallback, neutrinos and gravitational waves not directly
supplying this charge. Thus \(\dot N_{\rm acc}^{\rm dark}=0\) unless an explicit dark reservoir or portal is specified, while \(\dot N_{\rm sb}\) accounts for charge carried by outgoing scalar radiation. Hereafter the dark charge is held fixed and suppressed from the notation.

A simple estimate for the available binding-energy power is
\begin{equation}
	\begin{split}
	&\dot E_{\rm bind}(t)\simeq
	\alpha\,\frac{G M_{\rm tot}(t)\dot M_{\rm acc}(t)}{R_{\rm eff}(t)}\\
	&R_{\rm eff}(t)\sim \max\!\big(R_b(t),R_\phi(t)\big),
	\label{eq:Ebind}
	\end{split}
\end{equation}
with \(\alpha=\mathcal{O}(0.1\text{--}1)\).
The angular momentum of the dark core evolves according to
\begin{equation}
\dot J_\phi(t)=
\eta_J\,\dot M_{\rm acc}(t)\,\ell_{\rm acc}(t)
-\dot J_{\rm GW}(t)
	-\dot J_{\rm sb}(t),
	\label{eq:Jdot}
\end{equation}
where \(\ell_{\rm acc}\sim \sqrt{GM_{\rm tot}R_{\rm circ}}\) is a characteristic specific angular momentum of fallback material, with circularization radius $R_{\rm circ}$. The coefficient $0\leq\eta_J\leq1$ denotes the fraction actually transferred to the dark component.  In the absence of a portal or dissipative baryon--dark interaction, baryonic fallback remains a separate fluid and neither $\eta_J=1$ nor growth of the dark Noether charge is justified. The scalar-burst torque is modeled as proportional to the scalar-burst power:
\begin{equation}
	\dot J_{\rm sb}(t)=
	\chi_J\,\frac{\dot E_{\rm sb}(t)}{\Omega_\phi(t)},
	\label{eq:Jsb}
\end{equation}
where $\Omega_\phi$ is the dark-core rotation frequency and $\chi_J=\mathcal{O}(1)$ encodes the efficiency of
spin shedding.
To model re-landing, define the instantaneous excess above the rotating stability curve,
\begin{equation}
	\Delta M(t)\equiv M_\phi(t)-M_{\max}\!\left(J_\phi(t)\right),
	\label{eq:Delta}
\end{equation}
and prescribe scalar-burst losses by
\begin{equation}
	\dot E_{\rm sb}(t)=
	\begin{cases}
		0, & \Delta\le 0,\\[4pt]
		\kappa\,\Delta M(t)\,c^2/t_{\rm cool}, & \Delta>0,
	\end{cases}
	\label{eq:Esb}
\end{equation}
where \(\kappa=\mathcal{O}(1)\), \(t_{\rm cool}\) is the characteristic dark-core cooling time associated with scalar-burst relaxation, while \(t_{\rm fb}\) is the characteristic fallback-accretion timescale. The condition \(t_{\rm cool}\ll t_{\rm fb}\) expresses the requirement that excess mass-energy be shed faster than fallback reloads the remnant. This provides a toy feedback rule that drives the remnant back toward the stable branch whenever it overshoots it. The re-landing gap $\Delta M_{\rm reland}>0$ can be expressed as

\begin{equation}
	\Delta M_{\rm reland}\equiv \Delta M(t_{\rm pre})
	=
	M_\phi^{\rm pre}-M_{\max}\!\left(J_\phi^{\rm pre}\right).
	\label{eq:Deltam_reland}
\end{equation}
A compact way to encode the success of this channel is through a re-landing probability,
\begin{equation}
	P_{\rm reland}
	=
	P\!\left[
	M_\phi^{\rm pre}-\Delta E_{\rm rad}/c^2
	\le
	M_{\max}(J_\phi^{\,f})
	\right],
	\label{eq:Preland}
\end{equation}
where \(M_\phi^{\rm pre}\) is the dark-core mass at the onset of the radiative relaxation phase, \(J_\phi^{\,f}\) is the final angular momentum after cooling and \(\Delta E_{\rm rad}\) includes gravitational-wave, neutrino and scalar-burst losses. The key point is that seeded collapse is a conditional remnant channel: a horizonless GC9 endpoint is possible only if cooling and burst losses are efficient enough to satisfy Eq.~\eqref{eq:stability}.
\begin{figure}[t]
	\centering
	\includegraphics[width=\columnwidth]{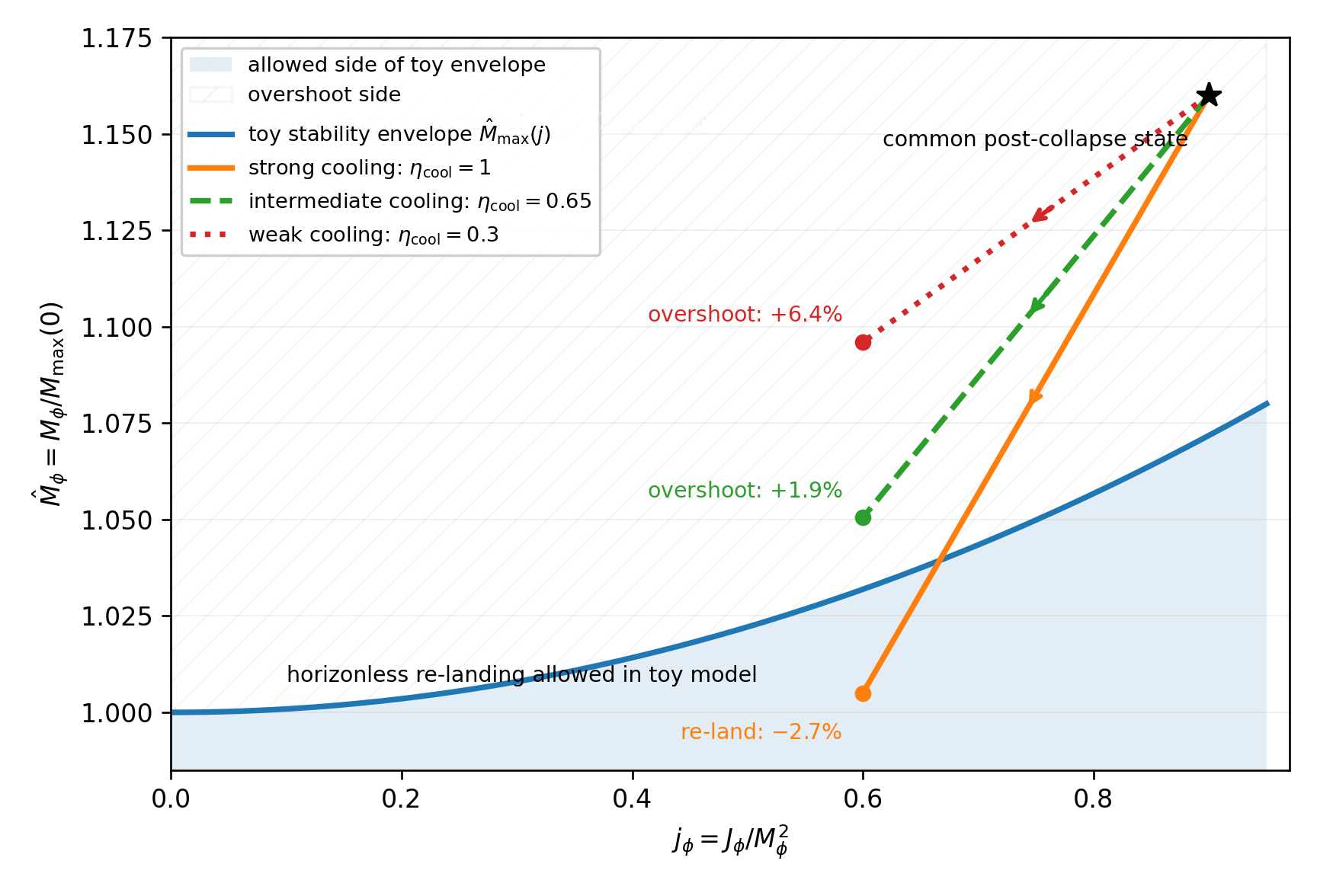}
	\caption{Toy-model seeded-collapse trajectories in the plane \((j_\phi,\hat M_\phi)\), with \(j_\phi=J_\phi/M_\phi^2\) and \(\hat M_\phi=M_\phi/M_{\max}(0)\). The solid curve is the assumed phenomenological stability boundary; the shaded region below it is the allowed side of this toy threshold, while the hatched region above it is an overshoot with no controlled GC9 endpoint. Arrows show the direction of evolution from the common post-collapse state. The parameter \(\eta_{\rm cool}\) multiplies the variable mass-loss term in Eq.~\eqref{eq:seeded_toy_trajectories}; it is not a first-principles radiative efficiency. Only the strong-cooling example crosses below the assumed envelope and re-lands.}
	\label{fig:gc9_seeded_relanding}
\end{figure}
To make the re-landing criterion explicit, we use the dimensionless mass \(\hat M_\phi\equiv M_\phi/M_{\max}(0)\) together with the physical spin parameter \(j_\phi\equiv J_\phi/M_\phi^2\). The illustrative rotating threshold is
\begin{equation}
 \hat M_{\max}^{\rm toy}(j_\phi)
 =1+0.08\left(\frac{j_\phi}{0.95}\right)^2,
 \qquad 0\le j_\phi\le0.95,
 \label{eq:seeded_toy_envelope}
\end{equation}
where the endpoint spin is aligned by construction with the heuristic GC9 mass-shedding proxy. This is not a computed rotating sequence. The three trajectories are parametrized by \(0\le t\le1\) as
\begin{align}
 j_\phi(t)&=0.90-0.30t,\\
 \hat M_\phi(t)&=1.16-\left(0.025+0.13\eta_{\rm cool}\right)t.
 \label{eq:seeded_toy_trajectories}
\end{align}
At their common final spin \(j_\phi=0.60\), the toy threshold is \(\hat M_{\max}^{\rm toy}=1.0319\). The strong-cooling trajectory \(\eta_{\rm cool}=1\) ends at \(\hat M_\phi=1.005\), about \(2.7\%\) below the boundary, and therefore re-lands within the toy model. The intermediate and weak examples end at \(1.0505\) and \(1.096\), respectively, corresponding to residual overshoots of approximately \(1.9\%\) and \(6.4\%\). They do not establish horizonless endpoints. The figure is consequently a transparent threshold illustration, not a prediction of the efficiency, trajectory shape, or endpoint distribution; those require a two-dimensional rotating GC9 sequence and a nonlinear collapse calculation.

Core collapse occurs on the dynamical timescale
\begin{equation}
	t_{\rm coll}\sim (G\rho_c)^{-1/2}\sim \mathcal{O}(1)\,{\rm ms},
	\label{eq:tcollapse}
\end{equation}
with \(\rho_c\) the central density \citep{Janka2012,Burrows2013}. The pre-existing dark seed, being a compact coherent configuration, responds on comparable or shorter timescales. Although dynamically negligible during the main-sequence phase, it can therefore act as a nucleation center for dark sector reconfiguration once relativistic compactness is approached.

\paragraph{Connection to failed supernovae and disappearing stars}
The observational consequence of seeded collapse is that some nominally failed supernovae may end not in black holes but in horizonless dark remnants. The Large Binocular Telescope (LBT) disappearing-star program has identified candidates consistent with massive stars that become dramatically fainter after weak transients, with N6946-BH1 the best-known example, while more recent James Webb Space Telescope (JWST) follow-up has highlighted the continued promise --- and ambiguity --- of the failed-SN channel \citep{Gerke2015,Adams2017,Adams2016,De2026M31,Beasor2026M31}. The same dark-core reconfiguration that helps the remnant re-land on the stable GC9 branch also
removes part of the post-collapse energy budget from the ordinary electromagnetic channel.

A fallback prescription can be written as
\begin{equation}
	\dot M_{\rm fb}(t)=
	\dot M_0\left(\frac{t}{t_0}\right)^{-5/3},
	\qquad
	t\gtrsim t_0\sim 10^2\text{--}10^4\,{\rm s},
	\label{eq:Mdotfb}
\end{equation}
where the exponent is the standard ballistic fallback value \cite{Chevalier}. If a fraction \(\eta_{\rm dark}\) of the accretion power is diverted into the dark sector, the bolometric luminosity becomes
\begin{equation}
	L_{\rm bol}(t)\simeq
	(1-\eta_{\rm dark})\,\epsilon_{\rm rad}\,\dot M_{\rm fb}(t)c^2.
	\label{eq:Lbol}
\end{equation}
If \(\eta_{\rm dark}\) is approximately constant at late times, the GC9 fallback tail preserves the same \(t^{-5/3}\) slope as the standard case but is shifted downward in normalization. The corresponding magnitude offset is
\begin{equation}
	\Delta m_{\rm tail}
	=
	-2.5\log_{10}(1-\eta_{\rm dark}).
	\label{eq:deltam_tail}
\end{equation}
Integrating the fallback-powered dark drainage gives
\begin{equation}
\begin{aligned}
E_{\rm dark,fb}
&\simeq\eta_{\rm dark}\,\epsilon_{\rm rad}
\int_{t_0}^{\infty}\dot M_{\rm fb}(t)c^2\,dt\\
&=\frac{3}{2}\,\eta_{\rm dark}\,\epsilon_{\rm rad}\,\dot M_0 t_0 c^2.
\end{aligned}
\label{eq:Edark_fb}
\end{equation}
Relative to a standard fallback-BH interpretation with \(\eta_{\rm dark}\approx 0\), the seeded-collapse channel therefore predicts suppressed late-time electromagnetic output whenever energy is efficiently drained into the dark core and its scalar-burst channel.

At the population level, the remnant formation rate takes the schematic form
\begin{equation}
	R_{\rm GC9\,rem}=f_{\rm seed}\,P_{\rm reland}\,R_{\rm failed},
	\label{eq:RBS}
\end{equation}
where \(R_{\rm failed}\) is the rate of disappearing or failed-SN candidates. The relevant observable combination is therefore not seed incidence alone, but the product \(f_{\rm seed}P_{\rm reland}\). Vera Rubin Observatory
 and continued JWST follow-up can constrain this combination statistically even without unique object-by-object identification \citep{Byrne2022}. In this sense, the
scalar-burst budget and the suppressed fallback luminosity are not independent add-ons, but direct consequences of the fact that the remnant must dynamically re-land on the same compact GC9 branch that underlies the Kerr-mimicking observables discussed earlier.

\begin{table}[t]
	\centering
	\scriptsize
	\renewcommand{\arraystretch}{1.10}
	\setlength{\tabcolsep}{2.4pt}
	\begin{tabular}{cccccc}
		\hline
		\(M_f\,[M_\odot]\) & \(\eta_{\rm dark}\) & \(\Delta m_{\rm tail}\,[{\rm mag}]\) & \(\delta_{\rm rel}\) & \(E_{\rm sb}^{\max}\,[{\rm erg}]\) & \(f_0^{(\ell=2)}\,[{\rm Hz}]\) \\
		\hline
		5  & 0.3 & 0.39 & \(10^{-5}\) & \(9\times10^{49}\) & \(1.22\times10^3\) \\
		5  & 0.5 & 0.75 & \(10^{-4}\) & \(9\times10^{50}\) & \(1.22\times10^3\) \\
		5  & 0.9 & 2.50 & \(10^{-3}\) & \(9\times10^{51}\) & \(1.22\times10^3\) \\
		10 & 0.3 & 0.39 & \(10^{-5}\) & \(1.8\times10^{50}\) & \(6.08\times10^2\) \\
		10 & 0.5 & 0.75 & \(10^{-4}\) & \(1.8\times10^{51}\) & \(6.08\times10^2\) \\
		10 & 0.9 & 2.50 & \(10^{-3}\) & \(1.8\times10^{52}\) & \(6.08\times10^2\) \\
		\hline
	\end{tabular}
	\caption{Seeded-collapse benchmark predictions for stellar-mass GC9 remnants. The late-time fallback-tail dimming is computed from
		\(\Delta m_{\rm tail}=-2.5\log_{10}(1-\eta_{\rm dark})\).
		The scalar-burst upper budget is estimated as
		\(E_{\rm sb}^{\max}\simeq \delta_{\rm rel} M_{\phi}^f c^2\),
		with \(\delta_{\rm rel}\equiv \Delta M_{\rm reland}/M_{\phi}^f\), $M_{\phi}^f$ the final remnant mass and $\Delta M_{\rm reland}\equiv M_{\phi}^{\rm pre}-M_{\max}(J_{\phi}^{\rm pre})$.
		The fundamental dark-core quadrupolar frequency is obtained from the compact-branch estimate
		\(M\omega_0^{(\ell=2)}\simeq 0.188320\),
		so that
		\(f_0^{(\ell=2)}\simeq \omega_0/(2\pi)\propto 1/M_{\phi}^f\).
		The same static background also has the axial spacetime QNM $M\omega_{\rm ax}=0.4638-0.09888i$, corresponding to $f_{\rm ax}\simeq2.99$~kHz at $5M_\odot$ and $1.49$~kHz at $10M_\odot$; these are mode frequencies, not merger excitation predictions.}
	\label{tab:seededcollapse_benchmarks}
\end{table}

Table~\ref{tab:seededcollapse_benchmarks} summarizes illustrative seeded-collapse benchmarks in the stellar-mass regime, using Kerr-mimicking observables computed before. The key point is that the GC9 channel produces a linked observational pattern: a vertical suppression of the late-time fallback tail controlled by \(\eta_{\rm dark}\), a finite dark-burst energy budget controlled by the re-landing gap \(\delta_{\rm rel}\) and a remnant spectrum containing internal modes in the few-\(10^2\)--\(10^3\) Hz range together with a static axial spacetime mode in the kHz range. Its excitation and rotating continuation require a future merger calculation.

\paragraph{Seed benchmarks and stellar backreaction.}
To remain invisible during ordinary stellar evolution, the seed fraction \(\epsilon\) in Eq.~\eqref{eq:eps} must remain small. For a representative progenitor with \(M_\star\simeq 20\,M_\odot\), three benchmark seed masses are displayed below.
\begin{center}
	\begin{tabular}{lccc}
		\hline
		Benchmark & \(M_{\rm seed}\) & \(\epsilon\) & Comments\\
		\hline
		B1 (micro-seed) & \(10^{-6}M_\odot\) & \(5\times10^{-8}\) & negligible gravity\\
		B2 (small seed) & \(10^{-3}M_\odot\) & \(5\times10^{-5}\) & still negligible\\
		B3 (macro-seed) & \(10^{-1}M_\odot\) & \(5\times10^{-3}\) & can steer collapse\\
		\hline
	\end{tabular}
\end{center}
The seed benchmarks refer to the stellar-remnant sector,
$\Lambda_T\simeq0.10$--$0.15$~GeV for
$M_{\max}\simeq5$--$10\,M_\odot$
(Sec.~\ref{sec:sector}, Table~\ref{tab:sectors}). At this
normalization the macro-seed B3 lies on the dilute branch with
$C\simeq0.3247\,(10^{-2})^{16/23}\simeq1.32\times10^{-2}$ and
$R\simeq11.2$~km --- well inside the progenitor iron core --- so the
seed can reside at the stellar center without perturbing hydrostatic
structure, while remaining compact enough to act as a nucleation
center once relativistic compactness is approached. Under the
supermassive keV normalization the same seed would instead have
$R\sim$~AU, larger than the helium core, and could not play this
role. Seeded collapse therefore selects stellar-sector GC9 objects
as seeds, and the remnant it produces inherits the stellar-sector
stiffness.
The B3 regime is the most relevant phenomenologically: such a seed can remain subdominant through most of stellar evolution while becoming important once the central baryonic core contracts. The perturbation to hydrostatic balance remains controlled by the additional potential term in Eq.~\eqref{eq:potential}; a detailed stellar-evolution treatment would require coupling this potential to a full stellar-evolution code, but in the present minimal framework we treat the \(\epsilon\ll 1\) regime as compatible with ordinary stellar evolution.

The compact-core feedback estimates below are therefore conditional on a pre-existing dark component already carrying approximately the mass and charge of the stellar-sector compact branch.  The dilute B1--B3 seed benchmarks do not dynamically grow into a $C\simeq0.3247$ pure GC9 core through gravity alone; without an external dark-charge reservoir their consistent endpoint is a hybrid two-fluid object.

\subsection{The shock revival problem}
\label{sec:ccsn}

This section addresses the long-standing shock-stagnation or \emph{shock
revival} problem in core-collapse supernovae (CCSNe) and asks whether a compact
GC9 component could provide an \emph{additional} feedback channel.  The absence
of an event horizon removes one specific sink---irreversible horizon
absorption---but it does not by itself make the core a reflecting boundary.
Baryonic material may transit the dark configuration, accumulate in a
baryonic component, or exchange energy with core and boundary-layer modes,
depending on microphysics that is not computed here.  The exploratory question
is therefore whether time-dependent accretion can excite such modes and return
a non-negligible fraction of the available accretion power to the surrounding
flow.
The correct claim at the present level is not that the mechanism is already demonstrated to solve CCSNe explosions, but that it provides a physically motivated extra source of heating and flow feedback that can lower the effective criticality threshold in otherwise difficult progenitors.
The delayed-neutrino mechanism, shock-stagnation problem and possible alternatives in core-collapse supernovae, from the perspective of neutrino heating, turbulence, multidimensional instabilities, rotation and alternative energy-injection channels, are reviewed extensively in \citep{BetheWilson1985,WoosleyJanka2005,Janka2012,MurphyBurrows2008,Burrows2013,Muller2016CCSN,Muller2025CCSN}; the failed-supernova and disappearing-star context relevant for our phenomenology is discussed in \citep{LovegroveWoosley2013,Gerke2015,Adams2016,Adams2017,Byrne2022,Beasor2024JWSTBH1}.

In the standard neutrino-driven mechanism, a massive star ($M\gtrsim 8M_\odot$) collapses when its iron core exceeds the Chandrasekhar mass. The core bounce launches a shock that initially propagates outward but rapidly stalls at $r\sim 100$--$200$ km due to energy losses from nuclear dissociation and neutrino cooling. The revival of this stalled shock requires sufficient heating in the gain region. A widely used criterion is the comparison of the advection time $\tau_{\rm adv}$ and the heating time $\tau_{\rm heat}$:
\begin{equation}
	\frac{\tau_{\rm adv}}{\tau_{\rm heat}} \gtrsim 1
	\quad \Longleftrightarrow \quad
	\text{runaway shock expansion.}
\end{equation}
In terms of neutrino luminosity $L_\nu$, one can write a critical curve (which resembles the Burrows--Goshy condition \cite{MurphyBurrows2008})
\begin{equation}
	L_{\nu,\rm crit}\propto \dot M^{\alpha} M_c^{\beta},
	\qquad \alpha,\beta=\mathcal{O}(1),	
\end{equation}
where $\dot M$ is the accretion rate, $M_c$ the central compact mass and the exact exponents depend on the analytic closure and transport assumptions. For high-compactness progenitors, $L_{\nu,{\rm crit}}$ often exceeds what realistic neutrino transport can supply.

In the standard picture, the central object is either a proto-neutron star or
a forming BH.  A proto-neutron star is already a dynamical, oscillating and
neutrino-emitting hydrodynamic object, whereas a forming BH becomes a genuine
absorbing boundary after horizon formation.  The relevant comparison is
therefore an ordinary proto-neutron-star engine versus a proto-neutron-star or
baryonic flow containing an additional GC9 component; a pure boson core is a
separate conditional endpoint.  The new component can in principle modify the
inner response in three ways: (i) energy need not be lost specifically through
horizon absorption, although baryons may still transit or accumulate; (ii)
some fraction of the infalling kinetic energy may excite coherent core or
boundary-layer modes; and (iii) a model-dependent fraction of that excitation
may be returned outward as mechanical, gravitational or baryonic-pressure
feedback.

We therefore parametrize the new channel as a pumping luminosity,
\begin{equation}
	L_{\rm pump}
	=
	\eta_{\rm pump}\,L_{\rm acc}
	\sim
	\eta_{\rm pump}\,\frac{G M_c \dot M}{R_c}
	\equiv \eta_{\rm pump}\,C_c\,\dot M c^2,
	\label{eq:Lpump}
\end{equation}
where $R_c$ is the boson-core radius, $C_c\equiv GM_c/(R_c c^2)$ is the core compactness and $\eta_{\rm pump}$ encodes the efficiency of mode excitation and re-emission. Because the compact GC9 branch computed in this work has
$C_c\simeq0.3247$ --- with the stellar-sector normalization
of Sec.~\ref{sec:sector}; a keV-normalized core of the same mass
would be dilute, $C_c\sim10^{-8}$, and dynamically irrelevant ---
the compactness factor is fixed for an assumed compact GC9 core, while core formation, excitation, release, transport and deposition efficiencies remain uncertain.
This candidate term is additional to ordinary neutrino heating and grows with $\dot M$; if the deposition efficiency is appreciable, it could improve the revival conditions. With the revival proxy being $\tau_{\rm adv}/\tau_{\rm heat}\gtrsim1$, if one adds pumping, the effective heating time becomes:
\begin{equation}
	\tau_{\rm heat}^{-1} \to \tau_{\rm heat,\nu}^{-1}+\tau_{\rm heat,pump}^{-1},
	\qquad
	\tau_{\rm heat,pump}\sim \frac{|E_{\rm bind,gain}|}{L_{\rm heat,pump}}.
\end{equation}
The total heating deposited in the gain region is then
\begin{equation}
\begin{aligned}
L_{\rm heat}^{\rm eff}&=L_\nu+L_{\rm heat,pump},\\
L_{\rm heat,pump}&=\epsilon_{\rm dep}L_{\rm rel}\leq L_{\rm pump}.
\end{aligned}
\end{equation}
and the modified runaway condition becomes
\begin{equation}
	L_\nu+L_{\rm heat,pump}\ge L_{\nu,{\rm crit}}(\dot M,M_c).
\end{equation}
Equivalently, the neutrino luminosity required for explosion is reduced to
\begin{equation}
	L_{\nu,{\rm eff}}=L_{\nu,{\rm crit}}-\epsilon_{\rm dep}L_{\rm rel}.
\end{equation}
A useful dimensionless measure of the relevance of this channel is
\begin{equation}
	\Pi_{\rm pump}\equiv \frac{L_{\rm heat,pump}}{L_{\nu,{\rm crit}}}
	=\frac{\epsilon_{\rm dep}L_{\rm rel}}{L_{\nu,{\rm crit}}}
	\leq \eta_{\rm pump}\,\frac{C_c\,\dot M c^2}{L_{\nu,{\rm crit}}}.
	\label{eq:PiPump}
\end{equation}
When $\Pi_{\rm pump}\gtrsim 0.1$, the pumping term is no longer a perturbative correction and can materially shift the explosion threshold; when $\Pi_{\rm pump}\ll 0.1$, the effect should be regarded as subdominant.

For compact boson cores with $R_c\sim(2$--$4)GM_c/c^2$, even a modest efficiency $\eta_{\rm pump}\sim10^{-2}$ can in principle reduce the critical luminosity by a non-negligible amount. This should be interpreted as a \emph{candidate feedback mechanism}: in addition to the assumed core compactness, the excitation, release, transport and deposition efficiencies are unknown and depend on mode excitation, damping, boundary-layer coupling and multidimensional flow geometry. If the required excitation, transport and deposition efficiencies are realized, the shock revival problem could be affected by:
i) preventing the BH engine switch-off;
ii) adding a new accretion-powered mechanical/gravitational feedback term;
iii) potentially modifying $\tau_{\rm adv}$ through boundary-mediated
turbulence or shock sloshing.

One may heuristically regard the GC9 boson core as a gravitational
resonator: time-dependent accretion can perturb the compact configuration
and excite its matter-supported modes. Part of the deposited energy could
subsequently be redistributed through scalar-field oscillations,
gravitational radiation, or hydrodynamic disturbances in the surrounding
baryonic flow. This analogy does not, by itself, imply efficient outward
energy transfer. The relevant characteristic timescales are
\begin{align}
	\tau_{\rm osc}
	&\sim
	\omega_0^{-1},
	&
	\tau_{\rm acc}
	&\sim
	\frac{R_c}{v_{\rm ff}},
	&
	\tau_{\rm rel}
	&\sim
	\frac{E_{\rm osc}}{L_{\rm rel}},
	\label{eq:shock_timescales_corrected}
\end{align}
where
\begin{equation}
	v_{\rm ff}
	\sim
	\sqrt{\frac{2GM_c}{R_c}},
\end{equation}
and \(L_{\rm rel}\) denotes the rate at which energy stored in the excited
core is released into channels capable of affecting the surrounding
system. It should not be identified with a gain-region heating luminosity
unless the corresponding transport and deposition efficiencies are
specified.

For the maximum-mass GC9 benchmark,
\begin{equation}
\begin{aligned}
M\omega_0&\simeq0.188320,\qquad
\frac{R_c}{M_c}\simeq3.079827,\\
C_c&\simeq0.324694.
\end{aligned}
\end{equation}
Consequently,
\begin{align}
	\frac{\tau_{\rm osc}}{M_c}
	&\simeq
	\frac{1}{M_c\omega_0}
	\simeq5.31,
	\\
	v_{\rm ff}
	&\simeq
	\sqrt{2C_c}
	\simeq0.806,
	\\
	\frac{\tau_{\rm acc}}{M_c}
	&\simeq
	\frac{R_c/M_c}{v_{\rm ff}}
	\simeq3.82,
\end{align}
and hence
\begin{equation}
	\frac{\tau_{\rm osc}}{\tau_{\rm acc}}
	\simeq1.39,
	\qquad
	\omega_0\tau_{\rm acc}\simeq0.72.
	\label{eq:shock_timescale_ratio}
\end{equation}
The GC9 benchmark therefore does not satisfy a hierarchy
\(\tau_{\rm osc}\ll\tau_{\rm acc}\). Instead, accretion perturbs the core
on a timescale comparable to its fundamental oscillation time. Such a
near-dynamical-timescale drive can in principle excite the core
nonadiabatically, but neither the excitation amplitude nor the fraction of
the mode energy returned to the baryonic flow follows from this comparison
alone.

For a core perturbation to contribute to shock revival, two distinct
conditions must be met. First, accretion must excite a non-negligible mode
energy,
\begin{equation}
	E_{\rm osc}
	=
	\epsilon_{\rm exc} E_{\rm acc},
	\qquad
	0\leq\epsilon_{\rm exc}\leq1,
	\label{eq:excitation_efficiency}
\end{equation}
where \(\epsilon_{\rm exc}\) depends on the time dependence and geometry of
the accretion flow. Second, a fraction of this energy must reach and be
deposited in the gain region before the material is advected through it.
Writing
\begin{equation}
	L_{\rm heat,pump}
	=
	\epsilon_{\rm dep} L_{\rm rel},
	\qquad
	0\leq\epsilon_{\rm dep}\leq1,
	\label{eq:deposition_efficiency}
\end{equation}
the relevant feedback condition is schematically
\begin{equation}
	\tau_{\rm feed}
	\equiv
	\tau_{\rm rel}+\tau_{\rm trans}
	\lesssim
	\tau_{\rm adv},
	\label{eq:feedback_timing_condition}
\end{equation}
where \(\tau_{\rm trans}\) is the propagation time from the compact core to
the gain region. This replaces the stronger and unsupported requirement
\(\tau_{\rm osc}\ll\tau_{\rm acc}\lesssim\tau_{\rm adv}\).

The effective gain-region heating rate can then be parameterized as
\begin{equation}
	\frac{1}{\tau_{\rm heat,eff}}
	=
	\frac{1}{\tau_{\rm heat,\nu}}
	+
	\frac{\epsilon_{\rm dep}L_{\rm rel}}
	{|E_{\rm bind,gain}|}.
	\label{eq:effective_heating_timescale_corrected}
\end{equation}
Similarly, a nonabsorbing dynamical inner configuration could alter the
residence time of matter in the gain region, but the sign and magnitude of
this effect are not known. We therefore write
\begin{equation}
	\tau_{\rm adv}^{G(2)}
	=
	\tau_{\rm adv}^{(0)}
	\left(1+\delta_{\rm adv}\right),
	\label{eq:advection_parameterization}
\end{equation}
where \(\delta_{\rm adv}\) must be determined from a hydrodynamic
calculation and is not assumed to be positive.

The corresponding runaway diagnostic becomes
\begin{equation}
	\frac{\tau_{\rm adv}^{G(2)}}
	{\tau_{\rm heat,eff}^{G(2)}}
	=
	\frac{\tau_{\rm adv}^{(0)}(1+\delta_{\rm adv})}
	{\tau_{\rm heat,eff}^{G(2)}}
	\gtrsim1.
	\label{eq:GC9_runaway_diagnostic}
\end{equation}
If \(\epsilon_{\rm exc}\), \(\epsilon_{\rm dep}\), and the released mode
power are appreciable, and if \(\delta_{\rm adv}>0\), the GC9 core could
lower the effective neutrino-heating threshold.

The conservative conclusion is therefore not that GC9 feedback
necessarily promotes shock revival, but that the computed core timescale
is comparable to the accretion-driving timescale and consequently permits
a potentially nonadiabatic coupling channel. Establishing whether this
channel deposits an astrophysically relevant amount of energy in the gain
region requires multidimensional radiation-hydrodynamic simulations with
the GC9 core implemented as a dynamical inner component.

\section{Phenomenological outlook III: cosmological assembly and the compact-object versus fluid crossover}
On cosmological and galactic scales, the scenario must be read in the broader context of compact dark matter, microlensing, halo heating, ultralight-wave phenomenology and compact-object population constraints \citep{Planck2018,Carr2020PBH,Tisserand2007EROS2,Alcock2000MACHO,MonroyRodriguez2014,Brandt2016EridanusII,Niikura2019HSC,Schive2014Cosmic,Ferreira2021UltraLight,SuarezRoblesMatos2014}.
\label{sec:cosmo}
A phenomenologically viable implementation of the GC9 scenario is not a single primordial compact-object mass function extending continuously from asteroid masses to supermassive galactic nuclei. Current lensing and non-lensing bounds make that interpretation unnecessarily rigid. A more realistic formulation is instead a hybrid hierarchy, \textit{i.e.} a classification ansatz like 
\begin{equation}
	\begin{split}
\psi_{\rm tot}(M,z)
=&
\psi_{\rm DM}^{\rm prim}(M)
+
\psi_{\rm NS+core}(M,z)
+\\
&\psi_{\rm GC9\text{-}BH}(M,z)
+
\psi_{\rm grow}(M,z),
\label{eq:psi_tot_hybrid_full}
\end{split}
\end{equation}
where \(\psi_{\rm DM}^{\rm prim}\) is the primordial compact DM component, \(\psi_{\rm NS+core}\) is the neutron-star remnant branch possibly containing a subdominant GC9 core, \(\psi_{\rm GC9\text{-}BH}\) is the compact branch replacing classical black hole remnants and \(\psi_{\rm grow}\) describes later mergers and accretion. 
The framework does \emph{not} remove neutron stars. The remnant branch that would classically terminate in a BH may be replaced by a pure GC9 boson star when the pre-existing or accreted dark-charge reservoir
already supports the target remnant mass. Under-charged collapses instead could produce a hybrid baryon--GC9 configuration and are classified within the
hybrid/NS+core sector at the present schematic level.
Accordingly, \(\psi_{\rm GC9\text{-}BH}\) denotes only the charge-sufficient subset of seeded-collapse events, it is not the outcome of generic small-seed collapse.

Throughout this section we use the common dimensionless mass-fraction
convention,
\begin{equation}
	\psi_i(M,z)\equiv
	\frac{1}{\rho_{\rm DM,0}}
	\frac{d\rho_i(M,z)}{d\ln M},
	\label{eq:psi_common_definition}
\end{equation}
whereas the kernels \(P_{\rm NS+core}\) and \(P_{\rm GC9\text{-}BH}\) below are instead probability densities per unit mass and satisfy \(\int dM\,P_i(M)=1\); they must therefore be converted to the convention
of Eq.~\eqref{eq:psi_common_definition} when entering \(\psi_i\).

\subsection{Primordial GC9 bulk in the asteroid window}

A first important observation is that GC9 boson stars are so compact that, for microlensing purposes, they behave as point lenses to excellent approximation. Writing
\begin{equation}
R_\star \simeq \frac{GM}{c^2 C},
\end{equation}
with representative GC9 compactness \(C\simeq0.3247\text{--}0.3371\), and comparing with the Einstein radius
\begin{equation}
R_E=\sqrt{\frac{4GM}{c^2}\frac{D_LD_{LS}}{D_S}},
\end{equation}
one finds
\begin{equation}
\frac{R_\star}{R_E}
=
\frac{1}{2C}\sqrt{\frac{GM/c^2}{D_{\rm eff}}},
\qquad
D_{\rm eff}\equiv \frac{D_LD_{LS}}{D_S}.
\label{eq:Rstar_RE_ratio_full}
\end{equation}

The point-lens property is normalization dependent.
Equation~\ref{eq:Rstar_RE_ratio_full} with $C\simeq0.3247$ presupposes that the primordial bulk lies near its own turning point, i.e.\ the asteroid-window
sector value
$\Lambda_T^{\rm bulk}\simeq
(A_{\rm GC9}M_{\rm Pl}^{3}/M_{\rm bulk})^{1/2}\simeq1.9$~PeV of
Table~\ref{tab:sectors}, for which
$R_\star\simeq1.4\times10^{-10}$~m and
$R_\star/R_E\lesssim10^{-14}$: the standard MACHO/PBH point-lens bounds then apply with excellent accuracy. The requirement is in fact mild. Using Eq.~\eqref{eq:diluteband}, $R_\star<R_E$ across
Galactic microlensing geometries ($D_{\rm eff}\sim{\rm kpc}$) demands only
\begin{equation}
	C\;\gtrsim\;\frac12\left(\frac{GM/c^{2}}{D_{\rm eff}}\right)^{1/2}
	\sim6\times10^{-16},
	\quad
	\Lambda_T^{\rm bulk}\gtrsim10^{-4}~{\rm GeV},
\end{equation}
a condition satisfied by roughly ten orders of magnitude at the sector value. By contrast, if the bulk shared the TON--618 keV stiffness, the same objects would be dilute, with $C\sim10^{-18}$
and $R_\star\sim4\times10^{4}$~km, giving
$R_\star/R_E\sim10^{2}$--$10^{3}$: such extended lenses have strongly finite-size-suppressed magnification and the standard point-lens exclusions would not apply as stated; the constraints would require a dedicated extended-lens re-derivation and would generically be weaker. Throughout this section we adopt the
sector-normalized compact bulk as the baseline, so that the point-lens bounds of \citep{Tisserand2007EROS2,Niikura2019HSC,Mroz2024OGLEHC,CarrKuhnel2020PBHReview,GreenKavanagh2021PBHReview}. apply directly.

For a monochromatic compact-object population, current surveys exclude a dominant DM fraction over broad mass ranges. In particular, EROS-2 excludes MACHOs as the dominant Milky-Way halo component over approximately
\begin{equation}
	6\times10^{-8}M_\odot \lesssim M \lesssim 15\,M_\odot,
\end{equation}
with subdominant allowed fractions even inside that interval \citep{Tisserand2007EROS2}. Subaru/HSC observations of M31 strongly constrain compact objects in the approximate window
\begin{equation}
	10^{-11}M_\odot \lesssim M \lesssim 10^{-9}M_\odot,
\end{equation}
once finite-source and wave-optics corrections are accounted for (below $\sim10^{-11}M_\odot$ the diffraction-suppressed signal degrades HSC's sensitivity \citep{Niikura2019HSC}), while recent OGLE high-cadence observations of the Magellanic Clouds push the exclusion of a dominant compact component into the planetary-mass regime, roughly
\begin{equation}
	1.4\times10^{-8}M_\odot \lesssim M \lesssim 1.3\times10^{-2}M_\odot,
\end{equation}
at the percent level or below \citep{Niikura2019HSC,Mroz2024OGLEHC}. Therefore, if the GC9 population were monochromatic, the statement that \emph{all} dark matter is made of boson stars would be viable only in narrow residual windows, not generically across the full mass spectrum \citep{CarrKuhnel2020PBHReview,GreenKavanagh2021PBHReview}.

Other microlensing surveys and modern reviews identify the asteroid-mass
window,
\begin{equation}
10^{-16}M_\odot \lesssim M \lesssim 10^{-12}M_\odot,
\end{equation}
as one of the residual regions in which a compact-object population may still
constitute a very large, and under some constraint sets potentially dominant,
dark-matter fraction
\citep{Tisserand2007EROS2,Niikura2019HSC,Mroz2024OGLEHC,CarrKuhnel2020PBHReview,GreenKavanagh2021PBHReview,Ballesteros2025PBH,Green2025NonParticleDM,Gorton2024AsteroidWindow}.

This motivates the primordial GC9 bulk
\begin{equation}
\begin{split}
\psi_{\rm DM}^{\rm prim}(M) &=
\frac{1}{\sqrt{2\pi}\sigma_{\rm bulk}}
\exp\!\left[
-\frac{\ln^2(M/M_{\rm bulk})}{2\sigma_{\rm bulk}^2}
\right], \\
& \qquad \int d\ln M\,\psi_{\rm DM}^{\rm prim}(M)=1,
\label{eq:psi_prim_full}
\end{split}
\end{equation}
with the benchmark choice
\begin{equation}
M_{\rm bulk}=3\times10^{-14}M_\odot,
\qquad
\sigma_{\rm bulk}=0.35.
\label{eq:bulk_choice_full}
\end{equation}
The rationale for these values is conservative. The central mass scale \(M_{\rm bulk}=3\times10^{-14}M_\odot\) is placed safely inside the middle of the asteroid window rather than near its edges, thereby reducing sensitivity to the precise location of the observational boundaries. The width \(\sigma_{\rm bulk}=0.35\) is chosen to be narrow enough that the distribution does not leak significantly into the more constrained neighboring mass ranges, but still broad enough to avoid an unrealistically monochromatic idealization.
Current lensing constraints leave residual asteroid-mass windows in which a
compact-object population can constitute a large and, depending on the adopted
constraint set, potentially dominant fraction of the dark matter
\citep{CarrKuhnel2020PBHReview,GreenKavanagh2021PBHReview,LehmannProfumoYant2018}.
For the finite-width lognormal benchmark of Eq.~\eqref{eq:psi_prim_full},
however, the present paper does not perform the integrated
extended-mass-function constraint analysis required to establish saturation of
the full dark-matter abundance.  We therefore use the unit normalization in
Eq.~\eqref{eq:psi_prim_full} as a population benchmark, not as a demonstrated
global-fit result; a subdominant normalization can be adopted without changing
the compact-star calculations.

\subsection{A remnant-dominated stellar-mass GC9 sector}

If one attempted to populate the stellar-mass compact-object sector directly through the primordial DM budget, the model would immediately run into not only lensing bounds but also dynamical-heating and CMB-accretion constraints. In particular, compact stellar systems in ultra-faint dwarfs disfavor dark matter composed entirely of compact objects above roughly \(5\text{--}10\,M_\odot\) over broad ranges \citep{Brandt2016EridanusII}, while CMB-accretion arguments likewise strongly pressure a dominant heavy compact DM component \citep{CarrEtAl2026PBHReview,Ballesteros2025PBH}. Therefore the stellar-mass GC9 branch should not be interpreted as part of the primordial DM bulk.

Treating the stellar-mass branch as an astrophysical remnant population avoids
assigning it to the primordial dark-matter budget, but its formation remains
conditional on the charge-sufficient seeded-collapse or dark-accretion
mechanisms discussed above.  The local binary black hole (BBH) merger rate
inferred from GWTC-4.0 is of order
\begin{equation}
R_{\rm BBH}\sim 14\text{--}26\ {\rm Gpc}^{-3}\,{\rm yr}^{-1},
\end{equation}
whereas the local core-collapse supernova rate is of order
\begin{equation}
R_{\rm CCSNe}\sim 7\times10^4\ {\rm Gpc}^{-3}\,{\rm yr}^{-1}.
\end{equation}
Thus only a tiny fraction of compact remnants need to enter merging binaries in order to account for the full observed LIGO-Virgo-KAGRA (LVK) stellar-mass ``black hole'' population \citep{GWTC42025,CCSNRate2025}. This is the main reason to interpret the stellar-mass GC9 sector as remnant-dominated rather than primordial DM-dominated.

\subsection{Remnant branching: neutron stars versus GC9 black hole replacements}

The remnant sector can be described by the kernel
\begin{equation}
\begin{split}
P_{\rm rem}(M\,|\,\xi_{2.5},Z,&\Omega_\star) =
p_{\rm NS}(\xi_{2.5},Z)\,P_{\rm NS+core}(M)
+ \\
&\left[1-p_{\rm NS}(\xi_{2.5},Z)\right]\,P_{\rm GC9\text{-}BH}(M),
\label{eq:Prem_full}
\end{split}
\end{equation}
where \(P_{\rm NS+core}\) is the neutron-star branch and \(P_{\rm GC9\text{-}BH}\) the branch replacing classical black hole remnants.

The branching probability is modeled by a smooth compactness/explodability law,
\begin{equation}
p_{\rm NS}(\xi_{2.5},Z)
=
\left[
1+\exp\!\left(
\frac{\xi_{2.5}-\xi_c}{\Delta_\xi}
\right)
\right]^{-1},
\label{eq:pNS_full}
\end{equation}
with the benchmark parameters
\begin{equation}
\xi_c=0.30,
\qquad
\Delta_\xi=0.05.
\label{eq:xi_choice_full}
\end{equation}
Equation \ref{eq:pNS_full} is a smooth logistic regularization of compactness-threshold prescriptions based on the O'Connor--Ott compactness parameter \(\xi_{2.5}\) \citep{OConnorOtt2011Compactness,Ertl2016Explodability}.
The values are chosen to represent a compact but not infinitely sharp transition between a neutron-star-producing regime and a BH-producing regime. The threshold \(\xi_c=0.30\) sits in the range commonly used in explodability-based remnant modeling as a proxy for the onset of difficult-to-explode progenitors, while \(\Delta_\xi=0.05\) prevents the model from becoming an unrealistic step function.

The neutron-star branch is represented by 
\begin{equation}
P_{\rm NS+core}(M)
=
\frac{1}{\sqrt{2\pi}\sigma_{\rm NS}M}
\exp\!\left[
-\frac{\ln^2(M/M_{\rm NS,0})}{2\sigma_{\rm NS}^2}
\right],
\label{eq:PNS_full}
\end{equation}
with
\begin{equation}
M_{\rm NS,0}=1.55\,M_\odot,
\qquad
\sigma_{\rm NS}=0.12.
\label{eq:NS_choice_full}
\end{equation}
The NS branch is modeled by a narrow lognormal kernel centered near the observed neutron-star mass scale, motivated by empirical NS mass distributions \citep{Ozel2012NSMasses,Kiziltan2013NSMassDistribution}.
The choice \(M_{\rm NS,0}=1.55\,M_\odot\) places the successful-explosion branch near the canonical observed neutron-star mass scale, while \(\sigma_{\rm NS}=0.12\) keeps the branch narrow enough to remain recognizably NS-like.

The black hole-replacement GC9 branch is modeled as
\begin{equation}
P_{\rm GC9\text{-}BH}(M)
=
\frac{1}{\sqrt{2\pi}\sigma_{\rm GC9}M}
\exp\!\left[
-\frac{\ln^2(M/M_{\rm GC9,0})}{2\sigma_{\rm GC9}^2}
\right],
\label{eq:PGC9BH_full}
\end{equation}
with
\begin{equation}
M_{\rm GC9,0}=8\,M_\odot,
\qquad
\sigma_{\rm GC9}=0.28.
\label{eq:GC9_choice_full}
\end{equation}
The GC9 black-hole-replacement branch is centered in the light stellar-remnant regime, motivated by core-collapse remnant mass calculations \citep{Sukhbold2016CoreCollapse}; the lognormal form itself is a phenomenological population kernel.
Here \(M_{\rm GC9,0}=8\,M_\odot\) is chosen to place the remnant branch directly in the light stellar-mass BH regime relevant to LVK observations, while \(\sigma_{\rm GC9}=0.28\) is deliberately broader than the NS branch to represent the wider spread expected from fallback-dominated collapse.

The remnant source is then
\begin{equation}
  \begin{split}
&\psi_{\rm rem}(M,z) = 
\frac{M^2}{\rho_{\rm DM,0}}
\int^z dz'\,
\left|\frac{dt}{dz'}\right|\times \\
&\int dM_{\rm prog}\,
\phi(M_{\rm prog},z')\,
\mathcal R_\star(M_{\rm prog},z')\,
P_{\rm rem}(M\,|\,\xi_{2.5},Z,\Omega_\star),
  \end{split}
\label{eq:psi_rem_full}
\raisetag{15pt}
\end{equation}
where \(\phi\) is the progenitor distribution, \(\mathcal R_\star\) the stellar-formation/core-collapse source term and \(\phi\,\mathcal R_\star\) is understood as a progenitor-collapse number rate per comoving volume, per unit progenitor mass and per unit
time. The factor \(M^2\) converts the remnant number distribution per unit mass into a dark-matter mass fraction per logarithmic mass interval.

\subsection{Viable GC9 cores inside neutron stars}

The framework does not eliminate neutron stars. Instead, it allows a neutron-star branch in which some remnants may contain a subdominant \(G(2)\) GC9 dark core. This is consistent with the broader literature on DM-admixed neutron stars, where a dense dark component can coexist with an ordinary baryonic neutron-star envelope while leaving the exterior phenomenology approximately neutron-star-like for sufficiently small dark fractions \citep{Grippa2025ReviewDMANS,Thakur2025FeasibilityDMANS,Koehn2024BosonicADMNS}.

The viability of the NS\(+\)core branch is controlled by the same constraints that govern ordinary neutron-star equations of state. First of all the maximum mass condition
\begin{equation}
M_{\max}^{\rm NS+core}\gtrsim 2\,M_\odot,
\end{equation}
in view of PSR J0740+6620, PSR J0348+0432 and PSR J1614$-$2230 \citep{Miller2021J0740Radius,Antoniadis2013J0348,Demorest2010J1614}.
Then a radius constraint, so that the branch must remain in the observational radius band of order \(12\text{--}14\) km for typical neutron-star masses \citep{Miller2021J0740Radius} and finally a tidal deformability argument: it must not generate tidal responses grossly inconsistent with current binary-neutron-star inferences \citep{Raithel2019GW170817Review,Koehn2024BosonicADMNS}.
These conditions motivate the effective requirement
\begin{equation}
f_{\rm core}\equiv \frac{M_{\rm core}^{\rm GC9}}{M_{\rm NS}^{\rm tot}},
\qquad
f_{\rm core}\ll 1,
\label{eq:fcore_full}
\end{equation}
so that the GC9 contribution remains perturbative on the neutron-star branch and becomes dominant only in the remnant channel that would otherwise form a classical BH.
In the present phenomenological treatment one may think of
\(f_{\rm core}\lesssim 0.01\text{--}0.1\), although the precise bound requires solving a two-fluid stellar model.

\paragraph{Binary-neutron-star post-merger remnants as a dark-charge selection test.}
Binary-neutron-star mergers provide a particularly sharp phenomenological
test of the GC9 framework because the conventional black-hole endpoint
does not require the creation of any new conserved charge, whereas a pure
GC9 remnant does in the approximately charge-conserving realization adopted
here. GW170817 was firmly identified as a binary-neutron-star coalescence
through its gravitational-wave signal and multimessenger counterpart
\citep{Abbott2017GW170817}. Its post-merger gravitational-wave signal was
not detected \citep{Abbott2017GW170817PostMerger}, so the nature of the
compact remnant was not measured directly. Multimessenger arguments and
remnant-classification studies nevertheless favor the formation of a
short-lived hypermassive neutron star followed by collapse to a black hole,
rather than either prompt collapse or a permanently stable massive neutron
star \citep{MargalitMetzger2017GW170817,
	RezzollaMostWeih2018GW170817,PuecherDietrich2024Remnant}.

GW190425 provides an even more massive test case. Its source-frame total
mass, \(M_{\rm tot}=3.4^{+0.3}_{-0.1}\,M_\odot\), is substantially larger
than that of known Galactic double-neutron-star systems; its component
masses are compatible with neutron stars, although the gravitational-wave
data alone cannot exclude one or both components being low-mass black
holes \citep{Abbott2020GW190425}. Under the binary-neutron-star
interpretation, numerical-relativity-based remnant classification favors
prompt collapse to a black hole
\citep{PuecherDietrich2024Remnant}. Dedicated searches have not detected a
post-merger gravitational-wave remnant from either GW170817 or GW190425
\citep{GraceWetteScott2024Postmerger}. Existing observations therefore
already contain systems for which
\[
{\rm NS}+{\rm NS}\longrightarrow{\rm BH}
\]
is a viable, and in specific analyses favored, standard interpretation,
although the high-frequency collapse and resulting black-hole ringdown
have not themselves been resolved.

The corresponding GC9 replacement obeys a stronger selection rule. Let
\(N_{D,1}\) and \(N_{D,2}\) denote the emergent dark Noether charges carried
by the two progenitors. Up to dark radiation and explicitly
charge-violating processes,
\begin{equation}
	N_D^{\,f}
	=
	N_{D,1}+N_{D,2}
	-N_{D,\rm rad}
	-\Delta N_D ,
	\label{eq:BNS_GC9_charge_bookkeeping}
\end{equation}
and a pure GC9 post-merger state is possible only if $	\left(M_f,J_f,N_D^{\,f}\right)
	\in {\cal D}_{\rm GC9}^{\rm stable}$,
with \(N_D^{\,f}\) sufficient to support the final several-solar-mass
configuration. Ordinary neutron stars with negligible dark charge
therefore cannot merge into a pure GC9 star merely through conversion of
baryonic binding energy into coherent dark-sector energy:
\[
{\rm NS}+{\rm NS}\not\longrightarrow{\rm pure\ GC9}
\]
within the minimal approximately charge-conserving realization. If each
neutron star instead contains a subdominant GC9 core, their dark charges
add during the merger and can form a more massive dark core; for the
fractions \(f_{\rm core}\ll1\) considered above, however, the natural
endpoint remains a hybrid baryon--GC9 remnant rather than a pure boson
star. A pure GC9 endpoint requires charge-sufficient progenitor cores,
accretion from a pre-existing macroscopic dark reservoir, or additional
dark-sector production physics beyond the minimal construction.

This distinction makes binary-neutron-star post-merger observations a
direct formation test of the framework. An event whose inspiral is
demonstrably neutron-star-like but whose post-merger remnant is
independently established to be black-hole-like could be reinterpreted as
a pure GC9 endpoint only if its pre-merger history supplied the required
dark sector. Conversely, a persistent baryonic remnant accompanied by
additional matter-supported dark-core modes would naturally belong to the
hybrid branch. Future detectors with improved sensitivity in the
kilohertz post-merger band can therefore probe not only horizon formation
versus a horizonless remnant, but also the distinction between
charge-sufficient pure-GC9 and hybrid NS--GC9 endpoints. Approximate dark
charge conservation thereby acts as a formation selection rule rather
than merely as an equilibrium assumption.

The preceding selection rule applies to the cold approximately
charge-conserving realization adopted in this work.  Since the underlying
\(E_6\to G(2)\times SU(3)_A\) gauge construction does not itself impose an
exact glueball-number \(U(1)\), a distinct possibility is that
number-changing dark reactions become efficient only in the hot,
strongly compressed merger phase and freeze out again in the relaxed
remnant.  Such a transient chemical-reprocessing channel could in
principle relax the pre-merger charge requirement, but would require a
finite-temperature nonequilibrium calculation of the dense \(G(2)\)
sector and is not assumed here.

\subsection{Dilute GC9 branch and the compact-object versus fluid crossover}
\label{subsec:GC9_dilute_crossover}
\label{subsec:gc9_dilute_crossover}

An important conceptual question is whether the low-mass GC9 sector should
always be regarded as a population of discrete compact objects or whether,
at sufficiently small mass, neighboring configurations could overlap and
behave as a continuous dark medium.  This question can be addressed using
the analytic dilute limit of the GC9 equation of state.

In the normalization
\begin{equation}
	\hat{\rho}\equiv\frac{\rho}{\rho_s},
	\qquad
	\hat{P}\equiv\frac{P}{\rho_s},
	\qquad
	\rho_s\equiv\Lambda_T^4,
\end{equation}
the exact GC9 Thomas--Fermi closure is
\begin{equation}
	\hat{\rho}(y)=2y+10y^9,
	\qquad
	\hat{P}(y)=8y^9.
	\label{eq:gc9_exact_parametric_eos}
\end{equation}
For \(y\ll1\), the density is dominated by the rest-mass contribution,
\begin{equation}
 \hat{\rho}=2y+\mathcal{O}(y^9),
\end{equation}
and hence $y=\hat\rho/2+\mathcal O(\hat\rho^9)$. Substitution into the
pressure gives
\begin{equation}
 \hat P=\frac{1}{64}\hat\rho^9+\mathcal O(\hat\rho^{17}).
\end{equation}
The dimensional dilute equation of state is therefore
\begin{equation}
 P\simeq K_9\rho^9,
 \qquad K_9\equiv\frac{1}{64}\rho_s^{-8}.
 \label{eq:gc9_dilute_polytrope_full}
\end{equation}
Thus the GC9 dilute branch is a Newtonian polytrope with $n=1/8$.
The Lane--Emden relation gives
\begin{equation}
 R\propto K_9^{1/23}M^{7/23}
 \propto\rho_s^{-8/23}M^{7/23},
\end{equation}
and
\begin{equation}
 C\propto K_9^{-1/23}M^{16/23}
 \propto\rho_s^{8/23}M^{16/23}.
 \label{eq:dilute_scalings_full}
\end{equation}
The effective Newtonian sequence consequently has no formal nonzero minimum
mass; toward smaller $M$, both $R$ and $C$ decrease. The continuation cannot
be extrapolated after the many-body fluid approximation fails.

For the geometrical-overlap estimate,
\begin{equation}
 R(M)=R_{\rm bulk}\left(\frac{M}{M_{\rm bulk}}\right)^{7/23},
\end{equation}
and $R(M_{\rm cross})=\ell_{\rm sep}(M_{\rm cross})$ gives
\begin{equation}
 M_{\rm cross}=R_{\rm bulk}^{69/2}M_{\rm bulk}^{-21/2}
 \left(f_{\rm GC9}\rho_{\rm DM}\right)^{23/2}.
 \label{eq:Mcross_full}
\end{equation}
The large exponents follow from $R/\ell_{\rm sep}\propto
M^{7/23-1/3}=M^{-2/69}$, so the relative overlap grows only very slowly
as the object mass decreases and the inferred crossover is extraordinarily
sensitive to the assumed radius normalization and environmental density.
For the benchmark values used here it formally gives crossover masses many orders of magnitude
below any credible many-body or continuum-condensate regime. We therefore do not assign physical significance to a numerical extrapolation of
\(M_{\rm cross}\); Eq.~\eqref{eq:Mcross_full} is retained only to demonstrate
that geometrical overlap is not reached within the domain in which the
GC9 fluid description is applicable.

At the benchmark mass, the mean separation is
\begin{equation}
	\ell_{\rm sep}(M_{\rm bulk})
	=
	\left(
	\frac{M_{\rm bulk}}
	{f_{\rm GC9}\rho_{\rm DM}}
	\right)^{1/3}.
\end{equation}
For \(f_{\rm GC9}=1\), this gives
\begin{equation}
	\ell_{\rm sep}^{\rm halo}\simeq29.7\,{\rm AU},
	\qquad
	\ell_{\rm sep}^{\rm cos}\simeq2.06\times10^3\,{\rm AU}.
\end{equation}
Thus, for the illustrative normalization \(R_{\rm bulk}=1\,{\rm AU}\),
\begin{equation}
	\frac{R_{\rm bulk}}{\ell_{\rm sep}^{\rm halo}}
	\simeq3.4\times10^{-2},
	\qquad
	\frac{R_{\rm bulk}}{\ell_{\rm sep}^{\rm cos}}
	\simeq4.8\times10^{-4}.
\end{equation}
The benchmark primordial population is therefore safely in the nonoverlapping, discrete-object regime. The illustrative normalization $R_{\rm bulk}=1$~AU is extremely conservative. Sector-normalized values range from $R_{\rm bulk}\simeq1.4\times10^{-10}$~m up to $\sim4\times10^{4}$~km under a fixed keV stiffness (referring to sector masses in
Table~\ref{tab:sectors}), in all cases far below
$\ell_{\rm sep}\simeq30$~AU. Since $M_{\rm cross}\propto R_{\rm bulk}^{69/2}$, realistic normalizations
push the overlap threshold to even more negligible values; the discrete-object conclusion is therefore
robust against the choice of sector.

This geometrical conclusion should not be confused with cosmological
coarse-graining.  A population of discrete, nonoverlapping GC9 objects may
still be represented as an effective collisionless fluid on scales much
larger than \(\ell_{\rm sep}\), just as an ordinary compact-object population
can be treated statistically.  The calculation above addresses whether the
individual condensates physically overlap, not whether their large-scale
stress--energy distribution admits a fluid description.

A mixed dark sector containing both diffuse glueball matter and compact GC9
objects is not excluded \emph{a priori}.  In the present analysis we examine
the sharper hypothesis that most of the dark-matter abundance resides in the
compact GC9 sector.  The \(n=1/8\) dilute-branch scaling shows that the benchmark asteroid-mass population remains a gas of well-separated
compact objects throughout the physically meaningful regime.

\subsection{Formation-scale interpretation of a frozen $\Lambda_T$}
\label{sec:lambdaT-frozen}

Sec.~\ref{sec:sector} identifies the sector dependence of $\Lambda_T$
as a phenomenological input, not a derived result, and notes that a
first-principles calculation would ultimately have to explain the
resulting mass dependence. We offer two complementary speculative consistency constructions; neither is a derivation from the microscopic $G(2)$ theory. The underlying statement to be explained is compact:
combining $M_{\max}=0.064157\,M_{\rm Pl}^3/\Lambda_T^2$ with the
turning-point radius scaling of Eq.~\eqref{eq:sectorLambda} means
that every viable object sits at its own maximum-mass turning
point, so explaining the benchmark assignment $\Lambda_T^{\rm bench}(M)$ is equivalent to explaining why
compact GC9 stars form preferentially near criticality.

\paragraph{Formation-temperature identification.}
If collapse occurs during radiation domination at dark-sector
temperature $T_D$, the horizon mass is the standard
\begin{equation}
M_H(T_D)\simeq0.30\,g_*^{-1/2}\,\frac{M_{\rm Pl}^3}{T_D^2},
\qquad
H(T_D)=1.66\sqrt{g_*}\,\frac{T_D^2}{M_{\rm Pl}},
\label{eq:horizon_mass}
\end{equation}
with $M_{\rm Pl}=1.2209\times10^{19}$~GeV, the same unreduced
convention used throughout Sec.~\ref{sec:sector}.  If the dark bath is decoupled, $T_D=\xi T_\gamma$ and the expansion rate is controlled by the total radiation density; expressed in terms of $T_D$, the horizon mass contains the additional factor $\xi^2/\sqrt{g_{*,\rm tot}}$.  The coefficient comparison below is therefore illustrative unless $\xi$ and the full thermal history are specified. Identifying
$\Lambda_T$ with the dark-sector temperature at collapse,
$\Lambda_T\equiv T_D(t_{\rm collapse})$, two dimensional-consistency checks motivate examining this identification. First, formally matching the normalizations of
$M_{\max}$ and Eq.~\eqref{eq:horizon_mass} exactly,
$0.064157=0.30\,g_*^{-1/2}$ would correspond to $g_*\simeq21.7$ under the simplifying identification $T_D=T_\gamma$: a numerically ordinary illustrative species count for a modest hidden sector, not a fitted prediction.
Second, as an order-of-magnitude check, evaluating Eq.~\eqref{eq:horizon_mass}
directly at the tabulated sector values of Table~\ref{tab:sectors},
with $g_*(T_D)$ set to its own value at that temperature rather
than the fixed $21.7$ above, reproduces the right order of
magnitude for the two extreme entries: at $\Lambda_T=1.3$~keV
($g_*\sim3$--$4$, appropriate to a late, mostly-decoupled bath)
one finds $M_H\sim1.6\times10^{11}M_\odot$, within a factor of
$\sim2.4$ of the tabulated SMBH (TON--618) mass
$6.6\times10^{10}M_\odot$; at $\Lambda_T=1.9$~PeV ($g_*\sim10^2$,
appropriate to a still-thermalized early bath) one finds
$M_H\sim1.3\times10^{-14}M_\odot$, within a factor of $\sim2.3$ of
the tabulated asteroid-window mass $3\times10^{-14}M_\odot$. That two illustrative choices of $g_*(T_D)$ land within an order-unity factor of the tabulated sector masses is an order-of-magnitude dimensional coincidence, not an independent prediction, because the tabulated $\Lambda_T$ values were themselves inferred from the target masses. Under this
identification, the coarse-graining coefficient $Z_{\rm IR}$ of
Sec.~\ref{sec:sector} acquires a candidate concrete meaning,
$(T_{\rm collapse}/\Lambda_{\rm conf})^4$, and for the primordial
bulk population the natural collapse epoch is $3\to2$ cannibal
freeze-out, computable from the number-changing-reaction machinery
already used in this paper
\citep{PappadopuloRudermanTrevisan2016,FarinaEtAl2016Cannibal,
ForestellMorrisseySigurdson2017Glueball}, making $\Lambda_T^{\rm bulk}$ a possible target of a future relic-abundance and collapse calculation rather than a prediction of the present work. We stress that the collapse
dynamics itself --- how a stiff condensing fluid actually falls
through horizon crossing into a bound configuration --- is not
modeled here and remains open.

\paragraph{Toy consistency test for a dynamical $\Lambda_T$.}
A natural worry is that if $\Lambda_T$ differs star to star, some
dynamical field must be setting it locally, in which case it should
also respond to each star's own density profile, undermining the
static TOV treatment used throughout this paper. The toy completion below shows that a simple adiabatically adjusting modulus of the chosen form does not preserve the GC9 closure.  This is not a no-go theorem for all collective fields; it motivates, but does not derive, an effectively frozen or metastable parameter on stellar timescales.

Write $V_{\rm GC9}(s)=m_G^2\,s_T\,f(s/s_T)=m_G^2s+(m_G^2/s_T^8)s^9$
using Eq.~\eqref{eq:gc9_reduced_f}; since $s_T=\Lambda_T^4/m_G^2$,
the nonic coefficient is $m_G^2/s_T^8=m_G^{18}/\Lambda_T^{32}$.
Promoting $\Lambda_T$ to the vacuum value of a modulus $\phi$
coupled through this same operator, plus an anomaly-matched
potential for $\phi$ itself of the type familiar from confining
effective Lagrangians
\citep{Schechter1980TraceAnomaly,MottolaVaulin2006,MigdalShifman1982},
gives the natural two-field completion
\begin{equation}
U(\phi,s)=\frac{m_G^{18}\,s^9}{\phi^{32}}+\lambda\,\phi^4,
\label{eq:dilaton_toy}
\end{equation}
where the second term is the leading piece of the log-running
dilaton potential that reproduces the trace anomaly of a confining
sector. Extremizing at fixed $s$,
\begin{equation}
\begin{aligned}
\frac{\partial U}{\partial\phi}
&=-\frac{32\,m_G^{18}s^9}{\phi^{33}}+4\lambda\phi^3=0,\\
\phi_*&=\left(\frac{8}{\lambda}\right)^{1/36}m_G^{1/2}s^{1/4}.
\end{aligned}
\end{equation}
and substituting back,
\begin{equation}
U_{\rm eff}(s)\equiv U(\phi_*,s)\propto m_G^2\lambda^{8/9}\,s.
\label{eq:dilaton_linear}
\end{equation}
For this specific toy potential, an adiabatically adjusting $\phi$ relaxes the nonic stiffness into an ordinary term linear in $s$ --- a mass renormalization --- and removes the extra pressure support responsible for the GC9 branch.  Preserving the static TOV closure in this particular completion would therefore require $\Lambda_T$ to be effectively hysteretic: frozen on stellar dynamical timescales and reset, if at all, only during formation or violent events rather than tracking the local density in real time.  Such hysteresis is an additional physical assumption, not a consequence already established by the one-field GC9 model.

Within this toy interpretation, the hierarchical-growth discussion of
Sec.~\ref{sec:hierarchical} sharpens into the following conditional,
falsifiable dichotomy. If
GC9--GC9 mergers reset $\phi$, the merger-relaxed population should
relax toward the single near-critical compactness
$C_{\max}^{\rm stable}\approx0.3247$ of
Sec.~\ref{subsec:GC9_first_order_results}, independent of formation
history --- testable via I--Love--Q-type universality relations
across the merger-remnant population. If mergers do \emph{not}
reset $\phi$, the per-sector ceilings of Table~\ref{tab:sectors}
are permanent relics of formation, consistent with the reading of
the supermassive sector as a late-time product of hierarchical
assembly rather than a primordial component, already adopted in
Sec.~\ref{sec:hierarchical}.

\paragraph{Comparison class.}
Two existing frameworks might seem to already cover this ground; it
is worth being explicit about why GC9's $\Lambda_T$ is neither.
Dilaton/glueball effective Lagrangians
\citep{Schechter1980TraceAnomaly,MottolaVaulin2006,MigdalShifman1982}
describe a genuinely dynamical modulus; Eq.~\eqref{eq:dilaton_linear} shows that the simple adiabatic toy realization used here would not reproduce the fixed-parameter GC9 closure. Density-dependent bag constructions instead build the
density-dependence into the Lagrangian by construction, as an
explicit function of the local density within a single star. GC9's
$\Lambda_T$ is neither: it is a fixed Lagrangian parameter,
constant within a given star, that varies only \emph{across} the
population as a quenched relic of each object's formation history
--- the frozen-modulus toy argument above illustrates one possible way of treating it this way, subject to the additional metastability assumption.

\subsection{Hierarchical growth and supermassive GC9 objects}
\label{sec:hierarchical}

The highest mass compact object sector is generated dynamically through $\psi_{\rm grow}(M,z)$,
which schematically accounts for GC9--GC9 mergers, baryonic accretion and dynamical assembly in dense stellar systems and galactic nuclei. At this stage \(\psi_{\rm grow}\) is not computed from a merger tree, it denotes the unresolved high-mass population-synthesis contribution.
This is phenomenologically essential: a substantial primordial heavy compact tail would immediately run into the combined lensing, dynamical-heating and accretion constraints discussed above. Therefore the supermassive GC9 sector should be interpreted as a late-time product of the same bosonic matter sector, not as a primordial DM component.

Sector normalization imposes a nontrivial consistency condition on this channel. An object driven by mergers or accretion above the
$M_{\max}(\Lambda_T)$ of its own sector has no stable GC9 branch on
which to re-land, so the gravitational-cooling regulation of
Secs.~\ref{sec:mergers} and \ref{sec:SB} no longer applies. Continued horizonless growth then
requires additional physics, such as an evolution of the effective
stiffness, re-condensation into a new coherent branch or an independently
populated higher-mass sector; without such a mechanism the present model
supplies no controlled endpoint. Hierarchical growth from the asteroid
bulk to stellar, massive and supermassive scales is therefore consistent
only if the effective stiffness itself evolves along the assembly history
--- for instance through an environment- or density-dependent
$Z_{\rm IR}$, or through re-condensation of the coherent state during
violent mergers --- or if each sector is populated by an independent
formation channel and grows only up to its own ceiling.

As a phenomenological benchmark rather than a derived abundance solution, the
hybrid GC9 census posits two population levels on galactic scales.  First, a
low-mass primordial compact GC9 component supplies the bulk normalization in
the illustrative unit-abundance benchmark (and may instead be subdominant if
extended-mass-function constraints require it).  Second, a heavy compact-object
population is generated astrophysically and subsequently concentrated into
binaries, dense stellar environments and galactic nuclei. As anticipated in \citep{Masi2024G2Resurgence}, according to the fact that baryons can affect the DM halo through their own gravity and by modifying the gravitational potential of the galaxy, the tiny primordial GC9 dark stars can be swept away from the galactic center via stellar feedback and DM heating \citep{DMheat1,DMheat2,DMheat3,DMheat5}, producing the "cored halo" we observe today \citep{DMheat4,DMheat6}.
 
In this benchmark sense the scenario attempts to replace diffuse dark matter, primordial black holes and classical black-hole final states by a single underlying bosonic sector, but it does so through different population channels rather than through a single primordial mass spectrum.

\paragraph{Connection with JWST little-red-dot compact seeds.}
A particularly relevant observational development is the emergence of JWST little-red-dot (LRD) sources as a population of compact, red, broad-line objects at high redshift \citep{Matthee2024LRD,Kocevski2024LRDSample,Durodola2024LRDOvermassive,Wang2025RUBIESLRD}: their physical interpretation is still under active discussion: in many analyses they are associated with
faint or obscured active galactic nuclei, while in others their unusual colors and compactness motivate more exotic early-growth scenarios.  In particular, the strongly lensed LRD Abell~2744--QSO1 at \(z=7.04\) has recently been reported to contain a directly measured compact central mass of order \(5\times10^{7}M_\odot\), inferred from a Keplerian gas rotation curve, with a conservative ratio \(M_{\rm BH}/M_\star>2\) \citep{Juodzbalis2026LRDDirectMass}. This has been interpreted as evidence for a massive black-hole seed caught during an early accretion phase. This observation does not distinguish a classical horizon from a horizonless compact object.  At the radii probed by gas kinematics it mainly measures the enclosed point mass.  Therefore a GC9 object would be degenerate with a black hole whenever
\begin{equation}
	R_{\rm GC9}\ll R_{\rm dyn},
	\qquad
	v_{\rm gas}^2(r)\simeq \frac{G M_{\rm GC9}}{r},
	\qquad r\gtrsim R_{\rm dyn},
	\label{eq:LRD_pointmass_degeneracy}
\end{equation}
where \(R_{\rm dyn}\) denotes the smallest physical scale resolved by the dynamical reconstruction.  Equation~\eqref{eq:LRD_pointmass_degeneracy} should be read as a point-mass degeneracy condition, not as a detection criterion for GC9 stars.  Its relevance is instead conceptual: LRDs sharpen the early-seed problem by showing that compact masses of order \(10^{7}\)--\(10^{8}M_\odot\) may already dominate their baryonic hosts at very high redshift.  In the present framework this is naturally associated with the growth channel \(\psi_{\rm grow}(M,z)\), in which massive and supermassive GC9 candidates arise from an initially lighter compact dark sector through mergers, accretion and dynamical assembly rather than from an unconstrained primordial heavy tail. 

\paragraph{Dense LRD environments, obscured nuclei and compact-object assembly.}
Recent studies sharpen the possible connection between Little Red Dots
and the massive compact-object sector from complementary directions.
Pacucci, Hernquist and Fujii show that, if the compact stellar masses
inferred for LRDs are approximately correct, their extreme nuclear
densities can drive rapid mass segregation and runaway stellar collisions,
producing very massive objects of order \(10^{4}\,M_\odot\) on sub-Myr
timescales \citep{PacucciHernquistFujii2025LRD}. Conversely, Pacucci,
Ferrara and Kocevski show that an initially massive direct-collapse seed
embedded in a dense atomic-cooling environment can sustain a long-lived,
strongly reprocessed accreting phase whose UV/optical spectrum reproduces
several characteristic LRD properties, including weak X-ray emission,
compact morphology and Balmer features
\citep{PacucciFerraraKocevski2026LRD}. These results identify LRD
environments as possible sites of rapid central-mass assembly and show
that their observed spectra can be controlled largely by gas reprocessing
well outside the microscopic central boundary.

A complementary qualification is provided by Rinaldi et al., who identify
at \(z=2.0145\) an LRD-like nucleus embedded in an extended spiral host.
When synthetically displaced to \(z=7\), cosmological surface-brightness
dimming renders the host undetectable, while stacking 99 high-redshift
LRDs reveals faint diffuse emission around their compact nuclei
\citep{RinaldiEtAl2026BeyondDot}. Thus the observed LRD size need not be
identified with the physical size of either the central compact object or
its complete baryonic host. These studies are formulated in terms of
black-hole or AGN evolution and do not constitute GC9 formation
mechanisms. Their relevance here is phenomenological: LRDs may provide
dense, rapidly evolving and strongly reprocessed environments containing
an unresolved central mass. A GC9 realization would have to reproduce
comparable assembly timescales and nuclear emission while additionally
satisfying the dark-sector charge, sector-normalization and stable-branch
conditions required for a massive GC9 configuration. 

We stress that this is an observational degeneracy rather than a detection claim.  LRD gas kinematics constrain the compact mass scale and
its growth history, while the distinction between a Kerr black hole and a
horizonless GC9 boson star remains tied to horizon-scale observables:
photon-region structure, tidal response, surface or reprocessing
effects, late-time relaxation, internal bosonic modes and possible
post-merger scalar-burst relaxation.  For this reason LRDs should be
included as phenomenological targets of the cosmological GC9 sector, but
not used as a primary validation of the model.

\subsection{Minimal viability statement}

The statements in this subsection define a deliberately sharp benchmark hypothesis, not a demonstrated cosmological solution or a global fit.  In particular, the all-dark-matter normalization, complete black-hole replacement and transitions between different $\Lambda_T$ branches remain conditional inputs.

The interpretation of this minimal hybrid population bookkeeping is: i) all cosmological dark matter is stored in primordial asteroid-mass GC9 compact objects; ii) neutron stars remain viable and may host a modest GC9 dark core; iii) the stellar-mass compact-object branch otherwise interpreted as black holes contain pure GC9 remnants for charge-sufficient seeds or dark accretion, while under-charged events remain hybrid
baryon--GC9 remnants;; iv) the massive and supermassive GC9 sectors are generated later by hierarchical growth.

In this sense, the framework can be parameterized so as to replace both free DM gas and classical BH final states, conditional on the uncomputed formation, constraint and branch-transition mechanisms, without requiring the neutron-star branch to disappear. The last explicit bookkeeping step at the present stage is therefore the decomposition
\begin{equation}
		\begin{split}
\psi_{\rm tot}(M,z)&
=
\underbrace{\psi_{\rm DM}^{\rm prim}(M)}_{\text{all-DM primordial bulk}}
+
\underbrace{\psi_{\rm NS+core}(M,z)}_{\text{neutron-star branch}} + \\
& + \underbrace{\psi_{\rm GC9\text{-}BH}(M,z)+\psi_{\rm grow}(M,z)}_{\text{BH replacements and hierarchical descendants}}.
	\end{split}
\end{equation}
A full confrontation with data requires the next layer of modeling, namely calibration of the branch probability \(p_{\rm NS}\) on stellar-evolution grids, an explicit dark-core fraction model for the NS branch and a numerical population synthesis for \(\psi_{\rm grow}(M,z)\).
\paragraph{GW candidate S251112cm.}
A useful recent example of why the low-mass compact sector is relevant is the public-alert GW candidate S251112cm \citep{S251112cmGraceDB}. GraceDB classifies it as a sub-solar-mass compact binary coalescence candidate, with a false-alarm rate of approximately one per 6.2 years and the alert products assign essentially all the source-frame chirp-mass probability to the sub-solar regime \citep{S251112cmGraceDB}. At present, however, the event should be treated only as a suggestive motivation rather than as evidence for the GC9 scenario. If confirmed to be astrophysical, such a candidate would be difficult to accommodate within ordinary stellar-remnant channels and would therefore strengthen the broader case for exotic low-mass compact objects, including PBH \citep{VieiraEtAl2026S251112cmFollowup,PBHInterpretationS251112cm2026} or the primordial GC9 sector here discussed. S251112cm represents an illustrative example of the type of future observation that could probe the low-mass compact branch of the hybrid GC9 framework.
\paragraph{Connection with lunar-mass microlensing candidates.}
A recent high-cadence microlensing search toward the Large Magellanic Cloud (LMC) reported an hour-long event called Phoebe \citep{KeyEtAl2026AMPMII}, whose inferred lens mass is of order
\begin{equation}
M_{\rm Phoebe}\simeq 10^{-7}M_\odot ,
\end{equation}
approximately a few lunar masses.  The event was interpreted as a primordial black hole candidate because the optical-depth analysis favours a dark halo lens over ordinary stellar or planetary populations. The microlensing light curve measures the compact lens mass,
geometry and transverse velocity, but it does not directly determine whether the lens is a black hole, a free-floating planet or a horizonless compact object.  Therefore a sufficiently compact GC9 object
with radius much smaller than the Einstein radius,
\begin{equation}
R_{\rm GC9}\ll R_E,
\qquad
R_E=
\left[
{4GM\over c^2}
{D_L(D_S-D_L)\over D_S}
\right]^{1/2},
\end{equation}
would be observationally degenerate with a point-mass PBH at the level of the leading microlensing light curve. Here $D_S$ is roughly the Earth–LMC distance and $D_L$ is the Earth–lens distance. 
For a lunar-mass lens, this relation again holds only with sector
normalization: $\Lambda_T\simeq1$~TeV
(Table~\ref{tab:sectors}) gives $R_{\rm GC9}\simeq0.5$~m against
$R_E\sim5\times10^{5}$~km, so the point-lens degeneracy with a PBH is exact for all practical purposes. 

This possible lunar-mass microlensing candidate Phoebe does not exclude an asteroid-window compact DM component because it measures the lens mass at the time of observation, not the primordial mass at formation. In the GC9 dark sector the initial low-mass distribution evolves through rare mergers, capture inside early substructure and gravitational cooling, producing an extended coalescence-grown tail. Future high-cadence microlensing surveys toward the LMC, SMC, M31 and Galactic fields can test whether the short-timescale events trace the dark-matter halo rather than ordinary planetary populations.

\section{Conclusions}
\label{sec:conclusions}

The conclusions of the present work should be read in the context of the modern strong-field program, in which the Kerr paradigm has proven highly successful phenomenologically, while still leaving room for disciplined horizonless alternatives \citep{Abbott2022TestsGR,EHT2019M87I,EHTSgrA2022I,Cardoso2019BHmimickers,Barack2019BHSurvey,YunesSiemens2013}. 

In this work we have developed a unified framework in which BH-like astrophysical objects are reinterpreted as horizonless, self-gravitating configurations of a dark bosonic sector, specifically G(2) glueballs described by an effective complex scalar field. The central theoretical ingredient is the scale-separated $G(2)$-inspired Casimir-six third-cumulant (GC9) closure, in which the microscopic glueball mass fixes the small-field pole while the macroscopic stellar scale is controlled by an infrared condensate stiffness scale \(\Lambda_T\). In the Thomas--Fermi description this gives the analytic, causal and bag-free EOS
\begin{equation}
\hat\rho=2y+10y^9,\qquad \hat P=8y^9,\qquad c_s^2\to {4\over5}.
\end{equation}
Conditional on the Thomas--Fermi effective-fluid reduction, the resulting branch admits highly compact configurations with $C_{\max}^{\rm stable}=0.3247$ and a post-turning maximum $C_{\max}=0.3371$, avoids the flat-space Q-ball criterion, and has a maximum-mass scaling \(M_{\max}=0.064157 M_{\rm Pl}^3/\Lambda_T^2\) with no direct \(m_G^{-2}\) suppression. 

Once one demands compatibility with supermassive compact objects such as TON--618 while retaining ultraheavy microscopic constituents, any viable construction must separate the collective stellar scale from the particle mass.  A large quartic coefficient is one way of diagnosing the failure of the ordinary weak-coupling one-scale parametrization, but it is not by itself a unique naturalness discriminator because the same quartic can be rewritten with an independent collective scale. Within the restricted candidate set studied here, GC9 is retained because its $c_s^2\to4/5$ stiffness produces the required compactness and tidal suppression while remaining causal, smooth, Q-ball safe and bag free; this is a phenomenological selection, not a uniqueness theorem or a first-principles derivation from $G(2)$.
Solving the TOV structure equations, we find that the conditional
GC9 fluid branch supports highly compact configurations at any
single target mass once $\Lambda_T$ is normalized to the
corresponding sector through
$M_{\max}=0.064157\,M_{\rm Pl}^{3}/\Lambda_T^{2}$. At fixed
$\Lambda_T$ the compact, Kerr-comparison band spans only a factor of
a few below $M_{\max}$, with $C\propto M^{16/23}$ on the dilute
branch; the cross-sector phenomenology of Secs.~V--VII therefore
rests on sector-dependent stiffness values, ranging from the keV
scale (supermassive sector) to the PeV scale (asteroid window), as
summarized in Sec.~\ref{sec:sector}.
The TON--618 scale corresponds to \(\Lambda_T\simeq1.3\text{--}1.6\,\mathrm{keV}\), while the microscopic glueball mass enters the condensate field scale \(s_T=\Lambda_T^4/m_G^2\) rather than the maximum mass. Indeed, after Thomas--Fermi rescaling none of the reduced equilibrium, tidal, slow-rotation or static-QNM observables determines $m_G$; that quantity remains an ultraviolet prior entering gradient control and microscopic excitations. This separation is algebraically consistent but the keV--ultraheavy hierarchy requires the computation of a many-body matching coefficient.

A second major result is that the compact branch is close to Kerr in a restricted controlled slow-rotation sense. The directly computed first-order combination is $\bar I C^{3/2}=0.9466$; a mass-shedding estimate is $j_{\max}^{\rm proxy}=\kappa_{\rm ms}\bar I C^{3/2}$ and therefore remains uncalibrated until a two-dimensional rotating GC9 sequence fixes $\kappa_{\rm ms}$. The second-order Hartle--Thorne matching estimate on the same reduced background gives
\[
\bar Q\simeq1.52158.
\]
At the representative slow-row spin \(j=0.328\), this corresponds to \(q_2=0.16395\) and \(\Delta q=-0.05620\). The corresponding off-Kerr waveform parameter is $\delta\kappa=\bar Q-1=0.52158$. Relative to the classic Barack--Cutler LISA EMRI benchmark, this deformation is larger than the quoted historical Fisher errors over the displayed mass range, although that comparison is only an order-of-magnitude sensitivity illustration and not a statistical significance or a dedicated GC9 forecast.

The controlled ISCO diagnostic shifts by $1.238\%$ (same-order comparison) to $1.347\%$ (against the exact Kerr circumferential radius) relative to Kerr at the slow-row spin, with the formal $\sim9\%$ high-spin value retained only as a formal fixed-denominator extrapolation, not as a prediction. By contrast, the stellar surface satisfies $R/M=3.08$ at the maximum-mass benchmark and remains above $3M$ throughout the radially stable branch, while the post-turning maximum-compactness configuration lies below $3M$. The present branch therefore has no external photon ring and the formally continued light-ring/eikonal values cannot be interpreted as shadow or QNM predictions. The current Kerr-mimicker claim is limited to the mass scale, slow multipoles, exterior ISCO and tidal observables, together with a distinguishable static axial QNM. Sec.~\ref{subsec:GC9_mimicry_scope} sets out what this costs observationally. Because the object is optically surfaceless as well as horizonless, no central brightness depression has yet been demonstrated. The static axial response survives the absence of an eikonal light ring: the $\ell=2$ Regge--Wheeler barrier peaks at $3.281M$, outside the surface, and the direct static axial calculation gives $M\omega=0.4638-0.09888i$ with $Q=2.345$ against $2.100$ for Schwarzschild. Its damping scale is black-hole-like, but its real frequency is $24.1\%$ above the Schwarzschild value at the same mass. A direct comparison with GW250114 gives the observational central value $M\omega_{220}\simeq0.5330-0.0824i$ in the LVK full-signal reconstruction; matching it would require corrections of approximately $+14.9\%$ in the real frequency and $+20.0\%$ in the damping time relative to the static GC9 result. Repeating the static calculation for the isotropic GC6--9--12 benchmark gives $M\omega\simeq0.463-0.097i$, leaving requirements of the same order, about $15\%$ in frequency and $18$--$20\%$ in damping time. The neighboring positive cumulants therefore do not provide a static cure; compatibility remains an explicit target for the uncomputed rotating coupled spectrum, not a demonstrated catalog-level fit. The absence of a light ring, monotonic axial potential and low-$Q$ mode exclude the specific axial trapping mechanism for the static endpoint, not complete nonlinear or rotating instability. A static surface-truncation model would map $R/M=3.0798$ to an illustrative continuum-fitting value $a_\ast\lesssim0.77$, but this is conditional rather than a hard ceiling because baryonic matter can penetrate the optically dark configuration. On the imaging side, the orbital angular velocity peaks at $R_\Omega=3.0068M$ and the associated apparent scale is $99.93\%$ of the Schwarzschild critical impact parameter; whether this produces a dark central depression remains a GRMHD and radiative-transfer question.

The dynamical extensions are more qualitative than the equilibrium
calculation.  In the successful horizonless substitution channel explored
here, gravitational cooling and re-landing are candidate regulators of
post-merger relaxation because horizon absorption is absent; they are not
guaranteed outcomes of horizonlessness.  Scalar bursts and matter-supported
internal oscillations are therefore treated as two possible manifestations of
that conditional post-merger channel. The burst energetics inferred in the compact branch are moderate rather than catastrophic. Because the one-field zero-temperature potential remains positive and bag-free, it contains no additional vacuum minimum of that restricted potential; this does not exclude density-, temperature-, composition- or collective-order-parameter transitions, nor does it guarantee relaxation through scalar radiation. We have shown that the burst energy can be estimated as $E_{\rm sb}\sim (\eta-1)f_c M_f$,
where $\eta$ quantifies the transient overcompression and $f_c$ is the core mass fraction. This process leads to a redistribution of the merger energy budget, producing an apparent deficit in gravitational-wave emission relative to the Kerr expectation. Such a feature could provide an observational handle only after a numerical merger waveform and a concrete visible-sector coupling or gravitational conversion mechanism are supplied. The Cowling analysis suggests a discrete spectrum of internal bosonic-core oscillations. 

We have also explored two broader astrophysical consequences of replacing an absorbing horizon by a dynamical bosonic core. First, we formulated a seeded-collapse channel in which compact GC9 remnants may emerge from stellar collapse only when a pre-existing dark seed is present, naturally connecting the framework to failed-supernova and disappearing-star searches. Second, we outlined a possible shock-revival feedback channel in
core-collapse supernovae, in which accretion onto a system containing a GC9
component may excite coherent core or boundary-layer modes and, if the
coupling and deposition efficiencies are non-negligible, return part of the
available accretion power to the surrounding flow.  Horizonlessness alone does
not guarantee this response.  The mechanism remains schematic and requires
validation in multidimensional radiation-hydrodynamic simulations.

On cosmological scales, we have outlined a conditional benchmark scenario in which the dark sector is composed of G(2) glueballs forming a hierarchical population of compact objects. In this picture, primordial seeds grow through accretion and mergers into stellar-mass and supermassive boson stars, while a fraction of the population populates galactic halos as asteroid-window primordial GC9 dark objects. The absence of additional vacuum minima avoids a first-order transition built into the potential. We have outlined parameter regions that are not obviously excluded by microlensing, dynamical and current gravitational-wave bounds 
The scenario may provide a route to early supermassive compact objects, potentially including high-redshift sources observed by JWST, rather than a demonstrated explanation of them. This cosmological interpretation is still only a minimal viability framework rather than a full global fit. Its role is to show how the GC9 scenario can be organized as a broader compact-object census, conditional on formation and branch-transition mechanisms, without forcing all dark matter, all remnants and all galactic nuclei into a single rigid primordial channel. It also clarifies the observational program ahead: microlensing, halo heating, sub-solar compact-object searches, neutron-star demographics, stellar-remnant populations and gravitational-wave catalogs all become part of the same test of the GC9 hypothesis. 

Taken together, these results support a coherent picture. The GC9 \(G(2)\) glueball framework is designed to remain close to Kerr in selected slow-rotation observables while replacing the interior by a regular bosonic medium and, conditionally, replacing horizons and singularities by a regular macroscopic dark-matter state with its own dynamics, oscillations and feedback channels. In this sense, the framework does not merely propose another BS potential: it provides a structured, $G(2)$-inspired and lattice-matchable effective compact-object scenario linking non-Abelian dark-sector microphysics, strong-gravity phenomenology and multi-messenger astrophysics. The same construction admits a broader GC6--9--12A connected-cumulant and anisotropic hierarchy, with the fully analyzed GC9 closure recovered exactly for $a_6=a_{12}=\lambda_A=0$ and $a_9=1$. This extension is quoted only as a future lattice- and stability-testing program and does not modify the GC9 benchmark conclusions.

At the same time, several key steps remain open. A definitive comparison with Kerr will require: (i) full rotating GC9 equilibria, so that timelike and null geodesics can be computed without a slow-rotation exterior continuation; (ii) dedicated ray tracing of a surface outside $3M$, rather than conversion of an internal formal light-ring radius into a shadow proxy; and (iii) a fully coupled spacetime--matter perturbation analysis yielding the complex rotating QNM spectrum and mode-mixing structure. On the astrophysical side, merger simulations, seeded-collapse calculations and core-collapse supernova modeling will be needed to turn the present branch-level estimates into predictive waveform and transient signatures.

Within those limits, however, the main conclusion is clear: GC9 boson stars made of \(G(2)\) dark glueballs provide a concrete horizonless compact-object candidate worthy of further numerical testing. They reproduce a subset of leading Kerr diagnostics--most clearly the slow-row ISCO and an order-unity quadrupole--while predicting finite tidal response, a shifted static axial QNM and matter-supported internal oscillations. 
Future observations across gravitational-wave astronomy, horizon-scale imaging, X-ray timing and transient surveys will determine whether nature realizes such a bosonic alternative to the classical BH picture.

\begin{acknowledgments}
The use of OpenAI ChatGPT (GPT-5.6 Sol) in manuscript
preparation is disclosed in the reproducibility appendix, in accordance with current APS policy. The author is solely responsible for the scientific content and conclusions.
\end{acknowledgments}

\appendix
\section{Second-order Hartle--Thorne computation of the GC9 quadrupole}
\label{sec:gc9_second_order_quadrupole}

To estimate the quadrupole moment of rotating GC9 boson stars beyond purely phenomenological scaling arguments, we implemented the second-order Hartle--Thorne formalism using the explicit equations given in the supplementary material of \citep{c3jx-5487}. The goal of this construction is to compute the mass quadrupole directly from the slow-rotation expansion, using the same background stellar model and EOS employed in the TOV analysis.

We begin from a static equilibrium solution of the TOV equations for the given GC9 closure. We first reconstructed the maximum-mass background from the corresponding central density $\rho_c(M_{\max})$, then solved the first-order frame-dragging equation and finally integrated the regular interior $\ell=2$ perturbations $h_{22}(r)$ and $k_{22}(r)$.
This determines the zeroth-order functions
\begin{equation}
	M(r),\qquad p(r),\qquad \rho(r),\qquad \nu(r),
\end{equation}
together with the stellar radius \(R\) and total mass \(M\).

At order \(O(\Omega)\), we solve the standard Hartle frame-dragging equation for the angular-velocity perturbation \(\bar\omega(r)\). Matching the interior solution to the asymptotically flat exterior form
\begin{equation}
	\bar\omega_{\rm ext}(r)=\Omega-\frac{2J}{r^3},
\end{equation}
yields the total angular momentum \(J\). This step is numerically robust and provides the input source terms for the second-order sector.

\paragraph{Second-order \(\ell=2\) sector.}
At order \(O(\Omega^2)\), the quadrupolar rotational deformation is described by the \(\ell=2\) Hartle--Thorne system. Adopting the notation of \citep{c3jx-5487}, the relevant interior variables are the metric perturbations \(h_{22}(r)\), \(k_{22}(r)\), \(m_{22}(r)\), together with the auxiliary pressure-displacement sector encoded through the algebraic \(\xi_{22}\) relation. Rather than integrating only a reduced subsystem, we use the full gauge-consistent second-order variable set, including the displacement contribution, in order to avoid spurious normalization and gauge artefacts.

\paragraph{Regular-center expansion.}
A crucial ingredient is the exact regular-center expansion. A naive direct integration of the second-order equations from \(r\approx 0\) is numerically unstable, because the \(\ell=2\) system is stiff and contains a homogeneous mode that contaminates the physical regular solution. We therefore use the regular-center series from the supplementary formalism to construct the interior solution in a basis adapted to the center:
\begin{equation}
	\mathbf{y}_{22}(r)
	=
	\mathbf{y}_{22}^{(0)}(r)
	+
	A_{22}^{\rm int}\,\mathbf{y}_{22}^{(1)}(r),
\end{equation}
where \(\mathbf{y}_{22}^{(0)}\) is one regular interior basis solution and \(\mathbf{y}_{22}^{(1)}\) is a second linearly independent regular solution. The coefficient \(A_{22}^{\rm int}\) is fixed by the surface match.

\paragraph{Exterior matching and extraction of \(Q\).}
Outside the star, the second-order \(\ell=2\) metric is written in the standard Hartle--Thorne exterior basis. The exterior solution contains the integration constant \(C_{22}^{\rm ext}\), which controls the quadrupole deformation. The key practical step is to match the interior and exterior solutions linearly at the stellar surface \(r=R\), enforcing continuity of the second-order metric functions. In practice, we impose the surface match using \(h_{22}(R)\) and \(k_{22}(R)\), and then verify consistency using \(m_{22}(R)\). This determines simultaneously the interior amplitude \(A_{22}^{\rm int}\) and the exterior constant \(C_{22}^{\rm ext}\).

Once \(C_{22}^{\rm ext}\) is known, the mass quadrupole follows from the Hartle--Thorne exterior formula
\begin{equation}
	Q \equiv M_2
	=
	-\frac{J^2}{M}
	-\frac{8}{5}M^3\,C_{22}^{\rm ext}.
	\label{eq:Q_HT_final}
\end{equation}
We also report the reduced quadrupole
\begin{equation}
	\bar Q \equiv -\frac{Q\,M}{J^2},
\end{equation}
which is particularly useful for comparison with Kerr, for which \(\bar Q_{\rm Kerr}=1\).
As a numerical consistency check, we verified that the interior $\ell=2$ system satisfies linear superposition to relative accuracy $10^{-8}$--$10^{-7}$, and that the surface matching residuals in $h_{22}$, $k_{22}$ and $m_{22}$ are negligible. We also performed an independent asymptotic extraction by fitting the large-$r$ behavior of the $\ell=2$ part of $g_{tt}$,
\begin{equation}
	g_{tt}^{(\ell=2)}(r)\simeq \frac{2Q}{r^3}P_2(\cos\theta)+O(r^{-4}),
\end{equation}
which yields values of $Q$ consistent with the matching result at the $10^{-5}$--$10^{-4}$ level.

\section{Relativistic Cowling system and boundary conditions}
\label{app:active_cowling}

For reproducibility, we state the convention used for the nonrotating polar-fluid estimate.  Write the static background as
\begin{equation}
\begin{aligned}
ds^2&=-e^{2\Phi_g(r)}dt^2+e^{2\Lambda_g(r)}dr^2+r^2d\Omega^2,\\
e^{-2\Lambda_g}&=1-\frac{2m(r)}{r}.
\end{aligned}
\end{equation}
where the subscript $g$ distinguishes the radial metric potential from the dimensionless tidal deformability.  With
\begin{align}
 \xi^r&=e^{-\Lambda_g}\frac{W(r)}{r^2}Y_{\ell m}e^{-i\omega t},\\
 \xi^\theta&=-V(r)\,\partial_\theta Y_{\ell m}e^{-i\omega t},\\
 \xi^\phi&=-V(r)\,\sin^{-2}\theta\,\partial_\phi Y_{\ell m}e^{-i\omega t},
\end{align}
the relativistic Cowling equations are
\begin{align}
 \frac{dW}{dr}&=\frac{1}{c_s^2}
 \left[\omega^2r^2e^{\Lambda_g-2\Phi_g}V+\Phi_g'W\right]
 -\ell(\ell+1)e^{\Lambda_g}V,
 \label{eq:active_cowling_W}\\
 \frac{dV}{dr}&=2\Phi_g'V-\frac{e^{\Lambda_g}}{r^2}W.
 \label{eq:active_cowling_V}
\end{align}
Regularity at the center requires
\begin{equation}
 W(r)=A r^{\ell+1}+O(r^{\ell+3}),
 \qquad
 V(r)=-\frac{A}{\ell}r^\ell+O(r^{\ell+2}),
\end{equation}
and the surface eigenvalue condition is the vanishing Lagrangian pressure perturbation,
\begin{equation}
 \omega^2e^{\Lambda_g(R)-2\Phi_g(R)}V(R)
 +\frac{\Phi_g'(R)}{R^2}W(R)=0.
 \label{eq:active_cowling_surface}
\end{equation}
The quoted $\ell=2$ roots were located by shooting from the regular center and bracketing zeros of the surface residual while checking the radial node count. Moving the numerical surface cutoff from $y=10^{-6}$ to $10^{-12}$ changes the displayed roots by less than $10^{-7}$ in $M\omega$.  The Cowling approximation freezes all metric perturbations; it therefore supplies real matter-mode frequencies only.  No independent convergence table or coupled spacetime--matter benchmark is supplied in this manuscript, so the extra displayed digits should be treated as numerical identifiers rather than established accuracy.

\section{Computational documentation and reproducibility manifest}
\label{app:computational_documentation}

This appendix records the common conventions, numerical sequence, boundary
conditions, benchmark inputs and precision policy used for the reported
computations.  The submission source package also contains a standalone
reproducibility guide, a machine-readable benchmark table and a short script
that verifies the algebraic reductions and derived numerical comparisons.
Those auxiliary files document the calculation but do not replace the full
radial profiles or solver source needed for independent high-precision
reproduction.

\subsection{Units, reduced variables and dimensional restoration}

We use $G=c=\hbar=1$.  For a potential
$V=\Lambda_T^4 f(y)$ with $y=s/s_T$ and
$s_T=\Lambda_T^4/m_G^2$, the density scale is $\Lambda_T^4$.  With the
Planck-mass convention used consistently in the TOV equations, the reduced
length and mass units are
\begin{equation}
 L_T=\frac{M_{\rm Pl}}{\Lambda_T^2},
 \qquad
 M_T=\frac{M_{\rm Pl}^3}{\Lambda_T^2},
\end{equation}
so that $r=L_T\hat r$, $m=M_T\hat m$, $\rho=\Lambda_T^4\hat\rho$ and
$P=\Lambda_T^4\hat P$.  All quoted compactnesses, Love numbers, reduced
moments, reduced quadrupoles and products $M\omega$ are invariant under this
rescaling.

For GC9,
\begin{equation}
 f(y)=y+y^9,
 \qquad
 \hat\rho=2y+10y^9,
 \qquad
 \hat P=8y^9,
\end{equation}
and
\begin{equation}
 c_s^2=\frac{72y^8}{2+90y^8}\longrightarrow\frac45.
\end{equation}
The checks $\hat P=0\Rightarrow\hat\rho=0$ and
$V(s)/s=m_G^2(1+y^8)\geq m_G^2$ establish the bag-free surface and the absence
of the standard canonical flat-space $Q$-ball condition.  The GC6--9--12
relations used for the outlook calculation are given in
Eqs.~\eqref{eq:GC6912_eos} and \eqref{eq:GC6912_sound_speed}.

\subsection{Static sequence and turning points}

For each central value $y_c$, the central pressure and density are obtained
from the parametric EOS and the TOV system is integrated from a regular-center
series until $P=0$.  The surface event therefore coincides with $y=0$.  A
numerically convenient implementation can use $y$ as the thermodynamic state
variable near the surface, avoiding inversion of the very soft dilute EOS.
The first maximum of $M(y_c)$ defines the maximum-mass stable benchmark used
throughout the paper; the later maximum of $C(y_c)$ belongs to the
post-turning continuation and is radially unstable by the turning-point
criterion.

The common benchmark manifest is
\begin{widetext}
\begin{center}
\begin{tabular}{lrrrr}
\hline\hline
configuration & $y_c$ & $M^{\rm red}$ & $R^{\rm red}$ & $C$\\
\hline
GC9 first mass maximum & 0.8730126 & 0.06415664 & 0.19759137 & 0.32469352\\
GC9 compactness maximum & 0.9811312 & -- & -- & 0.33711781\\
GC6--9--12 first mass maximum & 0.6827030 & 0.07205607 & 0.22304730 & 0.32305286\\
GC6--9--12A, $\lambda_A=0.3$ & 0.6783526 & 0.07391135 & 0.22534664 & 0.32798957\\
\hline\hline
\end{tabular}
\end{center}
\end{widetext}
The last row is an illustrative anisotropic background only.  It is not used
in the GC9 tidal, multipolar or QNM results.

\subsection{Tidal and first-order rotational calculations}

The Hinderer equation is integrated on the same GC9 TOV background, with the
regular central solution and the standard vacuum surface match.  The reported
maximum-mass values are
\begin{equation}
 k_2=0.02366722,
 \qquad
 \Lambda=\frac{2k_2}{3C^5}=4.372078.
\end{equation}
The first-order Hartle frame-dragging equation is integrated with a regular
central derivative and matched to
$\bar\omega_{\rm ext}=\Omega-2J/r^3$.  This yields
\begin{equation}
 \bar I=\frac{I}{M^3}=5.116358.
\end{equation}
For the controlled slow-row benchmark, $J^{\rm red}=1.3510944\times10^{-3}$
and $j=J/M^2=0.3282483$.  The number
$\bar I C^{3/2}\simeq0.95$ is only an uncalibrated mass-shedding proxy and is
not used as a constructed high-spin equilibrium.

\subsection{Second-order Hartle--Thorne quadrupole}

The regular interior $\ell=2$ system is integrated as a linear combination of
two regular basis solutions and matched at the surface to the asymptotically
flat exterior.  Continuity of $h_{22}$ and $k_{22}$ fixes the interior amplitude
and $C_{22}^{\rm ext}$; $m_{22}$ is retained as an independent consistency
check.  With
\begin{equation}
 Q=-\frac{J^2}{M}-\frac85 M^3C_{22}^{\rm ext},
 \qquad
 \bar Q=-\frac{QM}{J^2},
\end{equation}
the benchmark gives $C_{22}^{\rm ext}=0.0351243$ and
$\bar Q=1.52158$.  Interior linear superposition is satisfied at the
$10^{-8}$--$10^{-7}$ level, and the independent asymptotic $g_{tt}$ extraction
agrees at the $10^{-5}$--$10^{-4}$ level.

\subsection{Geodesic and null-orbit diagnostics}

ISCO quantities are evaluated in the Hartle--Thorne exterior at the controlled
slow-row spin and compared both with the same-order Kerr expansion and with the
exact Kerr circumferential radius.  The corresponding shifts are $1.238\%$ and
$1.347\%$.  Photon-region expressions are retained only as formal exterior
continuations whenever their radius lies inside the material surface.
For a static spherical background, internal and external circular null orbits
are checked using
\begin{equation}
 \chi(r)=\frac{3m(r)}{r}+4\pi r^2P_r(r)=1.
\end{equation}
At the GC9 maximum-mass point $\chi_{\max}=0.975563<1$.  For the illustrative
GC6--9--12A point, $\chi_{\max}=0.988204<1$.

\subsection{Static axial quasi-normal mode}

The axial master problem is Eqs.~\eqref{eq:axial_master} and
\eqref{eq:axial_potential}, with $Z\sim r^{\ell+1}$ at the center and a purely
outgoing solution at infinity.  The eigenvalue is a zero of the Wronskian
between an outward interior integration and an inward asymptotic integration.
Applying the upgraded machinery to Schwarzschild gives
$M\omega=0.37367168-0.08896232i$, reproducing the Leaver reference at the displayed precision.  The GC9 maximum-mass result is
\begin{equation}
\begin{aligned}
 M\omega&=0.463846-0.098881i,\\
 Q&=2.345,\qquad \tau/M=10.113 .
\end{aligned}
\end{equation}
with the audited matching variations at the $10^{-9}$ level and the outer-radius test $35M$--$50M$ changing the root by less than a few parts in $10^9$.

The isotropic GC6--9--12 background produces a small displacement,
reported conservatively as $M\omega\simeq0.463-0.097i$.  The calculation is
used only to establish that positive neighboring cumulants do not remove the
static discrepancy; its small shift is not assigned a statistically meaningful
sign.  No anisotropic GC6--9--12A QNM is quoted because its perturbation system
must be derived from the anisotropic stress closure.

\subsection{Cowling modes and root identification}

The active Cowling equations and boundary conditions are stated in
Appendix~\ref{app:active_cowling}.  Trial frequencies are integrated from the
regular center, the surface residual is bracketed and radial nodes are counted
to order the roots.  Moving the numerical surface cutoff from $y=10^{-6}$ to
$10^{-12}$ changes the displayed values by less than $10^{-7}$ in $M\omega$.
The roots remain real because the Cowling approximation freezes metric
perturbations; they are not complex ringdown QNMs.

\subsection{GW250114 conversion and discrepancy arithmetic}

The dimensionless central frequency is reconstructed from the measured
frequency, damping time and redshifted remnant time by
\begin{equation}
 M\omega=2\pi f_{220}t_{M_f}-i\frac{t_{M_f}}{\tau_{220}}.
\end{equation}
With $f_{220}=251.7\,\mathrm{Hz}$, $\tau_{220}=4.09\,\mathrm{ms}$ and
$t_{M_f}=0.337\,\mathrm{ms}$, this gives
$M\omega\simeq0.5330-0.0824i$ and $\tau/M\simeq12.14$.  Relative to static
GC9, passing through those central values would require approximately a
$14.9\%$ increase in the real part and a $20.0\%$ increase in $\tau/M$.
The rounded GC6--9--12 result leaves requirements of approximately $15\%$ and
$18\%$.  These are diagnostic targets for the rotating coupled spectrum, not a
catalog-level likelihood result.

\subsection{Precision policy and supplied files}

Analytic identities are quoted exactly.  Reduced TOV, Love and first-order
Hartle benchmarks retain six to eight digits as numerical identifiers for the
common sequence.  The second-order quadrupole retains the digits supported by
surface matching and asymptotic extraction.  Static GC9 QNM digits are supported by the upgraded logarithmic-derivative solver and the stated $10^{-9}$ matching checks; the GC6--9--12 shift is rounded because it
is below the current percent-level resolution.  Cowling extra digits identify
the numerical run and should not be interpreted as a full systematic-error
budget.

The accompanying reproducibility package contains: (i) this manuscript and its
bibliography; (ii) a detailed computation guide; (iii) a CSV benchmark manifest;
(iv) a script that verifies algebraic EOS relations, dimensional conversions,
quality factors and quoted percentage comparisons; and (v) a build-validation
report.  Full radial profiles and the complete production ODE/QNM solver are not
claimed to be publicly archived in this version.

\subsection{AI-assisted verification disclosure}

OpenAI ChatGPT (GPT-5.6 Sol) was used for algebraic cross-checks, code debugging and numerical-consistency audits. The tool was directed by explicit physical assumptions and numerical validation criteria. AI output was not treated as an authority: retained results were checked against the equations stated in the manuscript, numerical reruns and convergence tests where available and internal arithmetic consistency. The author retains full responsibility for the calculations, interpretations,
citations and final manuscript.

\section*{Data availability}

The source package accompanying this manuscript contains a computation guide,
a machine-readable benchmark manifest, a verification script for the analytic
reductions and derived numerical comparisons and the upgraded axial QNM solver
corresponding to the quoted $M\omega=0.463846-0.098881i$ benchmark.  The
remaining production scripts and high-resolution radial profiles belong to an
in-development numerical codebase that has not yet been separated into a
repository-ready public release.  They will be provided privately to editors,
reviewers and readers upon reasonable request.  The equations, normalizations
and boundary conditions needed to identify the calculations are stated in the
main text and appendices.
\bibliography{biblio_full_GC9}

\end{document}